\documentclass[rmp, reprint, aps, floatfix]{revtex4-1}
\usepackage{graphicx,amsmath,amssymb,bm}
\usepackage{array,multirow}

\newcommand{\twelve}{$^{12}$C}
\newcommand{\thirteen}{$^{13}$C}
\newcommand{\boldk}{\mathbf{k}}
\newcommand{\boldr}{\mathbf{r}}
\newcommand{\muorb}{\mu_\mathrm{orb}}
\newcommand{\DeltaSO}{\Delta_\mathrm{SO}}
\newcommand{\DeltaKK}{\Delta_{KK'}}
\newcommand{\boldA}{\mathbf{A}}
\newcommand{\boldB}{\mathbf{B}}
\newcommand{\boldC}{\mathbf{C}}
\newcommand{\gtensor}{\mathbf{g}}
\newcommand{\rs}{r_{\mathrm{s}}}
\newcommand{\gs}{g_{\mathrm{s}}}
\newcommand{\SSS}{\ensuremath{S} }
\newcommand{\SSSprime}{\ensuremath{S'} }
\newcommand{\AS}{\ensuremath{AS} }
\newcommand{\DeltaSAS}{\ensuremath{\Delta_{S,AS}} }
\newcommand{\DeltaASSprime}{\ensuremath{\Delta_{AS,S'}} }
\newcommand{\DeltaST}{\ensuremath{\Delta_{S,T}} }

\newcommand{\ie}{\textit{i.e.}}
\newcommand{\rrr}{\mathbf{r}}
\newcommand{\kkk}{\mathbf{k}}
\newcommand{\ppp}{\mathbf{p}}
\newcommand{\RRR}{\mathbf{R}}
\newcommand{\KKK}{\mathbf{K}}
\newcommand{\KKKp}{\mathbf{K'}}
\newcommand{\kappab}{{\bm{\kappa}}}
\newcommand{\taub}{{\bm{\delta}}}
\newcommand{\PsiAK}{\Psi_A^\mathbf{K}}
\newcommand{\PsiAk}{\Psi_A^\mathbf{k}}
\newcommand{\PsiBK}{\Psi_B^\mathbf{K}}
\newcommand{\PsiBk}{\Psi_B^\mathbf{k}}

\newcommand{\Hcv}{H_{\mathrm{cv}}}
\newcommand{\vF}{v_{\mathrm{F}}}
\newcommand{\sigmaone}{\sigma_1}
\newcommand{\sigmatwo}{\sigma_2}

\newcommand{\CINT}{INT} 
\renewcommand{\ie}{i.e.\ }

\begin{document}
\title{Quantum transport in carbon nanotubes}
\author{Edward A. Laird}
\affiliation{Department of Materials, Oxford University, Oxford OX1 3PH, United Kingdom}
\author{Ferdinand Kuemmeth}
\affiliation{Center for Quantum Devices \& Nano-Science Center, Niels Bohr Institute, University of Copenhagen, Universitetsparken 5, 2100 Copenhagen, Denmark}
\author{Gary A. Steele}
\affiliation{Kavli Institute of Nanoscience, Delft University of Technology, 2600 GA Delft, The Netherlands}
\author{Kasper Grove-Rasmussen}
\affiliation{Center for Quantum Devices \& Nano-Science Center, Niels Bohr Institute, University of Copenhagen, Universitetsparken 5, 2100 Copenhagen, Denmark }
\author{Jesper Nyg\aa rd}
\affiliation{Center for Quantum Devices \& Nano-Science Center, Niels Bohr Institute, University of Copenhagen, Universitetsparken 5, 2100 Copenhagen, Denmark }
\author{Karsten Flensberg}
\affiliation{Center for Quantum Devices \& Nano-Science Center, Niels Bohr Institute, University of Copenhagen, Universitetsparken 5, 2100 Copenhagen, Denmark }
\author{Leo P. Kouwenhoven}
\affiliation{Kavli Institute of Nanoscience, Delft University of Technology, 2600 GA Delft, The Netherlands }

\begin{abstract}
Carbon nanotubes are a versatile material in which many aspects of condensed matter physics come together. Recent discoveries have uncovered new phenomena that completely change our understanding of transport in these devices, especially the role of the spin and valley degrees of freedom. This review describes the modern understanding of transport through nanotube devices. 

Unlike in conventional semiconductors, electrons in nanotubes have two angular momentum quantum numbers, arising from spin and valley freedom. We focus on the interplay between the two. The energy levels associated with each degree of freedom, and the spin-orbit coupling between them, are explained, together with their consequences for transport measurements through nanotube quantum dots. In double quantum dots, the combination of quantum numbers modifies the selection rules of Pauli blockade. This can be exploited to read out spin and valley qubits, and to measure the decay of these states through coupling to nuclear spins and phonons. A second unique property of carbon nanotubes is that the combination of valley freedom and electron-electron interactions in one dimension strongly modifies their transport behaviour. Interaction between electrons inside and outside a quantum dot is manifested in SU(4) Kondo behavior and level renormalization. Interaction within a dot leads to Wigner molecules and more complex correlated states.

This review takes an experimental perspective informed by recent advances in theory. As well as the well-understood overall picture, we also state clearly open questions for the field. These advances position nanotubes as a leading system for the study of spin and valley physics in one dimension where electronic disorder and hyperfine interaction can both be reduced to a low level.
\end{abstract}
\maketitle

\tableofcontents

\section{Introduction and motivation}\label{introduction}
Carbon nanotubes are exceptional materials in many different ways. They are mechanically ultra-strong, the surface is perfectly clean, electrons move ballistically, and they vibrate like guitar strings with record-breaking quality factors. Moreover, by zipping nanotubes open one obtains the other wonder material, graphene. Together with C$_{60}$-buckyballs and diamond, these allotropes of carbon have a central position in nanotechnology. Many of their properties have been studied and reviewed in great detail (e.g.~\onlinecite{SaitoBook1998}).

Nanotube electronic transport properties have been studied since the mid-1990s, first in bulk and since 1997 using individual single-wall nanotubes~\cite{BockrathScience1997,TansNature1997}. Many of the basic transport properties were quickly discovered, including Coulomb blockade, Fabry-Perot interference, 1D electronic interactions, Kondo physics, spintronics effects, and induced superconductivity. These properties have all been comprehensively reviewed, with both theoretical (e.g.~\onlinecite{CharlierRMP2007}) and experimental focus (e.g.~\onlinecite{BiercukChapter2008}, \onlinecite{SchonenbergerSST2006})\footnote{The early generation of nanotube experiments that established basic quantum dot behavior was reviewed in~\cite{YaoTAP2001, NygardAPA1999}. Open devices and early attempts to analyse the quantum dot shell structure were described in~\cite{LiangARPC2005,SapmazSST2006}. Hybrid devices involving superconducting and ferromagnetic leads have been reviewed in~\cite{DeFranceschiNnano2010} and~\cite{CottetSST2006} respectively, while aspects pertinent to one-dimensional wires were addressed in~\cite{DeshpandeNature2010}. Coupled quantum dots were introduced in e.g.~\cite{BiercukChapter2008,SchonenbergerSST2006} whereas only recent reviews introduce spin-orbit interaction and valley physics~\cite{IlaniARCM2010, KuemmethMT2010}, which are the themes of this review.}. The general understanding in 2008 can be described as `consistent on a coarse scale'. On a fine scale the specific properties arising from residual disorder together with the specific, usually unknown, chirality of the nanotube under study were hampering a detailed description. On a coarse scale all nanotubes showed similar transport behaviour, but on a fine scale each experimentally studied nanotube was unique. 

An important technical advance was a device scheme in which the nanotube was not exposed to any fabrication chemicals, thereby retaining pristine material quality~\cite{CaoNmat2005}. Transport experiments on such ``ultra-clean'' nanotubes immediately showed more reproducible detail despite the still unknown chirality. Most importantly, the role of spin-orbit interaction was strikingly uncovered~\cite{KuemmethNature2008}. Although this spin-orbit interaction had already been predicted~\cite{AndoJPSJ2000}, it went unobserved and was therefore largely ignored until 2008. The experimental clarity revealed, however, that detailed understanding of quantum phenomena in carbon nanotubes has to include this effect.

The electronic orbits in nanotubes come in two flavors, known as the $K$ and $K'$ valleys, that roughly correspond to clockwise and counterclockwise motion around the nanotube. The resulting quantum states form interesting superpositions of spin up and spin down with the $K$ and $K'$ valleys. Our central aim is to present a coherent description of spin-orbit and $K-K'$ physics in carbon nanotubes. We present the theory on a conceptual level and make references to detailed calculations in the literature. More details of the theoretical background are given in Appendix~\ref{ap_isospin}.~We highlight experimental results that demonstrate the essential concepts most clearly. 

Spin-orbit and $K-K'$ physics and their experimental consequences are first described in Secs.~\ref{sec_basics}-\ref{sec_bandstructure} for nanotubes confined as single quantum dots. The quantum dot geometry allows for a precise, straightforward description of energy eigenstates, which can be probed with well-established techniques of Coulomb blockade spectroscopy. Double quantum dots increase the complexity, with quantum states now described by three numbers: spin (up or down), valley ($K$ or $K'$) and location (left or right). Since the occupancy of both quantum dots is so easily controlled by gate voltages, the double dot geometry provides for exquisite experimental control. Section~\ref{doubledots} describes spin-valley selection rules for tunneling, probed by Pauli blockade experiments. The experimental control in double dots is utilized further in Sec.~\ref{sec_qubits} describing the realization and operation of qubits employing various choice of basis states. 

Sections~\ref{sec_basics}-\ref{sec_qubits} make use of a simplified model with electron-electron interactions included as a capacitive charging energy. In Sec.~\ref{sec:open} we extend this picture to include interactions between quantum dot states and the continuum in the leads. Quantum dots strongly coupled to leads show  renormalization of the energy states as well as the formation of macroscopic coherence in a Kondo state. Sec.~\ref{sec:open} focuses on renormalization and Kondo effects in the specific context of spin-orbit and $K-K'$ physics.  In Sec.~\ref{sec_correlations} we consider interaction effects within quantum dots, which can be extraordinarily strong in the one-dimensional geometry of nanotubes. In quantum dots of a somewhat longer length, this leads to the formation of correlated Wigner molecules.   

\section{Basics of carbon nanotube devices}\label{basics}
\label{sec_basics}
\subsection{Structure of carbon nanotubes}
\begin{figure}
\center 
\includegraphics{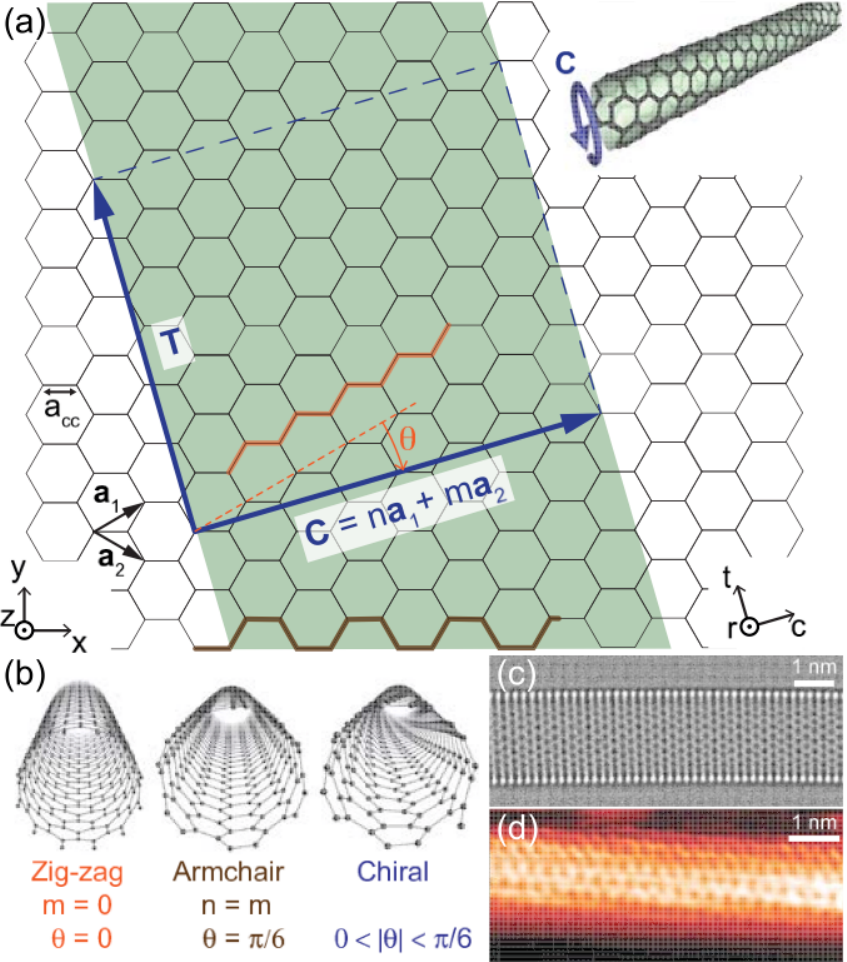}
\caption{\footnotesize{(Color online) Atomic structure of carbon nanotubes. (a) Derivation of nanotube structure from graphene. A single-wall nanotube is equivalent to a rolled-up graphene strip (shaded, with the direction of rolling chosen so that the printed pattern faces outwards). The chiral vector $\mathbf{C}$ spans the nanotube circumference (inset) and connects lattice sites that are brought together by rolling up. Chiral indices~$(n, m)$ completely define the nanotube structure. The unit cell of the nanotube (which is much larger than the unit cell of graphene) is outlined by dashed lines, and the unit vector~$\mathbf{T}$ is indicated. Graphene coordinates $(x,y,z)$, nanotube coordinates $(t,c,r)$ and the chiral angle $\theta$ are also marked. In this example,~$(n,m)=(6,2)$ and $\theta=13.9^\circ$. (b) Nanotubes are divided into three classes according to their chiral indices: Zig-zag, armchair or chiral. Zig-zag and armchair nanotubes are so called because of the shape of the edge formed by a cut perpendicular to the nanotube axis (see highlighted lines in (a)). These three nanotubes are (12,0), (6,6) and (6,4). (c,d) Nanotubes directly imaged by transmission electron microscopy ((c), a (28,0) zig-zag nanotube) and scanning tunneling microscopy~((d), an unidentified chiral nanotube). Adapted from~\onlinecite{ChurchillThesis2012,CharlierRMP2007,WarnerNmat2011,VenemaAPA1998}.}}
\label{structure}
\end{figure}

Carbon nanotubes consist of one or more concentric cylinders of graphene~\cite{SaitoBook1998}. Both multi-wall and single-wall carbon nanotubes (MWCNTs and SWCNTs) can be synthesized and measured, but in this review, we will discuss only SWCNTs. As well as being simpler, these are the most studied both experimentally and theoretically. 

The structure of nanotubes derives from the structure of graphene. A SWCNT is equivalent to a rolled-up strip taken from the two-dimensional honeycomb of carbon atoms that makes up a graphene sheet~(Fig.~\ref{structure}(a)). Since there are infinitely many ways of selecting a strip to roll up, there are correspondingly many different nanotube structures. Each structure is specified by its chiral vector $\mathbf{C}$, which connects lattice sites on opposite sides of the strip that are superposed by rolling up into a nanotube. A given structure is labelled by its chiral indices~$(n,m)$, which are the coordinates of the chiral vector $\mathbf{C}=n \mathbf{a}_1 + m \mathbf{a}_2$ in terms of the graphene basis vectors $\mathbf{a}_1,~\mathbf{a}_2$. From Fig.~\ref{structure}(a), $n$ and $m$ are integers; to ensure that the same structure is not labelled two different ways,~$m$ is conventionally taken in the range $-n/2 < m \leq n$. Instead of specifying~$(n,m)$, a nanotube can also be described by its diameter and chiral angle $\theta$, defined as the angle between~$\mathbf{C}$ and~$\mathbf{a}_1$. 

Two special cases are zig-zag structures ($m=0$) and armchair structures ($n=m$), so called because of the arrangement of atoms along a cut normal to the nanotube. Structures not in either category are called chiral~(Fig.~\ref{structure}(b)). Unlike armchair and zig-zag structures, chiral nanotubes lack inversion symmetry; the inversion isomer (with $\theta \rightarrow -\theta$) of an $(n,m)$ chiral structure is an $(n+m, -m)$ structure. From the Onsager-Casimir relations, the transport properties of isomer pairs are expected to be similar, but they may differ in their nonlinear conductance in the presence of electron-electron interactions and time-reversal symmetry breaking by a magnetic field~\cite{IvchenkoPRB2002,SanchezPRL2004,SpivakPRL2004,WeiPRL2005}. The differences between isomer pairs, well established in optical measurements~\cite{SamsonidzePRB2004,PengNnano2007}, are not discussed further here. Some structure parameters and their dependence on chiral indices are given in Table~\ref{tab:structureparameters}.

\begin{table}
   \centering
   \begin{tabular}{lll}
   \hline \hline
     Name \hfill 				& Symbol \hfill		& Value \hfill \\	\hline
     C-C bond length			& $a_\mathrm{CC}$	& 0.142 nm\\
     \vspace{-0.2cm} \\
     Graphene lattice constant			& $a$			& $\sqrt{3}a_\mathrm{cc}$ = 0.246 nm \\
     \vspace{-0.2cm} \\
     Graphene basis vectors			& $\mathbf{a}_{1,2}$	& $\left(\frac{\sqrt{3}}{2},\pm\frac{1}{2}\right)a$\\
     \vspace{-0.2cm} \\
    \parbox{3cm}{Graphene reciprocal lattice vectors}\hspace{0.1cm}	& $\mathbf{b}_{1,2}$	& $\left(\frac{1}{\sqrt{3}},\pm 1\right) \frac{2\pi}{a}$\\
     \vspace{-0.2cm} \\
     Graphene Dirac points\hspace{0.1cm}	& $\KKK, \KKKp$	& $\pm\frac{\mathbf{b}_2 - \mathbf{b}_1}{3}=\left(0,\mp 1\right)\frac{4\pi}{3a}$\\
          \vspace{-0.2cm} \\
     Chiral vector				& $\boldC$		& $n \mathbf{a}_1 + m \mathbf{a}_2$ \\
     						&				& ($m,n$ integer; $n>0$;\\
						&				&  $-n/2< m \leq n$) \\
	\vspace{-0.2cm} \\
     Chiral angle 			& $\theta$			& $\tan^{-1} \left(\frac{\sqrt{3}m}{2n+m}\right)$\\ 
          					&				& $\left(-\frac{\pi}{6}<\theta\leq\frac{\pi}{6}\right)$  \\

     \vspace{-0.2cm} \\
     Nanotube diameter		& $D$			& $a \sqrt{n^2 + m^2 + nm}/\pi$ \\
	 \hline \hline
   \end{tabular}
   \caption{Summary of structure parameters for an $(n,m)$ nanotube \cite{BaskinPR1955, SaitoBook1998}. Vectors are written with respect to the graphene coordinates~$(x, y)$ defined in Fig.~\ref{structure}.}
   \label{tab:structureparameters}
\end{table}

This structure is confirmed by atomic-resolution microscopy. Transmission electron microscopy images the entire cross section, allowing exact chiral indices to be deduced (Fig.~\ref{structure}(c)). Nanotubes on surfaces can be imaged by scanning tunneling microscopy (Fig.~\ref{structure}(d)), although because of the poor edge resolution, the precise chirality is usually undetermined. Both images confirm the atomic arrangement of Fig.~\ref{structure}(a), with the same atomic spacing~$a_\mathrm{CC} = 0.142$~nm as graphite.

Unfortunately, high-resolution microscopy is usually incompatible with transport measurements and the chiral indices of nanotubes in electronic devices are often unknown. A few experiments have combined transport measurements with structure determination by electron diffraction~\cite{KociakPRL2002, AllenPRB2011}. The structure can also be determined using optical Raman or Rayleigh spectroscopy, which is less invasive but does not always give unambiguous chiral indices~\cite{CaoPRL2004, HuangNL2005, DeshpandeScience2009}. Most of the results in this review will therefore be from nanotubes of unknown chirality; however, as discussed in the next section, the electronic properties of nanotubes are sufficiently independent of the chiral indices that most of the underlying physics can still be explored.

\subsection{Quantum dots}

\begin{figure}
\center
\includegraphics{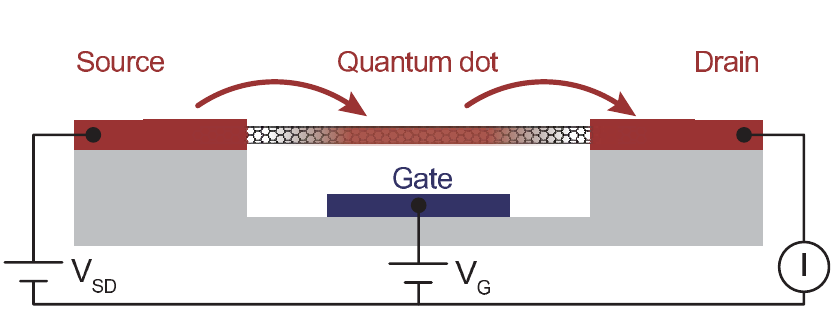}
\caption{\footnotesize{(Color online) Schematic of a basic quantum dot device. The device consists of a nanotube contacted by source and drain electrodes and capacitively coupled to a gate. Tunnel barriers to the source and drain, imposed through the combination of the gate potential and Schottky barriers, define a quantum dot. The current~$I$ through the device is measured as a function of bias voltage~$V_\mathrm{SD}$ and gate voltage $V_\mathrm{G}$. Both the number of electrons $N$ on the island and the dot energy levels can be adjusted by tuning $V_\mathrm{G}$. }}
\vspace{-0.3cm}
\label{CNTdot}
\end{figure}

A basic carbon nanotube electronic device is shown in Fig.~\ref{CNTdot}. The purpose is to allow measurement of the electrical current $I$ through a single nanotube~\cite{BockrathScience1997, TansNature1997}. To achieve this, the nanotube is contacted with metallic source and drain electrodes connected to an external circuit. A third electrode, the gate, coupled capacitively, allows the electrostatic potential to be tuned. Quantum dots are usually measured at low temperature ($\leq 1$~K) to suppress thermal smearing of transport features.

A nanotube naturally confines electrons to one dimension. In quantum transport experiments, it is common to add longitudinal confinement by introducing tunnel barriers. These barriers can be created by modifying the electrostatic potential using gate voltages, often taking advantage of Schottky barriers induced near the metal contacts in the nanotube~\cite{HeinzePRL2002, BiercukChapter2008}. The stretch of nanotube between the barriers where electrons are trapped is called a quantum dot. By studying the current through such a quantum dot as a function of bias, gate voltage, and other parameters such as magnetic field, the energy levels of electrons in the nanotube can be deduced. Quantum dot transport spectroscopy has been extensively reviewed e.g.\ in~\cite{KouwenhovenBook1997,KouwenhovenRPP2001, HansonRMP2007}. Basic concepts needed in this Review are explained in Appendix~\ref{secQuantumDots}.

\subsection{Fabrication challenges of gated quantum devices}

\begin{figure*}
\center 
\includegraphics[width=17.2cm]{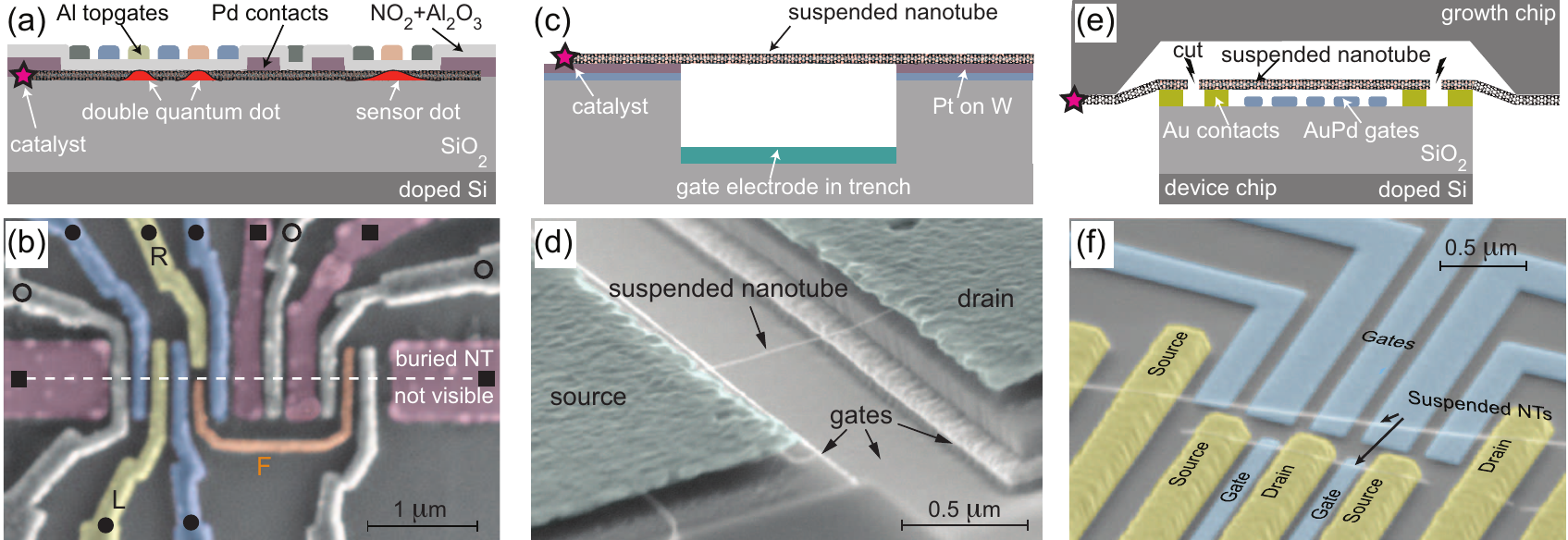}
\caption{\footnotesize{
(Color online) Schematics (a,c,e) and scanning electron micrographs (b,d,f) of devices fabricated by different methods.
(a,b)~Top gating: Nanotubes are located on a growth chip, and electrodes fabricated afterwards. Here a nanotube (not visible) is contacted by metal electrodes (marked $\blacksquare$) and covered by a thin gate oxide. Five gates ($\bullet$) control a double quantum dot, while a floating antenna~(F) allows charge sensing via a separate dot on the same nanotube. Other electrodes ($\circ$) are helper gates. Adapted from~\onlinecite{ChurchillPRL2009}.
(c,d)~Bottom gating: Trench, contacts, and gates are fabricated from inert materials before synthesis, and nanotubes grown across. Adapted from~\onlinecite{SteeleScience2009}.
(e,f)~Mechanical transfer: Suspended nanotubes are synthesized on a growth chip, while electrodes are patterned on a device chip. By stamping the chips together, a nanotube is transferred to the device. Electrical current can be used to cut the nanotube at specific places. In this complex two-nanotube device, five gates define a single or double quantum dot in the upper nanotube, while a pair of dots in the lower nanotube serve as independent charge sensors.
Adapted from \onlinecite{WaissmanNatNano13}.}}
\label{topandbottomgating}
\end{figure*}

The realization of clean and tunable quantum dots in carbon nanotubes is not straightforward. 
Unlike carriers in III-V heterostructures, which are separated from the crystal's surface by an atomically clean buffer layer, the nanotube's $\pi$-band is composed of atomic $p$-orbitals that stick out perpendicular to the surface (see Sec.~\ref{bandstructure}). Patterning of gate oxides, mechanical deformation, and contamination from fabrication chemicals can easily induce disorder and irreproducible device characteristics \cite{BezryadinPRL1998,ZhuAPL05}.

The characteristics of nanotube quantum dots depend on the bandgap, which varies widely between different nanotubes~(Sec. \ref{sec_SNMCNT}). Semiconducting nanotubes (bandgap $\gtrsim 0.1$~eV) often show poor transport characteristics at low carrier density and low temperature. Presumably, this arises from unintentional localization of carriers into disordered puddles, facilitated by the carriers' relatively large effective mass. Conversely, in quasi-metallic nanotubes~(bandgap $\lesssim 10$~meV), controlled creation of sufficiently opaque barriers by electrostatic potentials is difficult, presumably due to the small effective mass (Sec. \ref{SNMCNTtheory}). Kinks made by atomic force microscope~(AFM) manipulation or mechanical templating can be used to locally induce bandgaps and backscattering centers \cite{YaoNature1999,BozovicAPL2001,PostmaScience2001,ParkAPL2002,BiercukNL2004,StokesAPL2008}, resulting in addressable tunnel barriers and Coulomb blockade even at room temperature. A similar effect can occur unintentionally due to disordered mechanical deformations induced by fabrication~\cite{BezryadinPRL1998}. 

The largest experimental interest has been attracted by devices between these extremes (narrow-gap nanotubes). Tuneable tunnel barriers can then be induced rather easily by electrostatic gates. These nanotubes allow gate-controlled devices that do not uncontrollably break up into disordered puddles, yet their tunnel barriers remain tuneable over a wide range, even in the few-charge regime. Most devices can be classified according to whether gate fabrication occurs after nanotube growth (top gating), before growth (bottom gating), or on a separate chip (mechanical transfer method). 

\subsubsection{Top gating}
The simplest way to make devices is usually to fabricate electrodes on top of nanotubes. This allows complex devices with many kinds of contact material including normal metals, ferromagnets, and superconductors. After growth or deposition, suitable nanotubes are imaged, and the electrodes are patterned by electron-beam lithography and liftoff.
Early single-electron transistors were contacted in this way \cite{BockrathScience1997}, as were the first double quantum dots~\cite{MasonSci04}. Although cleanliness and fabrication-induced disorder are  a concern, devices fabricated this way have demonstrated ambipolar operation and discrete excited states \cite{BiercukNL05}, as well as charge sensing and pulsed gate spectroscopy \cite{BiercukPRB2006,GotzNL2008}.

Full control of a double quantum dot requires at least five gate electrodes, necessitating thin, high-dielectric constant gate oxides (e.g.\ atomic-layer-deposited aluminum or hafnium oxide) and densely packed gate arrays \cite{ChurchillPRL2009,ChurchillNPhys2009}. 
Such a device is shown in Fig.~\ref{topandbottomgating}(a,b), consisting of a fully tunable double quantum dot capacitively coupled via a floating gate to a charge-sensing single quantum dot on the same nanotube. Among other applications, these devices allow measurement of spin relaxation and dephasing (Sec.~\ref{qubits}). 
By selectively etching beneath the nanotube, suspended devices can also be fabricated \cite{LeturcqNphys09}.

\subsubsection{Bottom gating}
A drawback of top gating is that the fabrication process itself can introduce disorder in the nanotube. An alternative is to grow or deposit nanotubes over predefined electrodes, resulting in devices with improved control and cleanliness \cite{CaoNmat2005}. 
Early single quantum dots were realized by depositing nanotubes across Pt source and drain electrodes, using the Si/SiO$_2$ substrate as a backgate~\cite{TansNature1997}. 
Similar to graphene devices, where suspending the layer dramatically improved the mobility \cite{BolotinSSS2008,DuNNano2008}, suspended nanotubes often show near-ideal transport characteristics, indicating that much of the disorder arises from interactions with the substrate \cite{SteeleNNano09,IlaniARCM2010,JungNL2013}.

Motivated by the results of suspended single quantum dots as in Fig.~\ref{topandbottomgating}(d), more complex contact and gate arrays were developed that can be loaded into the nanotube growth furnace as the last step before cool down and measurements~\cite{KuemmethNature2008,SteeleNNano09}.
Although these devices were of high quality\footnote{Nanotubes that have never been in contact with solvents, resists, or a substrate are sometimes called ``ultraclean" \cite{DeshpandeScience2009, SteeleNNano09, PeiNnano12, pecker2013observation, WaissmanNatNano13, BenyaminiNatPhys2014}.} and resulted in new discoveries, the harsh conditions in the growth reactor greatly restrict the materials and design. The overall device yield is low because a nanotube must grow across contacts and gates by chance.

\subsubsection{Mechanical transfer}

Mechanical transfer attempts to benefit from the best of both approaches, achieving high gate tunability without  post-growth processing.
The device chip (without nanotubes) and the growth chip (with nanotubes suspended across trenches) are fabricated separately. 
Just before measurement, a single nanotube is transferred from growth chip to device chip using an aligned stamping process~\cite{WuNL2010,PeiNnano12}. By employing piezo-controlled scanning probe microscope manipulators, the transfer is possible in vacuum at cryogenic temperatures~\cite{WaissmanNatNano13}, allowing the cleanliness of the nanotube to be tested \emph{in situ}. A state-of-the-art example is shown in~Fig.~\ref{topandbottomgating}(e-f). 

\subsection{Nanotube synthesis and isotopic engineering}
For research applications nanotubes are readily synthesized in desktop furnaces, using chemical vapor deposition (usually from methane, ethanol or ethylene) in the presence of suitable catalysts~\cite{KongAPA1999,KuemmethMT2010}.
Unlike III-V devices such as GaAs double dots, in which all stable isotopes possess a nuclear magnetic moment, carbon nanotubes allow fabrication of devices with and without nuclear spins in the host material in a straightforward way.

Nanotubes synthesized from natural hydrocarbons consist of $~99\%$ \twelve\  and $\sim1\%$ \thirteen. 
By using isotopically purified $^{13}$CH$_4$ or $^{12}$CH$_4$, the isotopic composition can be tuned during growth. 
This not only affects the phonon modes (revealed by Raman spectroscopy \cite{Liu:2001eq}), but also  the electron spin properties, because
\thirteen\ possesses a nuclear spin $|\vec{I}|=1/2$, while \twelve\ has $|\vec{I}|=0$. 
As discussed in Sec. \ref{relaxationbynuclei}, a local spin impurity (such as~\thirteen) can flip the spin and/or valley of an electron.  

\section{Carbon nanotube bandstructure}\label{bandstructure}
\label{sec_bandstructure}

Just as the atomic structure of carbon nanotubes can be derived from that of graphene, the electronic band structure inherits from graphene many of its properties. However, the simple effect of being rolled up drastically modifies the band structure, leading to many effects that are not present in graphene. The most dramatic difference is the introduction of a bandgap, which allows electrons in nanotubes to be confined using gate voltages, but a variety of other subtle effects arise.

Briefly, the results are as follows. Although graphene is a semimetal, the formation of a nanotube leads to a confinement bandgap (a few hundred~meV) for two-thirds of the possible structures. These are known as semiconducting nanotubes. Most of the remaining nominally metallic nanotubes show narrow bandgaps~($\sim 10$~meV) due to a combination of curvature and strain. If the bandgap is undetectibly small, the nanotube is called quasimetallic, and a metallic nanotube is defined as one for which the bandgap is exactly zero.  More subtle details of the band structure become evident in a magnetic field, including a magnetic moment associated with the valley degree of freedom, and spin-orbit coupling that is much stronger than in graphene and arises from curvature. 

\begin{figure}
\center 
\includegraphics[width=8.6cm]{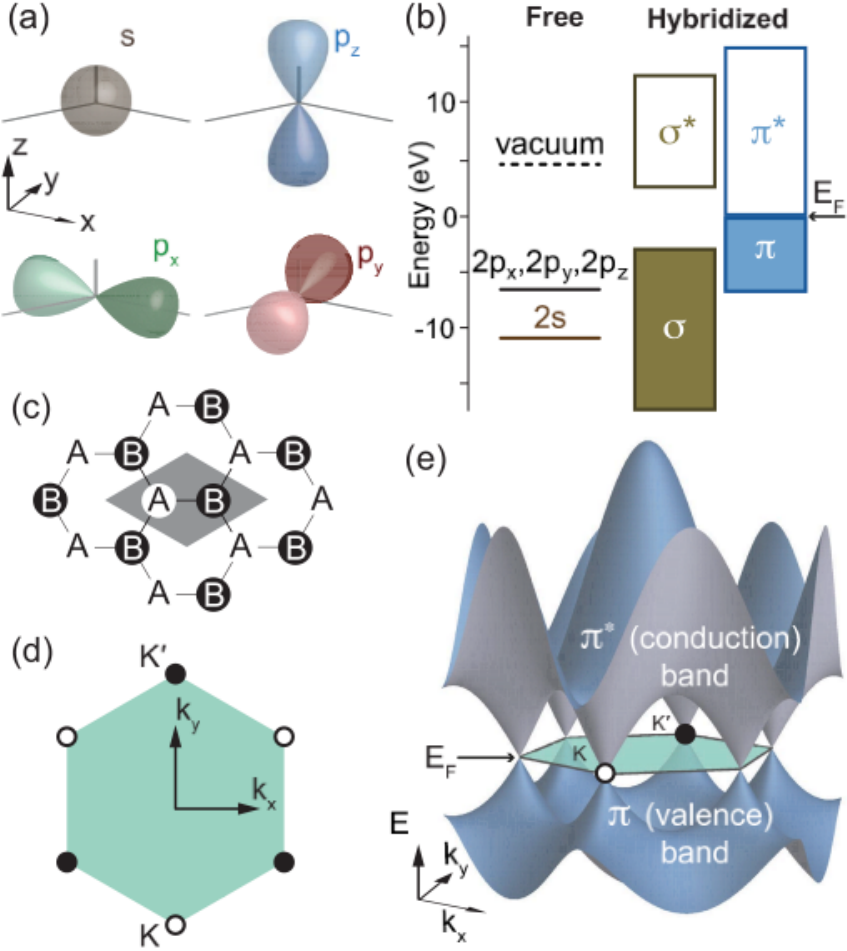}
\caption{\footnotesize{(Color online) (a) Electron orbitals of atomic carbon. Lighter (darker) colours denote regions where the $p$-orbital wave functions are positive (negative). Bond directions in graphene are indicated by grey lines. (b) Schematic energy levels of atomic (left) and $sp^2$ hybridized (right) carbon. Energies are referenced to $E_\mathrm{F}$, approximated as equal to the negative of the work function.~(c) Segment of graphene with the unit cell shaded and the A and B sublattices marked. (d) First Brillouin zone of graphene in reciprocal space, showing the six symmetry points, labelled $K$ or $K'$. (e) Energy bands ($\sigma$ bands omitted) of graphene close to the Fermi level, showing the six Dirac cones where $\pi$ and $\pi^*$ bands touch.}} 
\label{bandstructuregraphene}
\end{figure}

\subsection{From atomic carbon to graphene band structure}

To understand nanotube band structure, we begin with the energy levels of atomic carbon.
In a free atom, the six electrons occupy the configuration $1s^22s^22p^2$. The outermost atomic shell includes one spherically symmetric $s$-orbital and three $p$-orbitals $p_x$, $p_y$, $p_z$ (Fig.~\ref{bandstructuregraphene}(a)). Because of twofold spin degeneracy in each orbital, there are therefore eight states in the outermost shell of the atom, of which four are occupied. 

The $2s-2p$ energy splitting is small enough (less than a typical bond energy) that all four outermost orbitals can hybridize to form covalent bonds. For a given structure, the number of $2p$ orbitals that hybridize with the $2s$ orbital is determined by symmetry. In graphene, the $p_z$ orbital, oriented perpendicular to the plane, is odd under $z$ inversion and therefore cannot hybridize with the even-parity $2s$ orbital. No such symmetry protects the~$p_x$ and $p_y$ orbitals. This type of hybridization, in which an $s$-orbital is mixed with two $p$-orbitals, is known as $sp^2$ hybridization.

In graphene, these three orbitals further hybridize across neighbouring atoms in the crystal, forming a low-energy (bonding) band $\sigma$ and a high-energy (antibonding) band $\sigma^*$ (Fig.~\ref{bandstructuregraphene}(b)). Likewise, hybridization of the $p_z$ orbitals forms bonding and antibonding bands denoted $\pi$ and $\pi^*$, although with smaller bonding energy because the interatomic overlap is less. In undoped graphene, the electrons exactly fill the bonding bands, with three electrons per atom occupying $\sigma$ and one occupying $\pi$. The $\sigma$ band remains filled at all times and does not participate in transport. The electrical behaviour of nanotubes is therefore determined almost entirely by the properties of the $\pi$ and $\pi^*$ bands. 

Ignoring spin-orbit coupling, the $p_z$ orbitals do not hybridize with any of the lower-lying states, so the structure of the $\pi$ and $\pi^*$ bands follows simply from energy levels in the honeycomb graphene potential. Graphene consists of a rhombus unit cell with a two-atom basis~(Fig.~\ref{bandstructuregraphene}(c)), and has the hexagonal Brillouin zone shown in Fig.~\ref{bandstructuregraphene}(d). The corners of this hexagon in $k$-space are alternately labelled $K$ or $K'$. Because the three $K$ points are connected by reciprocal lattice vectors, by Bloch's theorem they correspond to equivalent electron states; likewise, the three $K'$ points are equivalent to each other, but not to the $K$ points. States close to the $K'$ point are time-reversal conjugates of those close to the $K$ point.

The band structure that arises from this potential (Fig.~\ref{bandstructuregraphene}(e)) has quite unusual properties~\cite{WallacePRL1947, SaitoBook1998, CastroNetoRMP2009}. Although there is no bandgap,  the $\pi$ and $\pi^*$ bands touch only at $K$ and $K'$, where the density of states is zero. Since the available electrons exactly fill the $\pi$ band, these points are where the Fermi level $E_\mathrm{F}$ intersects the band structure, so that undoped graphene is neither a true metal nor a true semiconductor, but a semimetal. Close to the Fermi surface, the dispersion relation is linear, with a slope that determines the Fermi velocity\footnote{This value is derived from numerical simulations of graphite~\cite{PainterPRB1970,TatarPRB1982,TrickeyPRB1992} and nanotubes~\cite{MintmirePRL1992}, which indicate $v_\mathrm{F}= 7.8-9.8 \times 10^5~\mathrm{ms}^{-1}$ (although interactions may renormalize the value significantly~\cite{KanePRL2004}), as well as nanotube STM density-of-states measurements~\cite{WildoerNature1998, OdomNature1998} and ballistic electron resonance experiments~\cite{ZhongNmat2008}, which give $v_\mathrm{F}= 7.9-8.7\times 10^5~\mathrm{ms}^{-1}$. }  $v_\mathrm{F}=\frac{1}{\hbar} |\nabla_{\boldk}E|\approx 8 \times 10^{5}~\mathrm{ms}^{-1}$. Expanding about the $K$ or $K'$ point by writing $\boldk=\KKK+\bm{\kappa}$ or $\boldk=\KKKp+\bm{\kappa}$, and defining $E_\mathrm{F}$ as the zero of energy, the dispersion relation for $|\bm{\kappa}| \ll |\KKK|$ is simply
\begin{equation}
E=\pm \hbar v_\mathrm{F} |\bm{\kappa}|,
\label{eq_dispersionrelation}
\end{equation}
where the $+$ sign applies to electrons and the $-$ to holes. Because this dispersion relation also describes massless Dirac fermions, the points where the bands touch are known as Dirac points, and the nearby bands as Dirac cones. The correspondence of electron states in a nanotube with solutions of a Dirac-like equation is explained in  Appendix~\ref{ap_isospin}. Although we use this correspondence only in a few places in this Review, it is theoretically convenient because it allows many effects on nanotube band structure to be derived as perturbations to the Dirac equation.

\subsection{Semiconducting, narrow-gap and metallic nanotubes}
\label{sec_SNMCNT}
\subsubsection{Theory: the zone-folding approximation}
\label{SNMCNTtheory}
\begin{figure}
\center 
\includegraphics{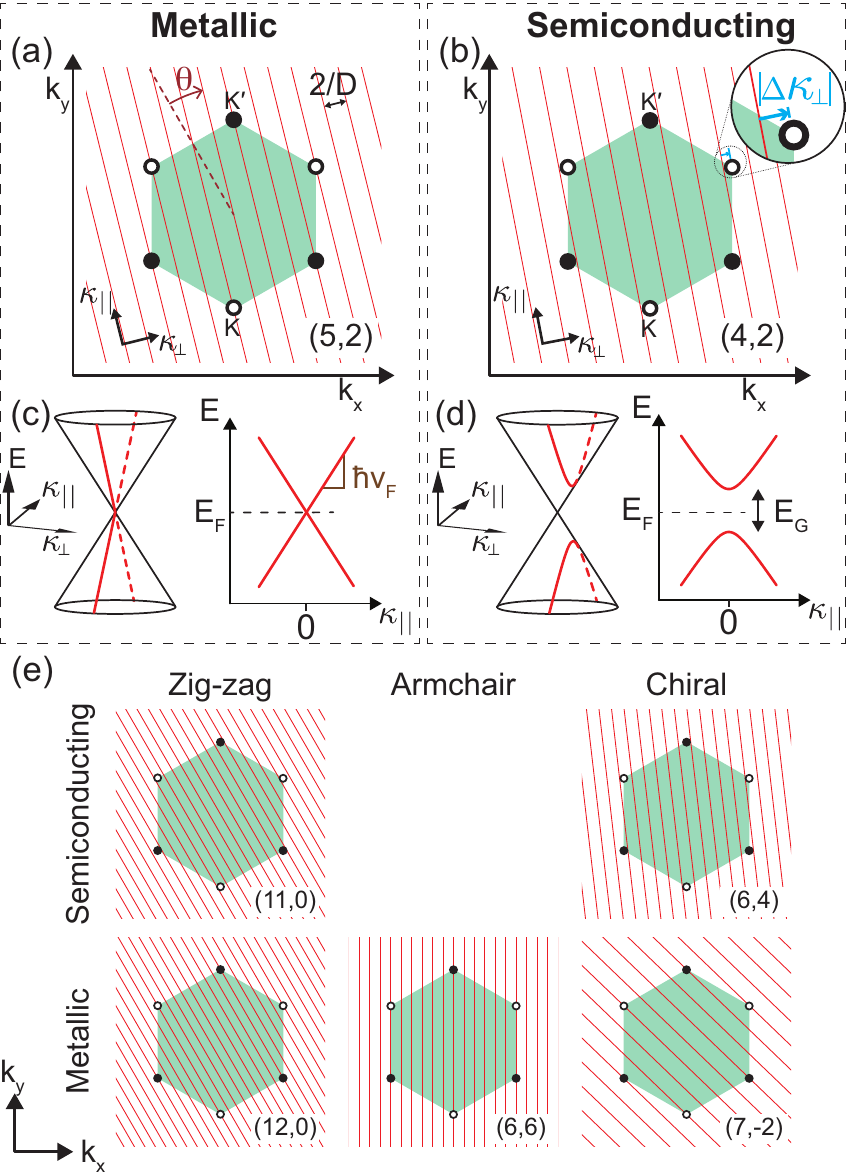}
\caption{\footnotesize{(Color online) The effect of periodic boundary conditions. (a,b) Requiring that the electron wave function be single-valued constrains $\mathbf{k}$ to lie on one of the quantization lines in reciprocal space corresponding to integer values of $k_c D/2$. If quantization lines intersect the Dirac points the nanotube is metallic (a), otherwise it is semiconducting (b), with minimum quantization line offset $|\Delta \kappa_\perp|=2/3D$. The $\{\kappa_\perp, \kappa_{||}\}$ axes in reciprocal space, corresponding to motion around or along the nanotube, are indicated. (c,d) Dispersion relations (in the lowest-energy one-dimensional band) close to a Dirac point for the two kinds of nanotube, showing how the offset gives rise to a bandgap. The Fermi level for undoped nanotubes is indicated. (e) Examples of quantization lines for several metallic and semiconducting structures of the three types shown in Fig.~\ref{structure}(b). Alone of the six combinations, armchair semiconducting nanotubes do not exist.}}
\label{quantizationkperp}
\end{figure}

Since the nanotube diameter is usually much larger than the interatomic spacing, the graphene band structure is to a good approximation unperturbed by rolling up into a nanotube except for the imposition of a periodic boundary condition \cite{HamadaPRL1992,SaitoAPL1992}. This is known as the ``zone-folding approximation''. 
The boundary condition to ensure single-valuedness is that $\boldk \cdot \boldC=2\pi p$, where $p$ is an integer, \ie the component of $\boldk$ perpendicular to the nanotube axis is $k_c=2p/D$. The allowed $\boldk$-values correspond to a series of lines in reciprocal space, known as quantization lines, running at an angle  $\pi/3 + \theta$ from the $k_x$ axis (Fig.~\ref{quantizationkperp}(a-b)).

The one-dimensional dispersion relation $E(\kappa_{||})$ is a cut along the quantization lines of the two-dimensional graphene dispersion relation. Since it is the branches closest to $E_\mathrm{F}$ that determine transport properties, we neglect the other branches. The nanotube bandgap depends on the minimum separation of the quantization lines from the Dirac points. There are two possible situations. If quantization lines run straight through the Dirac points (Fig.~\ref{quantizationkperp}(a,c)), then $E(\kappa_{||})$ is linear near~$\kappa_{||}=0$, giving zero bandgap and a metallic nanotube. However, if the lines bypass the Dirac points with separation $|\Delta \kappa_\perp|$, the situation is as shown in Fig.~\ref{quantizationkperp}(b,d). The dispersion relation gives a pair of hyperbolae with bandgap~$E_\mathrm{G}=2\hbar v_\mathrm{F} |\Delta \kappa_\perp|$, and therefore a semiconducting nanotube.



In the zone-folding approximation, the bandgap is determined by a simple rule: If $n-m$ is a multiple of three, the nanotube is nominally metallic~\cite{HamadaPRL1992,SaitoAPL1992}. Otherwise, it is semiconducting, with bandgap $E_\mathrm{G}=4 \hbar v_\mathrm{F}/3D\approx 700~\mathrm{meV}/D[\mathrm{nm}]$. In a collection of nanotubes with random chiral indices, semiconducting nanotubes will therefore outnumber metallic ones by approximately 2:1. Figure~\ref{quantizationkperp}(e) illustrates how the chiral indices determine whether the quantization lines intersect the Dirac points for various nanotube structures. Examples of both cases are shown for the three kinds of structure defined in~Fig.~\ref{structure}(b), with one exception: Zig-zag and chiral tubes can be either semiconducting or metallic, but all armchair nanotubes are metallic.

The nanotube structure also sets the electron dispersion relation and hence the effective mass. The equation of the hyperbola in Fig.~\ref{quantizationkperp}(d) is~\cite{ZhouPRL2005}:
\begin{equation}
E^\pm(\kappa_{||})=\pm \sqrt{\hbar^2 v_F^2 \kappa_{||}^2+E_\mathrm{G}^2/4}.
\label{eq_Ekpar}
\end{equation}
This low-energy dispersion relation is clearly electron-hole symmetric. This is a fragile symmetry, because any charge in the environment couples oppositely to electrons and holes, but it is sometimes reflected in data~\cite{JarilloHerreroNature2004}.

The effective mass arises from the curvature of the dispersion relation and for low energy ($|E^\pm(\kappa_{||})| \ll E_\mathrm{G}$) is:
\begin{equation}
\label{eq_meff}
m_\mathrm{eff}	= \hbar^2\left(\frac{d^2E}{d\kappa_{||}^2}\right)^{-1}\approx \frac{E_\mathrm{G}}{7.3~\mbox{eV}} \times m_\mathrm{e},
\end{equation}
where $m_\mathrm{e}$ is the free electron mass. A bandgap of~$100$~meV corresponds to effective mass $\sim 0.014m_\mathrm{e}$, smaller than that in many conventional semiconductors (e.g. in GaAs $m_\mathrm{eff}=0.067m_\mathrm{e}$). Because small $m_\mathrm{eff}$ leads to larger longitudinal level spacing, nanotubes with small $E_\mathrm{G}$ are often preferred for quantum dot experiments. 

\subsubsection{Valley as a robust quantum number}

\begin{figure}
\center
\includegraphics{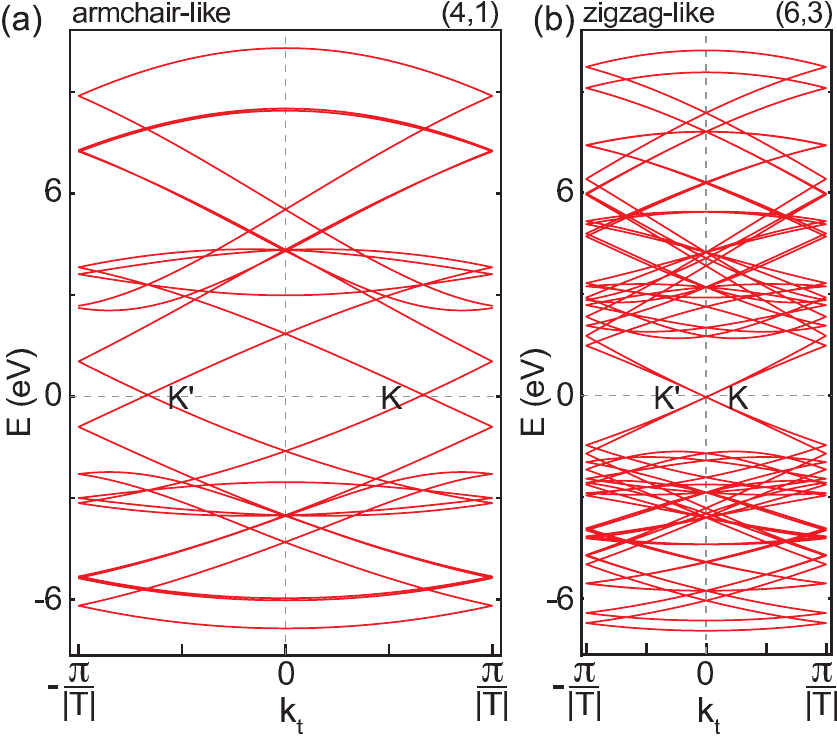}
\caption{\footnotesize{ (Color online) Robustness of the valley index in nanotubes. All subbands of the 1D dispersion relation (corresponding to different quantization lines in
Fig.~\ref{quantizationkperp}) are plotted in the first longitudinal Brillouin zone versus longitudinal wavevector $k_t$. All nominally metallic nanotubes can be classified as armchair-like or zig-zag like. For armchair-like nanotubes, the two Dirac points are separated in $k_t$; for zigzag-like nanotubes, they are separated in crystal angular momentum. Since all metallic nanotubes fall into one of these classes, valley is a good quantum number in a slowly varying potential. (a) Armchair-like (4,1) nanotube. (b) Zigzag-like (6,3) nanotube. Each spin-degenerate band is calculated using a graphene tight-binding model taking account of nearest-neighbor overlap integrals but without spin-orbit coupling. Only states near
$E=0$
participate in transport. Note that the Brillouin zone in (a) has been plotted wider than in (b), to reflect the different longitudinal length $|T|$ of the unit cell in real space.}}
\label{fig_valleyvalley} 
\end{figure}

Just as in graphene, the band structure in nanotubes is characterized by the distinct and time-conjugate valleys $K$ and $K'$.
In graphene the robustness of the valley quantum number is linked to the symmetries of the lattice. Mixing between valleys requires a large transfer of crystal momentum, and is therefore weak in a smoothly varying Coulomb potential.
This is less obvious in metallic nanotubes, because the two Dirac points sometimes remain well separated in momentum space, and sometimes they merge at $k_t=0$.
In fact, all metallic nanotubes (see~Fig.~\ref{fig_valleyvalley}) can be divided into two classes~\cite{SamsonidzeNanosNanot2003, MarganskaArxiv2014}: the Dirac points are either well-separated in longitudinal momentum space (such nanotubes are known as armchair-like metals), or collapse to the origin of the longitudinal Brillouin zone (zigzag-like metals). For chiral metallic nanotubes, this classification is possible by introducing a helical translational basis vector \cite{LundePRB05}.
For the zigzag-like metals, the two bands at $k=0$ are distinct by having different crystal angular momentum \cite{LundePRB05}, where the angular momentum is defined as the quantum number related to the rotation part of the helical symmetry \cite{WhitePRB1993}. In the armchair-like metals, the angular momenta are the same, but their longitudinal crystal momenta differ by $4\pi/3|T|$.
Consequently, in both cases valley-valley scattering is suppressed by a difference in crystal angular momenta or crystal longitudinal momenta.
Scattering within a valley may also require atomically sharp Coulomb scatterers or lattice imperfections, due to the spinor structure of the solutions to the Dirac equation, which differs between right movers and left movers~\cite{AndoJPSJ98a,AndoJPSJ98,RocheAPL2001,McEuenPRL99}.

Armchair-like and zigzag-like band structures are exemplified in Fig.~\ref{fig_valleyvalley}. The number of subbands equals the number of carbon atoms in the unit cell of the nanotube, spanned by $\mathbf{C}$ and $\mathbf{T}$ in Fig.~\ref{structure}(a). Each subband shown is two-fold degenerate due to spin, and arises from a mapping of the quantization lines in Fig.~\ref{quantizationkperp} into the 1D Brillouin zone of the nanotube.

The above discussion applies to narrow-gap nanotubes, \ie those that would be metallic in the zone-folding approximation but where other perturbations introduce a small bandgap~(Sec.~\ref{structural}). For nanotubes that are semiconducting even in the zone-folding approximation~\cite{WhitePRB1993, MintmireCarbon1995}, the situation is similar. This can again be seen using helical quantum numbers~\cite{WhitePRB1993, MintmireCarbon1995} as follows: All bands can be classified by their crystal angular momentum, which means that mixing of two bands with different crystal angular momentum is protected (as for zigzag-like metallic tubes). Mixing of bands with the same crystal angular momentum is suppressed by their difference in wavenumbers when folded onto the smaller translational Brillouin zone.

Local Coulomb scatterers can flip the valley index \cite{PalyiPRB10, BerciouxPRB2011},  and spin-carrying impurities can flip both spin and valley with comparable rates \cite{PalyiPRB09}. One example is hyperfine coupling to nuclear \thirteen\ spins, which can cause both spin and valley relaxation~(Sec.~\ref{relaxationbynuclei}). Another example is the local part of the electron-electron interaction, discussed in Sec.~\ref{correlations} and Appendix~\ref{ap_isospin}. In addition, electrical contacts can induce valley scattering due to valley mixing during tunneling~(Sec.~\ref{subsec:level_renormalization}).%

\subsubsection{Experiment}
\begin{figure}
\center 
\includegraphics{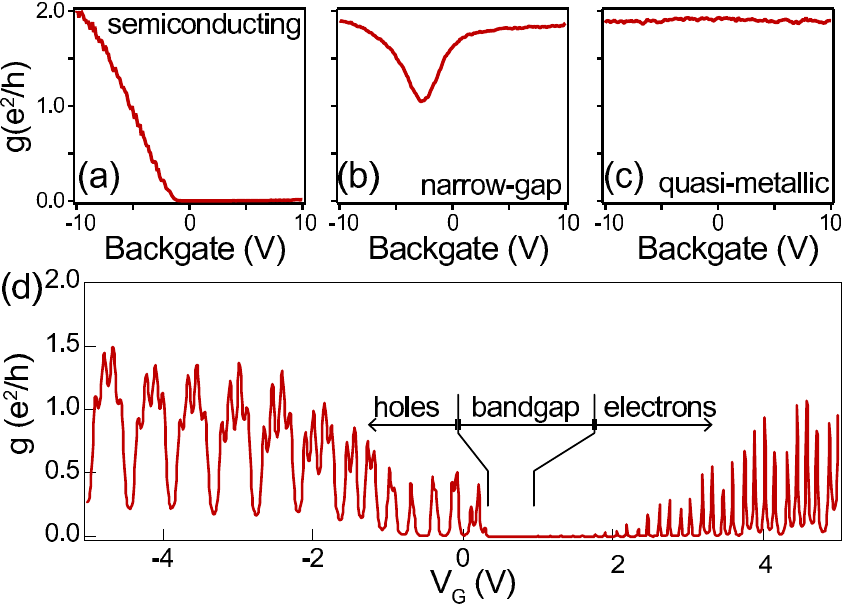}
\caption{\footnotesize{(Color online) Signatures of the bandgap in transport. (a-c) Room temperature conductance measurements as a function of gate voltage. A semiconducting nanotube (a) has $E_\mathrm{G}\gg k_\mathrm{B}T$, and can be tuned between a conducting state (Fermi level in the valence band) and an insulating state (Fermi level in the bandgap). Transport via the conduction band is not observed, because it would require a much higher gate voltage. (b) A narrow-gap nanotube ($E_\mathrm{G}\sim k_\mathrm{B}T$) shows transport via both conduction and valence bands. Tuning the Fermi level into the bandgap does not completely suppress current at room temperature. (c) A few nanotube devices ($<1$~\%) show no gate dependence of conduction. These could be truly metallic, although it cannot be excluded that this device in fact contains a bundle of nanotubes that screens the gate. (d) Conductance of a single narrow-gap nanotube at 300~mK. Transport is now completely suppressed in the bandgap, and the device can be tuned into electron or hole configurations by tuning $V_\mathrm{G}$. Adapted from~\onlinecite{ChurchillThesis2012, CaoNmat2005}.}}
\label{bandgapexpt}
\end{figure}

Nanotubes of different kinds can be distinguished experimentally by measuring the current as a function of~$V_\mathrm{G}$ at fixed $V_\mathrm{SD}$, as in Fig.~\ref{bandgapexpt}(a-c). The gate potential shifts the energy levels up or down and therefore tunes the position of the gap relative to $E_\mathrm{F}$. Tuning $E_\mathrm{F}$ into the bandgap suppresses the current. This can be seen in~Fig.~\ref{bandgapexpt}(a), where the Fermi level is shifted from the valence band (for $V_\mathrm{G}\lesssim 0$) to the bandgap (for $V_\mathrm{G}\gtrsim 0$), showing that the nanotube is semiconducting. A quasi-metallic nanotube, by contrast, is one with no dependence on $V_\mathrm{G}$ (Fig.~\ref{bandgapexpt}(c)), indicating $E_\mathrm{G}\ll k_\mathrm{B}T$, where $k_\mathrm{B}$ is Boltzmann's constant and $T$ is temperature.

Experimentally, the fraction of nanotubes showing quasi-metallic behavior at room temperature is very small ($\lesssim 1\%$) ~\cite{CaoNmat2005,ChurchillThesis2012}. More common is `narrow-gap' behavior (Fig.~\ref{bandgapexpt}(b)), where partial current suppression indicates a small bandgap $E_\mathrm{G}\sim k_\mathrm{B}T$ at room temperature~\cite{OuyangScience2001}. This interpretation is confirmed by low-temperature experiments~(Fig.~\ref{bandgapexpt}(d)), where precise measurements from Coulomb peak positions frequently give $E_\mathrm{G}\sim 10-100$~meV (e.g.\ in ~Fig.~\ref{bandgapexpt}(d), $E_\mathrm{G} =60 $~meV). To explain such small bandgaps from circumferential quantization alone  requires $D=7-70$~nm, which might be structurally unstable and is excluded by AFM topography measurements. More likely, nearly all nanotubes that the zone-folding model predicts should be metallic acquire narrow bandgaps by perturbations discussed in the next section.

\subsection{Structural origins of the narrow gap}
\label{structural}
\subsubsection{Theory}
\begin{figure}
\center 
\includegraphics{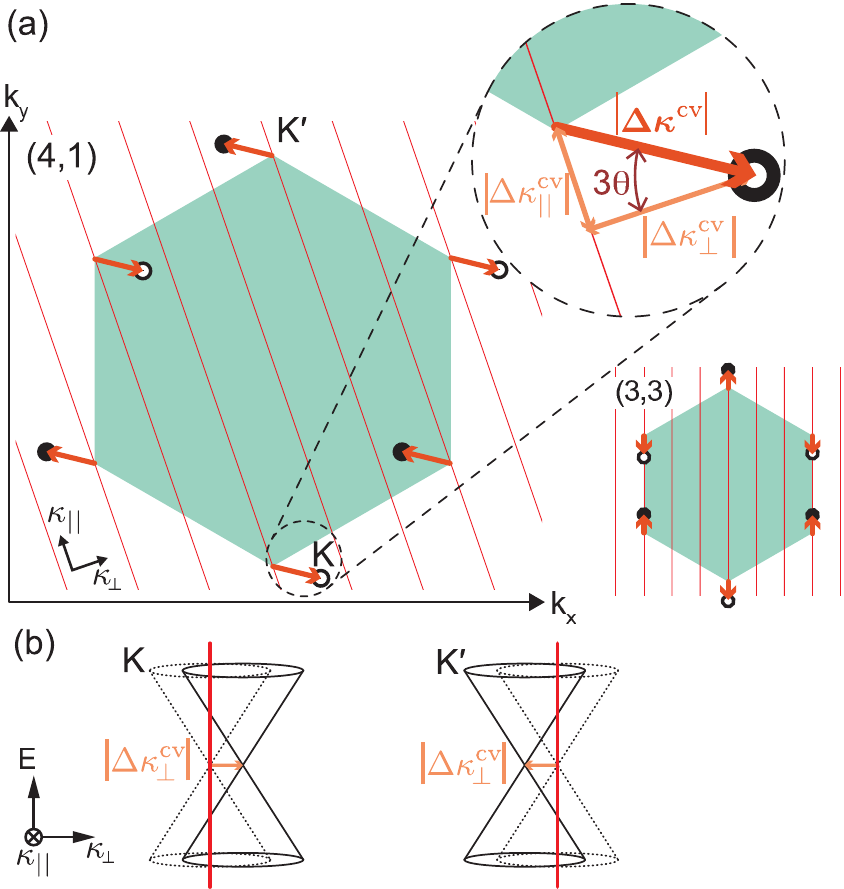}
\caption{\footnotesize{(Color online) Perturbation of the graphene band structure by the curvature in nanotubes. (a) Displacement of the Dirac points away from the corners of the Brillouin zone due to curvature in a (4,1) nanotube. For visibility, the shift has been exaggerated by a factor of 15. Top inset: Decomposition of the displacement vector $\bm{\Delta \kappa}^\mathrm{cv}$ near the $K$ point into components parallel and perpendicular to the nanotube axis. The shift is at an angle $3\theta$ to the nanotube circumference. Bottom inset: shift for an armchair nanotube. Because the shift is along the quantization lines, curvature does not lead to a gap in these structures. (b) Dirac cones close to $K$ and $K'$ valleys with (solid) and without (dotted) curvature effects, showing how horizontal shifts by~$\Delta \kappa_\perp^\mathrm{cv}$ open a bandgap in a nominally metallic tube.}}
\label{Curvature_theory}
\end{figure}

The zone-folding approximation assumes that the allowed electron states in nanotubes are exactly the same as their equivalents in graphene. Perturbations arise if the symmetry of the carbon bonds is broken by changing the overlap between adjacent electron orbitals. One unavoidable example is the curvature of the rolled-up sheet~\cite{BlasePRL1994}. This has two effects on the band structure. First, it leads to a small renormalization of the Fermi velocity by at most a few percent, which is insignificant in experiments~\cite{IzumidaJPSJ2009}. More importantly, it displaces the Dirac points in reciprocal space away from $K$ and $K'$~\cite{KanePRL1997, IzumidaJPSJ2009}, because it breaks the equivalence of the three bond directions. This shift is parameterized by a displacement vector $\mathbf{\Delta} \kappab^\mathrm{cv}$ (Fig.~\ref{Curvature_theory}(a)), and is opposite for $K$ and $K'$ because states in the two valleys are time-reversed conjugates of each other~\cite{CastroNetoRMP2009}. In semiconducting nanotubes, $|\mathbf{\Delta} \kappab^\mathrm{cv}|$ is much smaller than the offset $|\Delta \kappa_\perp|$ arising from quantization, and therefore has only a small effect. However, in nominally metallic nanotubes, the shift of the Dirac cones relative to the quantization lines introduces a bandgap $E_\mathrm{G}=2\hbar v_\mathrm{F} |\Delta \kappa_\perp^\mathrm{cv}|$, where  $\Delta \kappa_\perp^\mathrm{cv}$ is the component of $\mathbf{\Delta} \kappab^\mathrm{cv}$ perpendicular to the nanotube axis (Fig.~\ref{Curvature_theory}(b)). (The component parallel to the axis, $\Delta \kappa_{||}^\mathrm{cv}$, has no effect.)
This curvature-induced bandgap is always much smaller than the quantization energy difference. Unlike the quantization bandgap, it depends on the chiral angle. As shown in the inset of Fig.~\ref{Curvature_theory}(a), the vector $\mathbf{\Delta} \kappab^\mathrm{cv}$ points at an angle of $3\theta$ from the perpendicular. The curvature-induced bandgap is therefore proportional to $\cos 3 \theta$; it is calculated to be~\cite{KanePRL1997,ParkPRB1999,YangPRB1999,YangPRL2000, KleinerPRB2001,IzumidaJPSJ2009,KlinovajaPRB2011}:
\begin{equation}
E_\mathrm{G}^\mathrm{cv}\sim \frac{50~\mathrm{meV}}{D[\mathrm{nm}]^2}\cos 3\theta.
\end{equation}
For armchair nanotubes $\theta=\pi/6$~(Fig.~\ref{structure}(b)) and therefore $\cos 3 \theta=0$. These are the only nanotubes expected to be truly metallic, if no other perturbation is applied (see Appendix~\ref{ap_isospin} for more theoretical details).

A gap of similar magnitude can be opened by strain, in a way that also depends on nanotube chirality~\cite{HeydPRB1997,KanePRL1997,RochefortChemPhysLett1998,YangPRL2000}. A uniaxial strain $\epsilon$ has the same symmetry-breaking effect as curvature (namely to break~$C_3$), and therefore leads to a Dirac point displacement in the same direction with magnitude
\begin{equation}
|\mathbf{\Delta \kappab}^\mathrm{\epsilon}|=\frac{12 \zeta}{1+6\zeta}(1+\lambda)\epsilon/a_\mathrm{CC},
\label{straineqn}
\end{equation} 
where $\lambda \approx 0.2$ is the Poisson ratio and $\zeta \approx 0.066$ is a parameter related to the carbon-carbon bond force constants~\cite{NisoliPRL2007, HuangPRL2008}. A torsional strain $\gamma$ displaces the Dirac points by an amount $|\mathbf{\Delta \kappab}^\gamma|\approx\gamma$ at an angle $\pi/2-3\theta$ from the perpendicular~\cite{YangPRL2000}. The uniaxial bandgap is therefore proportional to $\cos 3 \theta$, while the torsional bandgap is proportional to $\sin 3 \theta$.  A third type of strain, nanotube bending, has no first-order effect on the bandgap for any structure~\cite{KanePRL1997}. Numerical estimates of these effects are given in~Table~\ref{tab:bandstructuresummary}.

\subsubsection{Experiment}
\begin{figure}
\center 
\includegraphics{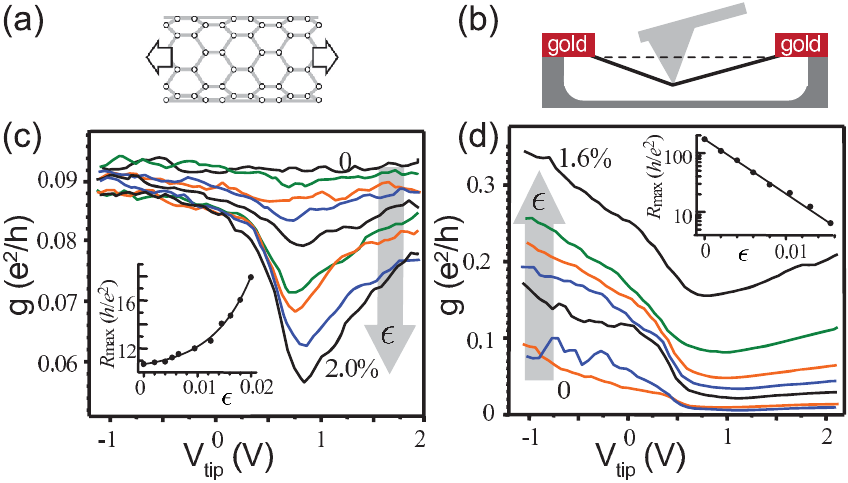}
\vspace{-0.5cm}
\caption{\footnotesize{(Color online) Contribution of strain to the bandgap. (a)~Distortion of the lattice by tensile strain. (b) The experimental setup. An AFM tip is used simultaneously to gate and tension a suspended nanotube. (c) Conductance of an initially metallic nanotube as a function of gate voltage measured for several values of strain $\epsilon$, showing increasing bandgap with $\epsilon$. Inset: Maximum resistance as a function of $\epsilon$, fitted assuming thermally activated conductance across a bandgap proportional to $\epsilon$. The fit yields the strain sensitivity $dE_\mathrm{G}/d\epsilon=35~\mathrm{meV}/\%$, where $\epsilon$ is expressed as a percentage. (d) Similar data for a semiconducting nanotube. In this case, the bandgap is found to decrease linearly with $\epsilon$, with fitted $dE_\mathrm{G}/d\epsilon=-53~\mathrm{meV}/\%$. Adapted from~\onlinecite{MinotPRL2003}.}}
\label{strain_expt}
\vspace{-0.4cm}
\end{figure}

The existence of a narrow gap in nominally metallic nanotubes was first shown in density of states measurements by scanning tunneling microscopy~\cite{OuyangScience2001}. For zig-zag nanotubes, a diameter dependence $E_\mathrm{G}=39~\mathrm{meV}/D[\mathrm{nm}]^2$ was found, in close agreement with theoretical expectations.  Isolated armchair nanotubes showed no bandgap, consistent with the expected $\cos 3\theta$ dependence (and implying negligible torsion for nanotubes lying on a surface). Many transport experiments have since found bandgaps of this order of magnitude, although usually without identifying the chirality. 

In transport measurements, quasi-metallic nanotubes typically show bandgaps a few times larger than expected from curvature alone, suggesting a significant contribution from strain. The sensitivity of the bandgap to uniaxial strain has been measured by using an AFM tip to apply tension to suspended nanotubes~\cite{MinotPRL2003}. By varying the applied force, it was possible both to induce a bandgap where none had been present before, and to decrease the bandgap in a semiconducting nanotube (Fig.~\ref{strain_expt}). From the variation of conductance with strain, it was possible to deduce $dE_\mathrm{G}/d\epsilon=+35~\mathrm{meV}/\%$ for the metallic and $dE_\mathrm{G}/d\epsilon=-53~\mathrm{meV}/\%$ for the semiconducting nanotube, both with unknown chirality, where $\epsilon$ is expressed as a percentage elongation of the nanotube. Both values are comparable with that expected from Eq.~(\ref{straineqn}), $dE_\mathrm{G}/d\epsilon=51~\mathrm{meV}/\% \times \cos 3\theta$. Similar results, including confirmation of the $\cos 3\theta$ dependence, have been obtained by optical methods~\cite{HuangPRL2008}.

\subsection{Longitudinal confinement and quantum dot energy shells}
\label{secShells}
\begin{figure}
\center 
\includegraphics{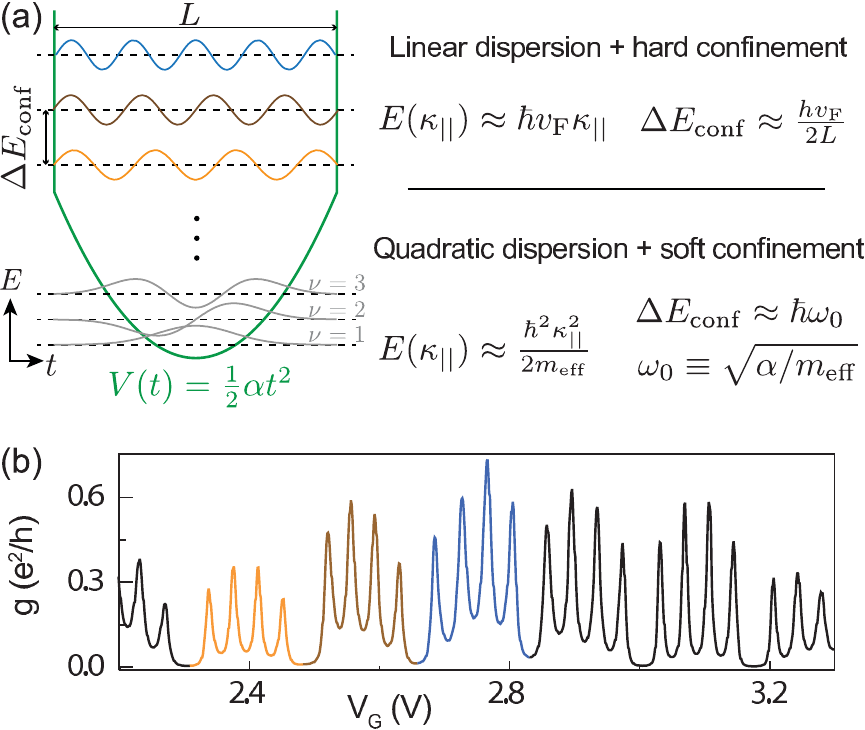}
\caption{\footnotesize{(Color online) (a) Schematic energy levels in the same quantum dot potential for two limits of dispersion and confinement discussed in the text. (Electron correlations beyond the constant interaction model, which change the potential depending on occupation, are not taken into account.) Superimposed on the level diagram are wave function envelopes $\psi(t)$ for several electron shells. Mode index $\nu$ is indicated for the first three shells. (b) Sequence of Coulomb blockade peaks measured in a quantum dot of length $L\approx 760$~nm~(adapted from~\onlinecite{SapmazPRB2005}). Peaks corresponding to three successive shells are colored to illustrate the connection to different wave functions. Although the four electrons that fill each shell occupy states of similar energy, extra energy $\Delta E$ is needed to populate a higher shell, leading to fourfold periodic peak spacing. (The connection between colored shells and particular longitudinal modes is only schematic, because absolute shell numbers cannot be deduced from this data.)}}
\label{FigConfinement}
\end{figure}

Different modes of the longitudinal wave function in a quantum dot of length~$L$ lead to different confinement energies $E_\mathrm{conf}$. The mode spectrum can be quite complicated, depending on the bandgap, boundary conditions, and interactions. The confinement can be classified as atomically sharp or non-sharp~\cite{McCannJPCM2004}, with the latter further subdivided into hard-wall or soft-wall cases depending whether the potential rises over a shorter or a longer distance than the dot length.
An additional complication arises from the fact that bound states are formed from right and left moving Dirac particles that do not necessarily have the same group velocity in the unconfined nanotube. In Fig.~\ref{fig_valleyvalley} right-movers and left-movers within a valley travel at different speeds. This effect arises away from $E_\mathrm{F}$ due to trigonal warping of flat graphene. If curvature of the nanotube is also taken into account~(not shown in Fig.~\ref{fig_valleyvalley}), an even stronger asymmetry arises. This happens already at $E_\mathrm{F}$ and directly affects how standing waves are constructed.
For example, if the confinement is sharp, then the discrete eigenstates of the quantum dot will be superpositions involving both valleys~\cite{IzumidaPRB2012}.
We mention two simple limiting cases (Fig.~\ref{FigConfinement}(a)). Electrons (or holes) with low enough energy sample only the region near the potential minimum where confinement is parabolic (``soft-wall confinement''). If the energy is also much less than $E_\mathrm{G}$, so that by Eq.~(\ref{eq_meff}) the electron behaves as a massive particle (e.g. in a sufficiently large few-electron dot), the spectrum is harmonic with mode spacing $\Delta E_\mathrm{conf}=\hbar \omega_0$, where $\omega_0$ is the harmonic frequency. Conversely, in a many electron-quantum dot the kinetic energy may both be large enough to reach the hard walls of the potential well and be in the linear part of the dispersion relation~Eq.~(\ref{eq_Ekpar}), so that the velocity is $v_\mathrm{F}$, independent of energy. The longitudinal modes then take on a sinusoidal form (Fig.~\ref{FigConfinement}(a)). The modes are again regularly spaced in energy, but now with $\Delta E_\mathrm{conf}=h v_\mathrm{F}/2L$~\cite{TansNature1997}. This regular spacing, first observed by~\cite{LiangPRL2002}, suggests that this picture is accurate in at least some real devices. If neither of these cases applies, or if the potential is strongly disordered, the mode spacing need not be regular. For example, when the electrons behave as massive particles in a hard-wall potential, the confinement energy is given by $E_\mathrm{conf}=\nu^2h^2/8m_\mathrm{eff}L^2$, with $\nu=1,2,3\ldots$.

A set of states with the same mode index $\nu$ is called a shell.  As explained in the next section, each state is characterized by twofold spin and valley quantum numbers, and thus the number of single-particle states per shell is four. In the so-called constant-interaction model (Appendix~\ref{ap_coulombspectroscopy}), the quantum dot states are filled in order of increasing energy, so that $\Delta E_\mathrm{conf}$ contributes to the Coulomb peak spacing only for every fourth electron. This is evident in the ground-state spectroscopy data of Fig.~\ref{FigConfinement}(b). The regular shell spacing shows that single-particle energy levels, in combination with the constant-interaction model, are a good approximation.

\begin{figure*}
\center 
\includegraphics[width=17.2cm]{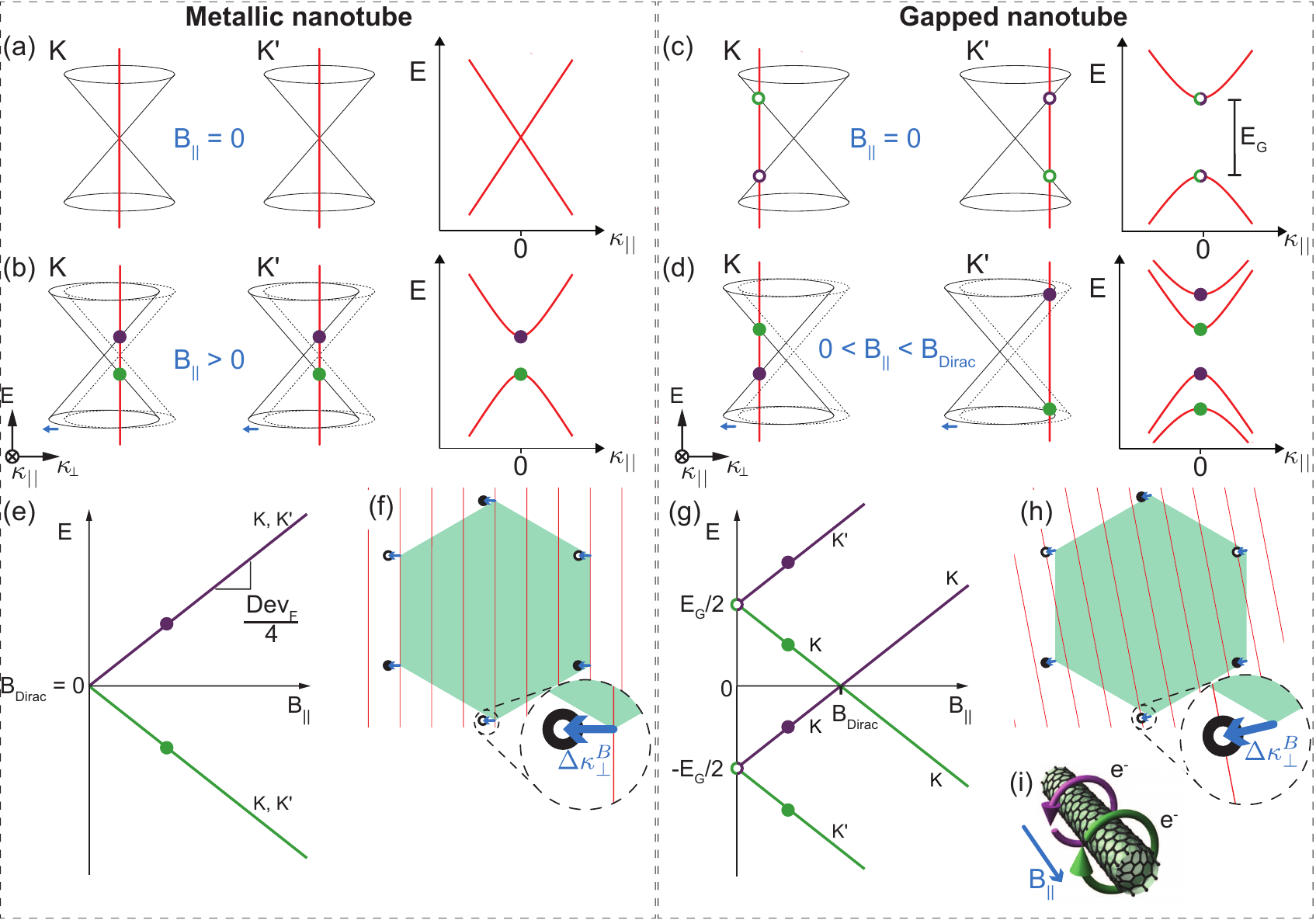}
\caption{\footnotesize{(Color online) Dependence of the band structure on parallel magnetic field. (a-b) Left: Dirac cones and quantization lines for a nanotube that is metallic at zero field, without (a) and with (b) magnetic field. Arrows in (b) mark the shift from zero-field (dotted) to finite-field (solid) Dirac cones. A field-induced horizontal shift $\Delta \kappa^B_\perp$ opens a bandgap between the conduction band (darker circles) and valence band (lighter circles). Right:  Corresponding one-dimensional electron dispersion relations. (c-d) The same plots for a nanotube with zero-field gap $E_\mathrm{G}$. A magnetic field shifts one Dirac point towards the quantization line  and one away, lifting valley degeneracy. (e) Conduction-band (darker line) and valence-band (lighter line) edges as a function of magnetic field for a metallic nanotube. In the conduction band both valleys increase in energy with field, corresponding to a negative magnetic moment $\mu_\mathrm{orb} =- Dev_\mathrm{F}/4$ or counterclockwise circulation. The valence band decreases in energy, corresponding to a positive magnetic moment and clockwise circulation. (f) Shift of the Dirac points perpendicular to the quantization lines for a zero-gap nanotube. (g) Band edges for a gapped nanotube. In the conduction band, $K(K')$ states move with positive (negative) magnetic moments. In the valence band, $K'(K)$ states move with positive (negative) magnetic moments. (h) Shift of the Dirac points for $B_{||} = B_\mathrm{Dirac}$, showing how one set of Dirac points is shifted onto the quantization lines. (i) Schematic of electron circulation directions corresponding to upmoving and downmoving states in (e) and (g).}}
\label{orbitalmoment}
\end{figure*}

\subsection{Orbital magnetic moment}
\label{SectionMuOrb}

Because each state in the $K$ valley has a time-reversed conjugate in the $K'$ valley, time-invariant perturbations such as curvature and strain do not break the degeneracy between them. This degeneracy can, however, be broken by a magnetic field. Intuitively, this can be seen by associating each state in the nanotube with a direction of circulation and hence a valley-dependent magnetic moment. This section shows how this orbital magnetic moment arises and is evident in the energy levels.


\subsubsection{Theory}
\label{sec_muorbTheory}
The orbital effect of a magnetic field $\mathbf{B}$ on an electron with charge $-e$ is captured by modifying the bare-electron Schr\"{o}dinger equation so that the momentum operator $\mathbf{p}$ is replaced by $\mathbf{p}+e\mathbf{A}$, where~$\mathbf{A}(\mathbf{r})$ is the vector potential and $\mathbf{B}=\nabla \times \mathbf{A}$~\cite{MerzbacherBook1998}. So long as $\boldA(\boldr)$ varies slowly on the scale of the lattice potential, the effect of this replacement on an electron confined in a closed loop is to add an Aharonov-Bohm phase to its eigenfunctions: if~$\psi_0(\boldr)$ is an eigenstate at $\boldA=0$, then
\begin{equation}
\psi_{\boldA}= \exp\left({ie\frac{\boldA(\boldr)\cdot \boldr}{\hbar}}\right) \psi_0
\label{eqnTransformPsi}
\end{equation}
is an eigenstate at finite $\mathbf{A}$ with the same energy~\cite{LuttingerPR1951, HofstadterPRB1976}. In other words, the finite-field dispersion relation $E_{\mathbf{A}}(\bm{\kappa})$ is related to the zero-field dispersion relation $E_{0}(\bm{\kappa})$ by:
\begin{equation}
E_{\boldA}(\bm{\kappa})=E_0(\bm{\kappa}+\bm{\Delta \kappa}_\perp^B).
\end{equation}
Here the field induced shift is $\bm{\Delta \kappa}_\perp^B = \Delta \kappa_\perp^B \mathbf{\hat{C}}$, with~\cite{AjikiJPSJ1993,LuPRL1995}
\begin{equation}
\Delta \kappa_\perp^B=\frac{eA}{\hbar}=\frac{eD}{4\hbar}B_{||}
\label{eq_kappaperpB}
\end{equation}
where $B_{||}$ is the component of $\mathbf{B}$ along the nanotube axis~$\mathbf{T}$, $A$ the corresponding component of $\mathbf{A}$, and $\mathbf{\hat{C}}$ is the direction of the chiral vector. The quantization condition, however, is unchanged.

The consequences for the band structure are shown in~Fig.~\ref{orbitalmoment}. For a true metallic nanotube (Fig.~\ref{orbitalmoment}(a-b)), the Dirac cones are shifted horizontally away from the quantization lines, opening a bandgap $E^B_\mathrm{G}=2\hbar v_\mathrm{F} |\Delta \kappa_\perp^B| = v_\mathrm{F}eD|B_{||}|/2$. If the nanotube already has a bandgap, the effect of the magnetic field is opposite for the two valleys~(Fig.~\ref{orbitalmoment}(c-d)). In one valley, the electron energy is initially reduced by $v_\mathrm{F}eDB_{||}/2$; in the other, it is increased by the same amount. At a field $B_{||}=B_\mathrm{Dirac}=E_\mathrm{G}/ev_\mathrm{F}D$, one of the Dirac cones crosses a quantization line and the bandgap vanishes. Increasing $B_{||}$ beyond $B_\mathrm{Dirac}$ causes the bandgap to increase again. For a true semiconducting nanotube, $B_\mathrm{Dirac}$ can be as large as $\sim 100$~T and is usually outside the experimental range, but for quasi-metallic nanotubes $B_\mathrm{Dirac}$ can be just a few tesla. Because the effective mass depends on bandgap~(Eq.~\ref{eq_meff}), $m_\mathrm{eff}$ can be tuned by magnetic field.

The ground-state energies are plotted in Fig.~\ref{orbitalmoment}(e-g) as a function of magnetic field. Each zero-field level is two-fold split, with slopes $dE/dB_{||}=\pm Dev_\mathrm{F}/4$. This linear splitting allows each state to be assigned a magnetic moment $\pm\mu_\mathrm{orb}$, which has a straightforward physical interpretation (Fig.~\ref{orbitalmoment}(i)): Electron states with positive (negative) magnetic moment correspond to clockwise (counterclockwise) circulation of electrons around the nanotube. In this interpretation, the direction of circulation for the first electron switches as the field is swept through $B_\mathrm{Dirac}$. A similar picture applies in the valence band. The orbital moment is related to the band structure by:
\begin{equation}
\mu_\mathrm{orb}\equiv\frac{dE}{dB_{||}}=\frac{eD}{4\hbar}\left|\frac{\partial E^\mathrm{Dirac}(\kappa_\perp,\kappa_{||})}{\partial \kappa_{\perp}}\right|,
\label{eqnPartial}
\end{equation}
where $E^\mathrm{Dirac}(\kappa_\perp,\kappa_{||})=\pm\hbar v_\mathrm{F} \sqrt{\kappa_\perp^2+\kappa_{||}^2}$ is the two-dimensional energy function describing the Dirac cone. For a low-energy electron ($\kappa_{||}\approx0$), this has the value:
\begin{equation}
\mu^0_\mathrm{orb}= Dev_\mathrm{F}/4.
\label{eqnmuorb}
\end{equation}

To emphasize the analogy with Zeeman spin splitting, an orbital $g$-factor $g_\mathrm{orb}\equiv \mu_\mathrm{orb}/\mu_B$ is sometimes defined, where $\mu_\mathrm{B}$ is the Bohr magneton. For a nanotube with a bandgap, the magnetic energy in a parallel field $B_{||}<B_\mathrm{Dirac}$ is then:
\begin{equation}
\label{eq_Emag}
E_\mathrm{mag}=\left(\mp g_\mathrm{orb}\tau + \frac{1}{2}g_\mathrm{s}s\right)\mu_\mathrm{B}B_{||},
\end{equation}
where $g_\mathrm{s}\approx 2$ is the spin $g$-factor, the $-(+)$ sign applies for electrons (holes), and the valley and spin quantum numbers are denoted by $\tau=\{+1,-1\}$ for $\{K,K'\}$ and $s=\{+1,-1\}$ for $\{\uparrow, \downarrow\}$, with the spin axis, along $t$, being parallel to the nanotube\footnote{Our convention for assigning valley labels is that conduction-band states decreasing (increasing) in energy with increasing $B_{||}$ are labelled~$K(K')$.}.

\subsubsection{Experiment}
\begin{figure}
\center 
\includegraphics{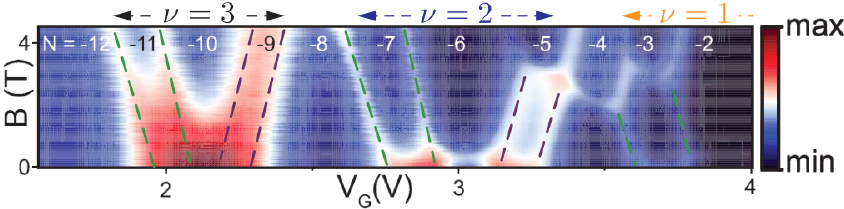}
\caption{\footnotesize{(Color online) Conductance of a nanotube quantum dot as a function of gate voltage and magnetic field, allowing ground-state spectroscopy of the first twelve hole states. (The first two peaks of the $\nu=1$ shell are not visible on this color scale.) Occupation numbers $N$ for holes 1-12 and shell numbers $\nu$ are indicated.  The orbital magnetic moment leads to a field-dependent shift of the conductance peaks, marked by dashed lines. From the slope of these lines, $\mu_\mathrm{orb}$ values of 0.90, 0.80 and 0.88~meV/T for shells 1-3 can be deduced, an order of magnitude larger than the spin magnetic moment. As well as the linear shift of peak positions, other features are seen: the complex network of lines above 2~T on the right-hand side reflects energy crossings between different shells, while the barely resolved low-field anticrossing at $V_\mathrm{G}\sim3$~V probably reflects spin-orbit coupling not recognized at the time (Sec. \ref{SectionSO}). Adapted from~\onlinecite{JarilloHerreroPRL2005}.}}
\label{magneticmomentexpt}
\end{figure}


The orbital energy splitting can be seen in Coulomb ground-state spectroscopy as a function of magnetic field (Fig.~\ref{magneticmomentexpt}), which shows the contribution $E_\mathrm{mag}$ to the single-particle energy levels $E_\mathrm{N}$~\cite{MinotNature04}. Ignoring spin-orbit coupling (to be discussed in the next section), the first four electrons fill the four lowest states in order of energy: $K{\downarrow}$, $K{\uparrow}$, $K'{\downarrow}$, $K'{\uparrow}$. Subsequent electrons must enter a higher shell of the dot, but repeat the fourfold filling sequence for spin-valley states. The expected pattern of ground state energies is therefore alternating pairs with positive and negative magnetic moments. Typical data is shown in~Fig.~\ref{magneticmomentexpt} for the first three hole shells. The measured magnetic moments, $\mu_\mathrm{orb}\sim 0.9$~meV/T, are of the expected magnitude for orbital coupling with $D\approx 4.5$~nm. In terms of orbital $g$-factors, this would correspond to $g_\mathrm{orb}\approx 16$, much larger than $g_\mathrm{s}=2$, qualitatively confirming the picture in the previous section. For a quantitative comparison with theory, an independent measurement of $D$ is necessary. This was achieved using an AFM for a nanotube with $D=2.6\pm 0.3$~nm, for which $\mu_\mathrm{orb}=0.7\pm 0.1$~meV/T was measured, in fair agreement with the value $\mu_\mathrm{orb}=0.52\pm 0.06$~meV/T expected from~Eq.~(\ref{eqnmuorb})~\cite{MinotNature04}.

\begin{figure}
\center 
\includegraphics{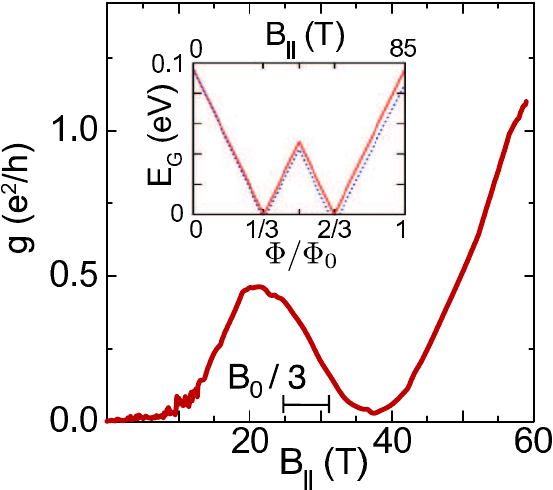}
\caption{\footnotesize{(Color online) Magnetoconductance of a semiconducting nanotube near the charge neutrality point, measured at 3.1~K. The observed maximum is slightly below the expected value (indicated by the bar labeled $B_0/3$). Inset shows calculated energy gap for a (95,15) semiconducting nanotube ($D=8.1$~nm) versus parallel magnetic field. Solid and dashed lines are without and with Zeeman effect. Adapted from~\onlinecite{JhangPRL2011}.}}
\label{fig_gapclosing}
\end{figure}

As seen from Fig.~\ref{orbitalmoment}, a nanotube that is semiconducting at zero field becomes metallic in a parallel field with magnitude $|B_{||}| = B_\mathrm{Dirac}$.  This peculiar metal-semiconductor transition is specific to the cylindrical form of nanotubes and in fact recurs periodically with every flux quantum, $\Phi_0 = h/e$ that threads the cross section. For a semiconducting nanotube, the gap closes twice per period, at flux equal to $\Phi\equiv\pi D^2B_{||}/4=\Phi_0/3$ and $\Phi=2\Phi_0/3$ where the open and filled circles respectively in Fig.~\ref{orbitalmoment}(h) cross quantization lines. The semiconducting gap reopens completely at $\Phi=\Phi_0$. Figure~\ref{fig_gapclosing} shows magnetoconductance of a semiconducting nanotube for which an AFM determined $D\approx 8$~nm. For this diameter, the expected $B_\mathrm{Dirac}$ is 27~T, which is accessible at dedicated facilities. The nanotube has low conductance at zero field when $E_\mathrm{F}$ is tuned into the gap. At 22~T the conductance is maximal, likely since the band gap is reduced to a smaller value. The band gap reopens to a maximum near 37~T before closing again as expected. The inset curve is calculated for $D=8.1$~nm and predicts gap minima at 27~T and 55~T. The observed conductance maximum, corresponding to the first gap closing, occurs at a somewhat lower field of 22 T, which is attributed to strain.

\begin{figure}[t!]
\center 
\includegraphics{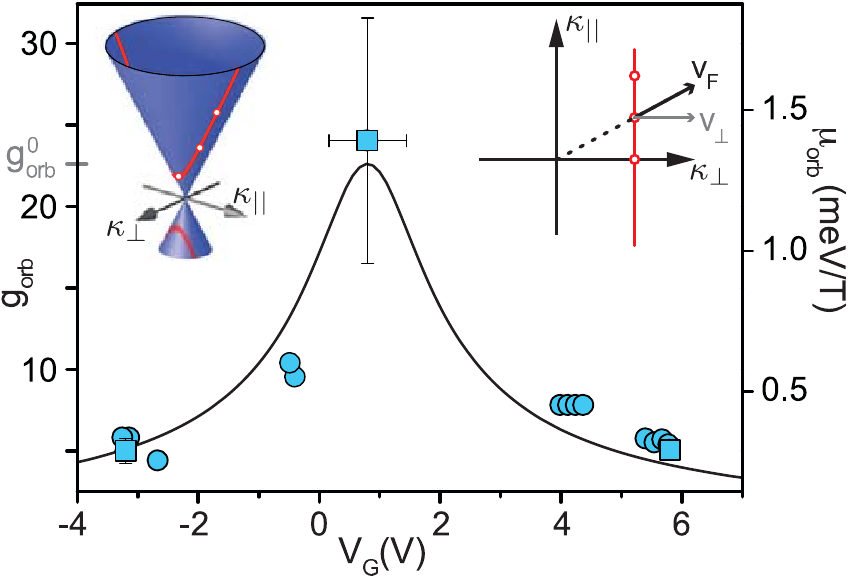}
\caption{\footnotesize{(Color online) Effect of confinement on magnetic moment. Main plot: Experimentally determined $g_\mathrm{orb}$ and~$\mu_\mathrm{orb}$ (points) as a function of gate voltage, together with a fit to~Eq.~(\ref{eqgorb})~\cite{JespersenPRL2011}. The fit assumes a constant dot length and linear dependence of $E_\mathrm{conf}$ on $V_\mathrm{G}$. Left inset: Dirac cone showing quantization line and points corresponding to~$\kappa_{||}$ values for three longitudinal energy levels. Right inset: View of the Dirac cone from above, giving a physical explanation for the reduction of~$\mu_\mathrm{orb}$. Regardless of $\kappa_{||}, \kappa_\perp$, the velocity (black arrow) of the electron is $v_\mathrm{F}$ directed away from the origin. Higher energy states, with larger $\kappa_{||}$, therefore have smaller perpendicular components~$v_\perp$ and therefore smaller $g_\mathrm{orb}$~\cite{JespersenPRL2011}.}}
\label{magneticmomentconfinement}
\end{figure}

Equation~(\ref{eqnmuorb}) assumed an electron with zero longitudinal momentum. For electrons confined in a quantum dot,~$g_\mathrm{orb}$ is reduced, for the following reason~\cite{JespersenPRL2011}. As seen from the insets of Fig.~\ref{magneticmomentconfinement}, the partial derivative in Eq.~(\ref{eqnPartial}) decreases with increasing~$\kappa_{||}$; the larger $\kappa_{||}$, the smaller the fraction of $v_\mathrm{F}$ directed around the nanotube and hence the smaller $\mu_\mathrm{orb}$.
Because of confinement the shells participating in transport consist of superpositions of states with $|\kappa_{||}| \neq 0$. (This can be seen for the sequence of shells in Fig.~\ref{FigConfinement}.) The total orbital moment therefore decreases with increasing confinement energy~(Fig.~\ref{magneticmomentconfinement}), with predicted scaling~\cite{JespersenPRL2011}:
\begin{equation}
g_\mathrm{orb}=\frac{g_\mathrm{orb}^0}{\sqrt{1+\left(\frac{2E_\mathrm{conf}}{E_\mathrm{G}^0}\right)^2}},
\label{eqgorb}
\end{equation}
where $g_\mathrm{orb}^0$ is the unconfined value derived from Eq.(\ref{eqnmuorb}),
\begin{equation}
g_\mathrm{orb}^0=\frac{ev_\mathrm{F}D}{4\mu_\mathrm{B}}
\label{eq_gorb0}
\end{equation}
and $E_\mathrm{G}^0$ is the bandgap at zero magnetic field without spin-orbit coupling. Figure~\ref{magneticmomentconfinement} shows a series of measured $g_\mathrm{orb}$ values as a quantum dot was tuned across the electron-hole transition using a gate voltage. As expected from Eq.~(\ref{eqgorb}), $g_\mathrm{orb}$ is maximized close to the transition, where electrons and holes can occupy the lowest-energy confined states, but reduced as the quantum dot occupation is increased. The data is well fit by Eq.~(\ref{eqgorb}) assuming $E_\mathrm{conf} \propto V_\mathrm{G}-0.8$~V, which is reasonable if the length of the quantum dot is independent of gate voltage.

The orbital moment deduced from Fig.~\ref{magneticmomentconfinement} corresponds by Eq.~(\ref{eq_gorb0}) to $D\approx 6$~nm. This is surprisingly large for the chemical-vapor-deposited nanotube used in the experiment, for which $D\lesssim 3$~nm is expected. Other experiments have measured a range of values for $g_\mathrm{orb}$; while some have obtained similarly large values~\cite{JarilloHerreroPRL2005, KuemmethNature2008, SteeleNcomm2013}, other results are consistent with smaller-diameter nanotubes~\cite{MinotNature04, MakarovskiPRB2007,ChurchillNPhys2009}. Although the discrepancy is not large, it is possible that expectations for either the nanotube diameter or the orbital $v_\mathrm{F}$ need to be revised. Measurements on devices with independently measured diameters would allow the discrepancy to be explored.

\subsection{Spin-orbit coupling}
\label{SectionSO}

On first consideration, it might be expected that carbon, as the second lightest of all semiconductors, should have negligible spin-orbit coupling. Indeed, spin-orbit coupling is comparatively weak in free carbon ($^3P_0-^3P_1$  splitting $\sim 2$~meV~\cite{KramidaWebsite2013}), and almost completely suppressed near the Dirac points in flat graphene~\cite{MinPRB2006,HuertasHernandoPRB2006}. However, it was realised by~\cite{AndoJPSJ2000} that the suppression relies on the symmetry of graphene. In a nanotube, this symmetry is broken by curvature, leading to a coupling up to a few~meV  between the spin and orbital moment of electrons. This coupling, first detected by~\cite{KuemmethNature2008}, is the key to controlling spins in nanotubes electrically. This section explains in detail how it arises and how it is measured.

\subsubsection{Origin of spin-orbit interaction in nanotubes}
\label{secSOtheory}

\begin{figure}
\center 
\includegraphics{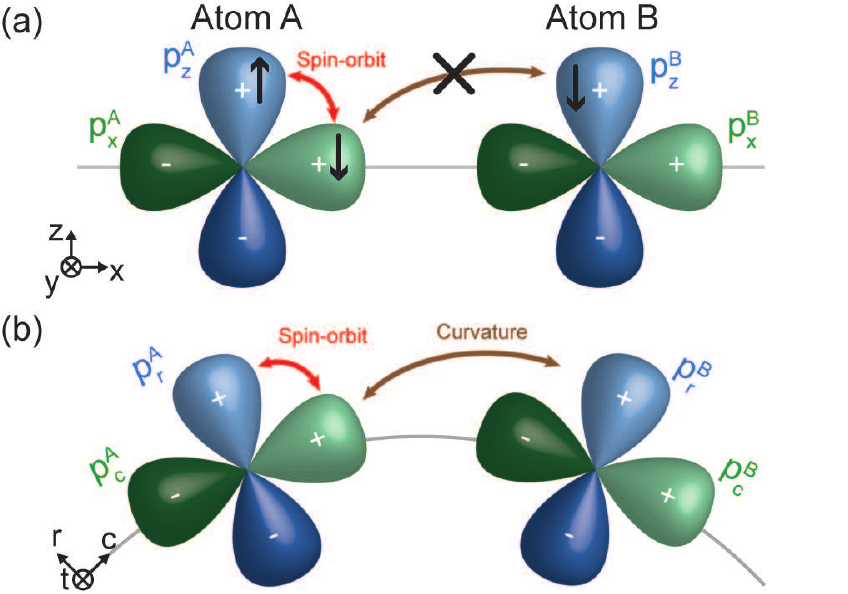}
\caption{\footnotesize{(Color online) How curvature enhances spin-orbit coupling. (a) Atomic spin-orbit coupling and interatomic hopping in flat graphene. The $p_z$ and $p_x$ orbitals of two adjacent atoms are shown, with the sign of the wave function in each lobe marked. Intra-atomic spin-orbit coupling mixes opposite-spin states involving different $p$-orbitals in the same atom, e.g.~$|p^A_z{\uparrow}\rangle$ and $|p^A_x{\downarrow}\rangle$. However, this does not mix spin states in the band structure, because hopping between different $p$-orbitals is forbidden by symmetry. (For example, the hopping contributions between $p^A_x$ and $p^B_z$ marked by a double arrow exactly cancel.) Consequently, there is no coupling between states such as $|p^A_z{\uparrow}\rangle$ and $|p^B_z{\downarrow}\rangle$ and therefore no bulk spin-orbit coupling. (b) In a nanotube, curvature breaks the up-down symmetry, meaning that direct hopping between different orbitals on adjacent atoms becomes possible. The combination of atomic spin-orbit coupling and interatomic hopping therefore mixes opposite-spin states on adjacent atoms (e.g. $|p_r^A{\uparrow}\rangle$ and $|p_r^B{\downarrow}\rangle$), leading to a spin-orbit splitting of the $\pi$ band.}}
\label{SO_fundamental}
\end{figure}

In atomic carbon, coupling between the total spin $\mathbf{S}$ and orbital angular momentum $\mathbf{L}$ adds a term to the Hamiltonian:
\begin{equation}
H_\mathrm{SO}^\mathrm{atomic}=\Delta_\mathrm{SO}^\mathrm{atomic} \mathbf{L \cdot S}
\end{equation}
where $\Delta_\mathrm{SO}^\mathrm{atomic}$ is the atomic spin-orbit strength (From the atomic $^3P_0-^3P_1$ splitting quoted above, $\Delta_\mathrm{SO}^\mathrm{atomic}\sim 4$~meV.) The effect of this coupling is to mix single-particle states with opposite spin from different orbitals, such as $|p_z{\uparrow}\rangle$ and $|p_x{\downarrow}\rangle$. Whether this leads to spin-orbit coupling in the band structure depends on how it affects hybridization between orbitals in different atoms, which in turn depends on the crystal structure.

The contrasting situations in flat and curved graphene are illustrated in Fig.~\ref{SO_fundamental}, which shows the atomic orbitals for two adjacent atoms $A$ and $B$. Any effect on the band structure arises through the combination of intra-atomic spin-orbit coupling and inter-atomic hopping. In flat graphene (Fig.~\ref{SO_fundamental}(a)), symmetry forbids direct hopping from a $p_x$ state on one atom to a $p_z$ state on another because $p_x$ and $p_z$ orbitals have opposite parity under $z$ inversion. Therefore atomic spin-orbit coupling between e.g.~$|p^A_z{\uparrow}\rangle$ and $|p^A_x{\downarrow}\rangle$ states does not introduce any non-spin-conserving hybridization between~$|p^A_z{\uparrow}\rangle$  and $|p^B_z{\downarrow}\rangle$, and thus spin-orbit coupling in the $\pi$ band is second-order and in practice negligible.

This situation is changed in the presence of curvature, which breaks the $z$ inversion symmetry on which the above suppression relies (Fig.~\ref{SO_fundamental}(b)). To understand this, it is convenient to work in the curved coordinate basis $\{r,c,t\}$ labelling radial, circumferential and axial directions, so that the $\pi$ band is composed predominantly of hybridised $p_r$ orbitals. Since the $p^A_c$ and $p^B_r$ orbitals are not orthogonal, hopping between them is allowed, leading to an indirect hybridization between $|p^A_r{\uparrow}\rangle$ and $|p^B_r{\downarrow}\rangle$ and consequently a spin-orbit coupling in the $\pi$ band.

As a result of this spin-orbit coupling, the effective hopping matrix element between $p_r^A$ and $p_r^B$ now contains both a direct and a spin-flip term. The interference between these terms causes a spin precession about the $y$-axis, and a corresponding splitting of the two spin states within a given valley as though by a magnetic field $\mathbf{B}_\mathrm{SO}$ directed along the nanotube. The spin-orbit splitting is defined as the Zeeman splitting due to this field, $\Delta_\mathrm{SO}=g_\mathrm{s}\mu_\mathrm{B} B_\mathrm{SO}$.

\begin{figure}
\center 
\includegraphics{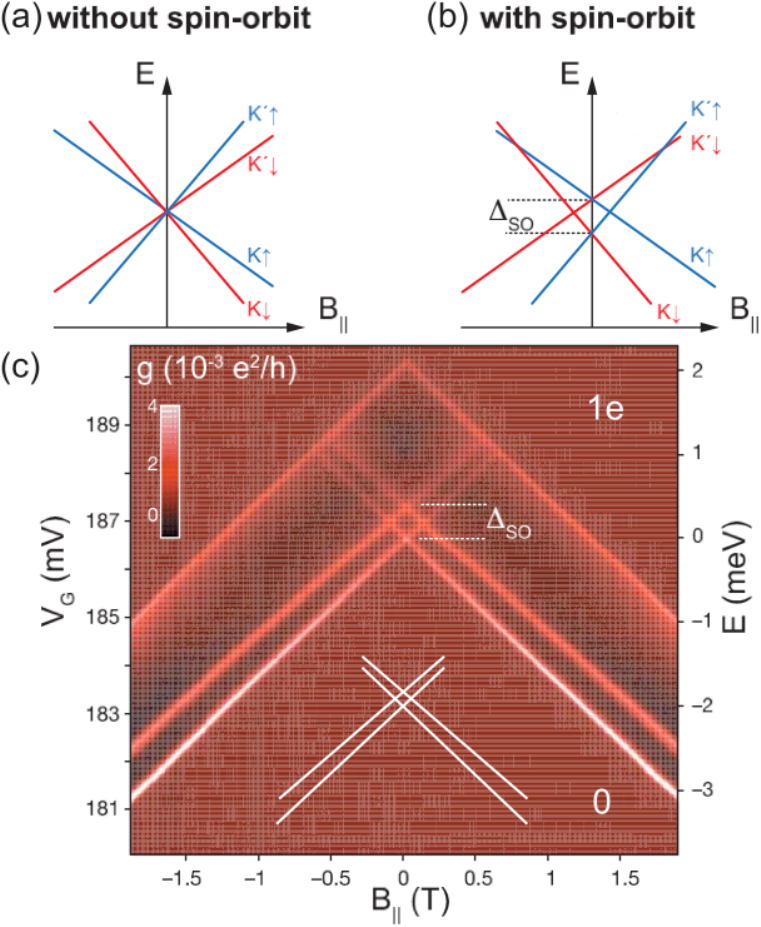}
\caption{\footnotesize{(Color online) Observation of spin-orbit coupling in a nanotube. (a,b) Expected spectra without and with spin-orbit coupling. (c) Conductance as a function of magnetic field and gate voltage across the 0-1e transition at $V_\mathrm{SD}=2$~mV, allowing high-bias spectroscopy of the lowest four one-electron states. From line slopes (inset), the spin and orbital magnetic moments can be deduced, allowing assignement of spin and valley quantum numbers. The data is clearly consistent with (b) rather than~(a). Adapted from~\onlinecite{KuemmethNature2008}.}}
\label{SO_evidence}
\end{figure}

Figure~\ref{SO_evidence}(a-b) show the consequences of spin-orbit coupling for the band edges. Without spin-orbit coupling~(Fig.~\ref{SO_evidence}(a)), the zero-field levels are four-fold degenerate, but are split in a magnetic field through a combination of Zeeman and orbital coupling. (This figure differs from Fig.~\ref{orbitalmoment}(g) by the inclusion of Zeeman spin splitting.) Spin-orbit coupling splits each four-fold degenerate level at $B=0$ into a pair of two-fold degenerate levels~(Fig.~\ref{SO_evidence}(b)); each element of the pair comprises a Kramers doublet, as required by time-reversal symmetry. The sign of~$\DeltaSO$ determines whether parallel or antiparallel alignment of spin and valley magnetic moments is favoured. For $\DeltaSO>0$, the magnetic moments of spin and valley of the lowest (highest) edge of the conduction (valence) band add, whereas they subtract for $\DeltaSO<0$. (see Fig.~\ref{SO_spectrumtheory} for examples.)
For $\Delta_\mathrm{SO}>0$, as drawn here, spin-orbit coupling favours alignment of the spin and valley magnetic moments. The lower doublet therefore comprises the states $\{K'{\uparrow}, K{\downarrow}\}$ for which both magnetic monts have the same sign, while the upper doublet comprises the states $\{K'{\downarrow}, K{\uparrow}\}$. 

\subsubsection{The discovery of nanotube spin-orbit coupling}

Figure~\ref{SO_evidence}(c) shows excited-state spectroscopy of the first electron shell of an ultraclean nanotube as a function of magnetic field~\cite{KuemmethNature2008}. The positions in gate voltage of the first four conductance peaks provide a map of the lowest four energy levels, as explained in Appendix~\ref{ap_specHighBias} (which also explains how to convert from gate voltage (left axis) to energy (right axis)). The dependence of the energy levels on magnetic field arises from the combination of valley and spin magnetic moments in each level, as in Fig.~\ref{orbitalmoment}(g), but taking account of the spin magnetic moment. From the line slopes, valley and spin quantum numbers can be assigned to each level (as in Fig.~\ref{SO_evidence}(a,b)).

The data are clearly consistent with Fig.~\ref{SO_evidence}(b) rather than Fig.~\ref{SO_evidence}(a). The key signature of spin-orbit coupling is the separation of the four spin-valley levels at zero field into two doublets. The magnitude of the splitting gives the spin-orbit coupling strength $\Delta_\mathrm{SO}=0.37$~meV, corresponding to $B_\mathrm{SO}=3.1$~T. From the spin and valley assignments deduced from the line slopes, it is clear that in this case spin-orbit coupling favors states with parallel spin and valley magnetic moments.


\subsubsection{Different types of nanotube spin-orbit coupling} 
\label{sec_SOdifferenttypes}

\begin{figure}
\center 
\includegraphics{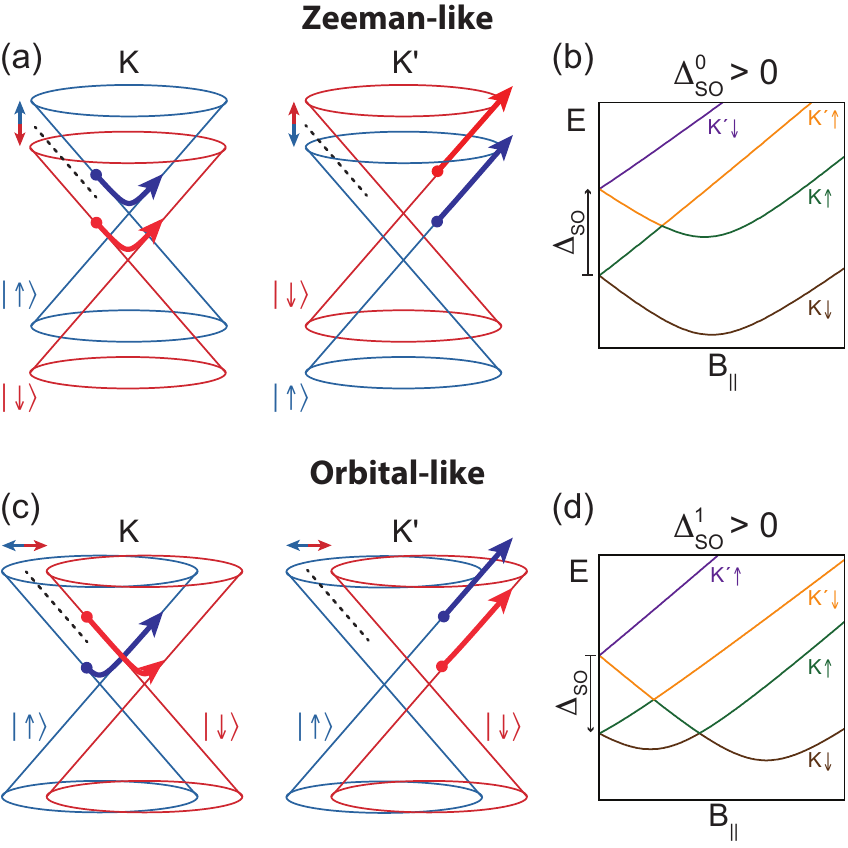}
\caption{\footnotesize{(Color online) Two types of spin-orbit coupling. (a,b) Zeeman-like coupling ~\cite{IzumidaJPSJ2009}  leads to a spin-dependent vertical shift of the band structure, equivalent to a Zeeman splitting (a) that is opposite in the two valleys.~(b) Energy levels as a function of $B_{||}$. Going beyond Fig.~\ref{SO_spectrumtheory}, a residual bandgap (induced for example by level quantization) is assumed, leading to rounding of the band minima. (c,d) Orbital-like coupling \cite{AndoJPSJ2000} leads to a horizontal shift of the band structure~(c) equivalent to a spin-dependent magnetic field coupling to $\mu_\mathrm{orb}$. (d)~Corresponding energy levels as a function of $B_{||}$. A signature to distinguish these two couplings comes from the minima in (b) and (d); whereas Zeeman-like coupling leads to minima in the first and second energy levels, orbital-like coupling leads to two minima in the first level. In (a) and (c), thick arrows indicate evolution of the state energies in the Dirac cones as $B_{||}$ is increased. Spectra (b) and (d) are calculated from~Eq.~(\ref{eq_Epmtaus}}) using $E_\mathrm{G}=4$~meV, $E_\mathrm{conf}=1$~meV, $\mu_\mathrm{orb}^0=0.9$~meV/T, $\Delta_\mathrm{SO}^0,\Delta_\mathrm{SO}^1=0$ or 2~meV.}
\label{SO_twotypes}
\end{figure}

Detailed calculations of the spin-orbit coupling reveal that there are actually two terms in the spin-orbit Hamiltonian, corresponding to Zeeman-like and orbital-like coupling~(Fig.~\ref{SO_twotypes}). The Zeeman-like contribution, characterized by a parameter $\Delta_\mathrm{SO}^0$, can be visualized as a vertical shift of the Dirac cones that is opposite for the two spin directions (Fig.~\ref{SO_twotypes}(a)). This is equivalent to an effective Zeeman shift of each spin state given by:

\begin{equation}
\Delta E^\mathrm{SO,Z}(\tau, s) = \Delta_\mathrm{SO}^0 \tau s,
\label{eqnSOZeeman}
\end{equation}
where $s$ and $\tau$ are defined at the end of Sec.~\ref{sec_muorbTheory}.

\begin{figure}
\center 
\includegraphics{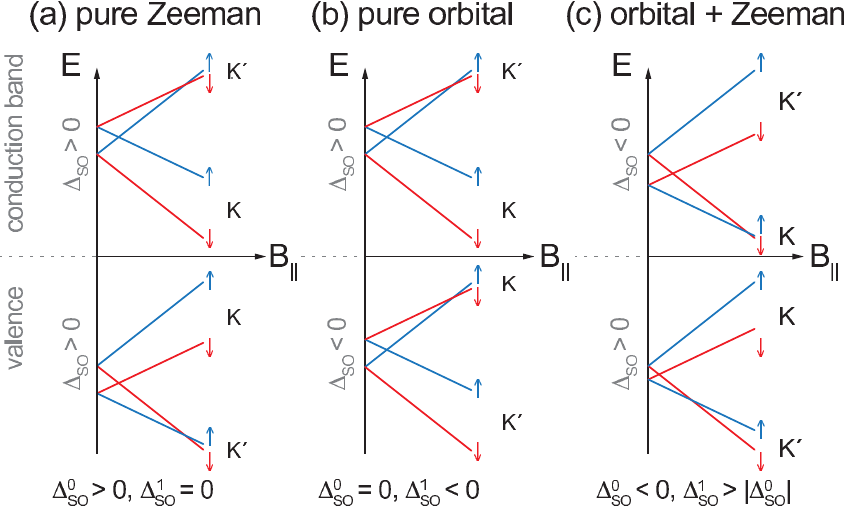}
\caption{\footnotesize{(Color online) Dependence of band edges on $B_{||}$ for different types of spin-orbit coupling for (a) Zeeman-like, (b) orbital-like  and (c) mixed spin-orbit coupling. At finite field, electron-hole symmetry is broken by orbital-like but not by Zeeman-like coupling; at zero field, a combination of both is required.}}
\label{SO_spectrumtheory}
\end{figure}

The orbital-like contribution to the Hamiltonian gives rise to a horizontal shift of the Dirac cones, similar to the curvature-induced shift discussed in Sec.~\ref{structural}, but opposite for the two spin directions~(Fig.~\ref{SO_twotypes}(c)). This is equivalent to a spin-dependent magnetic flux coupling to the orbital moment. The horizontal shift of each cone is: 
\begin{equation}
\Delta \kappa^\mathrm{SO,Orb}_\perp (s) = -s\frac{\Delta_\mathrm{SO}^1}{\hbar v_\mathrm{F}}.
\label{eq_SOOrbital}
\end{equation}
where $\Delta_\mathrm{SO}^1$ parameterizes the strength of the coupling. In this case, the hole energy levels are no longer simply the negative of the electron energy levels at finite field, and the electron-hole symmetry is broken~\cite{KuemmethNature2008}. 

Equations~(\ref{eqnSOZeeman}) and (\ref{eq_SOOrbital}) can be combined with the graphene dispersion relation Eq.~(\ref{eq_dispersionrelation}) and the magnetic field response Eq.~(\ref{eq_Emag}) to give the Hamiltonian in terms of the spin and valley quantum numbers, including spin-orbit and magnetic coupling: 
\begin{equation}
\label{eq_HSOexplicit}
\begin{split}
H=\left(\hbar \vF\left(\tau \Delta \kappa_\perp + s \frac{\Delta_\mathrm{SO}^1}{\hbar v_\mathrm{F}}\right)+\muorb^0B_{||}\right)\tau\sigma_1 
\\
+ \hbar \vF \kappa_{||} \sigma_2 + \tau s \DeltaSO^0 + \frac{\gs\mu_B}{2} \bf{s}\cdot\bf{B},
\end{split}
\end{equation}
where $\bf{s}$ is the spin operator (with eigenvalues $s=\pm 1$). Writing this Hamiltonian required the use of the  Pauli operators $\{\sigma_1, \sigma_2\}$, which act in a two-dimensional sublattice space describing the wave function amplitude on the two sublattices of Fig.~\ref{bandstructuregraphene}. These are described in more detail in Appendix~\ref{ap_isospin}. In Eq.~\ref{eq_HSOexplicit}, the horizontal and vertical shifts due to the two spin-orbit terms are evident.

The existence of two forms of spin-orbit coupling goes beyond the simple picture of Sec.~\ref{secSOtheory}. The orbital-like contribution can be understood as a Rashba-type coupling, arising from the broken reflection symmetry about the graphene plane. A curvature-induced displacement of the orbitals gives rise to a radial electric field, which circulating electrons experience as a magnetic field proportional to the azimuthal component of their momentum. It is similar to the Rashba coupling predicted by~\cite{KanePRL2005} for graphene in an electric field, and is equivalent to a horizontal shift of the dispersion relation of the form in Eq.~\eqref{eq_SOOrbital}.

\begin{table}
\vspace{-0.2cm}
\centering
\begin{tabular}{llrrl}
\hfill \hspace{0cm}&\hspace{0cm}&\hspace{0.8cm}&\hspace{0.8cm}&\hspace{0cm} 	\\
\hline \hline
Reference \hfill 					& $D$ \hfill		& \multicolumn{1}{c}{$\Delta_\mathrm{SO}^\mathrm{min}$}& \multicolumn{1}{c}{$\Delta_\mathrm{SO}^\mathrm{max}$} & \multicolumn{1}{c}{$\Delta_\mathrm{SO}$} \\
	        						&					&\multicolumn{2}{c}{(theory)}						&  \multicolumn{1}{c}{(expt)}			\\
									& nm				& \multicolumn{1}{c}{$\mu$eV} 						& \multicolumn{1}{c}{$\mu$eV} &
									\multicolumn{1}{c}{$\mu$eV}\\
\hline	\vspace{-0.2cm} \\
\cite{KuemmethNature2008}			&	5.0									& -120\hspace{0.2cm}	& 120\hspace{0.2cm}	&	370 (1e)			\\
									&										& -240\hspace{0.2cm}	& 0\hspace{0.2cm}	&	-210 (1h)			\\
\cite{ChurchillPRL2009}				&	1.5									& -400\hspace{0.2cm}	& 400\hspace{0.2cm}	&	170 (1e)			\\
\cite{JhangPRB2010}					&	1.5\footnote{AFM measurement}		& -800\hspace{0.2cm}	& 400\hspace{0.2cm}	&	$\pm2500$\footnote{Inferred from the magnetoresistance of an open CNT quantum wire.}		\\
\cite{JespersenNphys2011}			&	2.9\footnote{\label{note1}Derived from many-carrier regime but ignoring suppression of~$g_\mathrm{orb}$ by confinement~(Eq.~\eqref{eqgorb}). These values of $D$ are therefore likely to be underestimates.}															& -210\hspace{0.2cm}	& 210\hspace{0.2cm}	&	150 (many e)		\\
									&										& -410\hspace{0.2cm}	& 0\hspace{0.2cm}	&	75 (many h)		\\
\cite{LaiArXiv2012}		 			&	$1.3^\mathrm{c}$					& -920\hspace{0.2cm}	& 460\hspace{0.2cm}	&	$\pm220$ 			\\
									&										&						& 
									&(many e/h)	\\
\cite{SteeleNcomm2013}\\
\hspace{1cm}Device 1				&	7.2									& -80\hspace{0.2cm}		& 80\hspace{0.2cm}	&	3400 (1e)			\\
\hspace{1cm}Device 2				&	6.8									& -90\hspace{0.2cm}		& 90\hspace{0.2cm}	&	1500 (1e)			\\
\hspace{1cm}Device 3				&	4.1									& -150\hspace{0.2cm}	& 150\hspace{0.2cm}	&	-1700 (1e)\\
									&	3.7									& 0\hspace{0.2cm}		& -320\hspace{0.2cm}	&	
1300 (1h)			\\
\cite{CleuziouPRL2013}\\
\hspace{1cm}Device 1    			&      									&          	&	& -240     \\
									&										&			&			& (h, few shells)\\
\hspace{1cm}Device 2 &      &            		&&    -340 (1e,3e) \\
\cite{Schmid2013}				&			&			&&	-350 (17e)			\\
\vspace{-0.2cm} \\
\hline \hline
\end{tabular}
\caption{Summary of spin-orbit parameters measured in literature, compared with the largest theoretical values predicted by Eq.~(\ref{eq_SOtotal}). Unless otherwise noted, $D$ is measured from the orbital magnetic moment according to~Eq.~(\ref{eqgorb}). The spread of $\Delta_\mathrm{SO}$ values is large, but in many cases the measured values  considerably exceed the largest possible predictions or have the wrong sign. Other measurements (eg.~\onlinecite{PeiNnano12}) on the same devices are not included. Also not included are earlier measurements subsequently reinterpreted as indicating spin-orbit interaction with $\Delta_\mathrm{SO}\sim\Delta_{KK'}$~\cite{JarilloHerreroPRL2005,MakarovskiPRB2007}.}
\label{tab_SOparameters}
\end{table}

The Zeeman-like contribution also comes from a lack of reflection symmetry through the nanotube surface. In contrast to the orbital-like contribution, which is caused by the homogeneous part of the radial electric field, the Zeeman-like contribution comes from variation of the electric field within the graphene unit cell. In a tight-binding picture, this contribution can be thought of as curvature-induced spin-orbit scattering between next-nearest neighbours (e.g.\ from one $A$ site to another), whereas the orbital-like contribution comes from scattering between nearest neighbors (between an $A$ and a $B$ site). The perturbation theory leading to these separate effects is outlined in Appendix~\ref{ap_SO}.

The values of the coefficients  $\Delta_\mathrm{SO}^0$ and $\Delta_\mathrm{SO}^1$ depend on the structure of the nanotube. The theoretical values are estimated as:
\begin{eqnarray}
\label{eq_SOparam1}
\Delta_\mathrm{SO}^0 &\approx& -\frac{0.3 \mathrm{meV}}{D~[\mathrm{nm}]}\cos 3 \theta \hspace{0.6cm} \mbox{(Zeeman-like)}\\
\label{eq_SOparam2}
\Delta_\mathrm{SO}^1 &\approx& -\frac{0.3 \mathrm{meV}}{D~[\mathrm{nm}]} \hspace{1.6cm}\mbox{(Orbital-like)}.
\end{eqnarray}

These values are quite sensitive to the method of computation. The first calculations~(\cite{AndoJPSJ2000}, later refined by~\cite{YanikIEEE2004,HuertasHernandoPRB2006}) considered the modification of hopping amplitudes by atomic spin-orbit coupling which gives rise to the orbital-like contribution. Later work~\cite{IzumidaJPSJ2009} calculated the spin-orbit correction in more detail using a non-orthogonal tight-binding calculation that incorporated the spin degree of freedom and used four orbital states per atom. This work was the first to predict the Zeeman-like contribution. The parameters~$\Delta_\mathrm{SO}^0$ and $\Delta_\mathrm{SO}^1$ were estimated by~\cite{IzumidaJPSJ2009} in two ways: by nearest-neighbour tight-binding using density-functional theory potentials, and by fitting to a full numerical model, from which (Eqs. \ref{eq_SOparam1}-\ref{eq_SOparam2}) are taken. They have also been calculated by an extended tight-binding Slater-Koster method~\cite{ChicoPRB2009} and using density functional theory combined with atomic spin-orbit coupling and tight binding~\cite{ZhouPRB2009}. A similar calculation to \cite{IzumidaJPSJ2009} was performed by \cite{JeongPRB2009}, corroborating these results. The coefficients have also been estimated by~\cite{KlinovajaPRB2011}, who included the effects of external electric fields. These different methods differ quantitatively by a factor up to $\sim 3$, but the~$\cos 3\theta$ dependence is dictated by symmetry.


Combining Eqs.~(\ref{eq_Ekpar},\ref{eq_kappaperpB},\ref{eqnSOZeeman},\ref{eq_SOOrbital}) shows that both spin and valley are good quantum numbers in a magnetic field $B_{||}$ directed along the nanotube. The corresponding eigenenergies~\cite{JespersenNphys2011} are, again assuming a flat potential as in Eq.~\eqref{eqgorb}:
\begin{equation}
\begin{split}
E_{\tau,s}^\pm(B_{||})=\pm\sqrt{\left(-\tau\frac{E^0_\mathrm{G}}{2} + \mu^0_\mathrm{orb}B_{||}+ s \Delta_\mathrm{SO}^1\right)^2+E_\mathrm{conf}^2}
\\
+s\tau\Delta_\mathrm{SO}^0+\frac{sg_\mathrm{s}\mu_\mathrm{B}{B_{||}}}{2},\hspace{2cm}
\end{split}
\label{eq_Epmtaus}
\end{equation}
where $E^0_\mathrm{G}$ is the bandgap at zero field without spin-orbit coupling and the upper (lower) sign refers to the conduction (valence) band. In the limit $E^0_\mathrm{G} \gg \muorb^0|B_{||}|, |\Delta_\mathrm{SO}^1|$, this gives for the combined zero-field splitting: 
\begin{equation}
\Delta_\mathrm{SO}=2\left(\Delta_\mathrm{SO}^0 \mp \Delta_\mathrm{SO}^1 \frac{g_\mathrm{orb}}{g_\mathrm{orb}^0}\right).
\label{eq_SOtotal}
\end{equation}
The sign of $\Delta_\mathrm{SO}$ can be deduced from the spectrum as follows~(Fig.~\ref{SO_spectrumtheory}): If the two $K'$ states converge with increasing field and eventually cross (as in Fig.~\ref{SO_evidence}), then $\Delta_\mathrm{SO}>0$; if it is the $K$  states that cross, then $\Delta_\mathrm{SO}<0$~\cite{BulaevPRB2008,KuemmethNature2008}. 

\begin{figure}
\center 
\includegraphics{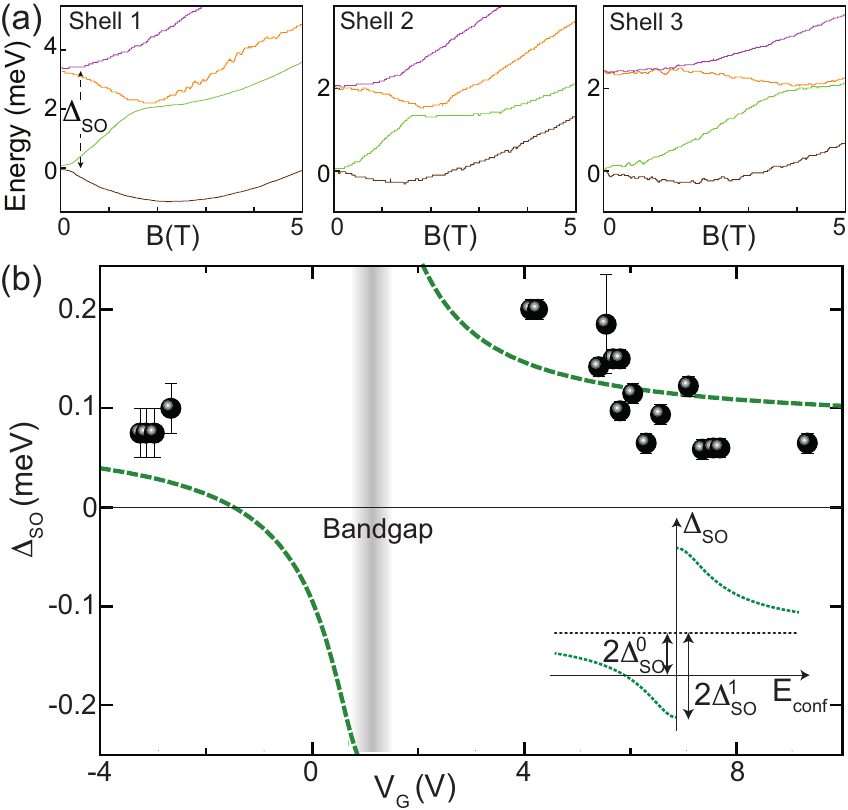}
\caption{\footnotesize{(Color online) Experimentally distinguishing orbital-type and Zeeman-type spin-orbit coupling contributions. (a) Measured magnetic field dependence of the first twelve electron energy levels, split across three shells, in a narrow-gap nanotube, derived from low-bias spectroscopy. (The vertical offsets of each trace, which this experimental technique leaves undetermined, have been chosen so that adjacent levels align.) The levels correspond more closely to Fig.~\ref{SO_twotypes}(b) than Fig.~\ref{SO_twotypes}(d) (and the colors have been chosen accordingly), indicating predominantly Zeeman-like coupling in this device. (b) $\Delta_\mathrm{SO}$ as a function of $V_\mathrm{G}$ in a separate device. The dashed lines are fits to theory taking account of the different dependence of the two contributions on orbital moment. Inset: Expected dependence of $\Delta_\mathrm{SO}$ on $E_\mathrm{conf}$, showing the strengths of the two contributions. $\Delta^0_\mathrm{SO}$ and $\Delta^1_\mathrm{SO}$ parameterize the Zeeman-like and orbital-like contributions respectively. Adapted from~\onlinecite{SteeleNcomm2013,JespersenNphys2011}.}}
\label{SO_twotypesExpt}
\end{figure}

From Eqs.~(\ref{eq_SOparam1},\ref{eq_SOparam2},\ref{eq_SOtotal}), four predictions can be derived:
\begin{enumerate}
\item{Spin-orbit coupling depends on chirality and diameter, hence different devices should display different coupling.}
\item{The different terms lead to different behaviour when fields comparable to  $B_\mathrm{Dirac}$ are applied (see Fig.~\ref{orbitalmoment}). This can be seen by plotting the first four energy levels as a function of $B_{||}$ across the Dirac point. For orbital coupling (and assuming a residual bandgap remains), the lowest-energy level has a pair of minima~(Fig.~\ref{SO_twotypes}(d)), corresponding to the Dirac points crossing the quantization lines for spin up and spin down. For Zeeman-like coupling, the two minima occur in the first and second energy levels, as in  (Fig.~\ref{SO_twotypes}(b)).}
\item{The orbital-like term contributes with opposite sign to $\Delta_\mathrm{SO}$ for electrons and holes, thereby breaking electron-hole symmetry (because the hole energy levels are no longer a mirror image of the electron levels.) The Zeeman-like term by itself preserves electron-hole symmetry~(Fig.~\ref{SO_spectrumtheory}(b)).}
\item{The orbital contribution leads to a smaller energy shift for higher energy shells in the same way that an orbital magnetic field does (Sec.~\ref{SectionMuOrb}). Thus $\Delta_\mathrm{SO}$ depends on density in the same way as~$g_\mathrm{orb}$~(Eq.~(\ref{eq_SOtotal}).}
\end{enumerate}

Evidence for prediction (1) comes from measurements on several devices, presumably with a distribution of structures.  The spin-orbit parameters are indeed found to take different values (Table~\ref{tab_SOparameters}), although it has not yet been possible to determine chiral indices in the same device for comparison. The spread of $\Delta_\mathrm{SO}$ is in fact larger than expected from Eqs.~(\ref{eq_SOparam1}-\ref{eq_SOparam2}), as discussed in~Sec.~\ref{sec_BandstructureOpenQuestions}.

Testing prediction (2) experimentally requires a device where $B_\mathrm{Dirac}$ lies at an accessible field. Figure~\ref{SO_twotypesExpt}(a) shows ground-state spectroscopy of the first four Coulomb peaks in a device where this was achieved~\cite{SteeleNcomm2013}. The pattern of energy levels clearly resembles~Fig.~\ref{SO_twotypes}(b) more than Fig.~\ref{SO_twotypes}(d), suggesting that the coupling in this device is predominantly Zeeman-like.

\begin{figure}
\center 
\includegraphics{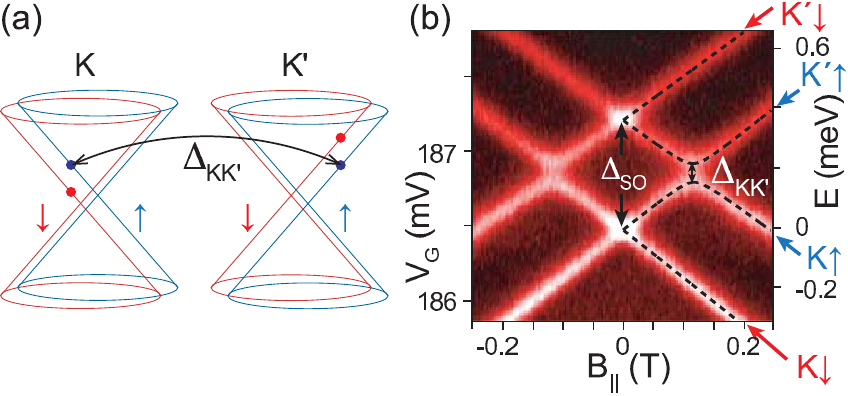}
\caption{\footnotesize{(Color online) (a) Example of a pair of states mixed by intervalley scattering, which couples states with equal spin in opposite valleys. (b) Zoom-in of Fig.~\ref{SO_evidence}. Intervalley scattering is manifest as an anticrossing between $|K{\uparrow}\rangle$ and $|K'{\uparrow}\rangle$ levels, with strength  $\Delta_{KK'}$. Adapted from~\onlinecite{KuemmethNature2008}.}}
\label{SO_intervalley}
\end{figure}

\begin{figure*}
\center 
\includegraphics{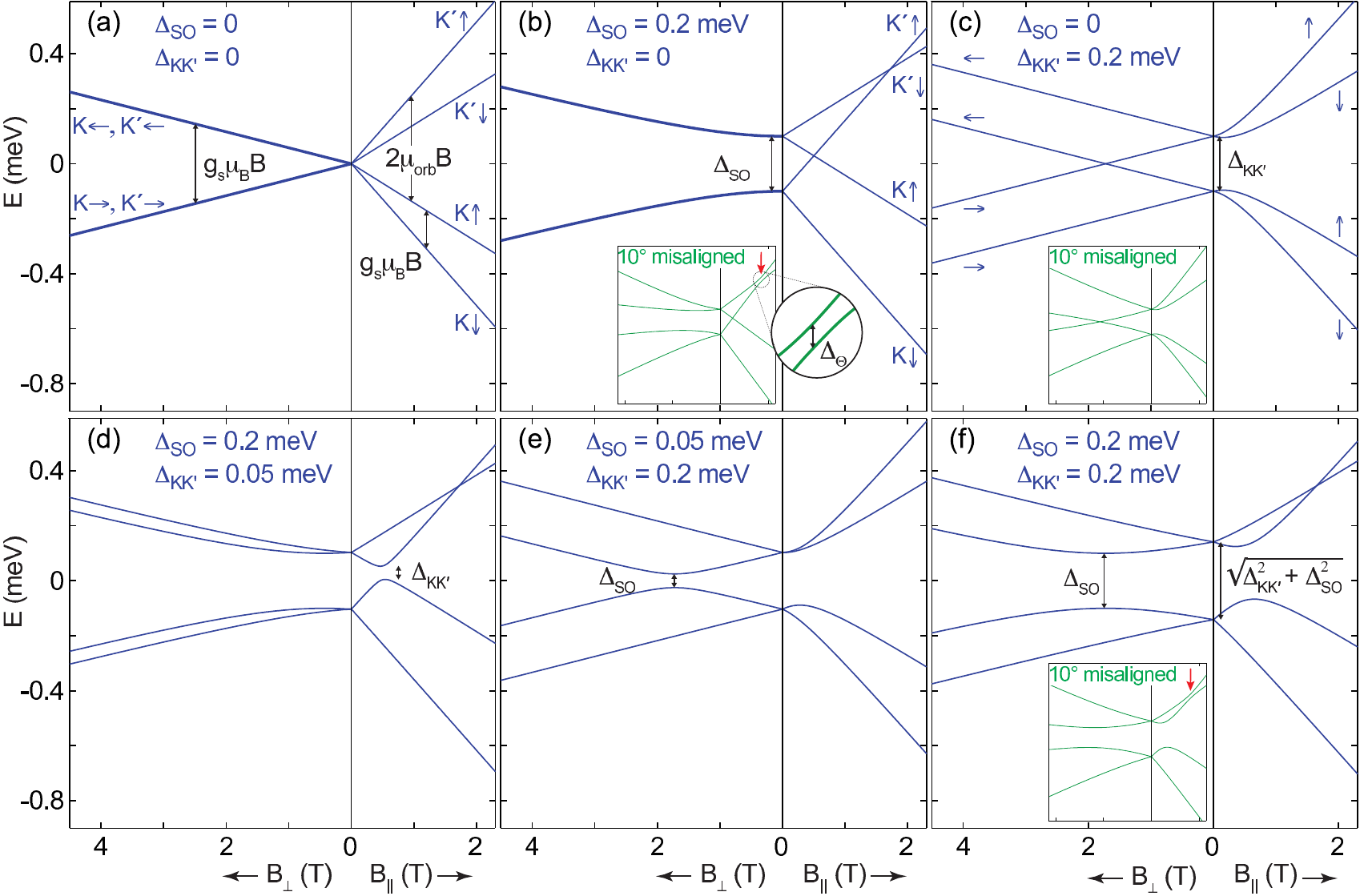}
\caption{\footnotesize{(Color online) (a-f) Single-particle energy levels plotted as a function of magnetic field in perpendicular and parallel directions for various strengths of spin-orbit coupling and intervalley scattering, taking $\mu_\mathrm{orb}=0.2$~meV/T. (a) No spin-orbit or intervalley scattering; (b) Spin-orbit only; (c) Intervalley scattering only; (d-f) combined spin-orbit and intervalley scattering. In selected panels, insets show the same spectra for field axes misaligned by $\Theta=10^\circ$ from the nanotube; this introduces an anticrossing (marked by arrows) between the two highest levels. Selected energy splittings mentioned in the text are marked. Double-thickness lines indicate degenerate levels.}}
\label{SO_combined}
\end{figure*}

Prediction (3) was confirmed in the first measurements by~\cite{KuemmethNature2008}, who found $\Delta_\mathrm{SO}$ to be different in both magnitude and sign for the first electron and first hole. Further confirmation, and a test of prediction~(4), was obtained in a different device where~$\Delta_\mathrm{SO}$ could be measured across several shells of electrons and holes~\cite{JespersenNphys2011}. As seen in~Fig.~\ref{SO_twotypesExpt}(b),~$\Delta_\mathrm{SO}$ decreases with higher $|E_\mathrm{conf}|$, qualitatively consistent with Eq.~(\ref{eq_SOtotal}) assuming an orbital contribution. For the shells measured in this device,~$\Delta_\mathrm{SO}$ did not change sign between electron and holes, providing further evidence of a Zeeman-type contribution.

\subsubsection{Uniform electric fields}
\label{sec_Rashba}
As well as intrinsic spin-orbit coupling from the nanotube structure, there is also predicted to be an extrinsic coupling due to electric fields~\cite{KlinovajaPRB2011}. This is a form of Rashba effect, and leads to a shift of the Dirac cones in $\kappa_{||}$ by an amount
\begin{equation}
\Delta \kappa_{||}^\mathrm{SO, R}=\frac{e E \xi}{\hbar v_\mathrm{F}}\tau s_\perp,
\end{equation}
where $E$ is the electric field (perpendicular to the nanotube) and $s_\perp$ is the spin component perpendicular both to the nanotube and to $E$. The parameter $\xi$, which governs the strength of this effect, is uncertain because it depends on several numerically calculated band structure parameters, but is estimated as $\xi\simeq 2 \times 10^{-5}$~nm~\cite{KlinovajaPRB2011}. This Rashba-like coupling has not yet been observed, but in principle allows for all-electrical spin manipulation~\cite{BulaevPRB2008, KlinovajaPRL2011}.

\begin{figure*}
\center 
\includegraphics{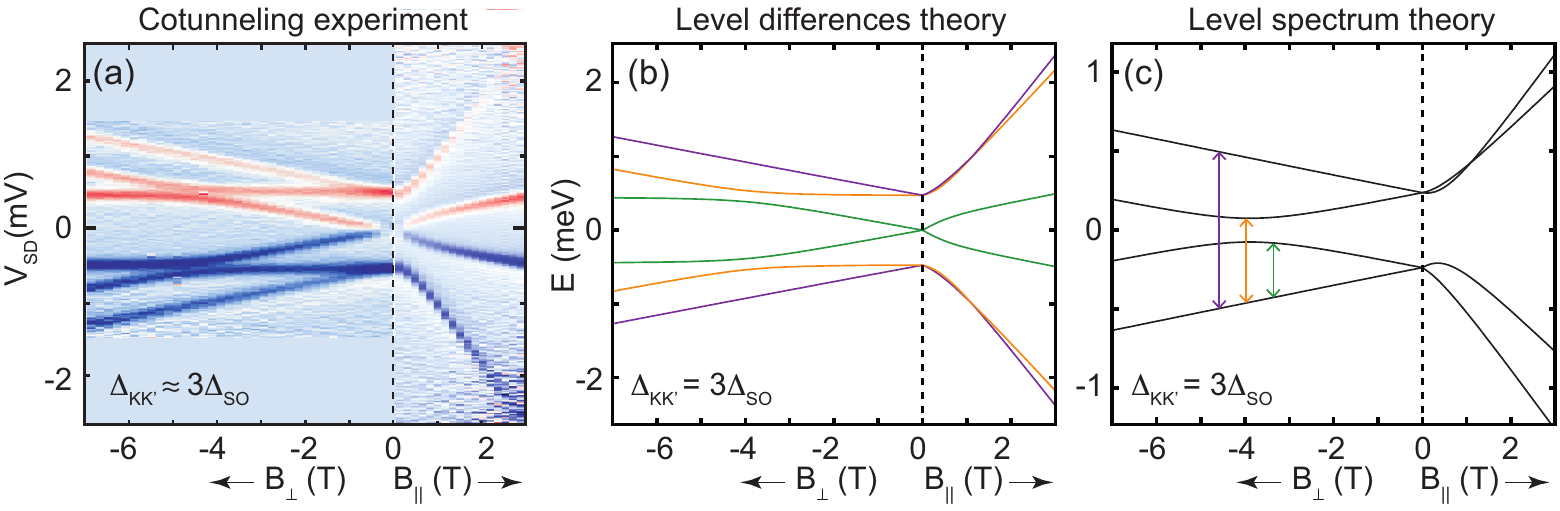}
\caption{\footnotesize{(Color online) Energy levels for comparable spin-orbit coupling and intervalley scattering. (a) Measurements of the four states of an electron shell by cotunneling spectroscopy (see text). The second derivative $d^2I/dV_\mathrm{SD}^2$ is plotted as a function of $V_\mathrm{SD}$ and magnetic field. (b) Calculated transition energies to the first three excited states assuming $\Delta_{KK'}=3\Delta_\mathrm{SO}$, showing good agreement with the peaks/dips in (a). (c) Energy levels corresponding to the transitions (marked with arrows) in (b). Adapted from~\onlinecite{JespersenNphys2011}.}}
\label{SO_combined_data}
\end{figure*}

\subsection{ Intervalley scattering}

In quantum dots (\ie confined states), magnetospectroscopy as in~Fig.~\ref{SO_evidence} allows measurement of a phenomenological parameter $\Delta_{KK'}$ that governs the strength of avoided level crossings between opposite valley states. This was observed by~\cite{KuemmethNature2008} in a suspended one-electron quantum dot for which $\Delta_{KK'}=65\mathrm{~\mu eV}$ was found (Fig.~\ref{SO_intervalley}(b)). Although this splitting indicates scattering between states in different valleys with the same spin, it does not reveal a specific mechanism.
Experiments performed on top-contacted nanotubes showed $\Delta_{KK'}$ as small as $25\mathrm{~\mu eV}$ \cite{ChurchillPRL2009}, and  \onlinecite{JespersenNphys2011} reported significant fluctuations within a device $\Delta_{KK'}=75-700\mathrm{~\mu eV}$ with no obvious correlation to gate voltage or occupation number. \onlinecite{Grove-RasmussenPRL2012} investigated the suppression of $K-K'$ mixing by application of a parallel magnetic field.
Although $\Delta_{KK'}$ has been used in other works as a empirical fitting parameter \cite{PeiNnano12,LaiArXiv2012}, the microscopic origin has not been investigated. In particular, all experiments involved finite tunneling to the source or drain electrode, and hence the intrinsic valley coupling in closed quantum dots has yet to be measured.

Historically, we suspect that electrical disorder on the scale of the interatomic spacing, leading to comparatively large $\DeltaKK$, was the main reason for the decade-long delay between the first nanotube quantum dots and the discovery of spin-orbit coupling. As evident below, $\Delta_{KK'}$ values larger than $\Delta_\mathrm{SO}$ obscure the signatures of spin-orbit coupling. It was only the development of low-disorder fabrication techniques that allowed such a delicate effect to be identified.

\subsubsection{Putting it all together}

Figure~\ref{SO_combined} shows calculated spectra in a single electron shell as a function of magnetic field in both perpendicular and parallel directions for a range of values of $\Delta_\mathrm{SO}$ and $\Delta_{KK'}$. This calculation proceeds by first working out the eigenenergies $E^\pm_{\tau,s}$ for a magnetic field directed along the nanotube and in the absence of disorder. Under the assumption that there is no mixing of electron and hole shells, \ie in the limit
\begin{equation}
E_\mathrm{G}^0\gg |\Delta_\mathrm{SO}^1|, \mu_\mathrm{orb}^0 |B_{||}|, 
\end{equation}
Eq.~(\ref{eq_Epmtaus}) reduces to:
\begin{equation}
E_{\tau,s}^\pm \approx E_0^\pm + s\tau \frac{\Delta_\mathrm{SO}}{2} + \left(\mp\tau g_\mathrm{orb}+\frac{1}{2}s g_\mathrm{s}\right) \mu_\mathrm{B}B_{||},
\end{equation}
where $E_0^\pm=\pm\sqrt{(E_\mathrm{G}^0)^2/4+E_\mathrm{conf}^2}$ is the energy without magnetic field or spin-orbit coupling.

Introducing disorder or tilting the field relative to the nanotube axis by angle~$\Theta$ (so that $B_{||}=B \cos \Theta$) mixes the eigenstates. In the basis $(K{\uparrow}, K'{\downarrow}, K{\downarrow}, K'{\uparrow})$, the Hamiltonian is:
\begin{equation}
\begin{split}
H = \left( 
\begin{array}{cccc}
E^\pm_{1,1} 	& 0 			& 0 			& \Delta_{KK'}/2 \\
0 				& E^\pm_{-1,-1}	& \Delta_{KK'}/2	& 0				\\
0				&\Delta_{KK'}/2	&	E^\pm_{1,-1}& 0				\\
\Delta_{KK'}/2	& 0				& 0				& E^\pm_{-1,1}	\\
\end{array}
\right)\hspace{1.2cm}
\\
\hspace{0.2cm}
+ \frac{g_\mathrm{s} \mu_\mathrm{B} B}{2}
\left(
\begin{array}{cccc}
0			 	& 0 			& \sin{\Theta}	& 0 			\\
0 				& 0				& 0				& \sin{\Theta}	\\
\sin{\Theta}	& 0				& 0				& 0				\\
0				& \sin{\Theta}	& 0				& 0				\\
\end{array}
\right).
\end{split}
\label{eq_SPH}
\end{equation}
Figure~\ref{SO_combined} shows the numerically calculated eigenstates of this Hamiltonian. With no spin-orbit or disorder (Fig.~\ref{SO_combined}(a)), the four states are degenerate at zero field, being split through a combination of orbital and Zeeman coupling. Pure spin-orbit coupling (Fig.~\ref{SO_combined}(b)) splits the zero-field quadruplet into two doublets; note that in small perpendicular field, Zeeman coupling is ineffective, because the spin states are locked to the valley, and valleys are not coupled. Pure disorder~(Fig.~\ref{SO_combined}(c)) suppresses orbital coupling at low field, but preserves the Zeeman splitting in both field directions.

\begin{table*}
   \centering
   \begin{tabular}{llll} 
   \hspace{4.8cm}	&   \hspace{4cm}	&   \hspace{5cm}	&   \hspace{3.5cm}	\\
   \hline \hline
     Quantity \hfill & Expression \hfill& Value \hfill & Reference \\	\hline
      	Confinement bandgap    	&  $4 \hbar v_\mathrm{F}/3D$ & 0.70~eV/$D$~[nm]  & \cite{CharlierRMP2007}\\
     \vspace{-0.2cm} \\
      	Curvature bandgap    	& $\sim\frac{\hbar v_\mathrm{F} a_\mathrm{CC}}{2D^2} \cos 3 \theta$ & $\sim50~ \mathrm{meV}/(D~[\mathrm{nm}])^2\times \cos 3 \theta$ & \cite{IzumidaJPSJ2009}\\
     \vspace{-0.2cm} \\
      	Strain bandgap    	& $\frac{2\hbar v_\mathrm{F}}{a_\mathrm{CC}}\frac{12\zeta}{1+6\zeta}(1+\lambda)\epsilon \cos 3 \theta$
      								&$51~\mathrm{meV} \times \epsilon~[\%] \times \cos 3 \theta$ & \cite{HuangPRL2008}\\
     \vspace{-0.2cm} \\
      	Torsion bandgap    	& $2\hbar v_\mathrm{F} \gamma \sin 3 \theta$  
      								& $0.018~\mathrm{meV}\times \gamma~[^\circ/\mu\mathrm{m}] \times \sin 3 \theta$ & \cite{YangPRL2000} \\
\vspace{-0.2cm} \\
      	Effective mass		& $m_\mathrm{eff} = E_\mathrm{G}/2v_\mathrm{F}^2.$ & $0.14 m_\mathrm{e} \times E_\mathrm{G}~[\mathrm{eV}]$
     \vspace{0.4cm} \\
       	Orbital magnetic moment    	& $\mu^0_\mathrm{orb}= Dev_\mathrm{F}/4$  & 0.20 meV/T $\times D$~[nm] & \cite{AjikiJPSJ1993}\\
       								& $g^0_\mathrm{orb}\equiv\mu^0_\mathrm{orb}/\mu_\mathrm{B}$										&$3.5\times D$~[nm]\\
     \vspace{-0.2cm} \\
         Zeeman magnetic moment 	& $\frac{1}{2}g_\mathrm{s} \mu_\mathrm{B}$  & 58~$\mu$eV/T& \\
     \vspace{0.4cm} \\
       	Spin-orbit coupling		& $\tau s(\Delta^0_\mathrm{SO} + \Delta^1_\mathrm{SO} \sigma_1)$  &  $\Delta^0_\mathrm{SO} \approx \frac{-0.3 \mathrm{~meV}}{D[\mathrm{nm}]}\cos 3 \theta$	& \cite{IzumidaJPSJ2009}\\
 								& (see Appendix~\ref{ap_isospin})     		  	& $\Delta^1_\mathrm{SO} \approx \frac{-0.3 \mathrm{meV}}{D[\mathrm{nm}]}$ & \cite{IzumidaJPSJ2009} \\
      						& 								  & $\Delta_\mathrm{SO}\equiv 2(\Delta_\mathrm{SO}^0 \mp \Delta_\mathrm{SO}^1 g_\mathrm{orb}/g_\mathrm{orb}^0)$ & \\				
     \vspace{-0.2cm} \\
Electric field spin splitting    			& $eE\xi$ & $\sim 20~\mu \mathrm{eV} \times E~[\mathrm{V nm}^{-1}]$ & \cite{KlinovajaPRB2011}\\      
      							
     \vspace{0.4cm} \\
	Intervalley scattering    	&$\Delta_{KK'}$& $\Delta_{KK'}\geq 60~\mu$eV (typical)& \cite{KuemmethNature2008}\\
	     \vspace{-0.2cm} \\
		Longitudinal mode spacing    	& \multirow{2}{*}{$\Delta E_\mathrm{conf} = \frac{h v_\mathrm{F}}{2L}$}& \multirow{2}{*}{$1.7~\mathrm{meV}/L~[\mu\mathrm{m}]$}& \multirow{2}{*}{\cite{TansNature1997}}\\      
		(high-energy limit)\\
 \hline \hline
   \end{tabular}
   \caption{Summary of nanotube quantum dot energy parameters. For the numerical values, representative estimates are given, based on experiments for the last two lines and theory elsewhere. A value $v_\mathrm{F}=8\times 10^5$~ms$^{-1}$ is assumed.}
   \label{tab:bandstructuresummary}
\end{table*}

In cases where both terms are finite but one dominates, the smaller parameter leads to anticrossings~(Fig.~\ref{SO_combined}(d-e)). Finally, if the two terms are of comparable finite magnitude, a complex spectrum emerges showing a mixture of effects~(Fig.~\ref{SO_combined}(f)). A small misalignment of the nanotube relative to the field axes, illustrated in the insets, introduces an anticrossing between the upper two levels. The magnitude of the anticrossing is $\Delta_\Theta=|\Delta_\mathrm{SO}|\tan{\Theta}$. 

This picture is confirmed in Fig.~\ref{SO_combined_data} for a device with comparable $\Delta_{KK'}$ and $\Delta_\mathrm{SO}$. The energy levels were measured by cotunneling spectroscopy, which maps out energy differences between ground and excited states~(see~\onlinecite{JespersenNphys2011} for discussion of this experimental technique). The resonant transitions appear as peaks or dips in $d^2I/dV_\mathrm{SD}^2$ whenever $eV_\mathrm{SD}$ is equal to the difference of energy levels. The measured transitions (Fig.~\ref{SO_combined_data}(a)) as a function of perpendicular and parallel field agree well with the predicted level differences~(Fig.~\ref{SO_combined_data}(b)) from the  spectrum (Fig.~\ref{SO_combined_data}(c)), calculated in the same way as in Fig.~\ref{SO_combined} assuming  $\Delta_{KK'}=3\Delta_\mathrm{SO}$.~In particular, the curvature of energy levels in parallel field due to $\Delta_{KK'}$ mixing and the anticrossing in perpendicular field due to $\Delta_\mathrm{SO}$ are seen. From similar data the parameters $\Delta_\mathrm{SO}$ and $\Delta_{KK'}$ can be measured precisely over a wide range of electron and hole occupation~\cite{JespersenNphys2011}.

We conclude this section by summarizing, with best numerical estimates, the various nanotube band structure parameters discussed in the text~(Table \ref{tab:bandstructuresummary}). Experimentally, one set of parameters usually suffices to characterize an entire shell, implying that these parameters are not strongly affected by addition of a few extra electrons. This is as expected from the constant-interaction model, which assumes all interactions can be parameterized by a single constant capacitance (Appendix~\ref{ap_coulombspectroscopy}).

\subsection{Open questions}
\label{sec_BandstructureOpenQuestions}

Comparison of theoretical and experimental spin-orbit coefficients shows serious discrepancies. As shown in Table~\ref{tab_SOparameters} different devices give unexpectedly large variation. With the diameters inferred from $\mu_\mathrm{orb}$, Eqs.~(\ref{eq_SOparam1}-\ref{eq_SOparam2}) predict that~$|\Delta_\mathrm{SO}|$ should range up to~$\sim 900~\mu$eV. Instead, values as large as $3.4$~meV have been reported, with several devices yielding results up to sixteen times larger than expected~\cite{JhangPRB2010,SteeleNcomm2013}. Other experiments have found $\Delta_\mathrm{SO}$ within the expected range~\cite{ChurchillPRL2009,LaiArXiv2012,CleuziouPRL2013}. Furthermore, in some cases the calculations even predict the wrong sign for both couplings $\Delta_\mathrm{SO}^0$ and~$\Delta_\mathrm{SO}^1$~\cite{KuemmethNature2008,JespersenNphys2011}.

This is clearly an open question for both theory and experiment. One explanation might be uncertainty in the tight-binding overlap integrals, which enter~Eqs.~(\ref{eq_SOparam1}-\ref{eq_SOparam2}) as empirical input parameters. Alternatively, electron interactions (Sec.~\ref{sec_shortrangeinteractions}) may play a role~\cite{RontaniPRB2014}. Another possibility is that some other symmetry-breaking between inside and outside the nanotube is responsible for the observed couplings, such as gate dielectric or adsorbates. For example, hydrogen adsorbed onto graphene is known to enhance the spin-orbit coupling~\cite{BalakrishnanNPhys2013}. In nanotube devices, adsorbed water affects the current-voltage characteristics~\cite{KimNanoLett2003}, although it is uncertain whether this is directly by modifying the band structure~\cite{NaAPL2005} or through another mechanism such as gathering ions from the environment~\cite{SungAPL2006}. 

A possibly related effect is that diameters inferred from measured $\mu_\mathrm{orb}$ are sometimes unexpectedly large. Whereas chemical-vapor deposited nanotubes are expected to have $D \lesssim 3$~nm, values of $\mu_\mathrm{orb}$ corresponding to $D\sim 5$~nm have been measured~\cite{MinotNature04,KuemmethNature2008,JarilloHerreroNature2004,JarilloHerreroPRL2005}. However, other devices yield values in the expected range~\cite{MinotNature04,MakarovskiPRB2007,ChurchillPRL2009,DeshpandeNPhys2008}. Measurements on nanotubes with known chirality should help clarify these discrepancies. Interestingly, the prediction of complete band closing at $B_\mathrm{Dirac}$~(Fig.~\ref{orbitalmoment}) is not borne out by experiments, which typically find minimal bandgaps $E_\mathrm{G} \approx 10-100$~meV. This hints at physics beyond the single-particle picture discussed here, such as formation of a Mott gap~\cite{DeshpandeScience2009} or an excitonic insulator~\cite{RontaniPRB2014}.

A separate set of questions addresses how to take advantage of the spin-orbit interaction. Ideas in this direction such as spin filtering and detection \cite{MazzaPRB2013, BrauneckerPRL2013}, generation of helical states \cite{KlinovajaPRL2011} and spin-orbit mediated spin control~\cite{FlensbergPRB2010, BulaevPRB2008} have already emerged (the latter is the topic of~Sec.~\ref{sec_nanotubeEDSR}).

\begin{figure}[]
\center 
\includegraphics{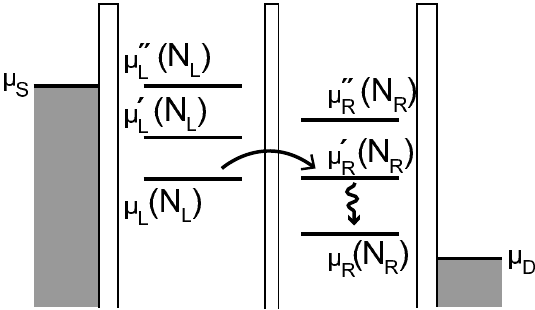}
\caption{\footnotesize{Transport through double quantum dots involves a dot-to-dot transition with a probability reflecting selection rules. The strictness of spin and valley selection rules depends on interactions that mix the spin or valley states, e.g. spin-orbit coupling, hyperfine interactions and disorder.}}
\label{fig_ddtransport}
\end{figure}

\section{Double quantum dots and Pauli blockade}\label{doubledots}
Transport through a single quantum dot involves an electron from the Fermi sea in one lead that tunnels via a discrete quantum state to an empty state in the other lead. 
For two dots in series an additional tunneling event occurs which involves a transition from one particular initial quantum state to a particular final state. This dot-to-dot transition is sensitive to selection rules, which determine the transition probability. The selection rules for nanotubes are based on the spin and valley quantum numbers. Whether they are obeyed in an experiment depends on to what extent spin and valley are good quantum numbers and how this is affected by spin-orbit coupling, hyperfine interaction or disorder. This sensitivity makes double quantum dots a versatile platform for studying quantum states and relaxation processes in nanotubes (Fig.~\ref{fig_ddtransport}).

\subsection{Role of bandgap and electron-hole symmetry in charge stability diagrams}

\subsubsection{Theory}

A double quantum dot formed by two dots in series defined within the same nanotube has similarities and differences to double dots defined in conventional semiconductors reviewed by \onlinecite{WielRMP02,Ihn2010}. Analogous to~Fig.~\ref{fig_dotschematic}, the device can be modelled by an electric circuit (Fig.~\ref{wielfig1}(a)). If each of the tunnel barriers is sufficiently opaque, $\Gamma_\mathrm{L,M,R} \ll E_\mathrm{C}$, then the charge within each dot is quantized and the number of electrons $N_\mathrm{L,R}$ can only change at specific gate voltages. A graph of the equilibrium charge configuration $(N_\mathrm{L},N_\mathrm{R})$ as a function of gate voltages is called a stability diagram (Fig.~\ref{wielfig1}(b)). The size of each region is a measure of the addition energy~(Eq.~\eqref{eqEadd}), horizontally for adding an electron in the right dot and vertically for an addition to the left dot. In most semiconductor dots either  the electron or hole region is accessible. In narrow-gap nanotubes the gate coupling can be sufficient to cross the bandgap, evident as larger honeycombs, and enter both regions. Because of the approximate electron-hole symmetry, a similar honeycomb pattern is expected in all regions of the stability diagram. Interestingly, the inter-dot tunnel barrier in the ($p,n$) and ($n,p$) regions is formed by a $pn$ junction. 

\subsubsection{Experiment}

\begin{figure}
\center 
\includegraphics{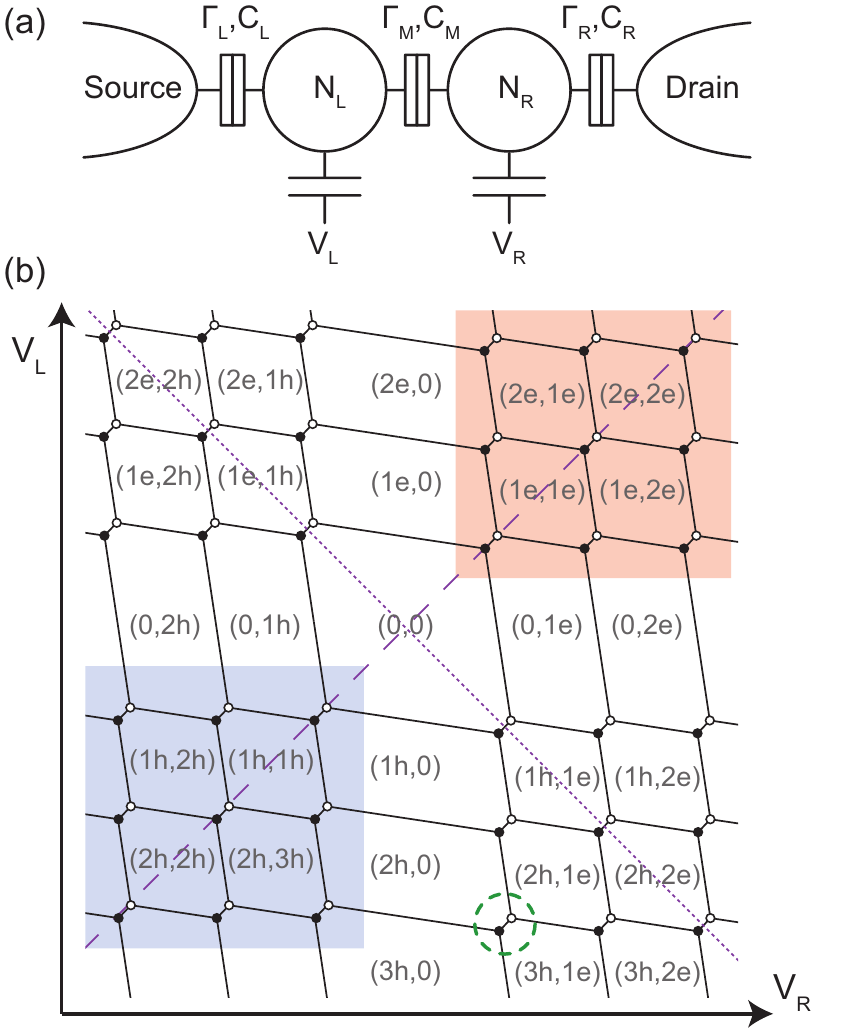}
\caption{\footnotesize{
(Color online) Charge stability with electrons and holes
(a) Circuit model of a series double quantum dot. Each of the three tunnel barriers is characterized by a tunnel rate $\Gamma_i$ and a capacitance $C_i$. 
(b) Stability diagram for a weakly tunnel-coupled double dot. Pairs of triple points define a honeycomb pattern.
The neutral Coulomb valley is largest, which can be understood by considering the $(1h,0)\rightarrow(1e,0)$ transition: adding two electrons to the left dot requires lowering its potential by an electrostatic energy (charging energy) and a kinetic energy (bandgap), $\Delta V_\mathrm{L}\propto 2E_\mathrm{C}+E_\mathrm{G}$.
Axes of approximate spatial and electron-hole symmetry are denoted by dashed and dotted lines respectively. 
The top right shaded region is most similar to conventional double dots, in which transport and occupation only involve electron-like carriers. 
Dashed circle marks a transition with particularly strong spin-valley blockade~(cf.~Figs.~\ref{Chap4SpinValleyExp_FK},~\ref{spinvalleyblockade}).}}
\label{wielfig1}
\end{figure}

\begin{figure}[]
\center 
\includegraphics[width=8.0cm]{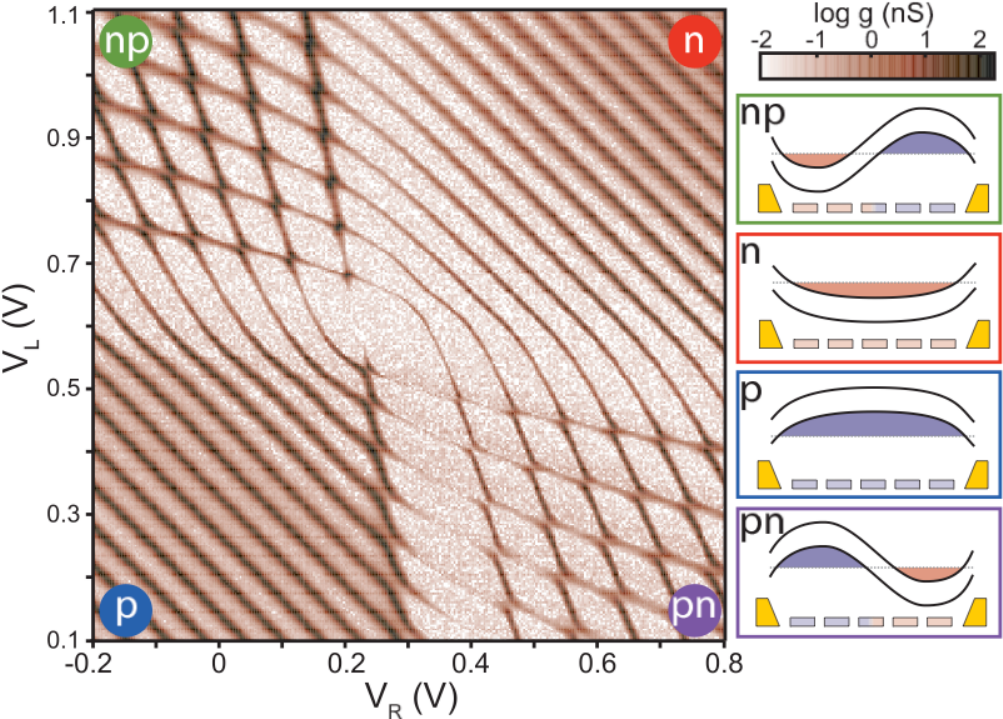}
\caption{\footnotesize{(Color online) Conductance through a narrow-gap nanotube suspended over five gate electrodes that allow independent control over $N_\mathrm{L}$, $N_\mathrm{R}$, $\Gamma_\mathrm{L}$, $\Gamma_\mathrm{M}$, and $\Gamma_\mathrm{R}$. 
For the gate configuration shown, the nanotube is neutral in the center of the plot $(N_\mathrm{L},N_\mathrm{R})=(0,0)$, forms a double quantum dot in the heteropolar regions (top left, bottom right), and forms a single quantum dot in the homopolar regions (top right for electrons, bottom left for holes). 
Insets visualize the charge distribution along the nanotube for electrons and holes.
Adapted from \onlinecite{WaissmanNatNano13}, device shown in Fig.~\ref{topandbottomgating}(f).
}}
\label{doubledot}
\end{figure}

The electron-hole stability diagram expected from~Fig.~\ref{wielfig1} is most easily observed in narrow-gap nanotubes. 
Full control over the basic parameters of a double quantum dot, namely charge occupation ($N_\mathrm{L}$, $N_\mathrm{R}$) and tunnel couplings ($\Gamma_\mathrm{L}$, $\Gamma_\mathrm{M}$, $\Gamma_\mathrm{R}$), requires at least five gate electrodes, and so the full charge stability diagram is at least five-dimensional.
A two-dimensional cut is shown in Fig.~\ref{doubledot}. Here the conductance $g$ through the device is plotted as a function of control parameters $V_\mathrm{R}$ and $V_\mathrm{L}$.  
Figure~\ref{doubledot} demonstrates that an actual device can show a stability diagram that is strikingly different from the diagram of Fig.~\ref{wielfig1}, characteristic of the weak-coupling regime. In Fig.~\ref{doubledot}, the middle gate voltage was intentionally chosen such that electron-hole double dots and single dots with electron or hole filling were demonstrated within the same device. In this regime, cotunneling processes give a significant contribution to transport, and hence boundaries between Coulomb valleys as well as triple points show up as conductance features.
Note both the spatial and electron-hole symmetry displayed by the data (mirror symmetry about the $+45^\circ$ and $-45^\circ$ diagonal respectively), attesting the cleanliness and tunability of suspended devices as in Fig.~\ref{topandbottomgating}(f).  

If the tunnel rates are too small to measure conductance, the charge stability diagram can be studied by charge sensing. Such capacitive sensing techniques are useful for the readout of pulsed-gate experiments on closed double dots and qubits \cite{ChurchillPRL2009}, as well as for investigating the quantum capacitance associated with electron interactions and correlations \cite{IlaniNPhys2006}.

Many other regimes are possible in double quantum dots. 
For example, 
a sizable  longitudinal level spacing in each dot can result in an overall eight-electron shell structure~\cite{JorgensenNatPhys08}, whereas
strong interdot tunneling lifts charge quantization within each dot, which can be interpreted as the formation of delocalized molecular states \cite{GraberPRB06}.
Devices with ambipolar charge stability and a high degree of tunability have been used to study many phenomenona, including Wigner crystallization \cite{pecker2013observation}, Klein tunneling \cite{SteeleNNano09}, and tunable electron-phonon coupling \cite{BenyaminiNatPhys2014}

\subsection{Spectroscopy of energy levels in bias triangles}

\begin{figure}
\center 
\includegraphics{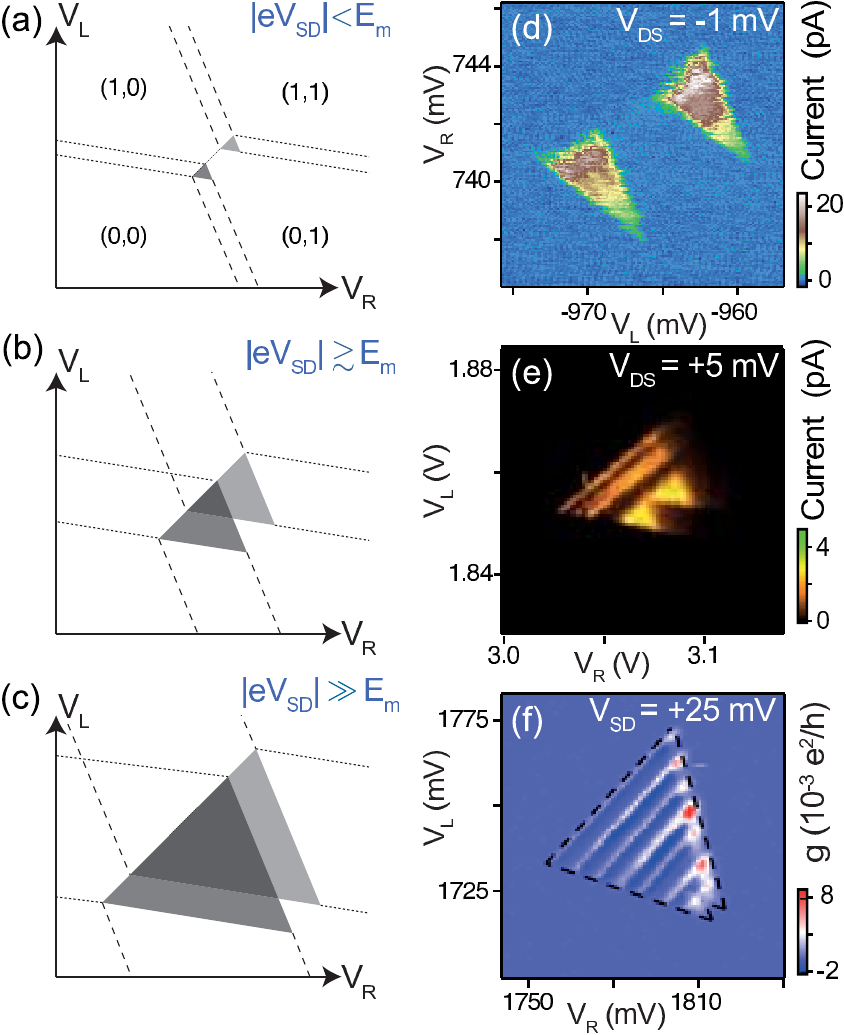}
\caption{\footnotesize{(Color online) 
(a-c): Dependence of bias triangles on $V_\mathrm{SD}$ (assumed negative). Dashed (dotted) lines indicate alignment of the electrochemical potential of the right (left) dot with the Fermi level in the right (left) lead. Sequential tunneling from source to drain is only allowed in the shaded regions, which expand with increasing $|V_\mathrm{SD}|$. The discrete density of states within each dot additionally restricts current (not shown). 
(d-f): Representative bias triangles for increasing $V_\mathrm{SD}$. Excited-state lines in (e) and~(f) identify $\Gamma_\mathrm{M}$ as the rate-limiting tunnel barrier. 
Adapted from \onlinecite{ChurchillThesis2012,SapmazNL06,JungNL2013}.}}
\label{biastriangles}
\end{figure}

Increasing the source-drain bias $V_\mathrm{SD}$  across a double dot allows non-linear conductance to be probed, providing spectroscopic information on the energy levels. Applying a bias large enough to overcome Coulomb blockade expands the triple points to finite-bias triangles. These triangles serve as a powerful experimental tool to reveal a variety of quantum effects in nanotubes. Figure~\ref{biastriangles}(a-c) shows their evolution with increasing $V_\mathrm{SD}$. Pairs of bias triangles start to overlap once $V_\mathrm{SD}$  becomes larger than the mutual charging energy $E_\mathrm{m}$ of the double dot\footnote{Details can be found in~\cite{WielRMP02} or~\cite{Ihn2010}.}. The finite bias breaks the left-right symmetry of a nominally symmetric device, yielding triangles pointing along the diagonal with a direction depending on the sign of~$V_\mathrm{SD}$. In panels (a-c) the sign of $V_\mathrm{SD}$ has been chosen so that electrons flow from the left contact (source) to the right contact (drain).

Under appropriate conditions, excited states (in either dot) are evident as discrete lines within a bias triangle. 
If interdot tunneling is the rate-limiting process, these lines appear parallel to the base of the triangles, but only if both tunnel rates to the leads remain at the same time smaller than level spacing and bias ($\Gamma_\mathrm{M}\ll \Gamma_\mathrm{L},\Gamma_\mathrm{R}< \Delta_\mathrm{ls},|eV_\mathrm{SD}|$).
From the line separation, measured from the base of the triangles, the corresponding excitation energies can be deduced. Examples from three different devices (Fig.~\ref{biastriangles}(d-f)) show the expansion of the bias triangles with $|V_\mathrm{SD}|$ and the appearance of excited-state lines\footnote{Figure~\ref{biastriangles} uses $V_\mathrm{DS}$ instead of $V_\mathrm{SD}$ for data with a different source-drain or left-right convention from Fig.~\ref{wielfig1}a.}. 

\subsection{Pauli blockade involving spin and valley}

\subsubsection{Motivation}

The dot-to-dot transitions in conventional double dots are strongly regulated by selection rules.
These selection rules arise from the Pauli exclusion principle, and can provide insight into the robustness of quantum numbers in the two dots and during interdot tunneling.
Since in nanotubes both spin and valley can form approximate good quantum numbers even in the presence of spin-orbit coupling, the manifestations of Pauli blockade and Pauli rectification are more complex than in conventional semiconductors.
In this section, we briefly review Pauli blockade in conventional semiconductors with only twofold spin degeneracy to establish useful terminology.
Next, we extend the model by adding twofold valley degeneracy to illustrate the persistence of Pauli blockade beyond spin blockade.
In order to make connection to actual nanotube experiments, we then discuss the main effects of spin-orbit coupling and electron-electron interactions on two-electron states within a quantum dot.
Finally, we present experimental evidence for Pauli blockade in nanotubes, and discuss the roles of spin, valley, and hyperfine coupling.

\subsubsection{State counting and Pauli blockade}

\begin{figure}[]
\center
\includegraphics{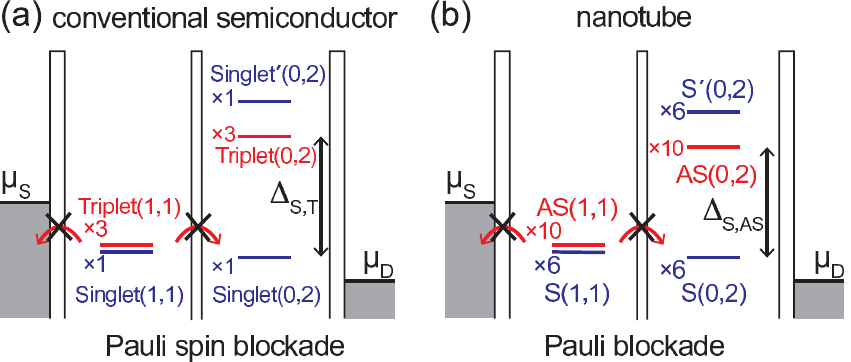}
\caption{\footnotesize{(Color online) Pauli rectification in conventional double dots and carbon nanotubes (Theory).
(a) In a conventional double dot without spin-orbit coupling, the spin-triplet (0,2) state involves an excited single-particle orbital of the right dot, and hence is higher than the spin-singlet (0,2) state by energy $\DeltaST$. Near the (1,1)-(0,2) degeneracy, one of the spin-triplet states Triplet~(1,1) can by chance become occupied. For sufficiently large bias ($\mu_\mathrm{S}-\mu_\mathrm{D}\gg k_\mathrm{B}T$), it is long-lived, resulting in a suppression of current compared with under opposite bias (not shown).
(b) If each dot participates with one longitudinal single-particle shell, then 16 different (1,1) states are possible in carbon nanotubes, colored here according to their longitudinal symmetry ($S$ or $AS$). If the right dot participates with the lowest longitudinal single-particle shell, then only six different (0,2) states are possible. Other (0,2) states are higher in energy by $\DeltaSAS$ (see Fig.~\ref{lowest02levels}). Similar to (a), sufficiently large bias $\mu_\mathrm{S}-\mu_\mathrm{D}$ can populate the long-lived (1,1) states (denoted $AS(1,1)$), leading to a suppression of current.
 }}
\label{STblockade}
\end{figure}

\begin{figure}[]
\center
\includegraphics{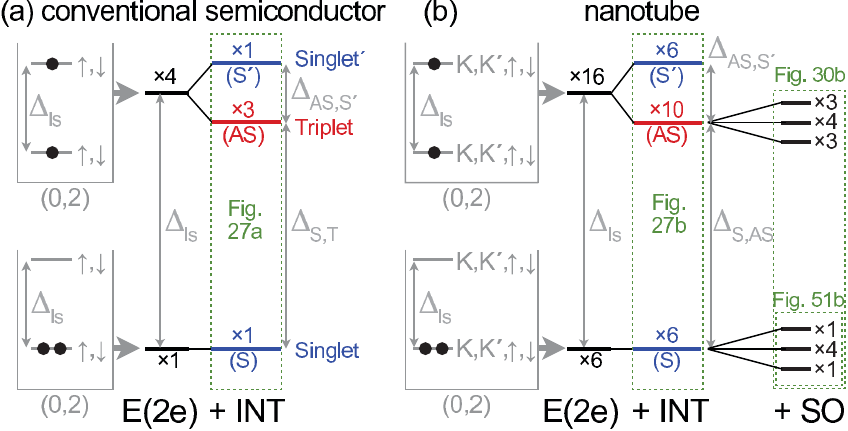}
\caption{\footnotesize{(Color online) Lowest quantum states of (0,2) for conventional semiconductors (a) and nanotubes (b) in the presence of interactions (\CINT) and spin-orbit coupling (SO).
For weak electron interactions ($\DeltaASSprime\ll \Delta_\mathrm{ls}$), the multiplet splitting $\DeltaST$ or $\DeltaSAS$ is approximately the single particle level spacing $\Delta_\mathrm{ls}$, and Pauli blockade between Triplet (1,1) and Singlet (0,2) or $AS(1,1)$ and $S(0,2)$ can be expected.
The states $S$, $AS$, $S'$ are part of the basis states of the matrix shown in Fig.~\ref{STblockadeMatrix}.
Spin-orbit interaction splits the two-electron multiplets in three. This is shown for $S$ and~$AS$, and further discussed in Fig.~\ref{TwoEBTheory}. Additional effects due to intervalley exchange are discussed in Fig.~\ref{theory_short_range}.
 }}
\label{lowest02levels}
\end{figure}

Pauli blockade is well established in conventional quantum dots containing a total of two electrons \cite{OnoScience2002, HansonRMP2007}. 
It relies on the fact that the (0,2) ground state is non-degenerate (it is a spin singlet) and is well separated in energy from the lowest spin triplet states, as illustrated in Fig.~\ref{STblockade}(a).
The spin triplet states are antisymmetric in their orbital degree of freedom with respect to electron transposition, and hence necessarily involve an excited single-particle state. 
Therefore, the energy cost $\DeltaST$ to form the Triplet\footnote{We capitalize Triplet and Singlet whenever we refer to specific spin-singlet and spin-triplet states indicated in Fig.~\ref{STblockade}.} (0,2) states, is approximately the single-particle level spacing in the right dot modified by electron interactions (cf. discussion of $\DeltaASSprime$ in Fig.~\ref{lowest02levels}(a)). This is in contrast to the Triplet (1,1) states, which for small interdot tunneling are nearly degenerate with the Singlet (1,1). 
If the (0,2) splitting~$\DeltaST$ is larger than temperature, then a Triplet (1,1) state cannot easily transition into the (0,2) state. If the applied source-drain bias is also larger than temperature, then the Triplet (1,1) is long-lived, and its occupation suppresses current flow due to Coulomb blockade. This effect is known as spin blockade, and manifests itself in current rectification (Pauli rectification).

In nanotubes, the two-electron spectrum is richer due to the two valleys. Ignoring spin-orbit coupling for clarity, it is schematically shown in Fig.~\ref{STblockade}(b). 
Anticipating  the breakdown of spin-singlet and spin-triplet terminology due to spin-orbit coupling, we have labeled the states according to the symmetry of the longitudinal quantum numbers under electron exchange (\SSS or \AS as defined below). As we will see, Pauli blockade is nevertheless possible in nanotubes. Even in the presence of spin-orbit coupling and small valley scattering, the projections of spin $s$ and valley $\tau$ in Eq.~(\ref{eq_SPH}) can remain (approximate) good quantum numbers, for example in the presence of a parallel magnetic field\footnote{Even in the absence of good quantum numbers Pauli rectification behavior can still occur, due to Kramers degeneracy. This was theoretically exemplified for double dots with strong spin-orbit coupling by \onlinecite{DanonPRB2009}.}. As in panel (a), interactions alter the spectrum, and the splitting $\DeltaSAS$ between the symmetric ground states and antisymmetric excited states is given by the level spacing modified by an interaction energy. 

Figure \ref{lowest02levels} shows a simple state-counting argument leading to the degeneracies indicated in Fig.~\ref{STblockade}(a-b) for the (0,2) states. For conventional semiconductors we consider two spin-degenerate levels separated by a level spacing $\Delta_\mathrm{ls}$. The non-degenerate ground state, Singlet (0,2), is formed by two electrons occupying the lower level (lower panel). If each level is occupied by only one electron, then four degenerate states are possible (upper panel). Exchange interactions (\CINT) result in an energy splitting $\DeltaASSprime$ between the Singlet$'$ (0,2) and Triplet (0,2) states, thereby reducing $\DeltaST$ slightly \cite{KouwenhovenRPP2001}.

The case of two fourfold degenerate nanotube shells is shown in Fig.~\ref{lowest02levels}(b). The lower shell can be occupied by two electrons in six ways, while 16 different states are possible with one electron in each shell\footnote{The multiplet $S^{\prime\prime}(0,2)$ representing two electrons in the upper shell is not shown in Fig.~\ref{lowest02levels}(b).}. Analogous to conventional semiconductor quantum dots and to lowest order, electron interactions (\CINT) split the 16 states into 10 longitudinal antisymmetric (lower energy) and 6 longitudinal symmetric states (higher energy). 
In the framework of first-order perturbation theory, this can be understood by calculating the (long-range) exchange integral associated with each two-electron basis state, and noticing that it differs between $\AS(0,2)$ states and $\SSSprime(0,2)$ states (Appendix~\ref{exlongrange}).
The complexity of the nanotube energy spectrum compared to conventional semiconductors is further revealed by turning on spin-orbit coupling within each multiplet (SO), as shown by the rightmost panel of Fig.~\ref{lowest02levels}(b) and discussed further below. 

State-counting arguments similar to those presented for (0,2) lead to $6+10=16$ symmetric and antisymmetric states in the (1,1) configuration.
Formally, this is accomplished by redefining the excited shell of the right dot as the lowest shell in the left dot: $AS(0,2)\rightarrow \AS(1,1)$, $\SSSprime(0,2)\rightarrow \SSS(1,1)$.
In the limit of vanishing interdot tunneling, interactions in the (1,1) regime can be neglected ($\DeltaASSprime\rightarrow0$).

\subsubsection{Symmetric and antisymmetric multiplets in nanotubes - theory}
\label{sub_SAStheory}

\begin{table}[]
 \centering
 \vspace{-0.45cm}
 \begin{tabular}{cc}
 \hspace{4.0cm} & \hspace{4.0cm}		\\
  \hline \hline
  \multicolumn{2}{c}{\textbf{Longitudinally symmetric states $S(0,2)$}}\\
  \multicolumn{2}{c}{$\tau s,\tau' s' \equiv |1\tau s\rangle_1 |1\tau' s'\rangle_2-|1\tau' s'\rangle_1 |1\tau s\rangle_2$}\\ 
  \hline
Designation $\tau s,\tau' s'$   		& Energy\\
\hline
$K{\downarrow}, K'{\uparrow}$	                            &	$-\Delta_\mathrm{SO}+E_\mathrm{C}$ \vspace{0.15cm} \\
$K{\downarrow}, K \uparrow$		                        &	\multirow{4}{*}{$E_\mathrm{C}$} \\
$K{\uparrow}, K'{\uparrow}$	                          	&  \\
$K{\downarrow}, K'{\downarrow}$                            & \\
$K'{\uparrow}, K'{\downarrow}$	                          	&\vspace{0.15cm}  \\
$K{\uparrow}, K'{\downarrow}$  	                        &	$\Delta_\mathrm{SO}+E_\mathrm{C}$		\\
  \hline
  \hline
  \multicolumn{2}{c}{\textbf{Longitudinally antisymmetric states $AS(0,2)$}}\\
      \multicolumn{2}{r}{$\tau s,\tau' s' \equiv |1\tau s\rangle_1 |2\tau' s'\rangle_2-|2\tau s\rangle_1 |1\tau' s'\rangle_2 $  }\\ 
   \multicolumn{2}{r}{$+|1\tau' s'\rangle_1 |2\tau s\rangle_2-|2\tau' s'\rangle_1 |1\tau s\rangle_2$  }\\ 
  \hline
  Designation $\tau s,\tau' s'$   	& Energy\\
\hline
$K{\downarrow}, K{\downarrow}$	                        &	  \\
$K{\downarrow}, K'{\uparrow}$	                        &	$-\Delta_\mathrm{SO}+E_\mathrm{C}+\Delta_\mathrm{S,AS}$                 \\\vspace{0.15cm}
$K'{\uparrow}, K'{\uparrow}$	                   		&							\\
$K{\downarrow}, K{\uparrow}$		                    &	\multirow{4}{*}{$E_\mathrm{C}+\Delta_\mathrm{S,AS}$}					\\
$K{\uparrow}, K'{\uparrow}$	                      	&							\\
$K{\downarrow}, K'{\downarrow}$	                    &                           \\\vspace{0.15cm}
$K'{\uparrow}, K'{\downarrow}$  	                    &	               	\\
$K{\uparrow}, K{\uparrow}$		                        &						\\
$K{\uparrow}, K'{\downarrow}$	                 &      	$\Delta_\mathrm{SO} + E_\mathrm{C}+\Delta_\mathrm{S,AS}$                     \\	$K'{\downarrow}, K'{\downarrow}$		\\
  \hline \hline
   \multicolumn{2}{c}{\textbf{Longitudinally symmetric states $S'(0,2)$}}\\
     \multicolumn{2}{r}{$\tau s,\tau' s' \equiv |1\tau s\rangle_1 |2\tau' s'\rangle_2+|2\tau s\rangle_1 |1\tau' s'\rangle_2 $  }\\ 
   \multicolumn{2}{r}{$ ~~~~-|1\tau' s'\rangle_1 |2\tau s\rangle_2-|2\tau' s'\rangle_1 |1\tau s\rangle_2$  }\\ 
  \hline
  Designation $\tau s,\tau' s'$   	& Energy\\
\hline
  $K{\downarrow}, K'{\uparrow}$	                            &	$-\Delta_\mathrm{SO}+E_\mathrm{C}+\Delta_\mathrm{S,AS}+\Delta_\mathrm{AS,S'}$ \vspace{0.15cm} \\
$
K{\downarrow}, K \uparrow$		                        &	\multirow{4}{*}{$E_\mathrm{C}+\Delta_\mathrm{S,AS}+\Delta_\mathrm{AS,S'}$} \\
$K{\uparrow}, K'{\uparrow}$	                          	&  \\
$K{\downarrow}, K'{\downarrow}$                            & \\
$K'{\uparrow}, K'{\downarrow}$	                          	& \vspace{0.15cm} \\
$K{\uparrow}, K'{\downarrow}$  	                        &	$\Delta_\mathrm{SO}+E_\mathrm{C}+\Delta_\mathrm{S,AS}+\Delta_\mathrm{AS,S'}$		\\
  \hline \hline

   \end{tabular}
  \caption{Two-electron states in the right quantum dot (normalization factors omitted). The three sections list quantum numbers for the three lowest two-electron multiplets ($\SSS,$ $\AS,$ and $\SSSprime$) at zero magnetic field. Explicit expressions for the states corresponding to particular quantum numbers in this notation are given below the table headers.
Each ket $|\nu \tau s\rangle_i$ represents a single-particle longitudinal/valley/spin state of the $i$th electron. The multiplets are classified as symmetric or antisymmetric according to their behavior under the interchange $\nu \leftrightarrow \nu'$.
  The energy splittings $\DeltaSAS$ and $\DeltaASSprime$ are defined in~Fig.~\ref{lowest02levels}. 
 }
  \label{tab_symmetricantisymmtric}
\end{table}

To gain more insight into the two-electron states in the presence of spin-orbit interaction\footnote{Related theoretical treatments are found in \cite{ReynosoPRB11, ReynosoPRB12, StecherPRB2010, WeissPRB2010}.}, we first consider the quantum numbers relevant for the lowest longitudinal symmetric multiplet $\SSS(0,2)$, i.e. two electrons occupying the lowest shell in the right quantum dot. The valley, spin (and longitudinal) quantum numbers are listed in Table \ref{tab_symmetricantisymmtric}, with the states organized according to their energies. The two-electron energy within the constant interaction model is found by adding the single-particle energies and a charging energy $E_\mathrm{C}$:
\begin{equation}
\label{eqKGR1}
E=E(\nu,\tau,s)+E(\nu',\tau',s')+E_\mathrm{C}
\end{equation}
where $\nu=\nu'=1$ specifies the lowest shell in the right dot and $\tau,\tau^\prime$ and $s,s^\prime$ are the valley and spin quantum numbers of the two electrons. 
The states are calculated by the method of Slater determinants, but taking account of the fact that $\nu$, $\tau$, and $s$ are coupled. 

Our conventions for labeling two-electron quantum states are stated below the three bold table headers. For example, consider the $S(0,2)$ state denoted $K{\downarrow},K'{\uparrow}$, which is spin- and valley-unpolarized. Written out explicitly, the energy and the state are\footnote{See Appendix~\ref{ap_isospin} for a more thorough derivation of the basis used.}:
\begin{eqnarray}
E^{S(0,2)}_{K{\downarrow}, K'{\uparrow}} & = &  -\Delta_\mathrm{SO}+E_\mathrm{C}\\
|\psi^{S(0,2)}_{K{\downarrow}, K'{\uparrow}}\rangle& =  &\frac{1}{\sqrt{2}}\left( |1K{\downarrow} \rangle_1|1K'{\uparrow}  \rangle_2-|1K'{\uparrow} \rangle_1|1K{\downarrow}\rangle_2 \right),\notag\\
\label{S(0,2)state}
\end{eqnarray}
where the subscripts on the right of~Eq.~(\ref{S(0,2)state}) refer to electron 1 or 2, and the three labels in a single-particle state $|\nu \tau s\rangle$ are the longitudinal shell, valley and spin quantum numbers associated with that electron. The states in Table \ref{tab_symmetricantisymmtric} are classified as longitudinally symmetric or antisymmetric according to whether the wave function remains the same or changes sign under the interchange $\nu \leftrightarrow \nu'$.
Thus the example in Eq. (\ref{S(0,2)state}) is symmetric ($\nu=\nu'$). 
We do not decompose $|\nu \tau s\rangle$ further into a product of longitudinal, valley, and spin wave functions, as this is strictly correct only at $B=0$ and in the absence of spin-orbit coupling\footnote{This decomposition can be illustrative~(e.g.\ in \onlinecite{pecker2013observation,PeiNnano12}), but is unsuitable for accurate calculations.}. In general, the longitudinal wave function depends on both $\tau$ and $s$~\cite{WeissPRB2010}. 

Next we consider two-electron states with one electron in the lowest shell ($\nu=1$) and one electron in the first excited shell ($\nu'=2$). By a similar procedure, sixteen distinct Slater determinants can be written. However, some of them are coupled by electron interactions (``\CINT" in Fig.~\ref{lowest02levels}(b)), resulting in the splitting $\DeltaASSprime$ and modifying $\DeltaSAS$ (see \ref{intin2e}). The eigenstates of the electron interactions are the ten antisymmetric states ($\AS(0,2)$) and six symmetric states ($\SSSprime(0,2)$) formed by linear combination of the Slater determinants. These eigenstates are listed with their energies in Table \ref{tab_symmetricantisymmtric}.

As for $\SSS(0,2)$, the energies at $B=0$ are found by adding the single-particle energies, charging energy, and interaction energy.
However, the expressions for the explicit states now contain four terms. 
For the spin and valley polarized states ($\tau s,\tau s$), these four terms reduce to a simple two-term Slater determinant, but for the remaining six $\AS(0,2)$ and six $\SSSprime(0,2)$ states this is not the case due to the interactions $\DeltaASSprime$.
This tells us that in general Eq.~(\ref{eqKGR1}) is too simple to predict the spectrum accurately. In particular, it does not take magnetic fields into account, nor differences in spin-orbit coupling and valley scattering between different shells, not to mention short or long range interactions discussed in Sec.~\ref{correlations}.
As an example, the energy and quantum state of the spin-valley unpolarized $\AS(0,2)$ state denoted by $K{\downarrow},K'{\uparrow}$ are:
\begin{eqnarray}
E^{AS(0,2)}_{K{\downarrow}, K'{\uparrow}} & = & -\Delta_\mathrm{SO}+E_\mathrm{C}+\DeltaSAS\\
|\psi^{AS(0,2)}_{K{\downarrow}, K'{\uparrow}}\rangle&  = &~\frac{1}{2}\left(|1K{\downarrow} \rangle_1|2K'{\uparrow}  \rangle_2-|2K{\downarrow} \rangle_1|1K'{\uparrow}  \rangle_2\right. \nonumber\\
& &~~\left.+|1K'{\uparrow} \rangle_1|2K{\downarrow}  \rangle_2-|2K'{\uparrow} \rangle_1|1K{\downarrow}  \rangle_2 \right).\notag\\
\label{ASstate}
\end{eqnarray}
The longitudinal antisymmetry of this state is easily seen by comparing terms in the same row of~Eq.~\eqref{ASstate}, while comparing between rows shows the symmetry of the spin-valley part. The spin-valley unpolarized $\SSSprime(0,2)$ state (also denoted $K{\downarrow},K'{\uparrow}$) involves the same single-particle states, but is obtained by changing the longitudinal symmetry, i.e.\ by changing the sign on the second and third terms in Eq.~(\ref{ASstate}). 

All relevant two-electron states in (1,1) can also be constructed from Table \ref{tab_symmetricantisymmtric}, simply by identifying $\SSS(1,1)$ with $\SSSprime(0,2)$, and $\AS(1,1)$ with $\AS(0,2)$. This works by assigning $\nu=1(2)$ to the lowest shell in the right (left) dot, and setting level spacing and interactions to zero. 
We call antisymmetric $(1,1)$ states ``blocked" states, because they vanish formally when setting $\nu=\nu'=1$. Physically, this means that these states cannot be converted into $(0,2)$ charge states without involving a higher orbital in the right dot or changing their spin or valley configuration.
The $(1,1)$ states generated in this fashion (along with the $(0,2)$ states) from Table \ref{tab_symmetricantisymmtric} are useful basis states for a tunnel-coupled double dot at $B=0$, assuming that the spin-orbit coupling in the two dots is identical
\footnote{In $(1,1)$, the splitting between symmetric states $\SSS(1,1)$ and antisymmetric states $\AS(1,1)$ vanishes for small interdot tunneling, $\DeltaASSprime^{(1,1)}\rightarrow0$, and hence a different basis would be used, appropriate for the regime $|\Delta_\mathrm{SO}^\mathrm{left}-\Delta_\mathrm{SO}^\mathrm{right}| \gg \DeltaASSprime^{(1,1)}$. Similarly, different orbital moments in left and right dot would require a different basis at sufficiently large magnetic field. }.

\begin{figure*}
\center
\includegraphics[width=13cm]{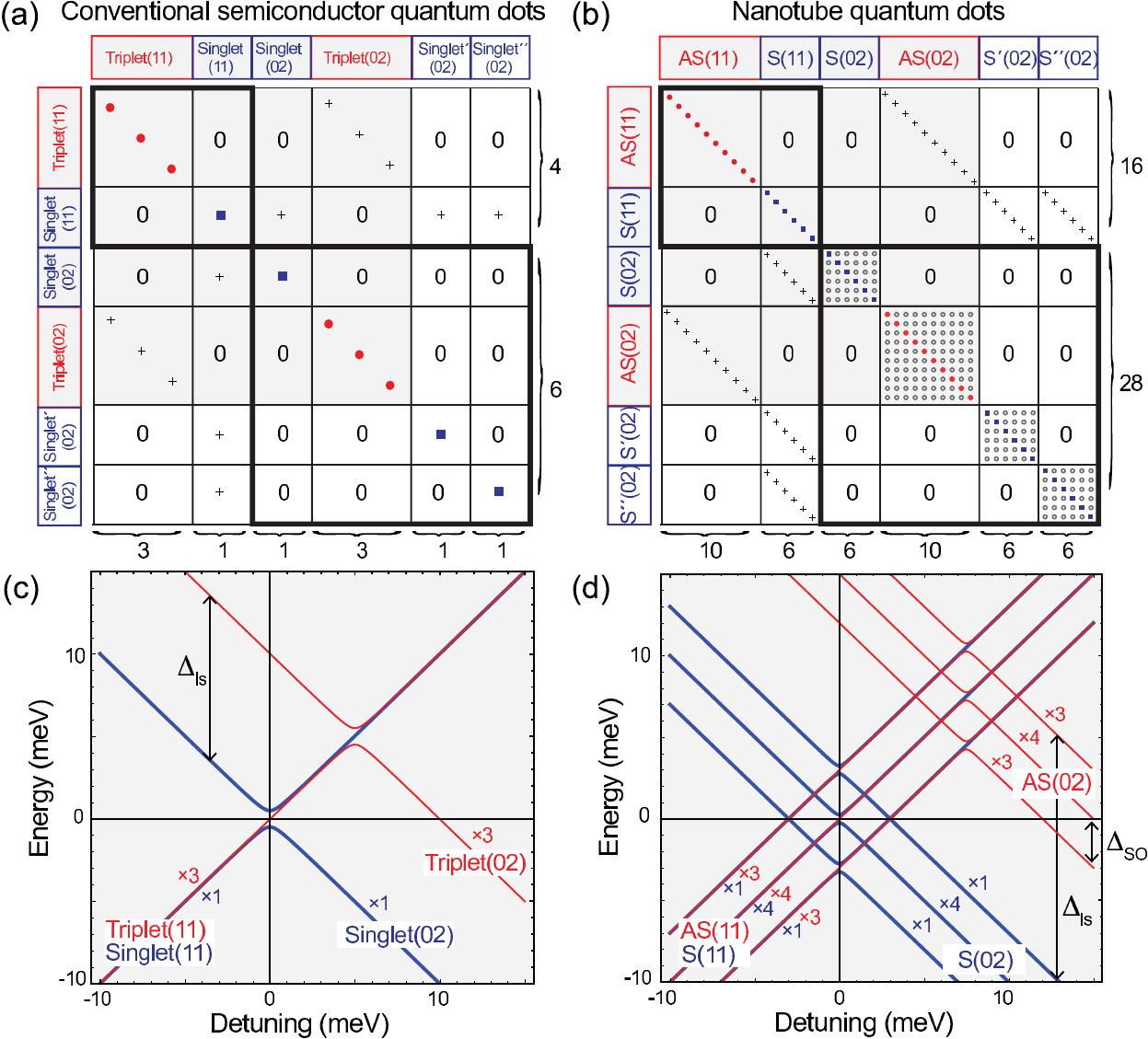}
\caption{\footnotesize{(Color online) 
(a) Matrix representing the two-electron Hamiltonian for a conventional semiconductor double dot, assuming a single orbital on the left and two orbitals on the right. The basis states are spin-singlet and spin-triplet states as indicated, and symbols indicate non-zero matrix elements.  The submatrices in the (1,1) and (0,2) subspaces are all diagonal, and interdot tunneling (+) leaves singlet and triplet states uncoupled. 
To lowest order, interactions modify the diagonal elements between Triplet (0,2) and Singlet$'$~(0,2) eigenstates ($\blacksquare$ and $\bullet$), corresponding to the interaction energy $\DeltaASSprime$ discussed in the text. 
For completeness, we include the higher-lying Singlet$''$ (0,2).
(b) The analagous matrix for a nanotube double quantum dot with spin-orbit interaction, assuming a single longitudinal shell on the left and two shells on the right. The basis states are those of Table \ref{tab_symmetricantisymmtric}, hence longitudinally symmetric $\SSS(1,1)$ and antisymmetric $\AS(0,2)$ states are uncoupled. 
Most submatrices are still diagonal, but interactions are expected to result in significant off-diagonal matrix elements ($\circ$) within the (0,2) sectors. For example, Sec.~\ref{correlations} will show that valley backscattering leads to off-diagonal elements between spin-valley unpolarized states within the $\SSS(0,2)$ subspace. Without going into details of the interactions, all off-diagonal elements marked $\circ$ are in principle allowed, as drawn here. In some cases, detailed analysis shows that particular interaction matrix elements vanish.
(c) Energy versus detuning of the shaded states in (a), neglecting interactions. Zero detuning corresponds to the degeneracy between the ground states of (0,2) and (1,1). 
Avoided crossings due to interdot tunneling occur for states with the same symmetry. The degeneracy of the levels is indicated. Numerical parameters in this plot were chosen for illustration, and differ in actual devices.
(d) The same plot for a nanotube, assuming identical spin-orbit coupling ($\Delta_\mathrm{SO}$) in all shells.
Interdot tunneling now leads to anticrossings between states with the same longitudinal symmetry. The effect of interactions on $\SSS(0,2)$ and $\AS(0,2)$ is neglected in this plot; see discussion of Figs.~\ref{TwoEBTheory}, \ref{twoelectronsExpt}, and \ref{theory_short_range} for details. }}
\label{STblockadeMatrix}
\end{figure*}

Having introduced all relevant (1,1) and (0,2) multiplets, a useful overview of the states and their mutual coupling is obtained by setting out the Hamiltonian in matrix form. Figure \ref{STblockadeMatrix} shows schematic matrices for (a) a conventional semiconductor double dot in the spin-singlet/spin-triplet basis, and (b) a nanotube double quantum dot in the basis of the longitudinal symmetric/antisymmetric multiplets. These matrices can be divided into submatrices coupling manifolds of particular symmetry. In conventional semiconductors all submatrices are diagonal even when interactions are included. Tunnel coupling between states of the same singlet/triplet character is reflected (+ symbols) in the submatrices between (1,1) and (0,2) states. In nanotubes, the overall structure is similar, but the number of states in each multiplet is increased and the diagonal elements now include spin-orbit coupling. Again, states with identical symmetry are coupled by diagonal tunnel matrices, and weak interactions (denoted $\DeltaASSprime$ above) appear as diagonal elements that shift the energies of multiplets with respect to each other. However, off-diagonal elements are allowed in the (0,2) multiplets ($\circ$ symbols), although arguments can be made that these are small (Appendix \ref{ap_SO}). Interactions within the $\SSS(0,2)$ multiplet (not included in Table \ref{tab_symmetricantisymmtric}) appear as diagonal and off-diagonal matrix elements, and are further discussed in Sec.~\ref{correlations}.


Plots of energy versus detuning for the low-energy states (Singlet, Triplet and $S,AS$) are shown in Fig.~\ref{STblockadeMatrix}(c-d) neglecting interactions ($\DeltaASSprime\rightarrow0$). The more complex spectrum for the nanotube is clearly revealed. Valley mixing (assumed zero in this figure) would lead to additional avoided crossings (off-diagonal elements in the matrix) between states with different valley quantum numbers. The non-avoided crossings between the $\AS(1,1)$ and $\SSS(0,2)$ states signify long-lived (1,1) states that give rise to Pauli blockade. 

Tunneling from a blocked $\AS(1,1)$ state to a $\SSS(0,2)$ state requires a change of the longitudinal symmetry. For some states this may simply involve dephasing between the left and right single-particle states. For other states it also necessitates a change in quantum numbers of at least one electron.
In particular, we distinguish lifting of Pauli blockade by:
\begin{itemize}
  \item Dephasing only
  \item Valley flips
  \item Spin flips
  \item Spin and valley flips
\end{itemize}
Inspection of all ten $AS$ states reveals that a single flip in one of the dots can result in an unblocked double dot configuration. For some states it involves a spin flip, for others a valley flip (see examples in \ref{appendixtwoelectronstatesPauli}).
However, these single flip processes do not conserve energy because of spin-orbit coupling, and therefore cannot lift the blockade. 
Therefore, simultaneous spin-and-valley flips can become the rate-limiting process (``spin-valley blockade").

To investigate which type of blockade is observed in experiments, careful identification of the involved states and quantum numbers is required. This is best facilitated by application of a  magnetic field. 
In Fig.~\ref{TwoEBTheory}(a), the energies of the two lowest single-particle shells are plotted against parallel magnetic field. The two-electron magnetic moments are obtained by summing the one-electron magnetic moments (Fig.~\ref{TwoEBTheory}(b)). As an example, the filled circles in~Fig.~\ref{TwoEBTheory}(a) indicate the two single-particle energies summed to give the energy of the $\SSS(0,2)$ state $K{\downarrow},K{\uparrow}$ as in Fig.~\ref{TwoEBTheory}(b). Similarly, the empty circles indicate the two-single-particle energies combining to give the $\AS(0,2)$ state denoted $K{\downarrow},K{\downarrow}$.
The field dependence of the remaining states is found in a similar fashion.  
The states can be divided into pairs of  valley-polarized, spin-polarized and spin-valley unpolarized states.
Although all degeneracies of the two multiplets are lifted at finite field, and spin and valley quantum numbers can in principle be assigned based on the observed magnetic moment\footnote{This requires that interactions have lifted the degeneracy with the $\SSS'(0,2)$ multiplet not included in Fig.~\ref{TwoEBTheory}(b).}, the spectroscopic intensity associated with each state can be very different, depending how exactly the spectrum is measured.
The solid lines in Fig.~\ref{TwoEBTheory}(b) indicate the available two-electron states within the $\SSS(0,2)$ and $\AS(0,2)$ multiplets, given that one electron occupies the $K{\downarrow}$ state. These two-electron states are relevant when measuring the addition spectrum given that the first electron is in its finite-field ground state $K{\downarrow}$.

\subsubsection{Symmetric and antisymmetric multiplets in nanotubes - experiment}

\begin{figure}[]
\center
\includegraphics{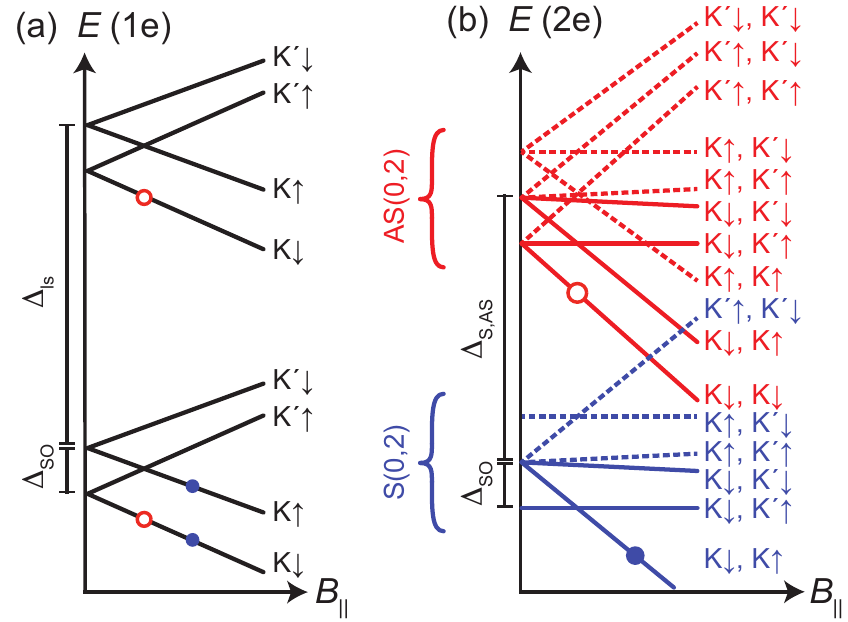}
\caption{\footnotesize{(Color online) Magnetic field dependence of the 16 lowest (0,2) states (cf. Fig.~\ref{lowest02levels}b).
(a) Spectrum of the two lowest single-particle shells of the right dot, versus parallel magnetic field $B_{||}$. There are six ways to fill the lowest shell with two electrons, each corresponding to a symmetric state in (b). The magnetic moment of each combination is simply the sum of the involved single-particle spin and orbital magnetic moments, shown here for two examples (colored dots).
(b) Magnetic field dependence of the ten antisymmetric and six symmetric states. 
States that involve the ground state of the lowest shell, $K{\downarrow}$, are plotted with solid lines.
}}
\label{TwoEBTheory}
\end{figure}

\begin{figure}[]
\center
\includegraphics{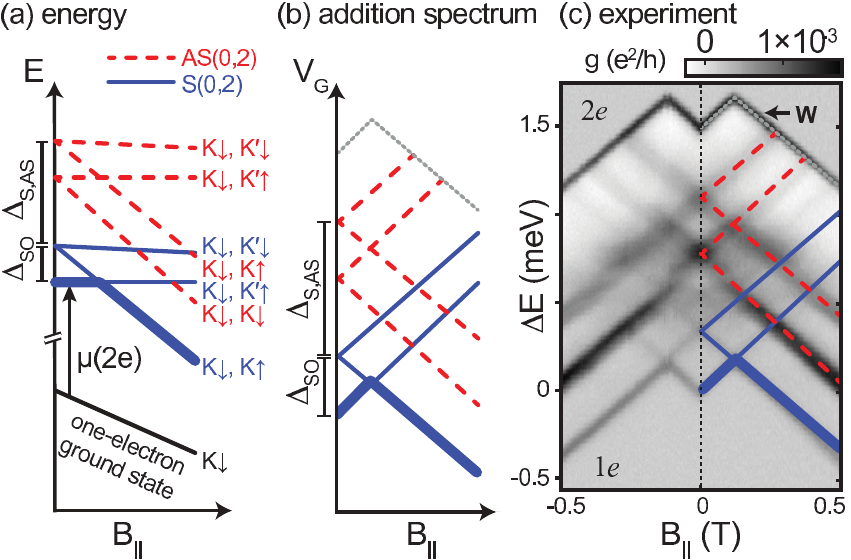}
\caption{\footnotesize{(Color online) Addition spectroscopy of two-electron states in a nanotube single quantum dot, assuming that the one-electron dot starts in its ground state $K{\downarrow}$.
(a) Two-electron spectrum $E(2e)$ from Fig.~\ref{TwoEBTheory}b as a function of parallel magnetic field. 
States that do not involve $K{\downarrow}$ are omitted.
(b)~Expected $1e$-$2e$ addition spectrum from (a).
(c)~Measured addition spectrum, with lines from (b) superimposed. The topmost conductance feature (w) results from the finite bias window of 1.7 mV used during measurement (see Fig.~\ref{CNTspectroscopy}).
The separation between solid and dashed lines is surprisingly small (0.85 meV) compared to the single-particle level spacing (7.8 meV), indicating that $\DeltaSAS\ll\Delta_\mathrm{ls}$ due to electron-electron interactions. 
Data from \onlinecite{pecker2013observation}.}}
\label{twoelectronsExpt}
\end{figure}

The two-electron spectrum can be measured by high-bias spectroscopy near the (0,1)-(0,2) transition. Assuming that the device starts in the (0,1) ground state~($K{\downarrow}$), then exactly three states of the symmetric two-electron multiplet can be reached and four states of the antisymmetric multiplet (Fig.~\ref{TwoEBTheory}(a)). All other two-electron states would require a higher-order process, in which the incoming electron also promotes the resident electron to change its spin or valley quantum numbers.
The expected addition spectrum is shown in Fig.~\ref{twoelectronsExpt}(b), and the corresponding data in~Fig.~\ref{twoelectronsExpt}(c). Although the qualitative agreement is quite good, the data shows a surprisingly small multiplet splitting $\DeltaSAS=0.85$ meV, significantly less than the measured single-particle level spacing of~7.8~meV. This is due to strong electron-electron correlations (Sec.~\ref{correlations}), that differ drastically from the weak exchange interactions typical in GaAs dots. This is of practical importance because a small splitting $\DeltaSAS$ makes it difficult to observe Pauli blockade in nanotubes. 

\subsubsection{Pauli blockade in nanotubes - experiment}

Generally, interdot transitions that are forbidden by spin or valley selection rules are of particular interest, because measurements of leakage current or double-dot charge state then illuminate spin dynamics and spin-valley relaxation processes. We return to this topic in~Sec.~\ref{qubits}. Although aspects of Pauli blockade in nanotubes have been observed by multiple groups, this phenomenon is experimentally less generic than in GaAs. A detailed understanding of the data is complicated by the large number of states involved (see Fig.~\ref{STblockadeMatrix}), and often a lack of knowledge of critical device parameters such as differences in intervalley scattering or spin-orbit coupling between left and right dots, and the strength of electron-electron correlations. All these were neglected in Fig.~\ref{STblockadeMatrix}.

Even the simplest manifestation of Pauli blockade, namely Pauli rectification in a DC transport experiment, can be obscured by other effects such as strong electron-electron correlations.
In Fig.~\ref{STblockade} we outlined the double dot energy levels for the $(1,1){\rightarrow}(0,2)$ transition and noted the importance of sufficiently large $\DeltaSAS$. In the ($n,n$) and ($p,p$) regimes, $\DeltaSAS$ is limited by the level spacing from the longitudinal quantization $\Delta_\mathrm{ls}$, and likely significantly reduced by correlation effects $\DeltaASSprime$.
In order to keep $\DeltaSAS$ as large as possible, the bandgap of the nanotube can be used as an effective large ``level spacing", making the observation of Pauli blockade more robust against interaction effects.
The transition between $(3h,1e)$ and $(2h,0)$ (dashed circle in Fig.~\ref{wielfig1}(b)), is one example where the level spacing is enhanced by the band structure gap, $E_\mathrm{G}$.

\begin{figure}[]
\center
\includegraphics{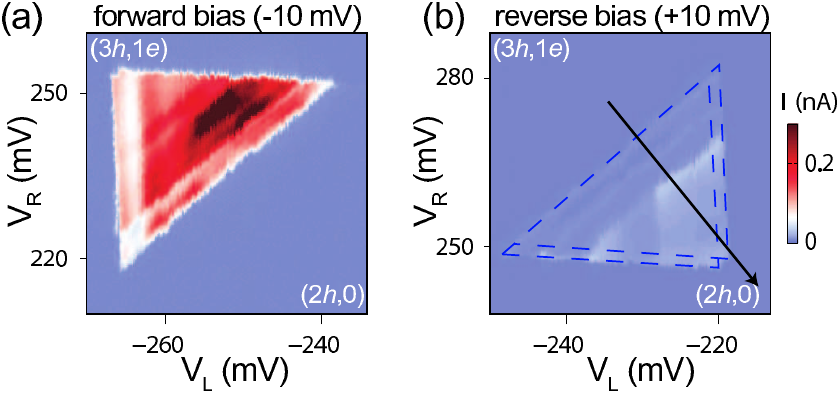}
\caption{\footnotesize{(Color online) Pauli rectification (experiment). Current in the bias triangles measured in forward (a) and reverse (b) direction near the $(3h,1e)$-$(2h,0)$ transition.
The asymmetry due to Pauli rectification is clearly observed, and persists up to $\pm 10$~meV, i.e.\ detunings much larger than $\Delta_\mathrm{SO}$ (=1.6~meV in this device).
The arrow defines the detuning axis used in Fig.~\ref{spinvalleyblockade}(c). Adapted from \onlinecite{PeiNnano12}.}}
\label{Chap4SpinValleyExp_FK}
\end{figure}

As shown in Fig.~\ref{Chap4SpinValleyExp_FK}, rectification behavior for such a transition is observed for detunings as high as the applied bias voltage ($\pm10$ mV). This is larger than the spin-orbit splitting and estimated level spacing in this device \cite{PeiNnano12}.
This observation of strong current suppression up to high bias can be linked to the advantageous use of the band gap (see Fig.~\ref{spinvalleyblockade}(a)).


In the blocked bias triangle of Fig.~\ref{Chap4SpinValleyExp_FK}(b), a small increase of leakage current is observed at a detuning of approximately $2\times\Delta_\mathrm{SO}=3.2$~meV.
This can be interpreted as a weak lifting of Pauli blockade, but a quantitative understanding of this leakage current, and identification of the corresponding relaxation rates, has not been reached. We speculate that it is necessary to include interaction effects beyond the constant interaction model to explain such features.

\subsubsection{Spin-valley blockade}
\label{subsubspinvalleyblockade}

Information about the role of spin and valley quantum numbers in the Pauli rectification of Fig.~\ref{Chap4SpinValleyExp_FK} can be obtained by applying a parallel magnetic field. If this induces an energy-level splitting larger than the interdot tunnel coupling or the intervalley scattering, 
we expect orbital and Zeeman couplings to restore valley and spin quantum numbers within the two-electron states. In turn we can associate these with the quantum numbers of single-particle levels.

\begin{figure*}[]
\center
\includegraphics{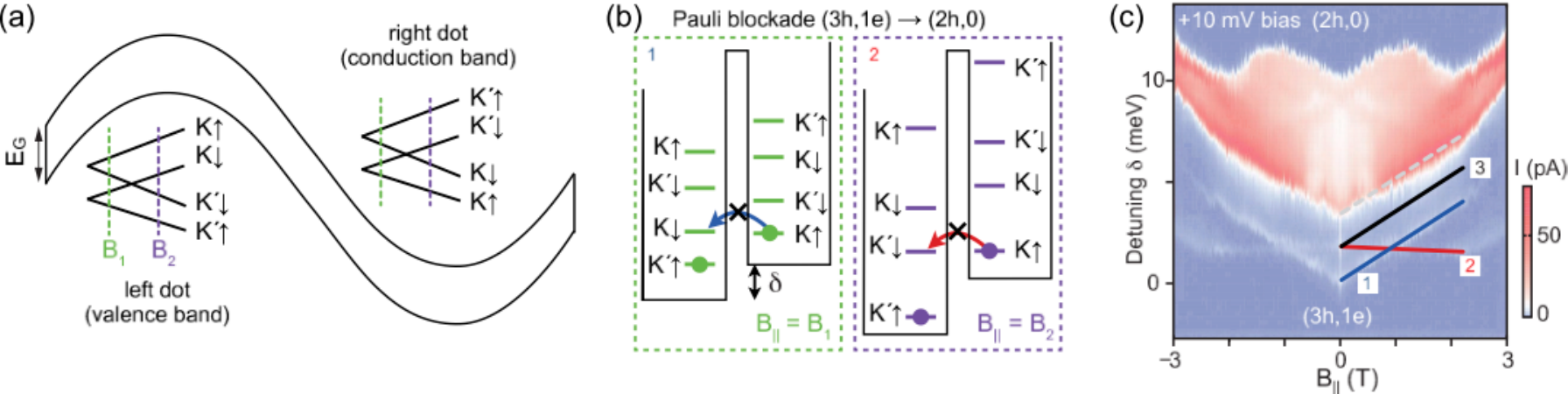}
\caption{\footnotesize{(Color online) Different levels of Pauli blockade involving spin and valley quantum numbers for the charge transition (3h,1e) $\rightarrow$ (2h,0).
(a) Single-particle electron levels versus magnetic field, sketched for the left and right sides of a $pn$ double  dot.
(b) At low field (marked $B_1$ in (a)) the ground state-to-ground state transition requires a spin-flip. At higher field (marked $B_2$) the ground state transition requires a spin-flip \emph{and} a valley-flip. The detuning $\delta$ between left and right dots is marked. Due to the different magnetic moments of the single particle states, the value of~$\delta$  at which each transition becomes resonant depends on magnetic field.
(c) Measurement of the leakage current versus $B_{||}$. Some of the features can be identified within the single-particle representation of (b) (lines 1 and 2).
The slight reduction of the ground state leakage current above 1~T indicates that a spin-and-valley flip (line 2) is slightly less frequent than a spin-only flip (line 1), and is evidence for spin-valley blockade. The absence of a simple charge relaxation $(RK\uparrow) \rightarrow (LK\uparrow)$ (line 3), and the onset of leakage current at $\delta \sim 2\Delta_\mathrm{SO}$ (dashed line) are currently not understood.
Adapted from \onlinecite{PeiNnano12}.
}}
\label{spinvalleyblockade}
\end{figure*}

In Fig.~\ref{spinvalleyblockade}(c), the leakage current is measured as a function of detuning (defined in Fig.~\ref{Chap4SpinValleyExp_FK}(b)) and $B_\mathrm{||}$. 
The base of the bias triangle (detuning $\delta\sim0$) corresponds to the ground state-to-ground state transition between $(3h,1e)$ and $(2h,0)$.
Energy conservation imposes spin and valley selection rules for this transition, and therefore the leakage current provides information about the relaxation of these selection rules.

Figure \ref{spinvalleyblockade}(a) shows the assignment of single-particle quantum numbers to the highest (lowest) longitudinal shell in the valence (conduction) band of the left (right) quantum dot. These quantum numbers were inferred from the magnetic field dependence of the stability diagram\footnote{To avoid confusion, we show quantum numbers of electronic states, even for the valence band. Other publications may consider the absence of a $K'{\uparrow}$ electron in the valence band as a $K'{\downarrow}$ hole, due to conservation of angular momentum. Quantum numbers in \cite{PeiNnano12,SteeleNcomm2013} are not consistent with our identification.}. Based on the single-particle picture, the ground state of $(2h,0)$ is expected to make a transition from $K'{\uparrow},K{\downarrow}$ to $K'{\uparrow},K'{\downarrow}$ as a function of parallel magnetic field\footnote{Table \ref{tab_symmetricantisymmtric} should not be used to generate these states, because conduction band and valence band differ in their assignment of $K$ and $K'$ (Fig. \ref{SO_spectrumtheory}).}.
This implies that at low field the ground state-to-ground state transition requires a spin flip of the right electron, whereas at higher magnetic field it requires a spin flip and a valley flip (compare the two arrows in Fig.~\ref{spinvalleyblockade}(b)). 
Therefore, if valley is a good quantum number and conserved during interdot tunneling, then one expects the ground state leakage current at high field to be smaller than at low field. Indeed, this is seen in panel~(c) by comparing conductance features marked with line 1 and line 2.
We mention that the $(3h,1e)$ ground state is not a blocked state in the sense of~Sec.~\ref{sub_SAStheory}: We can write it as a Slater determinant of one $K'{\uparrow}$ electron in the left ($L$) orbital and one $K{\uparrow}$ electron in the right ($R$) orbital, $|LK'{\uparrow}\rangle_1|RK{\uparrow}\rangle_2-|RK{\uparrow}\rangle_1|LK'{\uparrow}\rangle_2$, which does not vanish under the substitution $R\rightarrow L$ (i.e. $\nu=\nu'$). 
We therefore expect that relaxation to the $(2,0)$ charge state is allowed without involvement of higher orbitals or change of spin or valley configuration. The absence of this charge relaxation (line 3 in panel (c)) is not understood.

In summary, this data suggests that both spin and valley can contribute to Pauli blockade, but a quantitative understanding of the leakage current and relaxation rates has not been reached. We speculate that several mechanisms contribute, such as disorder, hyperfine coupling, spin-phonon coupling, or bend and spin-orbit mediated relaxation.

\subsection{Lifting of Pauli blockade by hyperfine coupling}
\label{relaxationbynuclei}

In this section we investigate double dots where Pauli blockade is partially lifted due to hyperfine interaction between the electron spin and the \thirteen~nuclear spins. 

\subsubsection{Theory}

Hyperfine interaction with disordered nuclear spins, such as \thirteen\ isotopes, couples different spin states by flip-flop processes and, in addition, different valley states, because of the atomically sharp length scale. It is therefore expected that hyperfine coupling generically lifts Pauli blockade  and results in spin relaxation and spin dephasing processes.
This mechanism of spin relaxation was considered by \cite{SemenovPRB07}, whose numerical estimates predicted a spin relaxation time $\sim 1$~s. Relevant for quantum dot experiments in nanotubes, \onlinecite{YazyevNL08, FischerPRB09} inspected the role of dipolar and Fermi contact interaction in 
$sp$-hybridized nanostructures, resulting in an interesting interplay between isotropic and anisotropic hyperfine interactions. 

\begin{figure}[b]
\center 
\includegraphics[width=7cm]{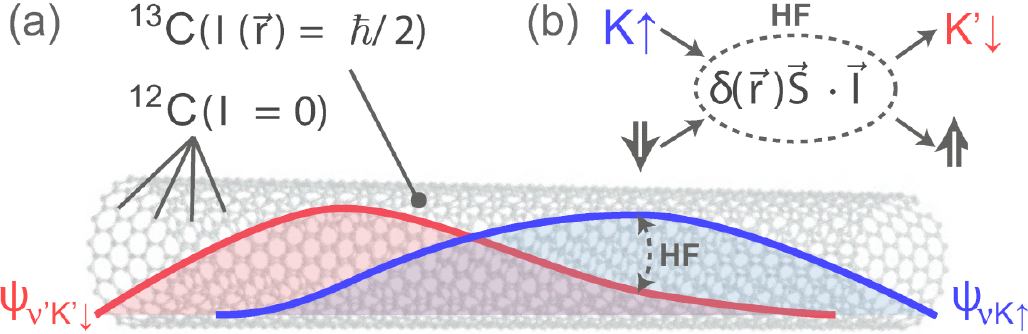}
\caption{\footnotesize{(Color online) 
(a) Illustration of two localized electronic wave functions with different valley,  longitudinal and spin quantum numbers. Hyperfine interaction with a \thirteen\ atom (HF) can scatter between them, because of the local nature of the scattering vertex~(b). This mechanism allows lifting of spin and/or valley-based Pauli blockade, if conditions of energy conservation are fulfilled. 
}}
\label{hyperfinetheory}
\end{figure}

In the tight-binding picture of \cite{PalyiPRB09}, the hyperfine interaction with \thirteen\ is modeled by a matrix element that arises on the site of each nuclear spin (Fig.~\ref{hyperfinetheory}). Because it acts locally at each atomic lattice site that contains a \thirteen, hyperfine interaction couples not only to the electron spin, but also mixes the valley index (cf. discussion of valley scattering in Fig.~\ref{fig_valleyvalley}).
The strengths of the valley-conserving and valley-mixing parts of the effective hyperfine coupling were estimated to have the same order of magnitude. Pauli blockade, even in its strongest form protected by spin and valley (Sec.~\ref{subsubspinvalleyblockade}), can therefore be lifted by the presence of \thirteen\ atoms. \onlinecite{PalyiPRB09} considered the situation where valley scattering is dominated by hyperfine coupling.
Ignoring spin-orbit coupling, they showed that the leakage current in the Pauli blockade regime of a double quantum dot is only strongly suppressed if both valley and spin splittings are larger than the hyperfine coupling. 

The other limit, where valley scattering is dominated by disorder-induced valley scattering, was considered subsequenctly~\cite{PalyiPRB10}. Ignoring hyperfine coupling, but assuming strong SOI (larger than the splittings due to disorder and interdot tunneling), it was predicted that the current in the Pauli blockade regime can show a dip at low fields. Although similar to experimental data discussed in Fig.~\ref{Chap4LeakagePeaks}d, the amplitude of the predicted dip is orders of magnitude smaller than observed. In the theory, the low-field dip occurs because a difference in the valley coupling splittings diminishes the matrix element for tunneling, while at high fields the valley mixing is suppressed.


\subsubsection{Experiment}

\begin{figure}[]
\center 
\includegraphics{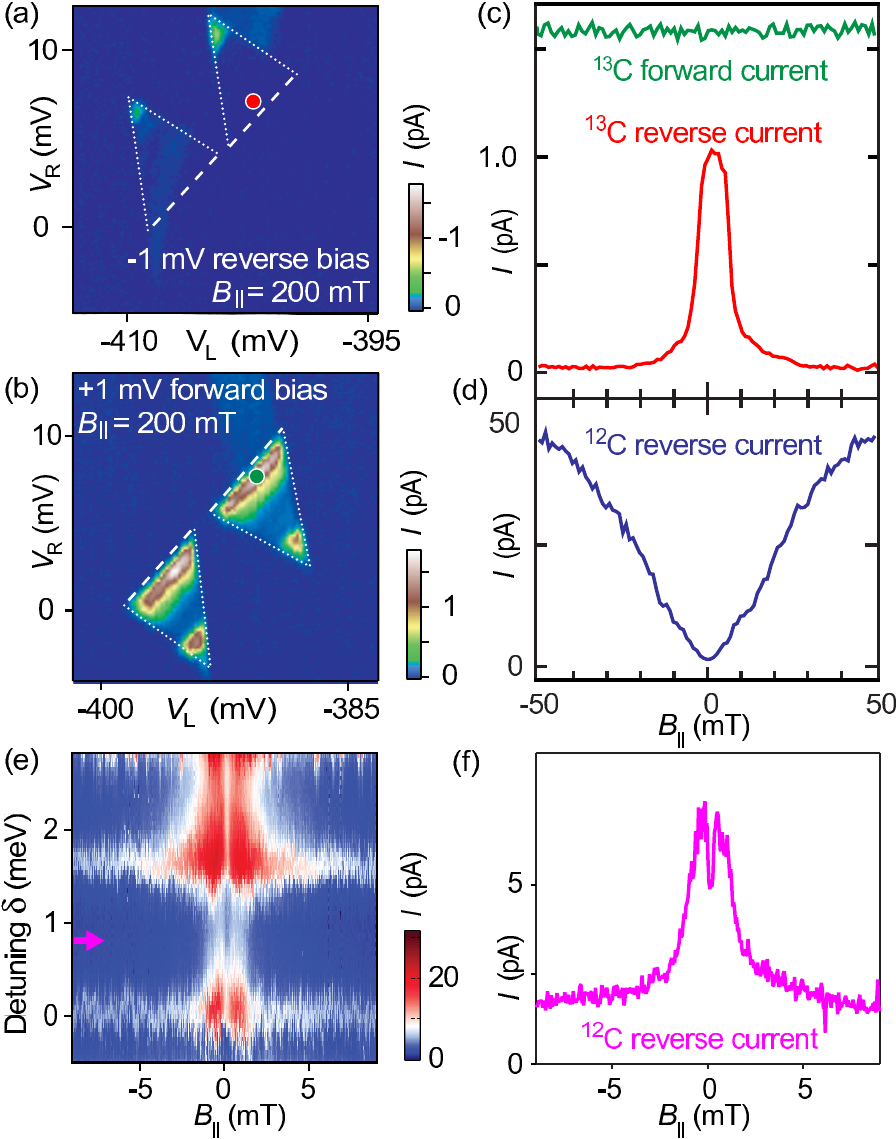}
\caption{\footnotesize{(Color online) 
(a) Current through a top-gated multi-electron double dot formed in a narrow-gap \thirteen\ nanotube. Near the base of the bias triangles (ground state-to-ground state transition) no current is observed for reverse bias. 
(b) Current at forward bias. 
(c) Magnetic field dependence of current in reverse and forward bias measured near base of a triangle (dots in (a,b)). The peak in reverse current at $B_{||}=0$ is attributed to hyperfine-mediated relaxation.
(d)~In a similar device formed in a predominantly \twelve\ nanotube, the opposite magnetic field dependence is observed: The reverse current shows a minimum at $B_{||} = 0$, presumably a consequence of Van Vleck cancellation.
(e)~Leakage current in a different device with natural abundance of \thirteen\ versus detuning and $B_{||}$. This peak, measured in the spin-valley blockade regime of Fig.~\ref{spinvalleyblockade}e, is narrower than in (c), possibly due to the smaller concentration of \thirteen. 
(f) Horizontal cut through~(e) at location of arrow, showing a small dip at $B_{||}=0$.
Panels (a-d) adapted from \onlinecite{ChurchillNPhys2009}, (e-f) from \onlinecite{PeiNnano12}.
}}
\label{Chap4LeakagePeaks}
\end{figure} 

Figure~\ref{Chap4LeakagePeaks} shows electron transport through weakly tunnel-coupled, Pauli blocked double dots. 
Panels (a-b) show the asymmetry in forward and reverse bias for a nanotube enriched with \thirteen, similar to the Pauli blocked \twelve\ device discussed in Fig.~\ref{Chap4SpinValleyExp_FK}. Figure~\ref{Chap4LeakagePeaks}(c) compares the magnetic field dependence of the reverse-bias leakage current near the base of the triangle with the forward current in the same tuning.
Whereas the forward current is independent of applied magnetic field (indicating that the rate-limiting tunnel barrier,~$\Gamma_\mathrm{M}$, is independent of magnetic field), the reverse leakage current is strongly suppressed only above a characteristic magnetic field scale, $B_\mathrm{C}\sim6$~mT (Fig.~\ref{Chap4LeakagePeaks}(c)).

\onlinecite{ChurchillNPhys2009} attributed the current peak at $B=0$ to spin relaxation via electron-nuclear flip-flops, similar to the situation in GaAs double dots and InAs nanowires \cite{KoppensSci05,PergePRB2010}. 
Because of the mismatch of electron and nuclear magnetic moments, these energy-conserving flip-flop processes are expected to be strongly suppressed once the difference in Zeeman splitting exceeds the strength of the hyperfine coupling $B_\mathrm{nuc}$. Estimating the number of nuclei, $N$, electron g-factor $g_\mathrm{e}\sim2 \gg g_\mathrm{nuc}$, and assuming uniform coupling to a Gaussian-distributed Overhauser field,  $g\mu_BB_{\rm nuc}=\mathcal{A}/\sqrt{N}$, an effective hyperfine coupling constant $\mathcal{A}\sim1$--$2\times 10^{-4}$~eV can be estimated. This is two orders of magnitude larger than predicted for nanotubes \cite{YazyevNL08,FischerPRB09} or measured in fullerenes \cite{PenningtonRMP96}. 

This puzzling result stimulated theoretical work by \onlinecite{CoishPRB2011}, who examined the role of thermally activated spin-flip cotunneling in lifting spin blockade. Their theory does not require large hyperfine coupling constants but predicts a peak width (set by temperature, here approximately $100$ mK) that is too large. Matching the width in Fig.~\ref{Chap4LeakagePeaks}(c) would require a temperature below $10$ mK for $g = 2$. This large value of the hyperfine interaction remains unexplained. It is, however, consistent with a subsequent measurement of the dephasing time in a \thirteen\ double-dot device \cite{ChurchillNPhys2009}, discussed in the next section.

A strikingly different field dependence of the leakage current through a Pauli-blocked \twelve\ double dot\footnote{i.e.\ with natural abundance \thirteen.} is shown in Fig.~\ref{Chap4LeakagePeaks}(d): The current shows a minimum at $B=0$, whose width depends on the interdot tunnel coupling. Such behavior was seen in both~\twelve\ and \thirteen\ devices, particularly for stronger interdot tunnelling. This is at first sight consistent with the predictions of \onlinecite{PalyiPRB10}, which give a peak or a dip depending on specific device parameters. However, the observed ratio of low and high field currents was~50, rather than~1.5 as predicted\footnote{The observed peak width in the regime of Fig.~\ref{Chap4LeakagePeaks}c cannot be explained by this theory either, because it did not depend on interdot tunneling, whereas the predicted peak width does.}. 

An alternative explanation of the large dip is spin relaxation mediated by phonon and spin-orbit interaction. Because spin-orbit coupling is even under time reversal, one-phonon processes cannot mediate a coupling between time-conjugate states (so-called van Vleck cancellation, similar to electric dipole transitions). This leads to suppressed spin relaxation near $B=0$ as discussed in Sec.~\ref{subspinrelaxation} \cite{VleckPR40,KhaetskiiPRB01}. 
However, this theory does not lead to a good quantitative fit to the observed dips.


Figure~\ref{Chap4LeakagePeaks}(e) shows the leakage current of a different device, namely the one presented in Fig.~\ref{spinvalleyblockade}(c), zoomed in to low fields and small detuning. 
At first sight, the magnetic field dependence resembles that of Fig.~\ref{Chap4LeakagePeaks}(c), with a peak width that is approximately ten times smaller. Noting that this device had a natural abundance of \thirteen\ ($\sim$ 1 $\%$), and that the effective hyperfine coupling scales with the square root of the \thirteen\ concentration, this data corroborates with the hyperfine coupling measured by \cite{ChurchillNPhys2009}. On second sight, a small splitting in the leakage current is evident near $B=0$ (Fig.~\ref{Chap4LeakagePeaks}(f)), indicating that a full understanding of this system has not been reached yet. As discussed in \cite[supplement]{PeiNnano12}, the splitting of the peak could arise from combinations of exchange interaction, spin-orbit coupling, and hyperfine coupling. 


Looking beyond nanotubes, we note that dips \emph{and} peaks of the leakage current at $B=0$ have been observed and discussed in InAs nanowires \cite{PfundPRL2007,PergePRB2010} and silicon quantum dots \cite{LaiSciRep2011,YamahataPRB2012}.

\subsection{Open Questions}
\label{openquestionsdoubledots}
\begin{itemize}
\item The reproducibility of quantum properties, and the variability of device characteristics among nanotubes of identical chirality, has yet to be established. 
It is experimentally unverified to what extent right-handed and left-handed species (i.e. inversion isomers of chiral nanotubes) display the same properties. 
Similarly, the robustness of the valley index (isospin) for armchair-like and zigzag-like nanotubes has not been checked experimentally.
\item Although Pauli blockade has been observed in different device geometries by several groups, it is not as well established as in III-V double dots. Several nanotube double dots showed no Pauli blockade\footnote{See for example \cite{MasonSci04,GraberPRB06,JorgensenNatPhys08,SapmazNL06,JungNL2013}.}, which can be attributed to differences in dielectric surrounding and interaction effects or disorder.
\item Relaxation times associated with valley, spin, or combined spin-valley relaxation in single and double quantum dots have not yet been measured systematically, including as a function of magnetic field. Existing experiments are discussed in Sec.~\ref{sec_qubits}.
\item The type of hyperfine interaction (Fermi contact term vs dipolar) has not been studied experimentally. The unexplained strength of hyperfine interaction inferred from one study by \onlinecite{ChurchillNPhys2009} remains to be confirmed. The lifting of Pauli blockade near $B=0$ may have alternative explanations, but we are not aware of any that are consistent with experimental conditions. The short dephasing time measured in \thirteen\ devices \cite{ChurchillPRL2009} is consistent with a large hyperfine coupling, but may originate from mechanisms unrelated to hyperfine coupling. \onlinecite{LairdNnano2013} measured a comparatively short dephasing time in predominantly \twelve\ nanotubes.
\item In Appendix~\ref{appendixtwoelectronstatesPauli} we argue that Pauli blockade protected by ``spin and valley" strictly speaking does not exist. However, a single spin flip or valley flip does not conserve energy due to spin-orbit coupling, possibly making spin-and-valley flips the dominant relaxation process. This underlines the importance of understanding both spin and valley for any quantum device based on carbon nanotubes.
\end{itemize}

\section{Spin valley coherence}\label{qubits}
\label{sec_qubits}
By studying the decay of spin and valley states, we can use them as delicate probes of their environment.  In this section, we discuss different ways for these states to decay, and show how they can be used as quantum bits. We focus especially on the interactions of electron spins with phonons and with magnetic nuclei.

We will discuss three distinct decay processes of quantum states, known as  relaxation, dephasing and decoherence. Relaxation (characterized by time $T_1$) describes the equilibration of population between two quantum states. Dephasing (characterized by time $T_2^*$) describes loss of phase information in a quantum superposition. The main mechanism by which this happens is through fluctuations of the quantum energy splitting leading to accumulation of random phases. Decoherence describes the loss of phase information when slowly varying fluctuations are removed by dynamical decoupling. For the simplest decoupling scheme, Hahn echo~(Sec.~\ref{sec_coherentmanipulation}), this is characterized by a time $T_\mathrm{echo}$ which is generally longer than $T_2^*$. For fuller discussion of spin qubits in single and double quantum dots, see~\cite{HansonRMP2007, Ihn2010}, which extensively reference the many experiments in other materials (principally GaAs). 


\subsection{Spin and valley coupling to phonons}
\label{subspinrelaxation}

\subsubsection{Theory}

Because the mechanical motion of nanotubes perturbs the confining potential of quantum dots, it couples distinct electron charge states. Through spin-orbit interaction, spin-valley states are also coupled to mechanical motion. This is most clearly evident as a relaxation channel for spin-valley states; excited states can decay by phonon emission, with a rate that depends on the coupling strength and the phonon density of states.

There are four types of phonon mode in nanotubes: radial breathing, twist, longitudinal and bending modes~\cite{MarianiPRB2009}. These couple to spin-valley states through two general coupling mechanisms. Deformation-potential coupling perturbs the bandstructure and, combined with spin-orbit coupling, induces spin flips~\cite{BulaevPRB2008}. Deflection coupling changes the alignment of the nanotube to the magnetic field, thereby coupling spin and valley through the anisotropy of the valley magnetic moment~\cite{BorysenkoPRB2008, RudnerPRB2010}. Although deformation-potential coupling is present for all four mode types, deflection coupling arises only from bending modes. Nevertheless, deflection coupling is calculated to be the dominant mechanism for phonon-mediated spin relaxation at low energy~\cite{RudnerPRB2010}.

Considering these mechanisms, several statements can be made about the expected $T_1$ between different valley-spin states as a function of magnetic field:
\begin{enumerate}
\item{Relaxation between time-conjugate states is suppressed at low magnetic fields due to van Vleck cancellation~\cite{KhaetskiiPRB01}. This is a consequence of time-reversal symmetry, and applies to relaxation within a Kramers doublet.}
\item{Relaxation between non-time-conjugate states occurs fastest when they are close together in energy. The reason is that the dispersion relation for bending-mode phonons, $\omega(k)\propto k^2$, leads  to a density of states $dk/d\omega\propto 1/\sqrt{\omega}$ which is maximal at $\omega \rightarrow 0$. This is in contrast to higher-dimensional systems, where the density of states is constant or increases with energy~\cite{BulaevPRB2008,RudnerPRB2010}.}
\item{The relaxation rate between two states is a non-monotonic function of their energy splitting, owing to interference between different contributions to the electron-phonon coupling. Interference is predicted between contributions from discrete and continuous phonon modes, as well as due to the match or mismatch of phonon wavelength with the wavelength of a confined electron.  These interference oscillations should be evident in the dependence of $T_1$ on magnetic field~\cite{BulaevPRB2008}}.
\end{enumerate}

\subsubsection{Experiment}

\begin{figure}[]
\center
\includegraphics{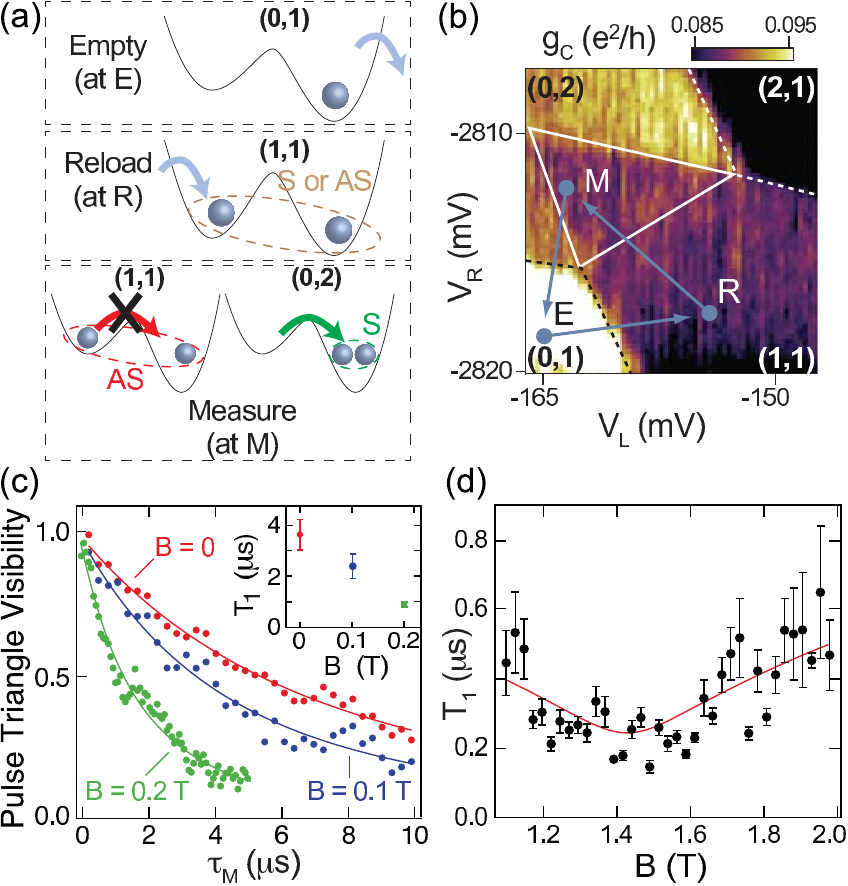}
\caption{\footnotesize{(Color online) Measurement of spin-valley $T_1$.
(a) Pulse cycle for measuring relaxation from AS to S states (see text). (b)~Stability diagram close to the (1,1)-(0,2) transition with pulse cycle $E\rightarrow R\rightarrow M \rightarrow E$ and $\tau_\mathrm{M}=0.5~\mu s$. Color represents time-average charge sensor conductance. Inside the ``pulse triangle'' (marked), Pauli blockade leads to an occupancy of (1,1) for metastable \AS states, yielding a reduced sensor conductance. (c)~Decay of pulse triangle visibility (points) as a function of $\tau_\mathrm{M}$, measured for three magnetic field values.  Lines are fits from which $T_1$ values (inset) are extracted. (d) Points: $T_1$ as a function of $B$ close to the upper spin-orbit anticrossing ($K'{\uparrow}-K'{\downarrow}$); Line: fit of the form $T_1 \propto \sqrt {\Delta E}$. Adapted from~\onlinecite{ChurchillPRL2009}.}}
\label{fig_churchillrelaxation}
\end{figure}

Spin-valley relaxation has been measured in the device of Fig.~\ref{topandbottomgating}(a,b). This device, fabricated from a  \thirteen\ nanotube, incorporates a double quantum dot and a nearby charge sensor, coupled via a floating coupling antenna, whose conductance is sensitive to the charge occupancy of the double dot. Relaxation is studied by preparing an \AS two-electron state, Pauli blocked in (1,1), and using the charge sensor to monitor the time to decay to an unblocked \SSS state~\cite{ChurchillPRL2009}.

The two-electron state is manipulated using a cycle of gate voltage pulses applied to gates L and R (Fig.~\ref{topandbottomgating}(b)), to switch the dot potentials between different configurations~(Fig.~\ref{fig_churchillrelaxation}(a,b)). The cycle~\cite{JohnsonNature2005} begins with the device configured at point E in gate space, where tunneling to the leads prepares the (0,1) configuration. The device is then pulsed to point R in (1,1), where an electron is reloaded into the left dot. Because $\DeltaSAS$ is small in the (1,1) configuration (left side of Fig.~\ref{STblockadeMatrix}(d)), the two-electron state after reloading can be either \SSS or \AS. For readout, the device is quickly pulsed to point M, corresponding to the right side of Fig.~\ref{STblockadeMatrix}(d), where the ground-state configuration is $\SSS(0,2)$. Here a symmetry-to-charge conversion occurs; if the prepared two-electron state was $\SSS$, the left electron will tunnel to the right dot, leading to (0,2) occupancy. However, for \AS states, Pauli blockade enforces occupancy (1,1). This persists for a time $\sim T_1$, until spin-valley relaxation\footnote{We define spin-valley relaxation as relaxation between an \AS and an \SSS state. As discussed in Sec.~\ref{sub_SAStheory}, this can involve a flip of spin, valley, both, or neither (i.e. dephasing only).} causes the \AS state to decay to an \SSS state, whereupon the device relaxes to (0,2) occupancy.

The time-average charge sensor conductance $g_\mathrm{C}$ is monitored with this pulse cycle applied continually. The duration $\tau_\mathrm{M}$ of the third step is chosen to be much longer than that of the others, so that $g_\mathrm{C}$ predominantly reflects the average occupancy at M. For $\tau_\mathrm{M}\ll T_1$, relaxation of the blocked states is negligible, resulting in a large admixture of (1,1) occupancy and corresponding reduced~$g_\mathrm{C}$ in the ``pulse triangle'' region of the stability diagram (Fig.~\ref{fig_churchillrelaxation}(b)). For $\tau_\mathrm{M}>T_1$, this admixture is reduced. By fitting the pulse triangle visibility (defined as the difference between measured $g_\mathrm{C}$ and the value expected for (0,2), normalized to unity at $\tau_\mathrm{M}\rightarrow 0$) as a function of $\tau_\mathrm{M}$, the time $T_1$ can be deduced~(Fig.~\ref{fig_churchillrelaxation}(c)).

As a function of magnetic field directed approximately along the nanotube, $T_1$ is observed to decrease initially, consistent with prediction (1) above (Fig.~\ref{fig_churchillrelaxation}(c) inset). However, $T_1$ shows a minimum at $B \approx 1.4$~T~(Fig.~\ref{fig_churchillrelaxation}(d)), where the two $K'$ states with opposite spin approach each other (as in Fig.~\ref{SO_combined}(d)). This is consistent with prediction (2) above, assuming that during step R an electron is sometimes loaded into a $K'$ state. Neglecting substrate interaction, the relaxation rate is expected to be proportional to the phonon density of states in the nanotube, giving $T_1 \propto 1/\sqrt{\Delta E}$, where $\Delta E$ is the energy difference between the two $K'$ states; taking the proportionality constant as a fit parameter and using the measured field misalignment and $\Delta_\mathrm{SO}$ for this device to calculate $\Delta E$, this prediction is found to be in good agreement with the data~\cite{ChurchillPRL2009}.

The interference oscillations of prediction (3) have not yet been reported. One reason may be that they are sensitive to the confinement potential. Whereas hard-wall confinement should give rise to sharp interference maxima of $T_1$ due to strongly varying overlap of electron and phonon wave functions with energy, soft confinement typical of few-electron devices is expected to lead to less pronounced maxima~\cite{BulaevPRB2008}.

\subsection{Hyperfine mixing of spin states}
\label{sec_hyperfinemixing}

\subsubsection{Theory}

As well as electron-phonon interaction, a major influence on spin evolution in semiconductors is hyperfine interaction with uncontrolled lattice nuclear spins~\cite{HansonRMP2007}.  An electron in a quantum dot interacts with all of the nuclei with which its wave function overlaps; this is equivalent to an effective Zeeman field  $B_\mathrm{nuc}$ that fluctuates slowly about zero due to nuclear spin diffusion. Approximating an equal overlap with all nuclei in the quantum dot, each root-mean-square component of this field is $B_\mathrm{nuc}=\sqrt{N_{13}}\mathcal{A}/N_{\mathrm{\Sigma}}\gs\mu_\mathrm{B}$, where $\mathcal{A}$ is the hyperfine constant and $N_{\mathrm{\Sigma}}$ is the number of nuclei in the dot of which $N_{13}$ are \thirteen. As discussed in Sec.~\ref{doubledots}, the local nature of the hyperfine interaction can cause spin relaxation. Here, we explain that this uncontrolled field also constitutes a major source of spin dephasing~\cite{TaylorPRB2007}, and its strength can be deduced by measuring $T_2^*$.

\subsubsection{Experiment}
\label{hyperfineexperiment}

\begin{figure}[t]
\center
\includegraphics{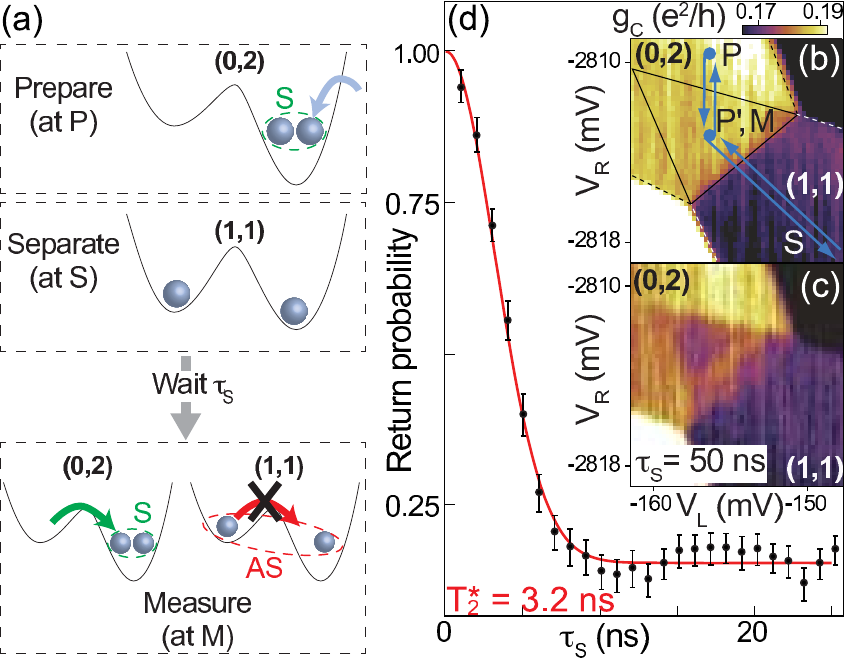}
\caption{\footnotesize{(Color online) Measurement of $T_2^*$ in a \thirteen\ double quantum dot. (a) Pulse cycle to measure mixing between \SSS and \AS states (see text). (b) Stability diagram close to the (1,1)-(0,2) transition, measured via time-averaged charge-sensor conductance. Gate settings at the three steps of the pulse cycle are indicated P, S, and M. (The step at P$'$, not discussed here, was inserted to reduce pulse overshoot.) The triangle marks the region where \AS states are Pauli blocked. (c) Stability diagram with pulses applied and $\tau_\mathrm{S}=50$~ns. The triangle of reduced conductance indicates mixing of \SSS and \AS states leading to higher probability of (1,1) during the measurement part of the cycle. (d) Points: Return probability as a function of $\tau_\mathrm{S}$, deduced from $g_\mathrm{C}$. Line: Gaussian fit giving $T_2^*=3.2$~ns. Adapted from~\onlinecite{ChurchillPRL2009}.}}
\label{dephasing}
\end{figure}

Electron spin dephasing was studied in the same device of Fig.~\ref{topandbottomgating}(b). This was synthesized using 99\% \thirteen, making hyperfine effects particularly strong. Dephasing was detected through the mixing of \SSS and \AS states with the two electrons in separate quantum dots~\cite{ChurchillPRL2009}. Because the random hyperfine field is in general different between the dots, the separated spins precess at different rates, mixing \SSS and \AS states.

The pulse scheme for this experiment~(Fig.~\ref{dephasing}(a-c)) first configures the device in (0,2) at point P, where large~$\DeltaSAS$ causes tunneling to the leads to prepare an \SSS state. By tilting the potential into (1,1) (point~S, corresponding to left of Fig.~\ref{STblockadeMatrix}(d)), the electrons are separated for a time $\tau_S$, during which precession in the hyperfine effective field can mix \SSS and \AS states. Finally, the gate voltages are pulsed back towards a (0,2) configuration (point M, corresponding to right of Fig.~\ref{STblockadeMatrix}(d)) for measurement. As in Fig.~\ref{fig_churchillrelaxation}, \SSS states relax to (0,2) occupancy, whereas \AS states remain blocked in (1,1). From the sensor conductance $g_\mathrm{C}$ in the measurement configuration, averaged over many cycles, the probability of return to (0,2), and hence the degree of $\SSS$-$\AS$ mixing during the separation step can be deduced~\cite{PettaScience2005}.

As a function of $\tau_\mathrm{S}$, the return probability $P(\tau_\mathrm{S})$ decays with characteristic time $T_2^*=3.2$~ns (Fig.~\ref{dephasing}(d)), saturating at a value $P(\infty)\approx 0.17$.  Attributing the observed $T_2^*$ solely to the difference of hyperfine effective field between dots, the effective hyperfine field is given by~$B_\mathrm{nuc}=\hbar/\gs\mu_\mathrm{B}T_2^*=1.8$~mT. This is within a factor~2 of the value deduced from Fig.~\ref{Chap4LeakagePeaks} ($B_{\rm nuc}=3.5\mathrm{~mT}$). However, this apparent agreement again suggests a hyperfine constant $\mathcal{A}$ two orders of magnitude larger than expected theoretically~\cite{YazyevNL08,FischerPRB09}. 

The long-$\tau_\mathrm{S}$ saturation value of the return probability $P(\infty)$ should reflect the level diagram of~Fig.~\ref{STblockadeMatrix}(d). Assuming a large longitudinal level spacing, so that only one shell in each dot needs to be considered, the (1,1) configuration allows 16 states, but the lowest manifold in the (0,2) configuration only six. If mixing is fully incoherent, the saturation probability will then be $P(\infty)=6/16=0.375$. If mixing is coherent, $P(\infty)$ will generally be higher\footnote{For example, in conventional semiconductors incoherent mixing gives $P(\infty)=1/4$ but coherent mixing in a random Overhauser field gives $P(\infty)=1/3$ or $1/2$ depending on magnetic field~\cite{TaylorPRB2007}.}. The case of a clean nanotube ($\DeltaKK=0$ but including spin-orbit coupling) was analyzed by~\cite{ReynosoPRB11}, who calculated the value for a range of specific cases. Depending on whether the system is prepared in its ground state, whether passage through the anticrossings in~Fig.~\ref{STblockadeMatrix}(d) is adiabatic, and depending on the strength of the magnetic field, $P(\infty)$ can be enhanced as high as unity. Including valley mixing makes the situation even more complicated because $\DeltaKK$ may differ between the dots. This gives rise to new avoided crossings in the level scheme of Fig.~\ref{STblockadeMatrix}(d). The speed at which these crossings are passed, set by the detuning sweep rate, is of critical importance~\cite{RibeiroPRL2013,RibeiroPRB2013}. In the simplest case, where the ground state is always prepared and the first crossing is adiabatic, a value $P(\infty)=1/3$ is predicted, with corrections due to non-adiabaticity always positive~\cite{ReynosoPRB12}. The measured $P(\infty)=0.17$ is therefore lower than all theoretical predictions.

\subsection{Qubits}

\subsubsection{Qubit states and the Bloch sphere}
\label{sec_qubitstates}

\begin{figure}
\center 
\includegraphics{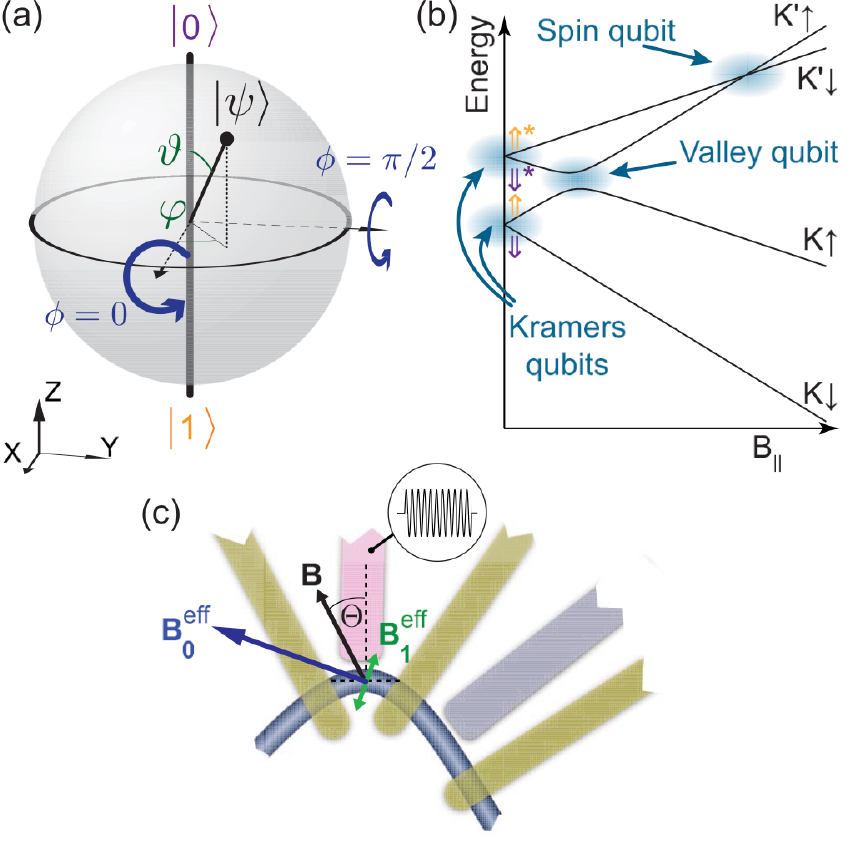}
\caption{\footnotesize{(Color online) (a) Bloch sphere representation of a generic qubit. The qubit state $|\psi\rangle$ is represented by a point with polar coordinates $(\vartheta,\varphi)$, so that north and south poles correspond to the basis states $|0\rangle$ and $|1\rangle$. In the rotating frame, with Cartesian coordinates $(X,Y,Z)$, the Rabi rotations driven by microwave bursts with phase $\phi=0, \pi/2$ are marked. (b) Four possible qubits in the spin-orbit coupled energy levels. (c) Qubit device, driven by an oscillating gate voltage. The applied and effective magnetic fields are indicated. Adapted from~\onlinecite{FlensbergPRB2010}.}}
\label{fig_qubittypes}
\end{figure} 

A quantum bit, or qubit, is a two-level system that can be controlled in a quantum coherent way~\cite{NielsenBook2000}. An intuitive way to represent the state of a qubit is as a point on the surface of the Bloch sphere (Fig.~\ref{fig_qubittypes}(a)). With two orthogonal states of the qubit (for example two spin states) assigned as the basis states $|0\rangle$ and $|1\rangle$, any superposition can be written $|\psi\rangle=\cos \frac{\vartheta}{2}|0\rangle + e^{i\varphi} \sin \frac{\vartheta}{2}|1\rangle$, where the parameters $\vartheta$ and $\varphi$ are polar coordinates representing that state. Any unitary single-qubit operation then corresponds to a rotation about the origin.

How can such rotations be achieved? One technique is by resonant driving at a frequency corresponding to the energy splitting between qubit states~\cite{HansonRMP2007}. In the case of a spin qubit with gyromagnetic ratio $g$ in static magnetic field~$\boldB_0$, a time-varying field $\boldB_1 \cos (2\pi f t + \phi)$ at driving frequency $f=g\mu_\mathrm{B}B_0/h$ and perpendicular to $\boldB_0$ induces transitions with Rabi frequency $f_R=g \mu_\mathrm{B}|B_1|/2h$. In a reference frame rotating with frequency $f$ about the $Z$ axis, the qubit state then precesses at a rate $f_\mathrm{R}$ about an axis in the $XY$ plane set by the phase $\phi$ of the driving field. One can achieve arbitrary rotations by concatenating bursts with appropriate phases.

\subsubsection{Valley, spin and Kramers qubits}

This section identifies various two-level subspaces in the spectrum of a generic, spin-orbit coupled carbon nanotube that can serve as qubits, and discusses how they differ in terms of their quantum numbers, ease of operation, and immunity to electrical or magnetic noise. The reason for focussing on these particular subspaces is that the splitting can be made small enough to allow resonant qubit manipulation using microwave fields with frequencies in the range $f \lesssim 40$~GHz that can readily be generated experimentally.

At low field, either Kramers doublet can be used as a qubit subspace, with the two basis states denoted $\{\Uparrow, \Downarrow\}$ (for one carrier in the shell) and $\{\Uparrow^*, \Downarrow^*\}$ (for two carriers in the shell) to emphasize the isomorphism with a spin qubit. In a parallel magnetic field and neglecting disorder, these are the eigenstates $\{K'{\uparrow}, K{\downarrow}\}$ and $\{K'{\downarrow}, K{\uparrow}\}$, but for general field direction or with disorder, they become entangled states of spin and valley. Experimental manipulation of these qubits, known as Kramers qubits or valley-spin qubits, is described in Sec.~\ref{sec_nanotubeEDSR}-\ref{sec_coherentmanipulation}~\cite{RohlingNJPhys2012}. Kramers degeneracy guarantees that by reducing the magnetic field the qubit splitting can be made as small as desired.

Two other qubits can be defined that have not yet been realized experimentally. A pure spin qubit can be defined between the two states that cross at $B_{||}=B_\mathrm{SO}$, i.e.\ $K'{\uparrow}$ and $K'{\downarrow}$ in Fig.~\ref{fig_qubittypes}. For magnetic field aligned with the nanotube, the energy splitting vanishes at the crossing. However, even quite small misalignment~$\Theta$ leads to an appreciable splitting $\Delta_\Theta$. For example, with $\Delta_\mathrm{SO}=1$~meV and $\Theta=1^{\circ}$, the minimum driving frequency is $f=\Delta_\Theta/h\approx 4.2$~GHz. Alternatively a pure valley qubit can be defined at the anticrossing between states of the same spin ($K'{\uparrow}$ and $K{\uparrow}$ in Fig.~\ref{fig_qubittypes}). However, the minimum driving frequency, set by $\Delta_{KK'}$, can again be substantial, with a typical experimental value  $\Delta_{KK'}=60~\mu$eV leading to minimum $f=\Delta_{KK'}/h \sim 15$~GHz. More importantly, it may be difficult to control $\Delta_{KK'}$ experimentally.

\begin{table}
   \centering
   \begin{tabular}{|l|c|cccc|}
   \vspace{-0.4cm} 
   \hspace{1.6cm}			&   \hspace{1.2cm}	& \hspace{1.2cm}&	\hspace{1.2cm}	&	\hspace{1.2cm}	& \hspace{1cm}\\
   \hline \hline
Qubit						& $B_\mathrm{ac}$ 	& \multicolumn{4}{c}{$E_\mathrm{ac}$ combined with}\\
							&					&	Rashba 	& Bend 			& Disorder 	& Inhom.~$B$\\
\hline
Kramers						& $\checkmark$ 		& $\checkmark$	& \hspace{0.5cm}$\checkmark^{\mathrm(E)}$ & $\checkmark$	& $\checkmark$\\
Spin						& $\checkmark$ 		& $\checkmark$	& $\checkmark$	& 				& $\checkmark$\\
Valley						& 			 		&		 		& 				& $\checkmark$	& \hspace{0.1cm}$\checkmark$\footnote{if $B$ varies on sublattice lengthscale, e.g. from magnetic impurities.}\\
	 \hline \hline
   \end{tabular}
\caption{Comparison of qubit types from Fig.~\ref{fig_qubittypes}. Th expected resonant driving mechanisms are marked $\checkmark$; the mechanisms considered are oscillating magnetic field and an oscillating electric field $E_\mathrm{ac}$ combined with extrinsic Rashba spin-orbit coupling, a bend, inhomogeneous static disorder, and inhomogeneous static magnetic field. Only the bend-mediated Kramers qubit (marked (E)) has been clearly demonstrated experimentally~\cite{LairdNnano2013}.}
\label{tab_qubits}
\end{table}

These various qubits can be manipulated using time-varying electric and magnetic fields. Techniques for driving single-qubit operations have been developed extensively in GaAs quantum dots and many should be applicable to nanotubes~\cite{HansonRMP2007}. Several schemes have been proposed. The conceptually simplest is to use an  alternating magnetic field as in Sec.~\ref{sec_qubitstates}. This should work for the spin and Kramers qubits, but not for the valley qubit, because it has no magnetic moment in the perpendicular direction. Since time-varying magnetic fields are hard to generate in nanostructures,  schemes have been suggested based on time-varying electric fields. The common principle is that moving the electron back and forth leads  to an effective magnetic field mediated by spin-orbit coupling. For example, the Rashba-like coupling discussed in Sec.~\ref{sec_Rashba} is equivalent to a momentum-dependent perpendicular magnetic field and can be used to manipulate both spin and Kramers qubits~\cite{KlinovajaPRB2011}. The corresponding Rabi frequency is however rather low ($f_\mathrm{R}\sim 5$~MHz), making the driving inefficient in the presence of decoherence. A stronger coupling can be achieved in the presence of a bend, which gives a position-dependent effective magnetic field~\cite{FlensbergPRB2010} and mediates the only coherent control so far clearly achieved in nanotubes (see next section). Similarly, coupling to a true inhomogeneous magnetic field has been suggested~\cite{SzechenyiArXiv2013}. Another proposal relies on inhomogeneity of the disorder parameter $\Delta_{KK'}$. Although random, if this inhomogeneity is static it should allow driving of both Kramers and valley qubits~\cite{PalyiPRL2011, SzechenyiArXiv2013}. These possibilities are summarized in Table~\ref{tab_qubits}.

As well as the driving mechanism, important considerations are dephasing and decoherence, which limit the coherence time and reduce the fidelity of gate operations. In general, a qubit suffers decoherence through every channel by which it can be driven. As discussed in~Sec.~\ref{subspinrelaxation}-\ref{sec_hyperfinemixing}, the dominant decoherence sources in nanotubes are expected to be random time-varying electric fields (e.g.\ from gate noise and nearby charge switchers) and hyperfine coupling to~$^{13}$C spins, which acts as an effective time-varying magnetic field on individual lattice sites. The coherence properties depend on the strength and power spectrum of the various noise sources. 

\begin{figure}
\center 
\includegraphics{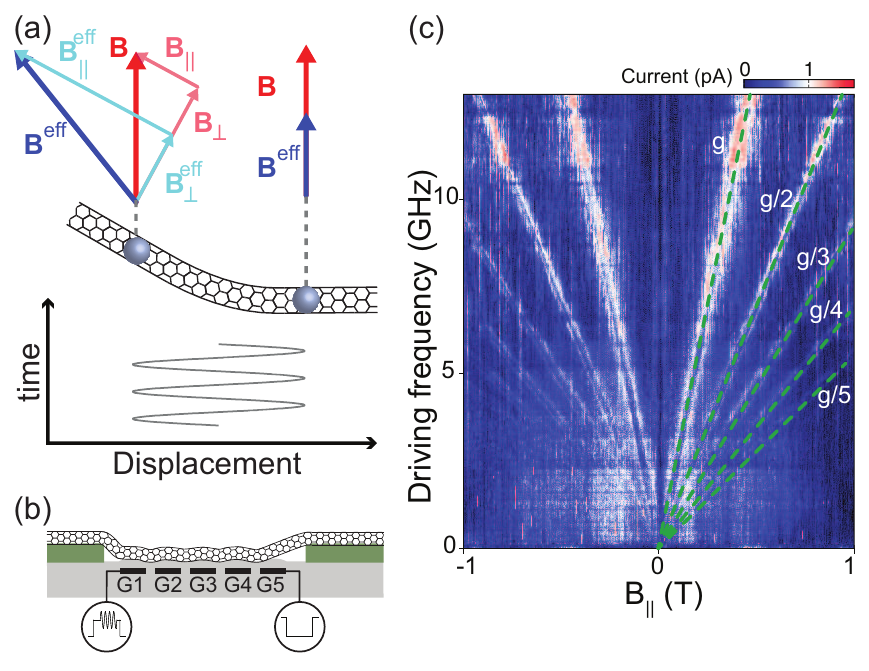}
\caption{\footnotesize{(Color online) (a) Principle of bend-mediated EDSR. An electron driven across a bend experiences a time-varying effective magnetic field which induces coherent qubit precession. (b) Schematic of bent nanotube double quantum dot.  Gates G1-G5 define the confinement potential and carry the time-dependent manipulation voltages. (c) Pauli blockade leakage current in a highly disordered device~($\Delta_{KK'}\gg\Delta_\mathrm{SO}$), showing resonance lines at $g\approx2$ and at subharmonics. To make the resonance clearer, the mean current at each frequency is subtracted. Adapted from ~\onlinecite{LairdNnano2013},~\onlinecite{PeiNnano12}.}}
\label{fig_EDSR}
\end{figure} 

\subsubsection{Electrically driven spin resonance in nanotubes}
\label{sec_nanotubeEDSR}

We now focus on the Kramers qubit, which has been experimentally demonstrated. Using bends to mediate qubit control was proposed by~\cite{FlensbergPRB2010} and realised by~\cite{LairdNnano2013}. It relies on the anisotropic splitting of the Kramers doublets with magnetic field (Fig~\ref{SO_combined}(d)). Each qubit can be regarded as an effective spin-1/2, with spin vector $\bm{s}^*$ whose components, just like those of the real spin, have eigenvalues defined as $\pm 1$. Unlike the real spin, the Zeeman splitting of this effective spin depends on field angle; the parallel and perpendicular components of the $\gtensor$-tensor are:
\begin{eqnarray}
g_{||}	&=& g_\mathrm{s} \mp \frac{g_\mathrm{orb}\Delta_\mathrm{SO}}{\sqrt{\Delta_{KK'}^2+\Delta_\mathrm{SO}^2}}\\
g_\perp	&=& \frac{g_\mathrm{s}\Delta_{KK'}}{\sqrt{\Delta_{KK'}^2+\Delta_\mathrm{SO}^2 }}.
\end{eqnarray}
where upper and lower signs correspond to starred and unstarred doublets respectively. The effective Zeeman Hamiltonian is then:
\begin{equation}
H_\mathrm{eff}=\frac{1}{2}\mu_\mathrm{B}\mathbf{s}^*\cdot \boldB_\mathrm{eff},
\label{eq_def_Beff}
\end{equation}
where $\boldB_\mathrm{eff}$ is an effective magnetic field, defined as the tensor product $\boldB_\mathrm{eff}\equiv\gtensor \otimes \boldB$,
about which $\mathbf{s}^*$ precesses. A geometric interpretation of~Eq.~(\ref{eq_def_Beff}) is shown in~Fig.~\ref{fig_EDSR}(a). When $\boldB$ is applied perpendicular to the nanotube (right side of figure), $\boldB_\mathrm{eff}$ is parallel to $\boldB$. However, when $\boldB$ is applied at an angle because the nanotube is bent (left side of figure), the parallel and perpendicular components couple differently, leading to a tilted~$\boldB_\mathrm{eff}$.

By applying a microwave electric field to a bent nanotube, a quantum dot can be driven back and forth across the bend, experiencing an effective magnetic field that contains both a static component and a perpendicular oscillating component\footnote{Because the confinement energy is usually much larger than the effective Zeeman splitting, to a good approximation the electron experiences the average $\boldB_\mathrm{eff}$ over the entire dot.}. Thus the electric field drives transitions between the two qubit states. Because these two states do not have the same spin, this is a form of electrically driven spin resonance (EDSR).

Detection of EDSR is by measuring the current through a double quantum dot configured in a Pauli-blocked configuration~(Fig.~\ref{fig_EDSR}(c)). With microwaves applied, a peak in current is observed at the resonance condition $f=g\mu_\mathrm{B}B/h$, indicating spin mixing by EDSR. The measured value $g\approx 2$ presumably arises because this device was highly disordered ($\Delta_{KK'}\gg\Delta_\mathrm{SO}$) and/or in the many-carrier limit, consistent with irregular Coulomb spacings seen in transport data (not shown). As well as the main resonance, a series of subharmonics at integer frequency fractions are seen, due to anharmonicity of the confinement potential or disorder~\cite{NowakPRB2012,SzechenyiArXiv2013a}. The resonant current increases with field as expected from~Eq.~(\ref{eq_def_Beff}) because~$\boldB_\mathrm{eff}$ and hence $f_\mathrm{R}$ becomes larger. 

\subsubsection{Qubit manipulation and characterization}
\label{sec_coherentmanipulation}

\begin{figure}
\center 
\includegraphics{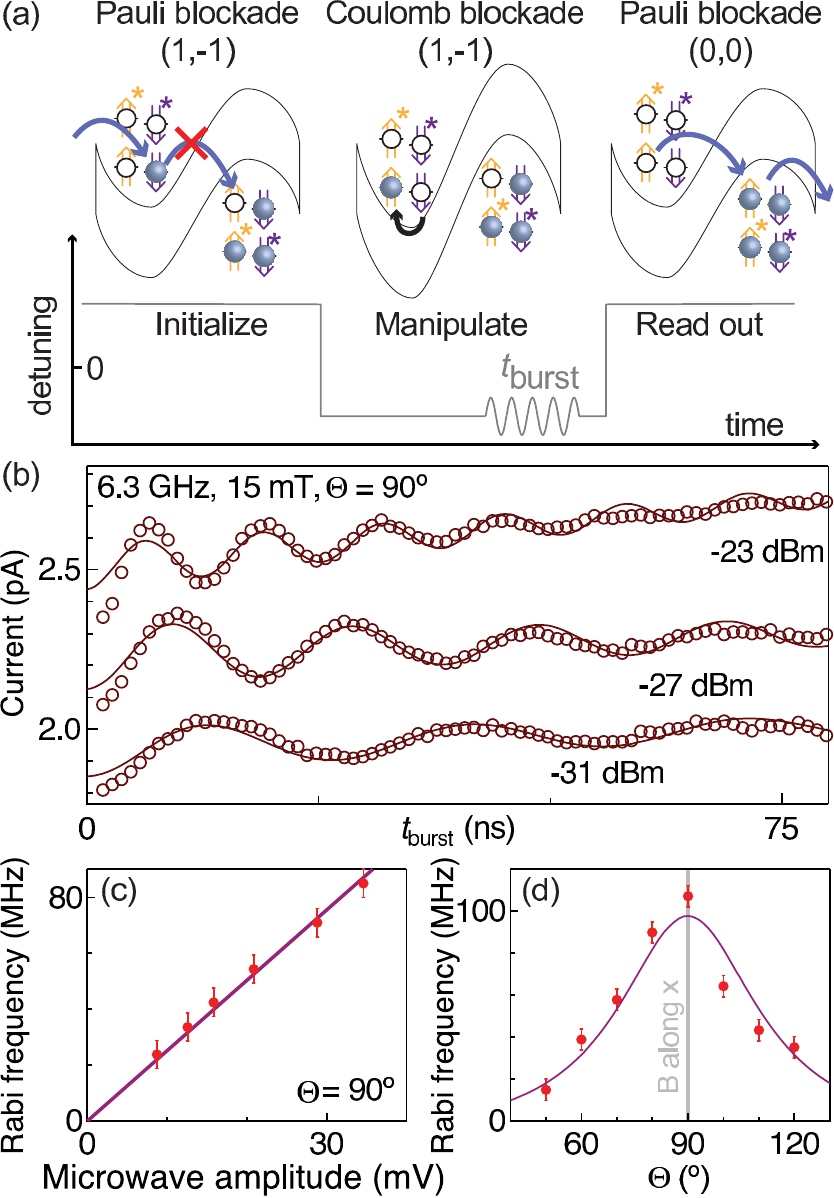}
\caption{\footnotesize{(Color online) (a) Pulse sequence used for coherent qubit manipulation. After initialization in a Pauli-blocked qubit state, the device is pulsed into Coulomb blockade to allow qubit manipulation and returned to Pauli blockade for readout. Electron tunneling occurs only if a qubit was flipped during the manipulation step. Repeated with a period $\sim 1~\mu$s, this leads to a current proportional to the qubit flip probability during the manipulation step. (b) Current (points) as a function of microwave burst duration for various applied powers, showing coherent Rabi rotations. The curves are fits to a model assuming a slowly varying random qubit detuning, due to e.g.~charge noise. Traces are offset for clarity. (c) Rabi frequency as a function of microwave amplitude, with fit showing the expected proportionality. (d) Rabi frequency at constant microwave amplitude and power, showing a dependence on field angle $\Theta$ consistent with bend-mediated EDSR. Adapted from~\onlinecite{LairdNnano2013}.}}
\label{fig_Rabi}
\end{figure}

The data of Fig.~\ref{fig_EDSR} shows only state mixing. Quantum coherence is demonstrated by measuring Rabi oscillations~\cite{HansonRMP2007}. This was achieved in a less disordered device using a pulsed measurement protocol that shuts off current while microwaves are applied~(Fig.~\ref{fig_Rabi}(a)). The scheme operates near the $(1,-1)\rightarrow(0,0)$ charge transition, where the qubit states participating in transport are $(\Uparrow,\Downarrow)$. Gate voltage pulses are used to adjust the detuning between two configurations, one Pauli-blocked where electron tunneling is selective on the qubit state, and one Coulomb-blocked where all tunneling is forbidden~\cite{KoppensNature2006}. The sequence has three stages. An initialization stage at a Pauli blockade configuration loads with high probability a parallel two-qubit state, (e.g. $\Downarrow \Downarrow$). The device is then configured in Coulomb blockade, where a qubit manipulation microwave burst is applied, possibly flipping one of the qubits. During this step, tunneling is energetically suppressed regardless of the spin state. Finally the configuration is returned to Pauli blockade. If no qubit flip occurred during the manipulation stage, the state remains blocked. However, if a qubit (in either dot) was flipped, tunneling will occur based on the overlap of the electron state on the left with the empty state on the right. Repeating this cycle many times, the time-average current is proportional to the qubit flip probability during the manipulation stage.

As a function of burst duration, this current is observed to oscillate (Fig.~\ref{fig_Rabi}(b)), indicating coherent rotations between qubit states at Rabi frequency $f_\mathrm{R}$. The fitted $f_\mathrm{R}$ is proportional to the driving microwave amplitude~(Fig.~\ref{fig_Rabi}(c)), consistent with a harmonic confinement potential and a smooth bend. The dependence of $f_R$ on field angle $\Theta$ is consistent with bend-mediated EDSR coupling ~\cite{FlensbergPRB2010} but not with e.g.~Rashba-mediated coupling, suggesting that the bend is indeed the dominant EDSR mechanism in this device.

\begin{figure}
\center 
\includegraphics{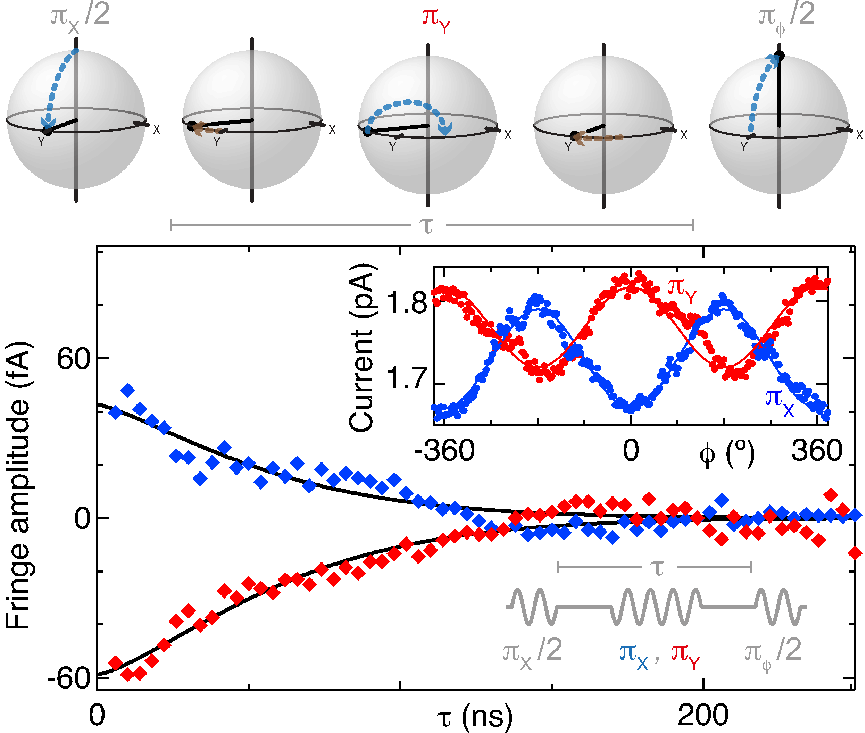}
\caption{\footnotesize{(Color online) Hahn echo amplitude (symbols) as a function of $\tau$ plotted for two phases of the central burst, together with fits (lines) of the form $e^{-(\tau/T_\mathrm{echo})^\alpha}$, yielding $T_\mathrm{echo}=65$~ns. Inset: example fringes as a function of $\phi$. The manipulation sequence is sketched at bottom right, and Bloch spheres along the top illustrate the resulting state evolution. Adapted from~\onlinecite{LairdNnano2013}.}}
\label{fig_Ramsey}
\end{figure} 

The qubit is characterized further by determining the coherence time $T_\mathrm{echo}$~\cite{HansonRMP2007} which characterizes how long a superposition can be preserved by the use of a Hahn echo pulse. 
The coherence time $T_\mathrm{echo}$ is measured by a Ramsey fringe experiment~(Fig.~\ref{fig_Ramsey}), which consists of (1) a $\pi/2$ rotation about $X$ to create a state on the equator; (2) a wait of duration $\tau$, with a $\pi$ rotation about $X$ or $Y$ inserted half way; (3) a~$\pi/2$ rotation with phase $\phi$. Neglecting decoherence, the three rotations interfere to give a qubit flip probability proportional to~$1\pm\cos\phi$~(Fig.~\ref{fig_Ramsey} inset). However, for $\tau \gg T_\mathrm{echo}$, phase information is lost during the wait step, and the qubit flip probability is 1/2 independent of $\phi$. By fitting the decay of fringe amplitude versus $\tau$, the decay time $T_\mathrm{echo} \approx 65$~ns is measured~(Fig.~\ref{fig_Ramsey}). This coherence time $T_\mathrm{echo}$ is distinct from the dephasing time $T_2^*$ discussed in the previous section because the $\pi$ rotation during step (2) makes it insensitive to slow fluctuations of the qubit splitting. The dephasing time of this qubit was measured by a similar method (not shown) and found to be $T_2^*\approx 8$~ns~\cite{LairdNnano2013}. 

Both $T_2^*$ and $T_\mathrm{echo}$ are quite short compared with some other semiconductor spin qubits, such as GaAs~\cite{PettaScience2005, GreilichScience2006, KoppensPRL2008, BluhmNPhys2011}, Si~\cite{MauneNature2012, PlaNature2012}, or diamond~\cite{LangeScience2010}, but similar to results in InAs and InSb nanowires~\cite{NadjPergeNature2010, BergPRL2013}. The measured $T_2^*$ is approximately consistent with  hyperfine dephasing, given the unexpectedly large coupling discussed in Sec.~\ref{hyperfineexperiment}~\cite{ChurchillNPhys2009}, but for $T_\mathrm{echo}$ to be limited in the same way, nuclear spin diffusion would have to be much faster than e.g. in GaAs. This would be surprising, given the one-dimensional geometry and low density of nuclear spins in a nanotube. Fuller consideration of mechanisms led to the tentative conclusion that $T_\mathrm{echo}$ and perhaps $T_2^*$ are limited by charge noise~\cite{LairdNnano2013}. To explain the ineffectiveness of echo at extending the coherence time (even with longer decoupling sequences), this charge noise would need a broad spectral range, in contrast to the $1/f$ spectrum expected for charge switchers in the substrate~\cite{CywinskiPRB2008}.

\subsection{Open questions}

Although the main results presented in this section are understood, there are still significant unresolved questions. In all these experiments, it is hard to convincingly identify the precise spin-valley states between which transitions occur. Whereas the experiment of Fig.~\ref{fig_churchillrelaxation} is sensitive to all forms of relaxation between \AS and \SSS states, which may be expected in general to have different rates, a single $T_1$ value appears sufficient to fit each decay curve in Fig.~\ref{fig_churchillrelaxation}(c). Likewise, in the experiment of Fig.~\ref{dephasing}, it is not clear how the \SSS and \AS state populations are redistributed by dephasing, and this is probably reflected in the unexplained $P(\infty)$ value discussed in Sec.~\ref{hyperfineexperiment}.


A related mystery comes from the measured EDSR spectra. Although in a disordered many-carrier device, expected resonances with $g\approx 2$ are observed~(Fig.~\ref{fig_EDSR}(c)), the spectrum in a cleaner device is complex and not understood~\cite{LairdNnano2013, LiPRB2014}. Whether the unexplained features relate to the short $T_\mathrm{echo}$ is not known.

An exciting area opened by this work is the possibility to combine the spin degree of freedom with mechanical and optical degrees in clean, suspended nanotubes.
There has already been progress in engineering quantized phonons in nanotubes and studying their interactions with the charge on quantum dots~\cite{SapmazNJPhys2005,SazonovaNature2004,LassagneScience2009,SteeleScience2009,HuttelNL2009,BenyaminiNatPhys2014}. 
Evidence for the discreteness of longitudinal stretching phonon modes, comes from Frank-Condon blockade in suspended nanotubes \cite{SapmazNJPhys2005,LeturcqNphys09}, discussed theoretically by \cite{SapmazPRB2003,FlensbergNJPhys2006,MarianiPRB2009}. By tuning the discrete phonon modes away from resonance with qubit splittings, long qubit lifetimes may be achievable. On the other hand, when the qubit splitting is nearly resonant with a discrete phonon mode, coherent energy exchange should be possible between them, in a solid-state analog of cavity quantum electrodynamics~\cite{PalyiPRL12}. Strong spin-phonon coupling in suspended nanotubes may also enable enhanced sensing of nanotube motion~\cite{OhmAPL12}. Finally, \onlinecite{GallandPRL2008,LiSciRep12} theoretically investigate spin-based mechanics and quantum optics. Using a combination of magnetic fields and optical pumping, they predict high-fidelity all-optical control of electron spins, phonon-induced transparency, and applications in quantum communication.

Finally, we compare this work with the large body of experiments on spin relaxation, dephasing, and diffusion in ensembles of single-walled carbon nanotubes measured via ESR and EDSR spectroscopy at higher temperatures $4-300\mathrm{~K}$ \cite{PetitPRB1997}. These techniques focus on resonances that appear within a few percent of $g=2$ at several Tesla. These results seem to contradict the understanding gained from quantum transport experiments because the $g$-factor is expected to be highly anisotropic in clean nanotubes. Therefore the debate whether such resonances reflect intrinsic spin properties of carbon nanotubes~\cite{KombarakkaranCPL2008,DoraPRL2008}, or defects~\cite{RicePRB2013}, has yet to be settled. Indeed, sufficiently purified nanotubes where removal of the catalyst was confirmed by TEM were found not to yield an ESR signal~\cite{ZakaACSn2010}. The level structure of Fig.~\ref{SO_combined} certainly provides no reason why randomly oriented assemblies of nanotubes should yield resonances at $g\approx2$. The only limit where $g\approx2$ may occur is for samples in which $|\DeltaKK|$ exceeds both $|\DeltaSO|$ and the orbital magnetic field splitting, which is not the case for clean, intrinsic nanotubes. 


\section{Valley physics in open quantum dots}
\label{sec:open}

\subsection{Transport in open regime}
\label{subsec:transportinopenregime}
Previous sections focused on closed quantum dots where low-transparency barriers ensure that transport occurs by sequential tunneling via strongly
confined quantum states. However, nanotube devices with a range of transparencies can be fabricated, allowing the transition from closed to open transport regimes to be studied
\cite{NygardArXiv2001,LiangARPC2005,CaoNmat2005,Grove-RasmussenPhysicaE2007}.
Valley and spin physics again play important roles for highly transmitting devices where the quantum dot states are hybridized with the Fermi seas in the leads. Here we provide an overview of transport mechanisms in open devices and then focus on 
phenomena involving valley physics: valley and SU(4) Kondo effects (Sec.~\ref{subsec:KondoeffectsinCNT}) and level renormalization (Sec.~\ref{subsec:level_renormalization}).

\begin{figure} 
\center
\includegraphics{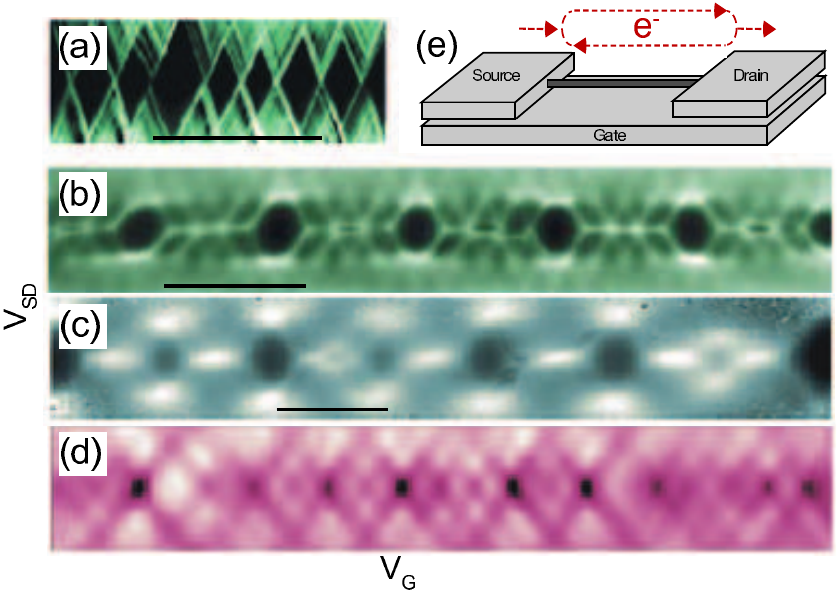} 
\caption{\footnotesize{(Color online) Quantum transport from closed to open devices. (a-d) Plots of ${\mathrm
d}I/{\mathrm d}V_\mathrm{SD}$ as a function of gate voltage $V_\mathrm{G}$ and bias $V_\mathrm{SD}$ for four nanotube devices with average conductances $g \sim 0.01, 0.5, 1.5,3$\,$e^2/h$ (schematic in (e)). All data was taken at 1.5~K and zero field. High (low) conductance is shown by light (dark) colors. Black bars indicate the gate voltage range for addition of four additional electrons. (e) Device schematic. Dashed paths indicate the origin of Fabry-Perot resonances induced by reflections at the contacts. Adapted from \onlinecite{LiangARPC2005}. }}
\label{fig:closed_to_open}
\end{figure}
Figure~\ref{fig:closed_to_open}(a-d) show transport data from four devices with varying contact transparency.
All devices exhibit metallic characteristics at room temperature (not shown) with average conductances ranging from around $0.01e^2/h$ to $3e^2/h$, i.e.\ approaching the
maximum conductance of $4e^2/h$ for a single nanotube.
The first, low conductance device (a) behaves as a closed quantum dot. 
 For the second device (b) with average $g\sim 0.5e^2/h$, i.e.\ close to the conductance quantum, the fourfold periodicity due to valley and spin degeneracy leads to clusters of four peaks (Sec.~\ref{secShells}). The enhanced background conductance reflects co-tunneling processes enabled by the stronger coupling.
 The most extraordinary features are the horizontal (gate independent) ridges of high conductance, e.g.\ occuring near zero bias in a large fraction of the Coulomb diamonds.
 These resonances are due to higher order tunneling processes, including 
 Kondo physics, described in the next section.

For the next devices (c-d) the quantum dot features are smeared out as the increased coupling to leads allows for charge fluctuations on the nanotube.
However, gate-periodic patterns remain and in the highly transmitting device (d) a distinct pattern of low-conductance lines dominates the spectroscopy plot. Here, in
the simplest picture, mode reflections at the contacts~(Fig.~\ref{fig:closed_to_open}(e)) give rise to interference in transmission, so-called
Fabry-Perot resonances \cite{LiangNature2001}.
The interference pattern appearing in bias spectroscopy plots is similar to universal conductance fluctuations (UCF) in other mesoscopic systems \cite{Ihn2010,Nazarov2009}. However,
the randomness that usually characterizes UCF is replaced by nearly perfect periodicity for one-dimensional, ballistic nanotube resonators.

\subsection{Spin, valley and SU(4) Kondo effect in nanotubes}
\label{subsec:KondoeffectsinCNT}

\subsubsection{Theory and background}
\label{subsubsec:Kondotheory}
\begin{figure*} 
\center
\includegraphics[width=16.2cm]{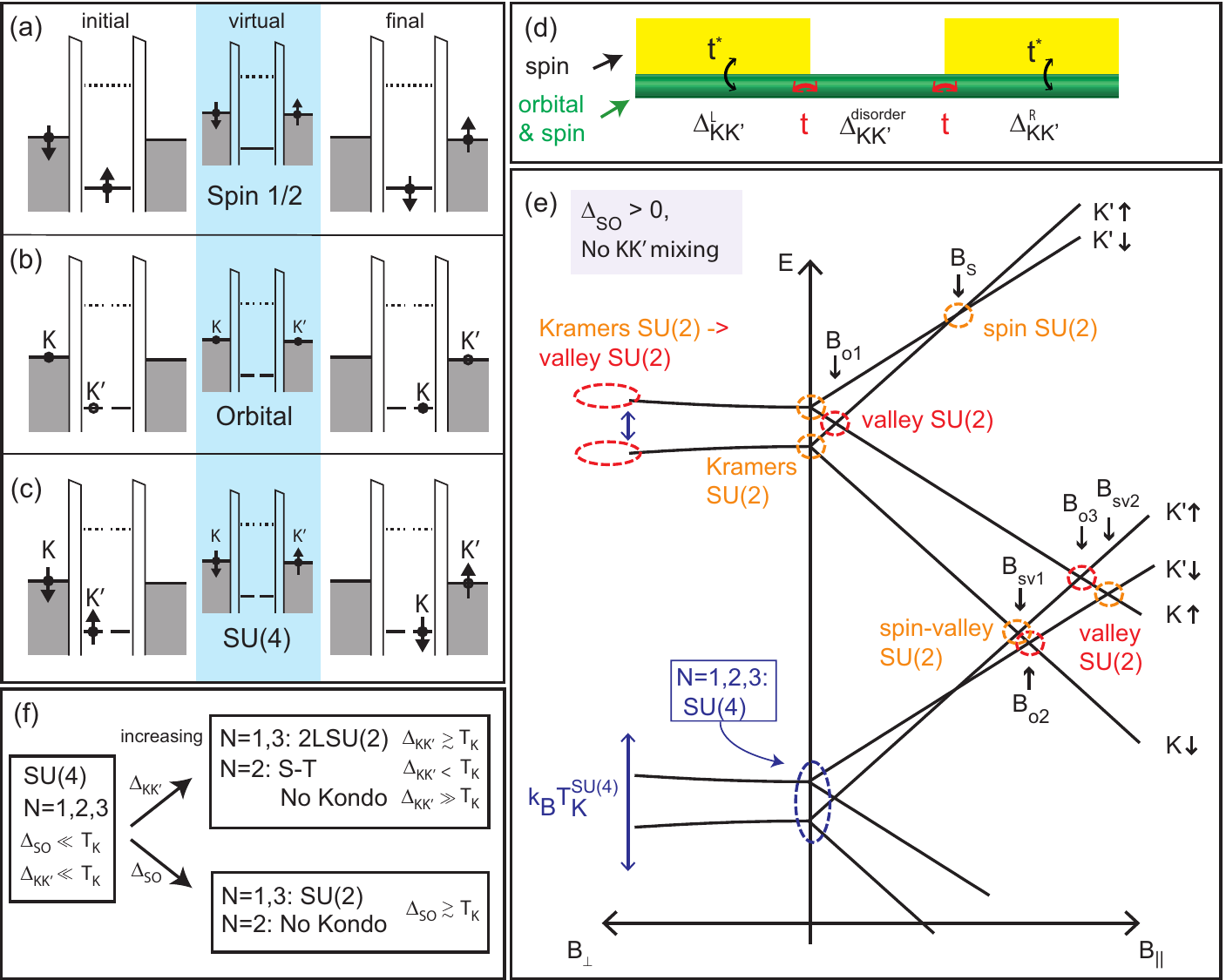}
\caption{\footnotesize{(Color online) Kondo physics in nanotubes due to spin and valley degrees of freedom. (a) Schematics of an elastic cotunneling process resulting in a
spin flip on the quantum dot. Such higher order processes give rise to the spin Kondo effect. (b) The Kondo effect can also occur if another quantum number is present in the
leads and the dot, e.g.\ the valley quantum number as shown here. (c) In nanotubes valley and spin quantum numbers may additionally give rise to the SU(4) Kondo effect
involving both degrees of freedom. (d) Device schematic showing valley mixing in the lead $(\DeltaKK^\mathrm{L,R})$ and the dot $(\DeltaKK^\mathrm{disorder})$. For SU(4) Kondo effects to be observed these should be small
and valley-conserving tunneling $(t)$ from nanotube leads is required. (e) Energy versus parallel and perpendicular magnetic field for two conduction-band shells
with spin-orbit interaction ($\DeltaSO>0$). Dashed ellipses and related $B$-fields indicate degeneracies resulting in Kondo phenomena, e.g.\ SU(4) and various SU(2) Kondo effects.
The SU(4) Kondo effect can be observed (lower shell) when the related energy scale is much larger than the zero-field (spin-orbit) splitting. (f) Flow diagram at $B=0$ for fillings $N=1-3$ showing that valley mixing reduces the SU(4) Kondo effect to two-level (2L), singlet-triplet (S-T) or no Kondo
effect. Similarly spin-orbit interaction reduces SU(4) Kondo effect to SU(2) or no Kondo effect. The intervalley mixing parameter $\Delta_{KK'}$ represents all relevant mixing
terms. Panels~(a-c) adapted from \onlinecite{Jarillo-HerreroNature2005}.
}}
\label{fig:Kondotheory}
\end{figure*}
Transport in nearly-closed quantum dots can be described in terms of first-order sequential tunneling. With stronger coupling to the leads higher order processes involving
virtual intermediate states become relevant (Fig.~\ref{fig:Kondotheory}(a)). Initially, the system is in a state of Coulomb blockade (left). Higher order fluctuations can
permit tunneling of the trapped electron to the right lead, while a second electron from the source enters the dot. Effectively, one electron charge, $-e$, has been
transferred from source to drain (right diagram) via an intermediate state (middle) that is classically forbidden due to energy conservation and Coulomb blockade. Such a process is called elastic co-tunneling \cite{AverinBook1992,PustilnikJPCM2004,Ihn2010}. In Fig.~\ref{fig:Kondotheory}(a) the spin on the dot is flipped as permitted by the
spin degeneracy of the level at zero field.
The non-trivial result of such higher order spin-flip transitions is the appearance of a new correlated ground state for the combined
lead-dot system. This is an effective singlet state between an unpaired spin and the Fermi sea of electrons in the leads. It forms a highly transmitting channel between source and drain at the Fermi energy, known as a Kondo resonance, leading to the breakdown of Coulomb blockade~\cite{Ihn2010,Nazarov2009,Heikkila2009}.

The Kondo state only exists at low temperatures where coherence is preserved. The transition temperature from Coulomb blockade to transport resonance is
denoted the Kondo temperature, $T_\mathrm{K}$. The energy scale $k_\mathrm{B}T_\mathrm{K}$ can be considered as the binding energy of the many-body singlet state formed from the quantum dot spin and a
screening cloud of electrons in the lead. Kondo resonances appear in~Fig.~\ref{fig:closed_to_open}(b) as ridges of high conductance near zero bias. The
corresponding charge states can be identified as the odd occupancy states where the quantum dot holds an unpaired electron. (The apparent
resonances in some of the even diamonds will be discussed below.) Several diagnostics can verify an underlying Kondo mechanism; the resonance should be suppressed
by increasing temperature, bias voltage or an external magnetic field that breaks the necessary level degeneracy.

\begin{figure}
\center
\includegraphics[width=8.6cm]{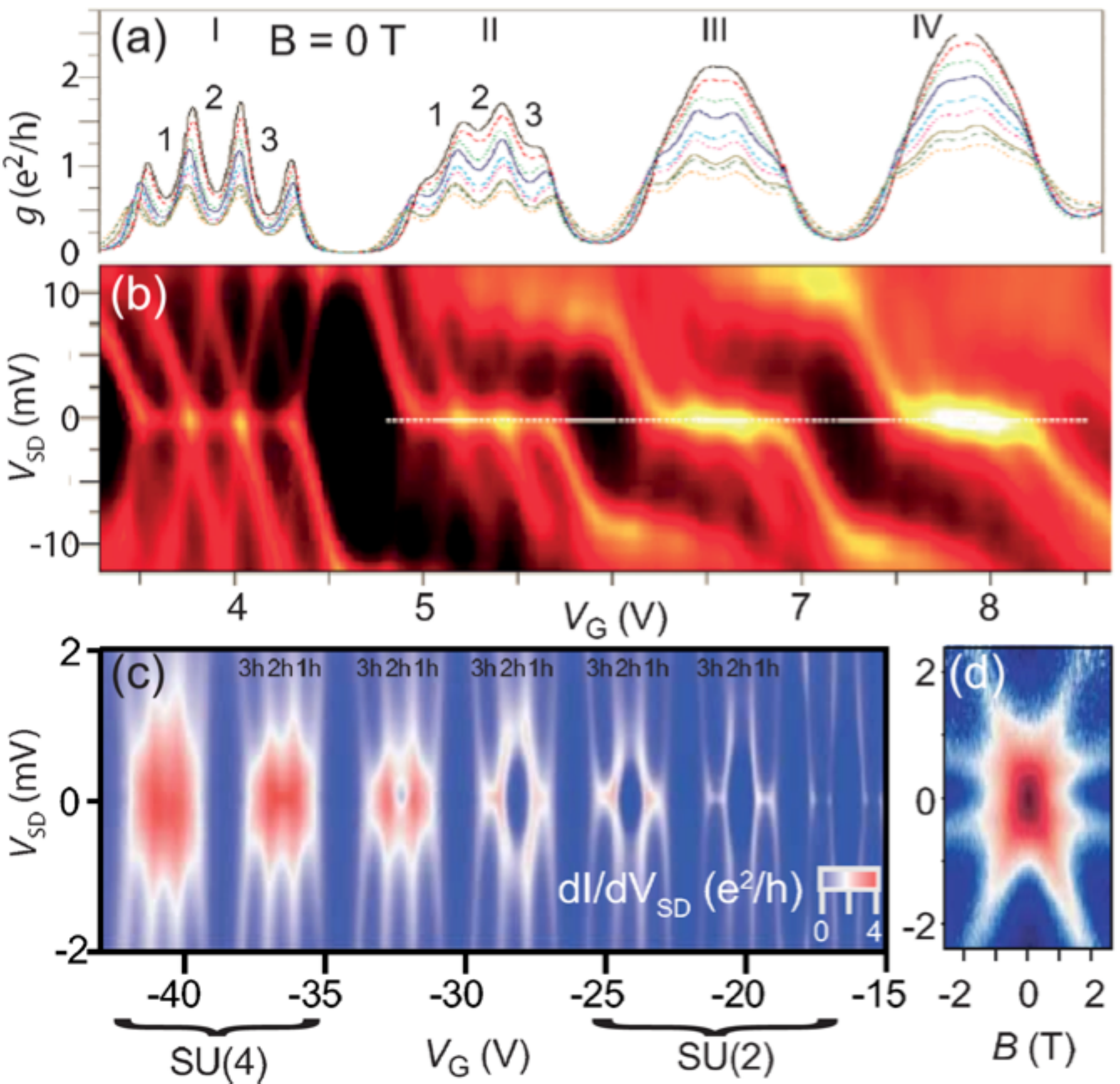}
\caption{\footnotesize{(Color online) Kondo physics near zero field. (a) Conductance versus gate voltage of a narrow-gap nanotube for four electron shells I-IV. For shells I and II the occupancy in each diamond is indicated. More positive $V_\mathrm{G}$ increases the tunnel couplings, tuning the dot from the Kondo to the mixed valence regime. Different traces correspond to temperatures from $1.3$\,K to $15$\,K. (b) Differential conductance in the same device as a function of $V_\mathrm{G}$ and $V_\mathrm{SD}$ at $3.3$\,K. (c) Similar data for a different device showing spin-orbit split SU(2) and SU(4) Kondo physics at low and high hole filling, respectively. (d) Magnetic field splitting of $N=1$ Kondo resonance into four peaks, indicating SU(4) Kondo physics. Panels (a-b) adapted from \cite{MakarovskiPRL2007}, (c) from \cite{CleuziouPRL2013}, and (d) from~\onlinecite{Jarillo-HerreroNature2005}.}}
\label{fig:Kondo_datazerofield}
\end{figure}

Quantum dot Kondo resonances were first discovered in two-dimensional semiconductor devices \cite{Goldhaber-GordonNature1998,cronenwettScience1998,SimmelPRL1999} but
observed shortly afterwards in carbon nanotubes \cite{NygardNature2000,LiangPRL2002,BuitelaarPRL2002KondoMWCNT, babicPRB2004}. For a more detailed account of Kondo physics
and the relation to quantum dots in general we refer to other reviews \cite{KouwenhovenPhysWorld2001,PustilnikJPCM2004,Grobis2006,ZarandPMag2006} and textbooks, e.g.~\cite{Ihn2010,Nazarov2009,Heikkila2009}. In all cases, the spin Kondo effect leads to an enhanced conductance at low temperatures. For nanotubes~$T_\mathrm{K}$ is typically 1-10~K.

We will not dwell on the details of the spin Kondo effect but rather point out that in principle any doubly (or higher) degenerate localized state with identical quantum
numbers in the leads could mediate Kondo-like resonances. The origin of the degeneracy does not need to be spin; for example, transport resonances
could be induced by an orbital level degeneracy on the dot (Fig.~\ref{fig:Kondotheory}(b)) \cite{Jarillo-HerreroNature2005}. In the context of nanotubes the valley degree of freedom comes to mind. Below we discuss such a valley Kondo effect.

Both the ordinary spin 1/2 Kondo effect and the two-fold valley Kondo effect reflect SU(2) symmetry. However, for ideal nanotube dots the concomitant existence of valley and
spin freedom leading to an approximate four-fold degeneracy could potentially lead to a Kondo effect described by the higher SU(4) symmetry class \cite{ChoiPRL2005}. This
situation is absent in most other quantum dots, which lack the spatial symmetry that naturally leads to valley degeneracy in nanotubes\footnote{SU(4)
Kondo physics has been studied in vertical quantum dots \cite{SasakiPRL2004}, parallel double quantum dots \cite{WilhelmPhysicaE2002, HolleitnerPRB2004,
OkazakiPRB2011,Keller2013} and single dopants \cite{TettamanziPRL2012}. In the vertical quantum dots, an orbital degree of freedom also exists in both the leads and the dot because they share similar symmetry. To study SU(4) physics in two parallel dots, however, requires a device with identical intra- and inter Coulomb energy.}. Figure~\ref{fig:Kondotheory}(c) shows an example of states involved in this scenario where both the valley and spin quantum numbers can be exchanged during
co-tunneling. In order to probe the SU(4) Kondo effect, it is essential that transitions between all four states are possible.

Even if valley degeneracy exists in the nanotube, this does not ensure that the valley quantum number is also present in the leads. Figure \ref{fig:Kondotheory}(d) shows
a nanotube quantum dot coupled to metallic leads. The double arrows indicate that electrons need to enter a nanotube lead segment ($t^*$) before tunnelling onto
the quantum dot ($t$) to allow for SU(4) Kondo physics since this effect requires valley-conserving tunneling. As discussed below, $K$-$K'$ mixing during tunneling ($\Delta_{KK'}^{\text{tunneling}}$), on
the dot ($\Delta_{KK'}^{\text{disorder}}$), and in the lead segments ($\Delta_{KK'}^{\text{L,R}}$) must all be weak, as must spin-orbit coupling.

To link the different Kondo phenomena to the nanotube level structure, the spectrum of two shells is plotted in Fig.\ \ref{fig:Kondotheory}(e), including spin-orbit interaction but not valley mixing. Dashed ellipses indicate degeneracies that can lead to Kondo physics. When the SU(4)
Kondo energy scale is much larger than the spin-orbit splitting, the SU(4) Kondo effect can be observed at zero field for occupations $N=1,2$ or 3 as depicted for the
lower shell. The opposite case of small SU(4) Kondo temperature is shown for the upper shell, resulting in SU(2) Kondo effects for the two
split Kramers doublets (dashed circles at zero field) \cite{FangPRL2008,FangPRL2010,GalpinPRB2010}.

Figure \ref{fig:Kondotheory}(e) also points to possible Kondo effects in a parallel magnetic field. Two intra-shell SU(2) Kondo effects are illustrated for the upper shell: a valley SU(2) Kondo effect at half-filling and a spin SU(2) Kondo effect for three electrons in the shell. Inter-shell valley Kondo effects can arise from level crossings between two different shells. Taking account of valley mixing within the shells modifies
the energy diagram (Fig.~\ref{SO_combined}), but does not qualitatively change the inter-shell valley degeneracies \cite{Jarillo-HerreroNature2005,Grove-RasmussenPRL2012}.
However, the intra-shell valley degeneracies for parallel and perpendicular field are split, and when valley mixing dominates, a spin-orbit split degeneracy (not shown)
emerges at finite perpendicular field giving rise to a singlet-triplet (S-T$_-$) Kondo effect \cite{NygardNature2000, PustilnikPRL2000, QuayPRB2007}. The zero-field SU(4) Kondo
effect survives as long as the Kramers doublet splitting due to valley mixing and spin-orbit coupling is much smaller than the Kondo temperature \cite{BordaPRL2003,
GalpinPRB2010,FangPRL2008}.

As shown in Fig.\ \ref{fig:Kondotheory}(e-f), the SU(4) Kondo effect can be observed for filling $N=1-3$ \cite{AndersPRL2008}, but only if the valley quantum numbers are
conserved during tunneling, i.e.\ no valley mixing. If this is not the case, the SU(4) Kondo effect reduces to two-level (2L) spin SU(2)
\cite{BusserPRB2011,BusserPRB2007,ChoiPRL2005,LimPRB2006} and singlet-triplet (S-T) \cite{IzumidaPRL2001,EtoPRL2000,SasakiNature2000} Kondo effects for $N=1,3$ and $N=2$,
respectively\footnote{The analysis in terms of singlet-triplet physics assumed zero spin-orbit interaction.}. For the 2LSU(2) Kondo effect, the Kondo temperature and maximum
conductance $g=2e^2/h$ at the center of the $N=1,3$ Coulomb diamond is the same as in the case of SU(4) Kondo effect, and the two effects are therefore to be distinguished
either by magnetic field spectroscopy \cite{Jarillo-HerreroNature2005,ChoiPRL2005} or by the overall shape of the $N=1-3$ conductance versus gate voltage
\cite{MakarovskiPhysicaE2008}. Similarly, in the center of the $N=2$ Coloumb diamond, both the singlet-triplet and SU(4) Kondo effects have the same theoretical maximum
conductance $g=4e^2/h$  \cite{Jarillo-HerreroNature2005}. However, for sufficiently strong valley mixing (or spin-orbit interaction), a zero-field Kondo effect is absent for $N=2$, as indicated in Fig.\ \ref{fig:Kondotheory}(f).

\subsubsection{Experiment}
\label{subsubsec:Kondoexp}
Unconventional Kondo physics in nanotubes was first experimentally studied in \cite{Jarillo-HerreroNature2005} and later by
\cite{JarilloHerreroPRL2005,MakarovskiPRB2007,MakarovskiPRL2007,MakarovskiPhysicaE2008,CleuziouPRL2013,Schmid2013}\footnote{Other types of unconventional Kondo phenomena
such as the two-impurity Kondo effect have been studied in nanotube systems \cite{ChangRepProgPhys2009, ChorleyPRL2012, BomzePRB2010}. Equipping nanotubes with
superconducting or ferromagnetic leads also give rise to new Kondo systems \cite{BuitelaarPRL2002, HauptmannNphys2008}, beyond the scope of this review.}.
For all but the last two experiments, the spin-orbit interaction was thought to be negligible and thus not included in the Kondo analysis
\cite{ChoiPRL2005,LimPRB2006,BusserPRB2007,AndersPRL2008,MizunoJPCM2009}. However, more recent theory includes spin-orbit interaction and reinterprets the
early data \cite{FangPRL2008,FangPRL2010,GalpinPRB2010}. Our aim is to present data relevant for the spectrum of Fig.~\ref{fig:Kondotheory}(e), starting with zero magnetic field Kondo physics
followed by finite field phenomena for one and two shells.
\begin{figure}
\center
\includegraphics{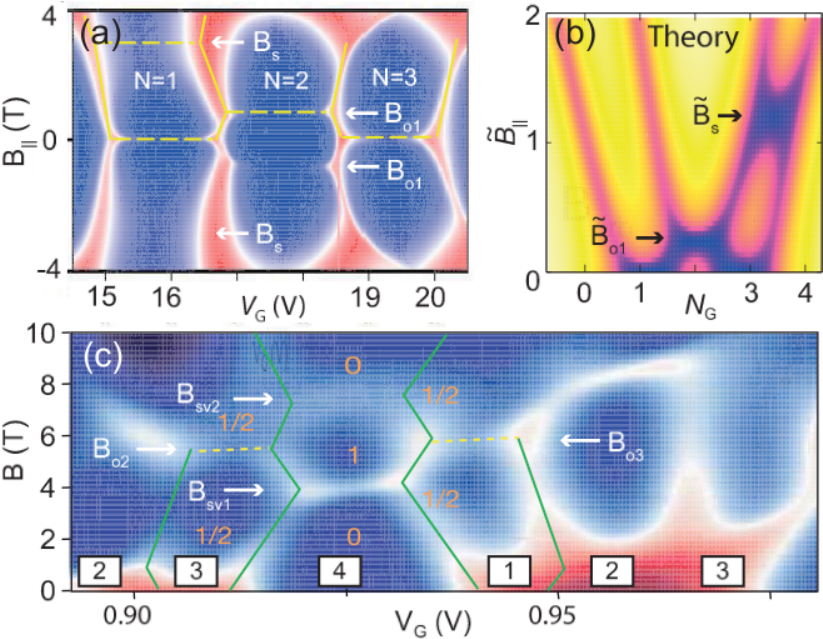}
\caption{\footnotesize{(Color online) Kondo physics at finite field. (a) Differential conductance versus gate voltage and parallel magnetic field for the first three electrons in
the conduction band. Dashed lines indicate faint Kondo resonances: Kramers (for $N=1,3$ at $B_{||}=0)$, valley (for $N=2$ at $B_{||}=B_\mathrm{o1}$) and spin SU(2) (for $N=1$ at $B_{||}=B_\mathrm{s}$).
(b) Theoretical linear conductance versus gate voltage filling (expressed as effective filling $N_\mathrm{G}$) and normalized parallel field $\tilde{B}_{||}$. The calculation assumed $|\DeltaSO|>T_\mathrm{K}^{\mathrm{SU}(4)}$
as in the top shell of Fig.~\ref{fig:Kondotheory}(e). (c) Differential conductance as a function of gate voltage $V_\mathrm{G}$ and parallel magnetic field $B$ at 30~mK. Boxed numbers indicate the occupation of the topmost shell and unboxed numbers indicate the ground state spin (0, 1/2, 1). Solid lines highlight the motion of the Coulomb blockade peaks.
Yellow dashed lines mark valley Kondo effects (compare white arrows to Fig.~\ref{fig:Kondotheory}(e)). Panels
(a-c) adapted from \onlinecite{CleuziouPRL2013}, \onlinecite{GalpinPRB2010} and \onlinecite{Jarillo-HerreroNature2005}, respectively.}}
\label{fig:Kondo_datafinitefield}
\end{figure}

The ideal devices for studying an SU(4) Kondo effect are quantum dots with tunable couplings to the leads. Such devices can be realized in narrow-gap nanotubes where the
conductance typically increases as carriers are added. Figure \ref{fig:Kondo_datazerofield}(a-b) shows the linear conductance $dI/dV_\mathrm{SD}$ versus gate voltage and the corresponding charge
stability diagram. A regular shell structure with zero-bias Kondo ridges for electron filling one, two and three is observed at low gate voltages. At stronger electrode coupling, the different charge states fully hybridize  \cite{MakarovskiPRL2007,MakarovskiPhysicaE2008} and a single merged peak is observed for the two rightmost
shells in Fig.~\ref{fig:Kondo_datazerofield}(a). In this mixed-valence regime the lifetime broadening is comparable to or larger than the charging energy and the
single-electron transport features of the stability diagram merge into broad ridges. The observed behavior in both the Kondo and mixed-valence regimes has been
reproduced by numerical renormalization group (NRG) calculations within an SU(4) Anderson model for various lead couplings, supporting this interpretation~\cite{AndersPRL2008}.

Figure \ref{fig:Kondo_datazerofield}(c) systematically studies the transition from SU(4) to SU(2) Kondo physics in a clean narrow-gap device
\cite{CleuziouPRL2013}. The crossover is measured across seven shells, thereby varying both the spin-orbit interaction (decreases with filling, see Fig.\ \ref{SO_twotypesExpt})
and the lead coupling (increases with filling). A transition from $\Delta_\mathrm{SO} > T_\mathrm{K}^\mathrm{SU(4)}$ to $\Delta_\mathrm{SO} < T_\mathrm{K}^\mathrm{SU(4)}$ versus hole filling, i.e. SU(2) to SU(4), can
thus be realized in accordance with the lower arrow in the flow diagram of Fig.\ \ref{fig:Kondotheory}(f).

The SU(4) Kondo effect was also reported by \cite{Jarillo-HerreroNature2005,JarilloHerreroPRL2005} studying the $N=1$ zero-bias peak. The temperature dependence (0.3\,K to 10\,K) of the conductance peaks was consistent with the expected empirical scaling formula \cite{Goldhbaber-GordonPRL1998} yielding high Kondo temperatures (around $10$\,K). Figure \ref{fig:Kondo_datazerofield}(d) shows the behavior of the Kondo resonance in a (parallel) magnetic field. The splitting into four peaks (ideally six if the Zeeman splittings are fully resolved) indicates SU(4) Kondo physics, since only two peaks would emerge if valley mixing induced by tunneling or disorder were dominant as for the 2LSU(2) Kondo effect \cite{ChoiPRL2005}. A zero-bias ridge observed for $N=2$ in the same work was analyzed in terms of singlet-triplet Kondo physics, but is also consistent with an SU(4) Kondo effect \cite{BusserPRB2007}.

Next we turn to Kondo phenomena at finite $B_{||}$ within a single shell (upper part of Fig.~\ref{fig:Kondotheory}(e)) \cite{FangPRL2008,FangPRL2010, GalpinPRB2010}. Recent experiments on the addition of
the first four electrons in a device with clearly identified spin-orbit interaction give hints of the valley and finite-field spin SU(2) Kondo effects \cite{CleuziouPRL2013}.
Figure~\ref{fig:Kondo_datafinitefield}(a) shows the linear conductance versus parallel magnetic field and gate voltage, with the electron filling and the possible Kondo ridges indicated. Compared to the model above, the finite field spin SU(2) Kondo effect at $B_\mathrm{s}$ for odd filling
appears for $N=1$ instead of $N=3$. (Whether it is the upper or lower two states that crosses at high field depends on carrier type and the sign of the spin-orbit
interaction (see Fig.~\ref{SO_spectrumtheory})). The $N=2$ valley Kondo effect at $B_\mathrm{o1}$ is clearly visible close to the Coulomb peaks, mimicking the behavior of the Kramers Kondo effect at zero field. Due
to small tunnel coupling in this device, valley or spin Kondo physics cannot be unambiguously identified from the data.

Similar data with strong Kondo ridges were already observed before the identification of nanotube spin-orbit interaction (see Fig.\ \ref{magneticmomentexpt}, shell around
$V_\mathrm{G} \sim 3$\,V), but the Kondo ridge at half filling and finite $B$ was interpreted in terms of singlet-triplet Kondo physics\footnote{Depending on the exchange
interaction, the triplet state occupying two different valleys may become the ground state. In a parallel magnetic field the singlet state occupying two down-moving valleys
thus crosses the almost degenerate triplet states, giving rise to a Kondo effect (assuming negligible Zeeman effect).}
\cite{JarilloHerreroPRL2005}. Later NRG modelling (Fig.~\ref{fig:Kondo_datafinitefield}(b)) suggested that the data could also be explained in
terms of the spin-orbit energy spectrum\footnote{While this is a likely interpretation, some caution should be applied since level crossings with a different shell
complicate the finite-field spin SU(2) identification.} \cite{GalpinPRB2010}.  The calculated color plot thus represents the ideal Kondo behavior for the upper shell in~Fig.~\ref{fig:Kondotheory}(e). Other experiments with dominating valley mixing ($\Delta_{KK'}>\Delta_\mathrm{SO}$) also identify the spin SU(2) Kondo effect at $B_{||}=B_\mathrm{s}$ \cite{JespersenNphys2011,Schmid2013}.

Finally we address the (inter-shell) valley Kondo effect. It can be found at particular fields ($B_\mathrm{o2},B_\mathrm{o3}$) that induce valley degeneracy but not spin
degeneracy (Fig.~\ref{fig:Kondotheory}(e)). Figure \ref{fig:Kondo_datafinitefield}(c) shows the (parallel) magnetic field evolution of the conductance for a
nanotube quantum dot that has strong Kondo resonances at zero field. At finite fields additional resonances appear at $B_\mathrm{o2}$ and $B_\mathrm{o3}$, consistent
with valley degeneracies inducing a spinless valley SU(2) Kondo effect. The appearance of this Kondo resonance indicates that tunneling preserves the valley symmetry,
which for this device is less obvious because of the metal deposited on top of the nanotube leads.

In several of the experiments above a Kondo effect arising from inelastic transitions gave rise to conductance peaks at finite bias. Such Kondo enhancement of inelastic cotunneling thresholds was observed and modelled in \cite{PaaskeNphys2006} while a recent experiment revealed that certain cotunneling thresholds are not observed in the strong coupling regime \cite{Schmid2013}.
The authors identify the two states involved in the relevant cotunneling processes as particle-hole symmetric\footnote{The work considers a shell with both spin-orbit
interaction and valley mixing. The eigenstates are therefore superpositions of all four basis states $|\tau \sigma \rangle$. A simple example of a particle-hole symmetric
pair is \{$|K{\downarrow}\rangle,|K'{\uparrow} \rangle$\}.} and show theoretically that these processes do not give rise to Kondo correlations (even though cotunneling is
allowed).

\begin{figure}
   \includegraphics{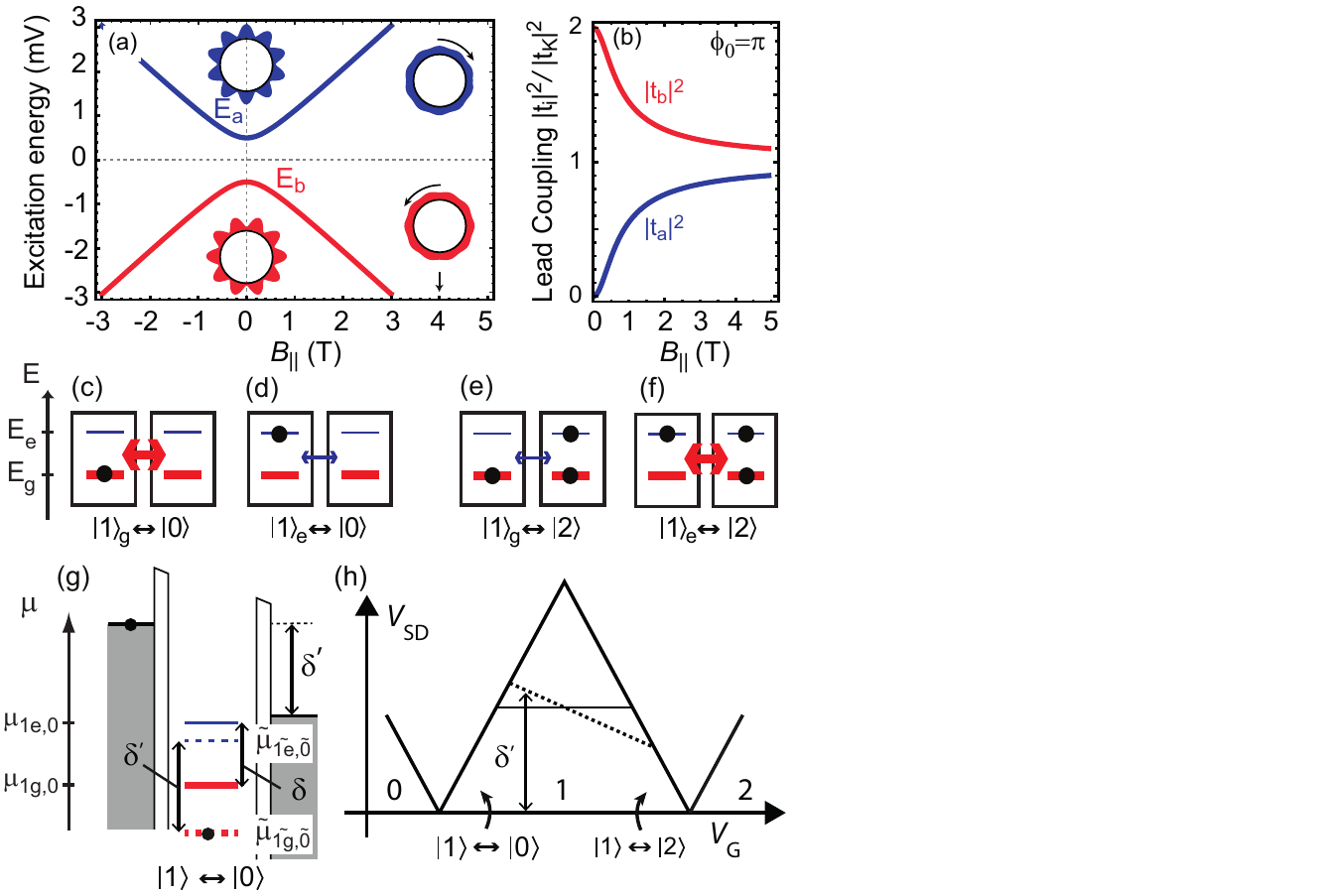}
  \caption{(Color online) Level renormalization. (a) Energy versus parallel magnetic field of two valley states with coupling $\Delta_{KK'}$ that splits the states at zero
  field. Insets show the electron density distributions at zero and high fields. (b) Lead couplings versus parallel magnetic field of the two eigenstates in (a). (c-f)
  Cotunneling processes for the two doublets of a singly occupied shell near the 1 to 0 (c-d) and 1 to 2 (e-f) charge degeneracy points. The thickness of the arrows indicates the tunnel couplings.
  (g) Electrochemical potentials $\mu_{1g,0} = E_{1g}-E_0$ (solid lines) and renormalized versions $\tilde{\mu}$ (dashed lines). Here $E_{1g}$ and $E_0$ are the energies of the one electron ground state ($1g$) and the empty-dot state ($0$). The lead potentials are set at the threshold for inelastic cotunneling; renormalization enhances the threshold energy $\delta '$ compared to the original doublet
  splitting $\delta$. (h)
Schematic charge stability diagram for the one-electron diamond. The resulting level renormalizations induce a gate-dependent inelastic cotunneling threshold (dashed) in
contrast to the single-particle prediction $\delta$ (solid line).
Adapted from \onlinecite{Grove-RasmussenPRL2012}.}\label{fig:renormtheory}
\end{figure}

\subsection{Level renormalization}
\label{subsec:level_renormalization}
The most visible consequence of increasing electrode couplings in Fig.\ \ref{fig:closed_to_open} is a broadening of the energy levels, but this hybridization is also
accompanied by level shifts. These shifts are named tunnel renormalization, since the effect stems from quantum charge
fluctuations (cotunneling events) that are particularly relevant when the tunnel couplings are large.
Nanotube quantum dots have turned out to be ideal to observe such level shifts, in particular when the two doublets are differently coupled to the leads
\cite{HolmPRB2008}.

\subsubsection{Theory}
\label{subsubsec:renormtheory}
We consider a spinless nanotube shell model without any internal couplings and assume metallic leads (no valley quantum number). Since $|K\rangle$ and $|K'\rangle$
states are time-reversed partners, the tunnel couplings are equal $t=|t_K|=|t_{K'}|.$
Introducing a complex valley mixing term $\Delta_{KK'}=|\Delta_{KK'}|e^{i\phi_0}$ results in new eigenstates 
which have phase dependent tunnel couplings $t(1\mp e^{i\phi_0})$ \cite{Grove-RasmussenPRL2012}.
Figure \ref{fig:renormtheory}(a) shows the energy versus parallel magnetic field for the new eigenstates that arise when the valley states are coupled. The schematic
insets show that the electron probability distributions for the two eigenstates are different at zero field and thus the tunneling amplitude may be different depending on
the exact site of a microscopic contact. This picture of tunneling into one particular atomic site with different probability for the two wave functions is probably too
simple for a real device but is adopted here as a minimal model. In Fig.\ \ref{fig:renormtheory}(b) the lead couplings for the two states are plotted for the case
$\phi_0=\pi$. At zero field the two new eigenstates have different couplings: the ground state doubles its coupling to the leads compared to the original states while the
excited state decouples. The model also makes an important prediction about the magnetic field dependence of the couplings. At high parallel magnetic fields the ground and
excited state couplings become equal, since the eigenstates are close to the original valley states $|K\rangle$ and $|K'\rangle$ whose probability distributions are equal at
high fields (Fig.\ \ref{fig:renormtheory}(a) inset).

\begin{figure}
  \includegraphics{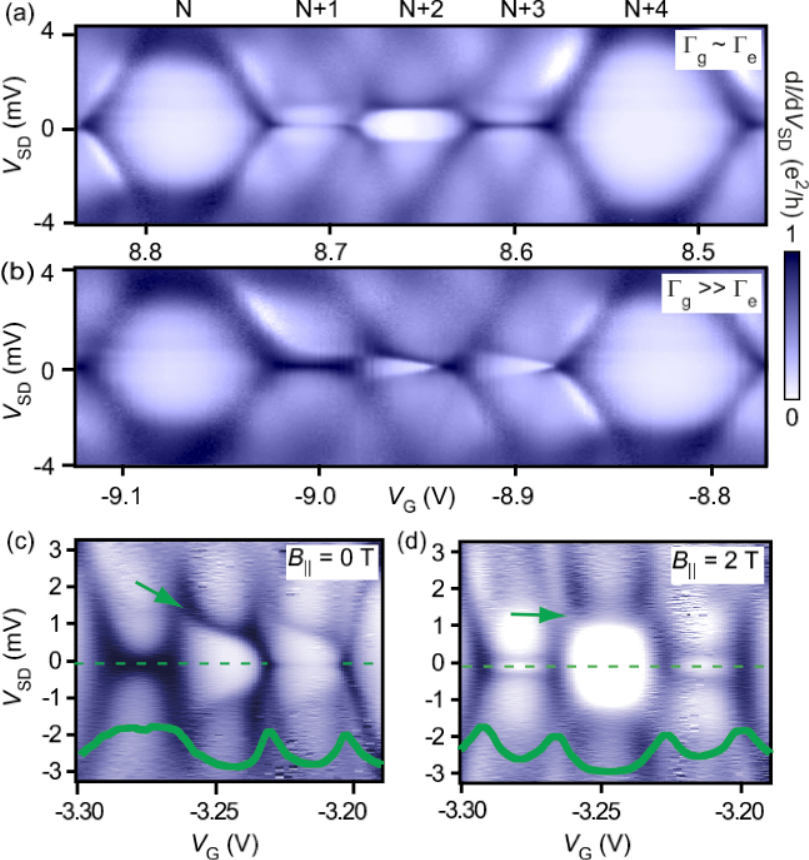}
  \caption{(Color online) Tunnel renormalization in nanotube stability diagrams. (a) Stability diagram for a shell with similar lead couplings to ground and excited states ($\Gamma_g\sim \Gamma_e$). Kondo ridges are observed for one and three electrons and inelastic cotunneling thresholds are gate-independent. Occupation numbers are marked above the plot. (b) Stability diagram
  for a different shell. Unlike in panel (a), the Kondo ridge appears only for the $N+1$ diamond, and the cotunneling thresholds in the $N+2$ and $N+3$ diamonds are gate-dependent. This is consistent with cotunneling provided that the doublet lead couplings are asymmetric ($\Gamma_g\gg\Gamma_e$ deduced from
  thresholds for $N+2$ and $N+3$). (c) Diagram at zero field for a different device. Again, a gate-dependent cotunneling threshold (arrow) indicates asymmetric couplings. (d)  At $B_{||}=2$\,T the threshold is gate-independent, consistent with equalization of the lead couplings by the magnetic field as discussed in the text. Adapted from \onlinecite{HolmPRB2008,Grove-RasmussenPRL2012}. }\label{fig:renormExp}
\end{figure}

We now examine the effect of the difference in lead-couplings on cotunneling. Consider two states with different tunnel couplings as shown in Fig.~\ref{fig:renormtheory}(c-d) for a singly occupied dot $|1\rangle$. The ground state experiences stronger cotunneling charge fluctuations via the zero state ($|1\rangle_g \leftrightarrow |0\rangle$) than the excited state does ($|1\rangle_e\leftrightarrow |0\rangle$) does.
The renormalization shift increases with the coupling of the levels and thus the more strongly coupled state has the larger energy shift (Fig.~\ref{fig:renormtheory}(g)). Consequently, the threshold for inelastic cotunneling transport will increase (from $\delta$ to $\delta '$) near the $|1\rangle  \leftrightarrow
|0\rangle$ charge transition (left side of diamond). In the stability diagram Fig.\ \ref{fig:renormtheory}(h) the threshold will correspondingly be shifted to higher bias.
In contrast, on the right side of the one-electron diamond, the dominant cotunneling processes involve the two-electron state ($|1\rangle  \leftrightarrow |2\rangle$). The
situation is opposite here since the excited state $|1\rangle_e$ experiences stronger fluctuations and level shifts than the ground state $|1\rangle_g$ (Fig.
\ref{fig:renormtheory}(e-f)). Thus the cotunneling threshold will be reduced compared to the original level in this gate range. The resulting stability diagram is sketched in
Fig.\ \ref{fig:renormtheory}(h). The models above are easily extended to include spin and higher fillings, but the conclusions remain the same: tunneling-induced level
renormalization can result in gate-dependent cotunneling features within the charge stability diamonds.

\subsubsection{Experiment}
\label{subsubsec:LevelrenExp}
Figure \ref{fig:renormExp}(a) shows a stability diagram with shell filling identified by the characteristic pattern of three small Coulomb diamonds followed by a large one and Kondo ridges for odd occupancies. However, Fig. \ref{fig:renormExp}(b) shows that this regular behavior is, not observed for all shells. In this case, the Kondo effect is only evident for one electron, and the inelastic cotunneling thresholds\footnote{For the one-electron case the inelastic cotunneling threshold is not visible due to the Kondo effect present at zero bias.} are seen to have a marked gate dependence for two and three electrons where arguments similar to the model above hold (compare Fig.\ \ref{fig:renormtheory}(h)). The absence of the Kondo effect for $N=3$ indicates that the ground-state doublet is more strongly coupled to the leads than the excited-state doublet. The different behavior for different shells, such as between Fig.\ \ref{fig:renormExp}(a) and (b), can be understood as due to valley mixing with different phases for each shell and tunnel renormalization.

The effective doublet splitting is set by both spin-orbit and valley coupling, $\delta=\sqrt{\DeltaSO^2+\Delta_{KK'}^2}$, while in the lead-coupling model
of Fig.\ \ref{fig:renormtheory}(b) the origin of the coupling asymmetry was the valley coupling alone. The validity of this model can be tested by
examining its prediction for the lead tunnel couplings in a magnetic field. Figure \ref{fig:renormExp}(c) shows a stability diagram of a shell whose behavior is similar
($\Gamma_g>\Gamma_e$) to that of Fig.\ \ref{fig:renormExp}(b).
The corresponding stability diagram in parallel magnetic field $B_{||}=2$\,T is shown in Fig.\ \ref{fig:renormExp}(d). As predicted qualitatively by the model, the gate
dependence of the cotunneling threshold disappears at large field where the lead couplings become equal, consistent with the observed widths of the Coulomb peaks within the shell (see cut in Fig.\ \ref{fig:renormExp}d). This observation indicates that the origin of the asymmetry is related to valley mixing \cite{Grove-RasmussenPRL2012}.

Renormalization effects have also been established in devices with ferromagnetic leads, where the difference in spin density of states of the electrodes leads to
different effective tunnel coupling \cite{MartinekPRL2003,MartinekPRB2005,PasupathyScience2004,HauptmannNphys2008}. Furthermore, if valley mixing originates from cotunneling
(not only disorder as described above) different gate-dependent shifts are predicted with the possibility of inelastic cotunneling lines crossing inside the Coulomb diamond \cite{KirsanskasPRB2012}.

\subsection{Open questions}
Several experiments have found features consistent with Kondo correlations originating from the additional valley degree of freedom. A complete study of the SU(4) Kondo effect would, however, call for control of the valley mixing parameters (Fig.~\ref{fig:Kondotheory}(d)), to tune the valley Kondo correlations. Valley mixing also plays a role for the inter-shell valley Kondo experiment, where incipient Kondo correlations enhancing finite bias features are analyzed~\cite{Jarillo-HerreroNature2005,PaaskeNphys2006}. Such correlations are generally more difficult to
quantify. Clearer experimental results may be possible in tunable gate-defined dots with weak valley mixing, where the leads are nanotube segments rather than metal.

\section{Correlated-electron effects}\label{correlations}
\label{sec_correlations}

\subsection{Introduction}

Electron correlation effects due to Coulomb interaction can be strong
in carbon nanotubes.  One reason is the one-dimensional
nature of the confinement: electrons in two or three dimensions can
minimize their Coulomb repulsion by moving out of each other's way.
Electrons in nanotubes do not have this freedom, and instead
tend to develop strong correlations.  Electrons in nanotubes also see
an environment with a low dielectric constant, which for suspended
nanotubes approaches the free-space value $\varepsilon = 1$.  This is in
contrast to electrons in semiconductors, where electric fields can be
screened by the large dielectric constant of the host material.  These
properties suggest that nanotubes are an interesting system for
studying electron correlations.

The strength of Coulomb interactions between particles with parabolic dispersion is
characterized by an interaction parameter $\rs$, defined as
the ratio of the average interparticle spacing $l$ to the Bohr radius $a_0$:
\begin{equation}
\rs = \frac{l}{a_0} = \frac{m_{\mathrm{eff}}e^2l}{\varepsilon \hbar^2} \approx
\frac{E_{\rm C}}{E_{\rm K}}
\end{equation}
Within a numerical constant of order unity, $\rs$ is also the ratio of
the Coulomb interaction energy $E_{\rm C} = e^2/\varepsilon l$ to the
kinetic energy $E_{\rm K} = \hbar^2 / m l^2$. At high densities
(small~$\rs$), kinetic energy dominates and the single-particle
approximation can be used for the electronic states. At low densities
(large~$\rs$), kinetic energy is quenched and the Coulomb interaction
dominates the physics.

For semiconducting nanotubes at sufficiently low densities, the Fermi
energy lies in the parabolic region of the dispersion relation
(cf.~Fig.~\ref{quantizationkperp}) and $\rs$ is a useful
characterization of the strength of electron interactions. A typical
Coulomb interaction energy in a nanotube for
$\varepsilon = 1$ and $l = 100$ nm is $E_{\rm C} = 13$~meV. To estimate the
kinetic energy, it is important to note that the effective mass is
strongly dependent on the bandgap. For a semiconducting nanotube with
a bandgap of 210~meV ($m_{\rm eff} = 0.029$), $E_{\rm K} = 260\ \mu$eV (Bohr
radius $a_0 \approx 2$~nm). This implies that two electrons in a 200
nm long quantum dot in such a nanotube corresponds to very strong
interactions ($\rs = 50$)\footnote{Taking $l \approx 100\ {\rm nm} =
  200\ {\rm nm}\ /\ 2$ electrons.}.

Advances in making clean suspended nanotubes have enabled the study of
quantum dots at very low density with very low electronic
disorder. Using these new devices, the question of electron
interaction and correlation effects in nanotubes is being
revisited from the ground up. In the few-electron regime in
the clean limit, a clear understanding of the simplest case of two
electrons is beginning to emerge.

\subsection{Interactions in two-electron nanotube quantum dots}
\label{intin2e}

The Coulomb interaction does not directly exert a force between spins.
For example, spin-spin exchange coupling does
not arise from a direct interaction between spins, but instead from a
combination of the Pauli exclusion principle with the Coulomb
repulsion between orbital wave functions.  Thus, to approach Coulomb
interaction phenomena, it is important to start by considering the
properties of nanotube electronic wave functions.

The spatial wave function in a nanotube has several degrees of
freedom: the position $x$ along the nanotube axis\footnote{For
  simplicity, in this section we use notation $x$ for the longitudinal coordinate instead of 
  $t$ as used elsewhere in the review.}, a subband index from the
quantization around the circumference, and a valley index $\tau$
specifying which of the two valleys ($K,K'$) is occupied by an electron. In
nanotubes, it is convenient to separate the $1/r$ Coulomb interaction
into long-range and short-range components. A natural length scale for
this separation is the nanotube diameter. This separation simplifies considerably the treatment of the Coulomb
interaction in nanotubes \cite{wunsch2009few}. A first approximation
is that neither the short-range nor long-range component mixes states
from different subbands: this is justified by the large subband
spacing, on the order of eV. A second approximation is that only the
short-range component of the Coulomb interaction mixes states from
different valleys: this is justified since intervalley scattering
requires a large momentum shift
\cite{mayrhofer2008spectrum,wunsch2009few,WeissPRB2010,secchi2013inter}. The
third approximation is that the envelope function $\psi(x)$ describing
the position of the electron along the nanotube axis is independent of
spin and valley. Although this is not true in general
(cf. Appendix~\ref{appendixsingleparticlestates}), it is a reasonable
approximation for smooth confinement potentials and large
quantum dots \cite{wunsch2009few}. Within these approximations, we can
treat the long-range and short-range components of the Coulomb
interaction separately: the long-range component couples to $\psi(x)$,
while the short-range component couples to the valley degree of
freedom.

Indications of Wigner correlations in multi-hole quantum dots were identified already in early works~\cite{DeshpandeNPhys2008}. In this review, we follow a pedagogical approach to the topic. 
We start with long-range interactions, presenting a pedagogical model
for the Wigner molecule in one dimension, reviewing calculations
performed for nanotubes, and discussing experimental results
demonstrating a two-electron Wigner molecule.  We then focus on
effects from the short-range interaction. At the end of the section, we give an outlook towards interactions in the many-electron and many-hole regimes.

\subsubsection{Long-range interactions and Wigner molecules}

Our   simple   model  focusses   on   the  spatial   two-electron
wave function,   assuming   that  antisymmetry   of  the   total
wave function is provided by  appropriate symmetries in the spin/valley
degree  of freedom.  Approximating the  dispersion  relation of  a
semiconducting  nanotube   by  a  parabola,   we  can  write   down  a
Schr\"odinger equation  with an effective  mass $m_{\rm eff}  = E_{\rm
  gap} / 2 v_F^2$:
\begin{equation}
\begin{split}
-\frac{\hbar^2}{2m_{\rm eff}}\frac{\partial^2 \Psi}{\partial x_1^2} 
-\frac{\hbar^2}{2m_{\rm eff}}\frac{\partial^2 \Psi}{\partial x_2^2}\hspace{4cm}
\\
+V_0(x_1)\Psi + V_0(x_2)\Psi+ V_{\rm C}(x_1,x_2)\Psi = E \Psi
\end{split}
\end{equation}
where $\Psi(x_1,x_2)$ is the two-electron spatial
wave function,~$V_0(x)$ the external confining potential, and~$V_{\rm
  C}(x_1,x_2)$ the Coulomb interaction between electrons. To restrict
ourselves to the long-range component of $V_{\rm C}$, we include a
cutoff in the $1/r$ Coulomb potential as follows:
\begin{equation}
V_{\rm C}(x_1,x_2) = 
\frac{1}{4\pi \varepsilon}
\frac{e^2}{\sqrt{(x_1-x_2)^2 + d^2}}
\label{eq:coulomb}
\end{equation}
with $d$ is chosen as the nanotube diameter.  It is instructive to
combine the confinement terms in a two-electron electrostatic
potential $V(x_1,x_2) = V_0(x_1) + V_0(x_2) + V_{\rm C}(x_1,x_2)$. The
problem of two interacting electrons in one dimension is then formally
equivalent to that of a single electron confined in a two-dimensional
potential~$V(x_1,x_2)$.

\begin{figure}
\center \includegraphics{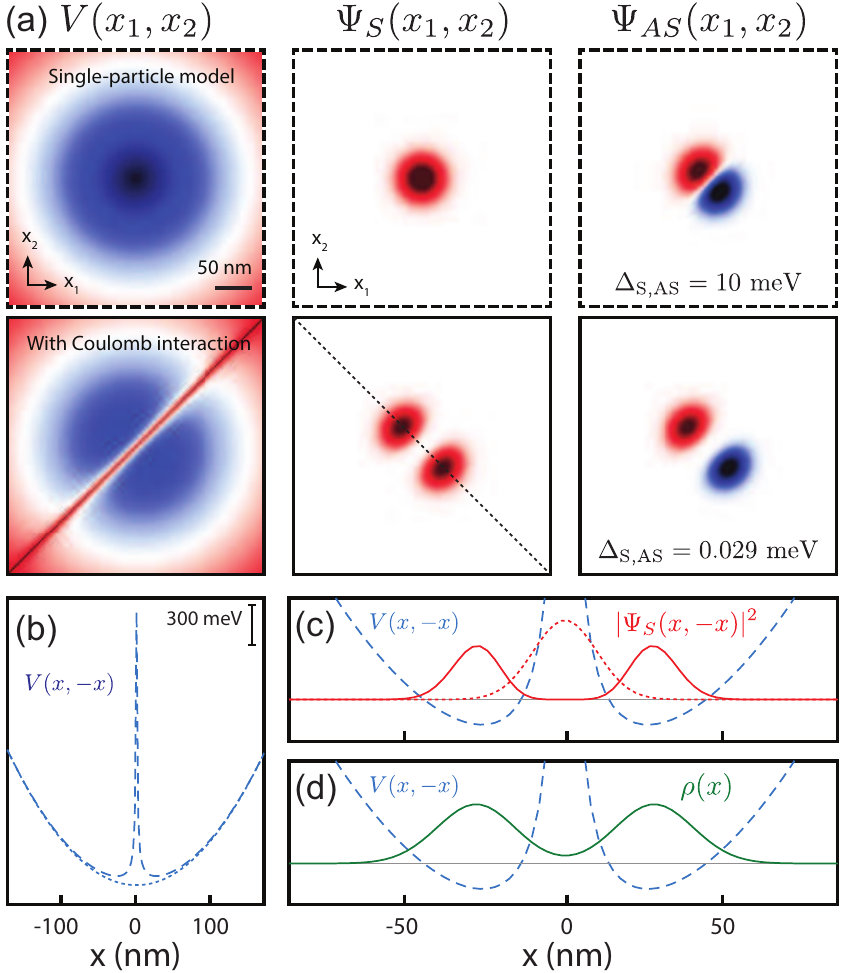}
\caption{\footnotesize{(Color online) Toy model for the two-electron
    Wigner molecule in one dimension. (a) Two-electron potential
    $V(x_1,x_2)$ (left panels), symmetric ground state wave function
    $\Psi_S(x_1,x_2)$ (middle panels), and antisymmetric excited state
    wave function $\Psi_{AS}(x_1,x_2)$ (right panels). Upper panels: no
    Coulomb interaction ($\varepsilon = \infty$), Lower panels:
    $\varepsilon = 1$.  (b) Linecuts of $V(x_1,x_2)$ along $x_1 =
    -x_2$ with (dashed) and without (dotted) Coulomb
    interactions. (c) Linecut of the ground state two-electron
    probability density $|\Psi(x,-x)|^2$ along $x_1 = -x_2$ in the
    single-particle model (dotted line) and the Wigner molecule
    (solid line). Dashed line: $V(x,-x)$. The suppression of
    $|\Psi(x,-x)|^2$ near $x=0$ is an indication of strong
    correlations.  (d) Electron density $\rho(x) = \int
    |\psi(x,x_2)|^2 dx_2$ in the Wigner molecule limit.}}
\label{simple_wigner}
\end{figure}

This problem can be solved on a desktop computer by exact
diagonalization
\cite{jauregui1993wigner,szafran2004spatial,balder2008modelling}.  The
results are shown in~Fig.~\ref{simple_wigner}, taking $m_{\rm eff} =
0.03\ m_\mathrm{e}$, (corresponding to $E_\mathrm{G} = 210$~meV and
$D \approx 3.3 $ nm) and $V_0(x) = \frac{1}{2}m_{\rm
  eff}\omega^2x^2$ with $\hbar \omega = 10$~meV (confinement length
$\approx$ 30 nm). These parameters correspond to $\rs \approx 18$
\cite{balder2008modelling}. The results with ($\varepsilon = 1$) and
without ($\varepsilon=\infty$) Coulomb interaction are shown in
Fig.~\ref{simple_wigner}(a). Including interactions (lower panels in
(a)), the Coulomb repulsion can be seen in $V(x_1,x_2)$ as a diagonal
line along $x_1=x_2$.  With these parameters, typical for nanotubes,
the Coulomb interaction dominates over the confinement potential. In
response, the single lobe of $\Psi_S(x_1,x_2)$ in the single-particle
model splits into two well-separated lobes, pushing the two electrons
away from each other to minimize Coulomb repulsion
(Fig.~\ref{simple_wigner}(c,d)).  In this state, the two-electron
probability density $|\Psi(x_1,x_2)|^2$ goes rapidly to zero along the
$x_1=x_2$ line, indicating the formation of a Wigner molecule. The
quantity $|\Psi(x,-x)|^2$ in Fig. \ref{simple_wigner}(c) can be viewed
as a two-particle correlation density\footnote{Integrating
  $|\Psi(x_1,x_2)|^2$ along the diagonal $x_1=x_2$
  in~Fig.~\ref{simple_wigner}(a) yields the two-particle correlation
  function $g(r) = \int |\Psi(r/2+X,-r/2+X)|^2dX$.}, and its
suppression near $x=0$ corresponds to the formation of a ``correlation
hole'' (Fig.  48c).

An important property of the Wigner molecule state is a strong
suppression of the splitting $\DeltaSAS$ between the spatially
symmetric ground state $\Psi_S(x_1,x_2)$ and the spatially
antisymmetric excited state $\Psi_{AS}(x_1,x_2)$. Without
interactions, $\DeltaSAS = \Delta_{\rm ls} = 10$~meV, where
$\Delta_{\rm ls} = \hbar \omega$ is the single-particle level
spacing. Including interactions, the splitting drops dramatically to
$\DeltaSAS = 29\ \mu$eV, smaller by a factor $\sim 300$. An
explanation can be seen by plotting the two-electron wave functions
$\Psi(x_1,x_2)$ (lower panels of Fig.~\ref{simple_wigner}(a)).  In the
Wigner molecule regime, the ground state wave function
$\Psi_S(x_1,x_2)$ deforms to minimize the Coulomb energy, gaining
kinetic energy and reducing $\DeltaSAS$ until the symmetric
ground state and antisymmetric excited state are nearly degenerate.

\begin{figure}[t]
\center \includegraphics[width=7cm]{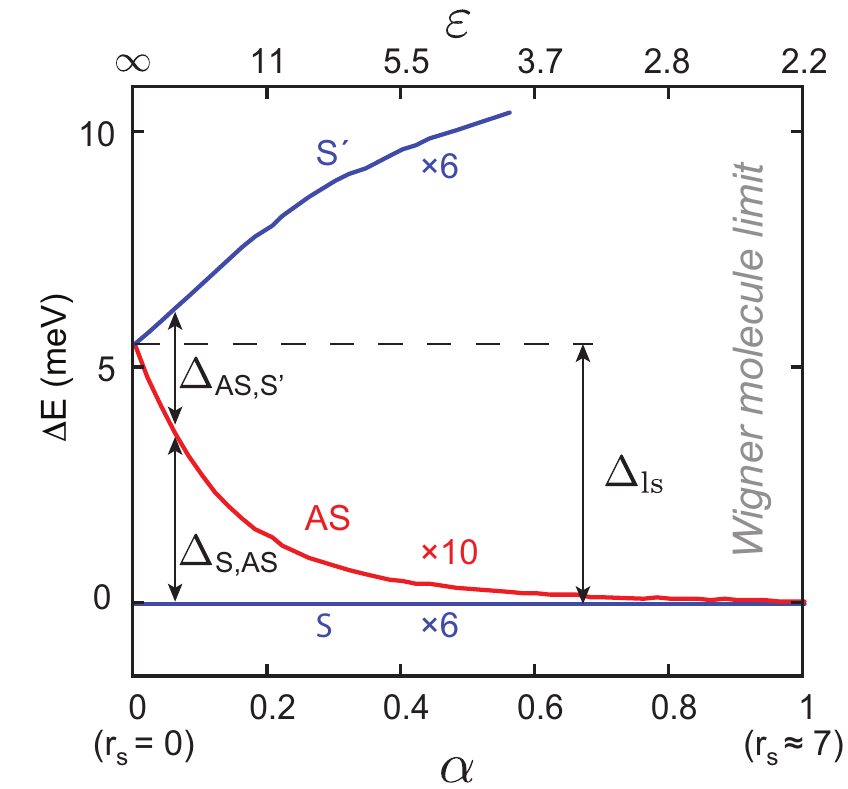}
\caption{\footnotesize{(Color online) Exact calculation for two
    electrons in a nanotube quantum dot.  Vertical axis: energies of
    the antisymmetric and symmetric excited multiplets relative to the
    symmetric ground multiplet.  Along the horizontal axis, the dielectric
    constant is changed to tune the interaction strength.  Numbers (6,
    10, and~6) indicate the degeneracy of each multiplet.  Spin-orbit
    coupling is not included ($\Delta_{\rm SO} = 0$) to illustrate only
    effects from the long-range interactions.  Increasing $\alpha =
    2.2/\varepsilon$, there is a transition from the single particle
    limit (left) to the Wigner molecule limit (right). In the Wigner
    molecule, the splitting $\DeltaSAS$ becomes
    exponentially small.  Adapted from \onlinecite{wunsch2009few}.}}
\label{theory_long_range}
\end{figure}

Until now, the discussion has been quite general and applies equally
to other one-dimensional systems. The relevance for carbon nanotubes
becomes clear when looking at the magnitude of such effects. Exact
diagonalization calculations for nanotubes
\cite{wunsch2009few,secchi2009coulomb,secchi2010wigner,secchi2012spectral,secchi2013inter,roy2009semiconducting,roy2012effective}
arrive at the same conclusions as our toy model: the two-electron dot
is strongly correlated, forming a Wigner molecule.  The results of
such a calculation are shown in Fig.~\ref{theory_long_range}.  The
input parameters are a 100~nm square-well confinement potential with a
50 meV barrier height, a nanotube diameter of 5 nm, a bandgap of 90~meV, and an effective mass of $0.009\ m_\mathrm{e}$.  To illustrate the transition from the single-particle limit to the Wigner
molecule limit, the strength of the Coulomb interaction is tuned by
changing the environmental dielectric constant $\varepsilon$,
parametrized by an effective fine structure constant $\alpha = e^2 /
\varepsilon \hbar v_F = 2.2 / \varepsilon$ (so that $\rs =
\frac{\alpha}{2.2}\frac{me^2l}{\hbar^2}$).  With no long-range
interactions ($\alpha=0$), the excited multiplet is 16-fold degenerate ($\DeltaASSprime = 0$) and separated from the ground-state multiplet
by the single-particle level spacing ($\DeltaSAS = \Delta_{\rm ls}$).
With strong interactions ($\alpha
= 1$), the antisymmetric multiplet is pushed down in energy relative
to the ground state, with $\DeltaSAS$ becoming exponentially
small. The two-electron state is deep in the Wigner molecule limit for
$\rs \sim 7$, and already strongly correlated for intermediate
$\rs$. Summarizing these predictions, Wigner correlations in a
two-electron quantum dot cause a collapse of $\DeltaSAS$.

\begin{figure}[t]
\center \includegraphics{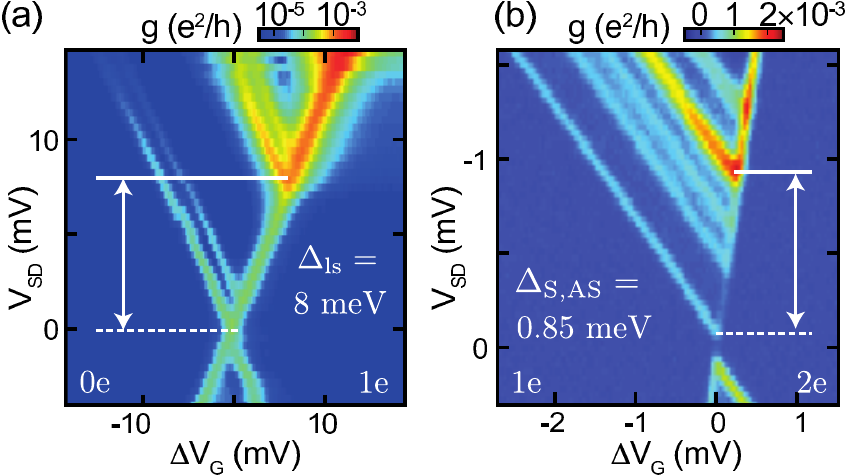}
\caption{\footnotesize{(Color online) Observation of a
Wigner molecule. (a)~Coulomb blockade spectroscopy of the 0-1 electron transition, from which a single-particle level spacing $\Delta_{\rm ls} = 8$ meV is extracted. (b)~Spectroscopy of the 1-2 electron transition: analysis of the magnetic field dependence of the levels allows identification of the excited state corresponding to splitting $\DeltaSAS$. The value $\DeltaSAS = 0.85$ meV is extracted, ten times smaller than   $\Delta_\mathrm{ls}$, indicating that the two-electron quantum dot is in the Wigner molecule regime with $\rs \approx 1.6$ ($\alpha \approx 0.5$ in Fig.~\ref{theory_long_range}). Adapted from \onlinecite{pecker2013observation}.}}
\label{wigner_exp}
\end{figure}

Experiments with clean nanotubes have provided clear evidence of
a Wigner molecule.  Specifically, Pecker {\em et.\ al} (2013) compared
a one-electron to a two-electron quantum dot in the same device (Fig.~\ref{wigner_exp}).  A crucial step was a careful analysis of the
magnetic field dependence of the excitation lines (Fig.~\ref{twoelectronsExpt}). This allowed the authors to
determine which splitting in~Fig.~\ref{wigner_exp}(a)
corresponds to $\Delta_{\rm ls}$, and
which splitting in Fig.~\ref{wigner_exp}(b) corresponds to $\DeltaSAS$.  They found~$\Delta_{\rm ls} = 8$~meV and $\DeltaSAS =~0.85$ meV.  The 10-fold suppression of $\DeltaSAS$ compared to
$\Delta_{\rm ls}$ indicates that the two-electron
quantum dot indeed forms a Wigner molecule.

\subsubsection{Short-range interactions and intervalley exchange}
\label{sec_shortrangeinteractions}

\begin{figure}
\center \includegraphics{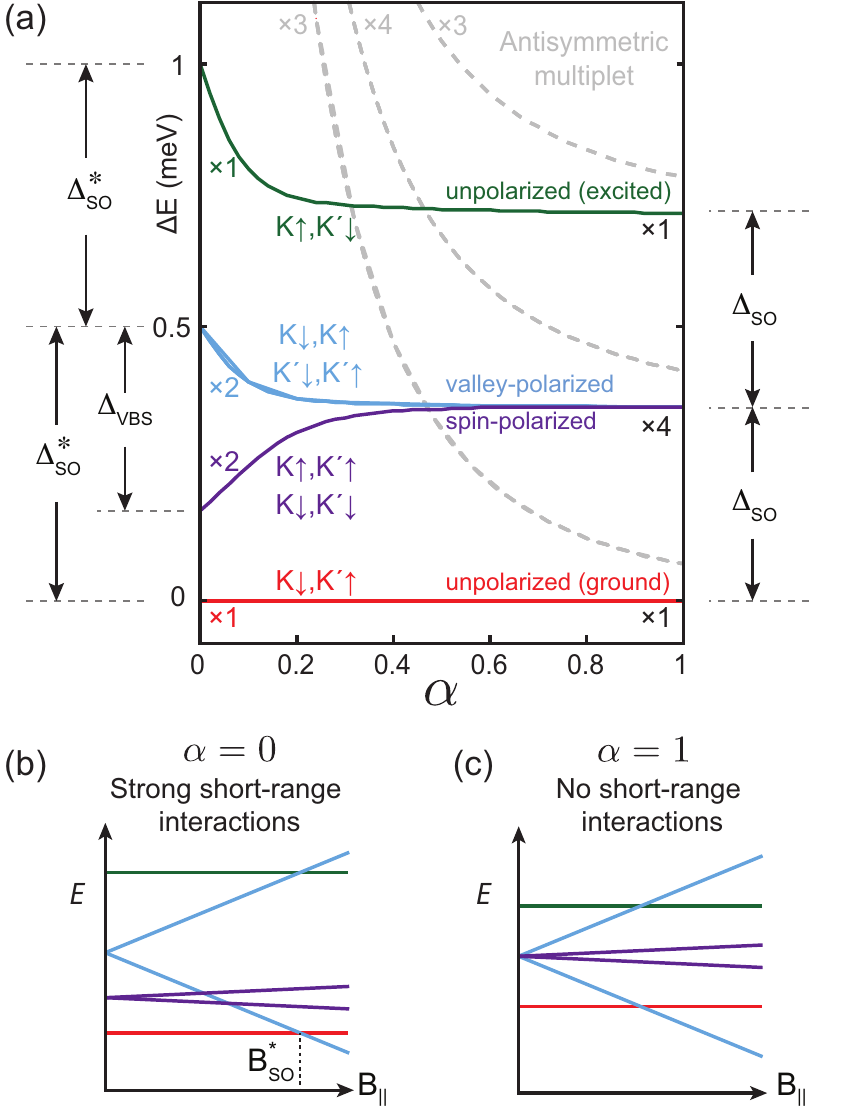}
\caption{\footnotesize{(Color online) Calculated two-electron quantum
    dot energies including short-range interactions. (a) Energies
    relative to the ground state of the spatially symmetric multiplet
    (solid lines) and the spatially antisymmetric multiplet (dashed
    lines).  The parameter $\alpha$ characterizes the
    strength of the long-range interactions: $\alpha = 1$, Wigner
    molecule regime; $\alpha = 0$: no long-range correlations.  The valley-polarized, spin-polarized, and unpolarized (ground and excited) levels are labelled, and the
    degeneracy of each set of lines is indicated.  For $\alpha = 1$,
    the dashed lines become degenerate with the solid lines, as in
    Fig.~\ref{theory_long_range} without spin-orbit coupling and
    short-range interactions.  For $\alpha = 0$, long-range
    correlations are suppressed, the single-particle wave functions
    begin to overlap and short-range interactions become
    stronger. Here, new splittings $\Delta_{\rm VBS}$ and $\Delta_{\rm
      SO}^*$ appear in the valley-spin structure of the multiplet.
    (b,c) Magnetic field dependence of the ground-state multiplet for
    $\alpha=0$ (b) and $\alpha=1$ (c). Short-range interactions in (b)
    break the degeneracy between the valley-polarized states and spin-polarized states at $B =
    0$. For $\alpha=1$, the valley/spin structure of the multiplet is
    the same as in the single-particle model, even though the
    longitudinal wave functions $\Psi(x_1,x_2)$ are highly
    correlated. States are labelled as in Table
    \ref{tab_symmetricantisymmtric}, except that the states used in
    the calculation are simplified such that they are independent of
    longitudinal~$\Psi(x_1,x_2)$. Panel (a) adapted from
    \onlinecite{wunsch2009few}.}}
\label{theory_short_range}
\end{figure}

In this section, we consider effects of the short-range Coulomb
interaction on the spectrum of a two-electron quantum dot. As seen in
the toy model, the long-range interaction distorts the longitudinal
wave functions, reducing the splitting $\DeltaSAS$ between the
symmetric and antisymmetric multiplets. Because the long-range
interaction does not couple to valleys, it does not change the
spin/valley level structure inside a multiplet. In contrast, the short-range Coulomb interaction can induce valley scattering, and thus
changes the splitting between different valley and spin states.

Several exact diagonalization calculations have been performed accounting for Wigner molecule effects, spin-orbit coupling, and
short-range Coulomb interactions
\cite{wunsch2009few,secchi2009coulomb,secchi2010wigner,secchi2012spectral,secchi2013inter}.
The resulting spectrum is shown in~Fig.~\ref{theory_short_range}.
In the calculation, $\alpha$ is set to zero to artificially
suppress long-range interactions while keeping the short-range on-site
energy\footnote{The Coulomb potential for the short-range interaction is
  taken with the same form as Eq.~\eqref{eq:coulomb}, but with
  cutoff $d$ chosen such than $V_{\rm C}(x_1 = x_2) = U_0$, where $U_0
  = 15$ eV is the charging energy associated with putting two
  electrons on the same $p_z$ orbital. Such a potential is also
  referred to as the Ohno potential \cite{mayrhofer2008spectrum}.} $U_0$
constant\footnote{Physically, this could be achieved by
  placing the nanotube close to a metallic gate or
  dielectric slab, such that the distance to the surface is large
  compared to the diameter but small compared to the nanotube
  length.}.  Figure \ref{theory_short_range}(a) shows the spectrum as a function of $\alpha$ at zero magnetic field. 
The colored (grey) lines
represent the \SSS ($\AS$) multiplet of Fig.~\ref{theory_long_range}. For $\alpha=1$, the dot is in the Wigner
molecule regime due to long-range interactions, and the grey lines in
(a) become nearly degenerate with the colored lines. Panels (b) and
(c) show the $B_{||}$ dependence of the spatially symmetric multiplet
for $\alpha = 0$ and $\alpha = 1$.

The influence of short-range interactions can be most
clearly seen for $\alpha = 0$ (Fig.~\ref{theory_short_range}(b)). The first effect is to lift the degeneracy of the spin-polarized and valley-polarized states.  In the single-particle model (and also at
$\alpha\gtrsim 1$, see below), these four states are degenerate at~$B=0$.  With strong
short-range interactions, the energy of the valley-polarized doublet is raised with respect to the spin-polarized doublet.
This is because a positive $U_0$ used in the calculation penalizes double occupancy of atomic sites, which by symmetry considerations does not occur for spin-polarized states (see Appendix \ref{appendixexchangeshortrange} for details).
This results in a new splitting $\Delta_{\rm VBS}$ within the $\SSS(0,2)$ multiplet, which can be viewed as an effective intervalley exchange energy\footnote{Note that Fig.~\ref{theory_short_range} plots the energies with respect to the ground state, explaining why spin-polarized states appear to be lowered with respect to valley-polarized states. The abbreviation~VBS denotes ``valley backscattering'', a name used for the term in the Coulomb scattering matrix that gives rise to an intervalley exchange splitting.}.
For the $\AS(0,2)$ multiplet, the effect of short-range Coulomb interaction is predicted to be much smaller \cite{secchi2013inter}, indicating that the longitudinal symmetry of the two-electron wave function plays an important role (Appendix~\ref{appendixexchangeshortrange}).

The second effect of the short-range interaction is seen in Fig.~\ref{theory_short_range} as an increase in the splitting between the unpolarized
ground state and the unpolarized excited state. The splitting is now equal to $2\Delta_{\rm SO}^* \equiv
2\sqrt{\Delta_{\rm SO}^2 + \Delta_{\rm VBS}^2}$ instead of $2\Delta_{\rm
  SO}$.  An important consequence is that the
two-electron ground state becomes valley-polarized at a higher
magnetic field than without short-range interactions (the crossing field $B_\mathrm{SO}^*$ is higher in (b) than in (c)). In the Wigner molecule regime
($\alpha=1$) the effects of short-range interactions are
suppressed: Due to strong correlations, the two electrons have little overlap irrespective of their longitudinal symmetry, thereby suppressing double occupancy of the
same atomic site.  Similar intervalley exchange effects have also been
seen in other calculations
\cite{mayrhofer2008spectrum,StecherPRB2010,secchi2009coulomb,secchi2010wigner,secchi2012spectral,secchi2013inter}.


\begin{figure}
\center \includegraphics{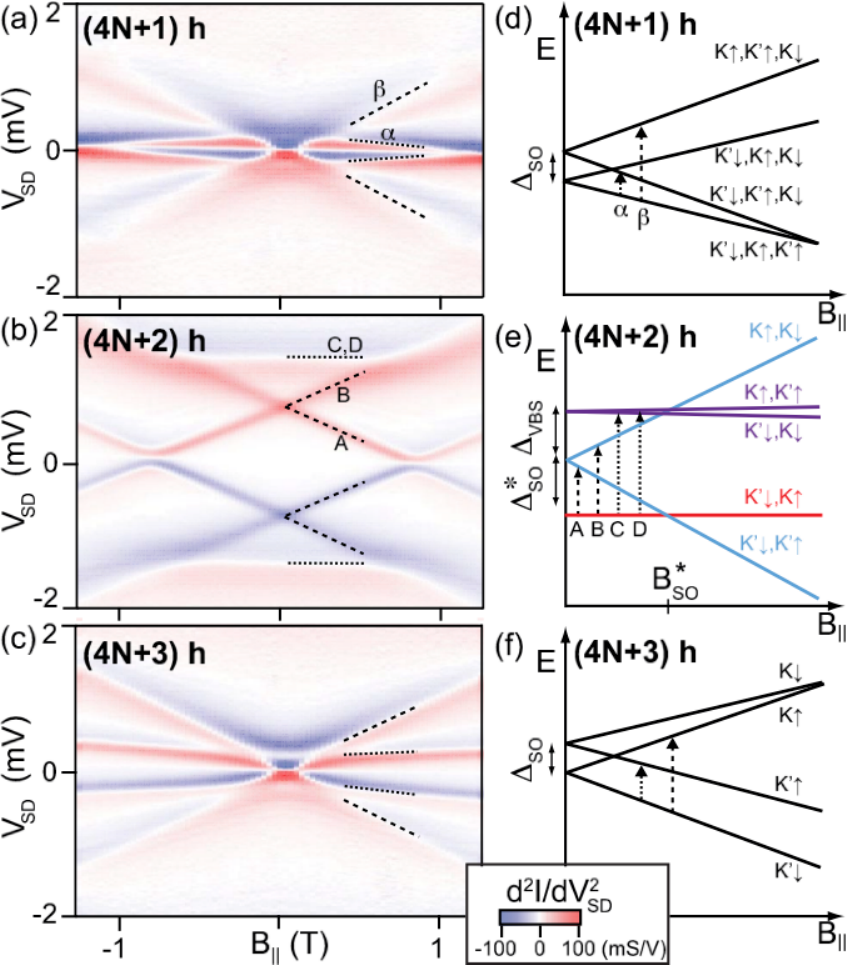}
\caption{\footnotesize{(Color online) Short-range Coulomb interactions
    in a many-hole quantum dot (estimated as containing 10-40 holes). (a-c) Cotunneling spectroscopy of
 of different charge configurations. (d-f) Expected spectra from a model with short-range Coulomb
    scattering. Quantum numbers of occupied electronic states in the valence band are indicated (cf.\ Fig. \ref{SO_spectrumtheory}). Lines in (a-c) measure length of the
    arrows in (d-f). Colors in (e) correspond to those used
    in Fig.~\ref{theory_short_range}. The signature of interactions
    can be seen by comparing the spectra of different charge configurations:
    the valley/spin spectra of the $4N{+}1$ and $4N{+}3$ configurations are consistent with those predicted by the single-particle model, while the
    spectrum of the $4N{+}2$ configuration is modified by
    short-range Coulomb interaction. Adapted from \onlinecite{CleuziouPRL2013}.}}
\label{short_range_experiment}
\end{figure}

Experiments by Cleuziou {\em et al.} 
demonstrated clear intervalley exchange effects in the spectra of
clean nanotubes in the many-hole regime (Fig.~\ref{short_range_experiment}). The key observation was that the $4N{+}1$
and $4N{+}3$ charge states showed a spectrum consistent with
shell-filling of the single-particle valley/spin levels, whereas a
different spectrum was observed for the $4N{+}2$ charge state \cite{CleuziouPRL2013}.  This difference between odd and even filling of the nanotube
multiplet is a clear signature that can only be
explained including short-range interactions.

The spectrum measured for $4N{+}2$ filling is similar to that of Fig.~\ref{theory_short_range}(b), except for the sign of the exchange
interaction. In particular, theoretical treatments of intervalley
exchange as outlined above predict that the valley-polarized states are
raised in energy above the spin-polarized states (as in Fig.~\ref{theory_short_range}(b)). In contrast, the
observed spectra~\cite{CleuziouPRL2013} show
the opposite (Fig.~\ref{short_range_experiment}(e)). This may
indicate that the exchange integral implied from the experimental
data has the opposite sign compared to that predicted by
theory so far. A similar observation was also reported for electrons
\cite{pecker2013observation}. The reason for this sign difference is
not understood, but suggests that the validity of theoretical approximations, as well as alternative mechanisms, should carefully be checked.

As discussed above, intervalley exchange is not expected to play a
significant role in the few-electron regime, since
the long-range Wigner correlations will suppress short-range Coulomb
interactions. The device of~Fig.~\ref{short_range_experiment}
was in the many-hole regime: a possible explanation of why
$\Delta_{\rm VBS}$ is so large is that the long-range Coulomb
interaction was screened by the holes in nearby
shells. However, a large $\Delta_{\rm VBS}$ was also reported in the
few-electron regime by both~\onlinecite{pecker2013observation} ($\Delta_{\rm VBS} = 0.2$ meV) and~\onlinecite{CleuziouPRL2013} ($\Delta_{\rm VBS} = 1.56$~meV calculated from the observed~$\Delta_{\rm SO}^*$).  
Estimates for uncorrelated states based on first-order perturbation theory predict~$\Delta_{\rm VBS}$ to be hundreds of $\mu$eV (Appendix~\ref{twoelectronexchange}). Taking into account correlations predicts a much smaller value, $1 - 10\ \mu$eV
\cite{wunsch2009few,secchi2009coulomb,secchi2010wigner,secchi2012spectral,pecker2013observation,secchi2013inter}.
  It is
an open question why $\Delta_{\rm VBS}$ is so large in these
experiments.

Finally, the large $\Delta_{\rm VBS}$ splitting seen in excited-state
spectroscopy has important implications for ground-state spectroscopy,
in which the magnetic field dependence of Coulomb peak position in
gate voltage is used to infer the spin-orbit splitting
\cite{ChurchillPRL2009,SteeleNcomm2013}. In particular, the magnetic
field where kinks occur in the ground-state chemical
potentials is now given by $B_{\rm SO}^* = \Delta_{\rm SO}^* / 2 \mu_{\rm orb}$
instead of by $B_{\rm SO} = \Delta_{\rm SO} / 2 \mu_{\rm orb}$. This raises the
possibility that the large spin-orbit interaction reported in
\cite{SteeleNcomm2013} extracted from the ground state energies could
be due to a large $\Delta_{\rm SO}^*$ and~$\Delta_{\rm VBS}$ instead of a large
$\Delta_{\rm SO}$. The intervalley exchange splitting required to match the experimental data, however, would be
$\Delta_{\rm VBS} \sim 3$~meV, even larger than the already unexpectedly large
values reported from excitation spectra
\cite{pecker2013observation,CleuziouPRL2013}.

\subsection{Beyond Wigner molecules: Correlation effects of many electrons in quantum dots}

The search for strong correlation effects with many electrons inspired some of the very first experiments on nanotubes. These early experiments were motivated by the predictions
of Luttinger liquid theory
\cite{luttinger1963exactly,imambekov2012one} and focused on power-law
behavior of the conductance
\cite{bockrath1999luttinger,YaoNature1999}. Detailed understanding
of these experiments was hampered, however, by the large disorder
present in these devices, leading to localization and Coulomb
blockade at low temperatures. Dynamical Coulomb blockade also leads
to power-law dependence on bias and temperature
\cite{ingold2005charge}. A clearer signature for a Luttinger liquid
would be spin-charge separation, giving spin and charge modes with
different velocities. Evidence for such spin-charge separation has
been seen in GaAs wires \cite{auslaender2005spin}, but has not been reported for nanotubes. 

Correlation effects for two electrons in a carbon nanotube quantum dot
are now fairly well understood theoretically, and well established
experimentally. A natural question is then: what happens if
more electrons or holes are added to the quantum dot?  One route could
be to probe the electron density itself in the quantum dot
using an STM \cite{ziani2013probing}. Another route is detailed
spectroscopy of suspended nanotube quantum dots in the multi-electron
and mult-ihole regime.  From an experimental point of view, there have
been suggestions of correlation effects in clean many-hole quantum
dots \cite{DeshpandeNPhys2008}. Studying the ground-state spin and
valley filling of a quantum dot as a function of hole number and
magnetic field, regions of strong spin and valley polarization were
observed that are not easily explained by a single-particle
picture.  A possible explanation of this strong spin and valley
polarization is the formation of a many-hole Wigner crystal in which
the kinetic energy is quenched by interactions, similar to
the suppressed $\DeltaSAS$ in the two-electron Wigner
molecule.  An exciting next step will be to use low-disorder
nanotubes with multiple gates to perform detailed spectroscopy with
tunable confinement, exploring the transition from the
well-established two-electron Wigner molecule to the regime where
many-electron Wigner crystals and Luttinger-liquid like correlations
may occur.
 
\subsection{Open Questions}

\begin{itemize}

\item Why is the observed intervalley exchange
  splitting $\Delta_{\rm VBS}$ so large? Even with a completely screened
  long-range interaction, calculations predict $\Delta_{\rm VBS} \lesssim
  \Delta_{\rm SO}$, while the opposite has been observed in some
  experiments.

\item Why does $\Delta_{\rm VBS}$ have the wrong sign in experiments? 
  A possible scenario is a superexchange mechanism for the short-range
  interaction. In superexchange, a net exchange of two electrons is
  achieved by two separate exchange processes with electrons in a
  third orbital. The third orbital could correspond to a state in a
  different shell, a state in the valence band, or a state in one of
  the higher subbands. Superexchange can have the opposite sign
  compared to direct exchange.

\item Why are not more devices deep in the Wigner molecule limit?
  Why is shell-filling theory so effective for describing so many
  devices?
  
\item Are there qualitatively new predictions from theories that take into account the dependence of the spatial wave function on $\tau$, $s$, and magnetic field (see Appendix~\ref{appendixsingleparticlestates})?

\item How do the well-understood strong correlations of the
  two-electron Wigner molecule extend to quantum dots with more
  electrons? Are there clear experimental signatures to look for?

\end{itemize}

\section{Conclusions and Outlook}\label{discussion}
The chirality is an unknown parameter in all the experiments discussed in this review. Although optical techniques have been developed to determine chirality on specially made, long nanotubes~\cite{LiuNnano2013,AmerNanoLett2013}, these characterizations are unfortunately incompatible with the short length of nanotube devices in the quantum regime. Therefore the growth of chirality-specific nanotubes remains an important challenge, although its realization may not be expected soon. Of course, several important (band-)structure parameters can be inferred from transport, such as the gap and the diameter. Uncertainties nevertheless arise since a measured gap may be affected by electron-electron interactions, or even mundane issues like surface coverage with water~\cite{EliasScience2009}. Such band-gap modifications are interesting subjects by themselves but then knowledge of the bare band gap, and thus the chirality, would be an indispensable input parameter.  

Fortunately, the general physics discussed in this review does not depend on the precise numerical values of the band-structure parameters. Most of this physics is a consequence of the simple fact that electrons in nanotubes live on a hexagonal lattice that is confined to a one dimensional, tubular geometry. We stress that in this sense nanotubes are unique solid-state structures. Given this structure the spin and valley phenomena described in this review represent general physics independent of the precise chirality. 

The inclusion of spin-orbit interaction completes our picture of the non-interacting, single-particle physics in nanotubes. Until 2008, this effect was widely thought to be negligible, but it has turned out to be highly relevant in many experiments. It is also important for applications involving coherent control of quantum states. The emerging picture of the role of interactions has also become much clearer in recent years since ultraclean devices have become available. We have described in detail the case of two interacting electrons. The study of many interacting electrons~\cite{DeshpandeNPhys2008} is a quest full of interesting challenges with its holy grail being formation of a long Wigner crystal. 

The theoretical understanding of the single-particle physics has been important to obtain a complete picture of the allowed spin-orbit terms.
There are still theoretical challenges relating to a quantitative microscopic understanding of the effects originating from the cylindrical geometry. As part of this, renormalization by long-range Coulomb interactions of the curvature-induced gap and the spin-orbit coupling has not been investigated.

Quantum states are best defined in closed systems and this review has therefore mainly focussed on quantum dots whose coupling to leads is weak (but large enough to allow a measurable current). Quantum dot states change when the coupling is increased, and qualitatively new phenomena can arise such as various Kondo effects and possibly even quantum phase transitions~\cite{MebrahtuNature2012}. Great potential for new experiments arises when the leads are given interesting properties. For example superconducting leads can induce superconducting correlations that are restricted by the special spin-valley quantum numbers of nanotubes. One can imagine Josephson junctions with the junction consisting of a nanotube with multiple bends. Experiments using superconducting contacts performed so far indeed indicate a rich research direction~\cite{DeFranceschiNnano2010}. Also in such hybrid nanotube devices, spin-orbit interactions may play a role, e.g. leading to modified Josephson currents \cite{ LimPRL2011}  or new detection schemes \cite{ BrauneckerPRL2013}. Superconducting contacts to ultraclean nanotubes have also recently been demonstrated~\cite{SchneiderSciRep2012}, opening the possibility of studying proximity-induced superconductivity with exceptionally low electronic disorder. Magnetic materials could be used as contacts to explore spintronics confined to one dimension~\cite{Sahoo2005}.

The field of quantum computing has been inspirational for the development of all kinds of qubit devices. This review discussed various types of nanotube. The coherent evolution in qubits can be used as an extremely sensitive probe of the environment. Nanotube qubits could be used to study in detail nuclear spins or mechanical vibrations. The holy grail here could be the realization of a coherent coupling between spin and motion. Another inspiration from the field of qubits is the use of nanotubes as on-chip charge sensors~\cite{BiercukPRB2006,GotzNL2008,ChurchillNPhys2009}, or as a component in circuit quantum electrodynamics~\cite{DelbecqPRL2011}. If long coherence times could be achieved, there is potential for a dramatically improved charge sensor~\cite{DialPRL2013}. Nanotubes have the advantage for sensing that short quantum dots show Coulomb blockade up to room temperature~\cite{PostmaScience2001}.

We hope this review inspires new generations of experiments taking advantage of the unique properties of carbon nanotubes. Similar to the evolution of quantum Hall physics in GaAs in the 1980s and 1990s, the remarkable quality of devices that can now be achieved using ultraclean techniques indicates that nanotubes will remain an exciting playground for exploring rich new physics for years to come.

\section{Acknowledgements}
We acknowledge many colleagues and collaborators, especially from Delft, Copenhagen, Harvard, Cornell, and Oxford. Especially we thank Ramon Aguado, Hugh Churchill, Thomas Sand Jespersen, Charles Marcus, Jens Paaske, Fei Pei, Emmanuel Rashba, Mark Rudner, Christoph Strunk, Stephan Weiss, and Floris Zwanenburg. We acknowledge support from the Carlsberg Foundation, the Danish Council for Independent Research, the Danish National Research Foundation, the Dutch National Science Foundation, NWO/FOM, the European Union, and the Royal Academy of Engineering.

\appendix
\section{Transport spectroscopy in quantum dots}
\label{secQuantumDots}
\label{ap_coulombspectroscopy}

Here we give a basic overview of quantum dots and transport spectroscopy, focusing on techniques by which data in this review is derived.

\begin{figure}
\center
\includegraphics{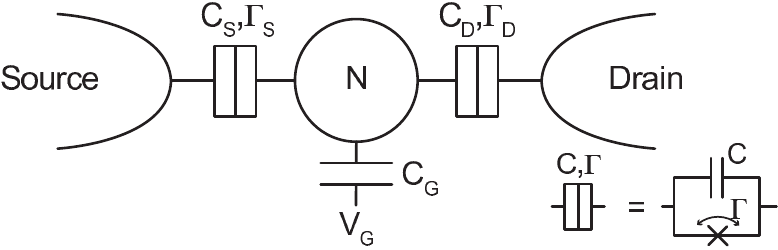}
\caption{\footnotesize{Electrical model of a single quantum dot: a conducting island is coupled with tunnel rates $\Gamma_\mathrm{S}$ and $\Gamma_\mathrm{D}$ to source and drain, and with capacitance $C_\mathrm{S}$, $C_\mathrm{D}$, and $C_\mathrm{G}$ to source, drain, and gates.}}
\label{fig_dotschematic}
\end{figure}

\subsection{Coulomb blockade and the constant interaction model}
An electrical schematic of a quantum dot as in~Fig.~\ref{CNTdot} is shown in Fig.~\ref{fig_dotschematic}. If both the tunnel rates to the leads $\Gamma_\mathrm{S,D}$ and the thermal energy $k_\mathrm{B}T$ are less than the charging energy $E_\mathrm{C}=e^2/C$, where $C$ is the total dot capacitance to the outside world, then the electron occupation becomes constrained to take an integer value. The equilibrium occupation can be adjusted by tuning~$V_\mathrm{G}$.

This quantization of the dot occupation strongly modifies the transport characteristics of this circuit. With all other parameters held fixed, any change of the dot occupation away from equilibrium increases the electrostatic energy of the system. This causes a suppression of the current known as Coulomb blockade.  However, for particular values of $V_\mathrm{G}$, the suppression can be lifted. This allows the energy levels of the quantum dot to be mapped out by measuring the device conductance as a function of $V_\mathrm{G}$ (for extensive reviews, see~\cite{KouwenhovenBook1997,KouwenhovenRPP2001, HansonRMP2007})

\begin{figure*}
\center
\includegraphics[width=17.2cm]{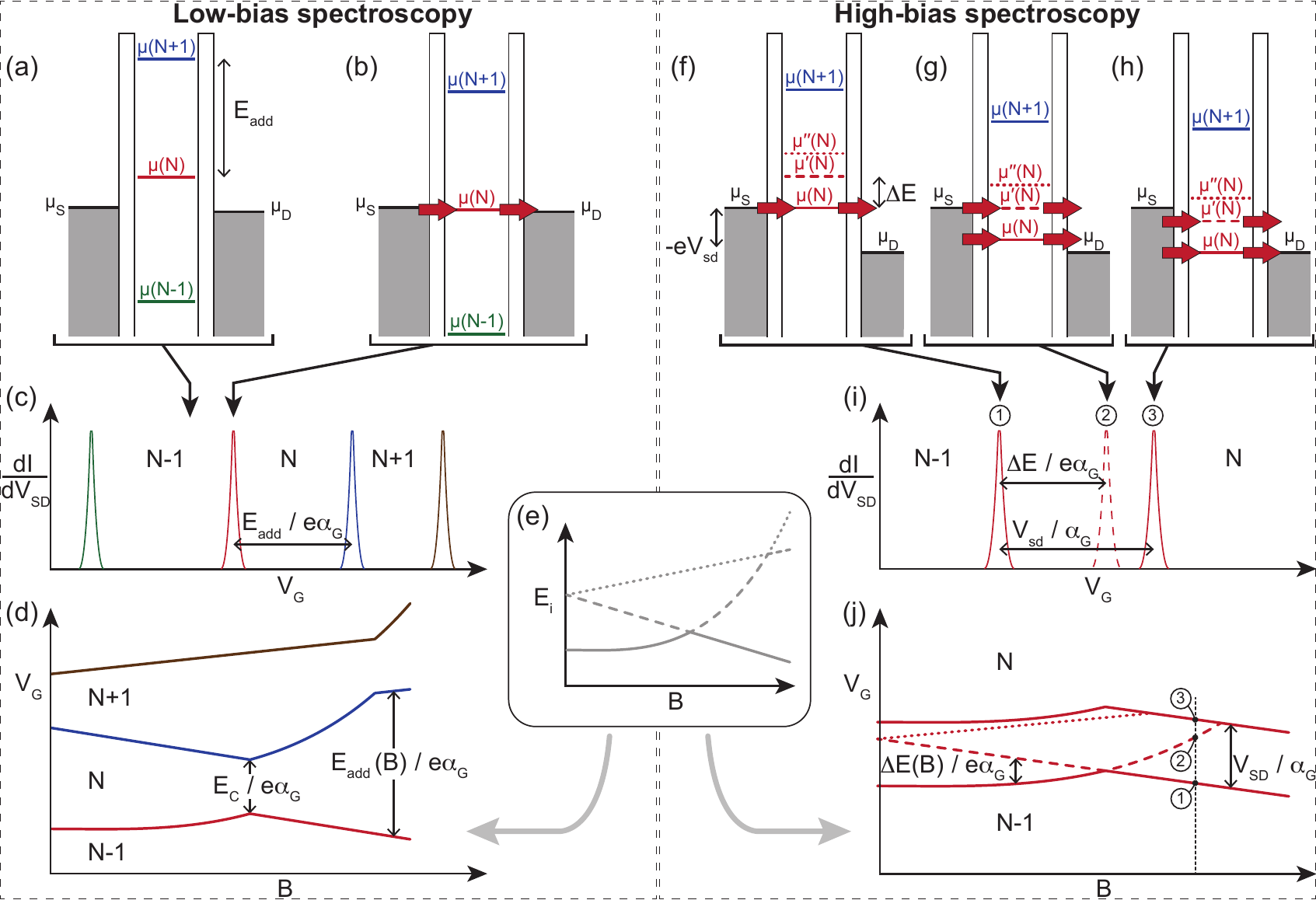}
\vspace{-0.2cm}
\caption{\footnotesize{(Color online) Two techniques of electron transport spectroscopy. In low-bias spectroscopy (a-d), at most a single quantum dot level falls within the bias window set by the lead electrochemical potentials. (a) With no ground-state level in the bias window, levels are filled up to $\mu_\mathrm{D}$, and the electron number is fixed (in this case, at $N-1$) due to Coulomb blockade. (b) With one of the levels $\mu(N)$ in the bias window, the electron number fluctuates between $N$ and $N-1$, leading to conductance through the dot. (c) Sweeping $V_\mathrm{G}$ moves each level in turn through the bias window, leading to a series of current peaks. From the peak spacing, the addition energy $E_\mathrm{add}$ for each electron number can be deduced, allowing the ground-state energy to be studied. (d) Sketch of transport peak evolution measured using this technique, where the energy levels $E_i$ are assumed to vary with magnetic field $B$ as shown in (e). In high-bias spectroscopy (f-j), more than one level can fall within the bias window, allowing excited-state energies to be measured. Sweeping $V_\mathrm{G}$ through the $N-1 \leftrightarrow N$ transition, a series of transport peaks appear in $dI/dV_\mathrm{SD}$. (f) Current first appears when $\mu(N)$ crosses $\mu_\mathrm{S}$. (g) When $\mu'(N)$ crosses $\mu_\mathrm{S}$, additional current can flow through an excited state of the dot. (h) The last peak in the series occurs when $\mu(N)$ crosses $\mu_\mathrm{D}$, leaving the bias window. (i) The corresponding series of transport peaks, with the spacings corresponding to $\Delta E$ and $V_\mathrm{SD}$ marked. When $\mu(N)<\mu_\mathrm{D}$, Coulomb blockade is reestablished with the dot in its $N$-electron ground state; there are therefore no peaks corresponding to states with excitation energies larger than $eV_\mathrm{SD}$, such as that marked by dotted line in (d-f). (j) Sketch of peak evolution measured using this technique for the same underlying energy levels as shown in (e). The $V_\mathrm{G}$ trace in (i) corresponds to the vertical dashed line in (j).}}
\label{CNTspectroscopy}
\end{figure*}

 Coulomb blockade is lifted whenever each step of electron tunneling through the device is energetically favourable. To understand how the observed conductance features relate to the energy levels of the device, we introduce the electrochemical potential $\mu(N)$ of the dot for occupation $N$, defined as the difference in energy~$U$ between $N$-electron and $N-1$-electron ground states:
\begin{equation}
\mu(N)=U(N)-U(N-1).
\end{equation}
Correspondingly, the electrochemical potentials of the leads are defined as the energy to add an additional electron at the Fermi level; with a bias $V_\mathrm{SD}$ applied to the source as in Fig.~\ref{CNTdot}(a),
\begin{eqnarray}
\mu_\mathrm{S}&=&E_\mathrm{F}-eV_\mathrm{SD} \\
\mu_\mathrm{D}&=&E_\mathrm{F}
\end{eqnarray}
where $E_F$ is the Fermi energy.

The electrochemical potential is in general related in a complicated way to the single-particle energy levels, because each electron added electrostatically perturbs the energies of the electrons already on the dot. However, the relationship becomes simpler in the \emph{constant-interaction model}, which makes two assumptions. First, all Coulomb interactions, both between electrons on the dot and between the dot and the environment, are parameterised by a single constant capacitance $C$, which is the sum of capacitances $C_\mathrm{S}, C_\mathrm{D}, C_\mathrm{G}$ to the source, drain and gate\footnote{In devices with more than one gate, additional capacitances must be added to the model in a straightforward way.}. Second, the single-particle energy levels are assumed to be independent of these interactions, and therefore not changed by adding additional electrons. Under these assumptions, the dot energy is:
\begin{equation}
U(N)=\frac{[-e(N-N_0)+C_\mathrm{S}V_\mathrm{S}+C_\mathrm{D}V_\mathrm{D}+C_\mathrm{G}V_\mathrm{G}]^2}{2C} + \sum_{i=1}^{N}E_i,
\end{equation}
where $N_0$ is the occupancy with no voltages applied (set by fixed charges in the environment, e.g.~substrate charges, and not necessarily quantized) and $E_i$ are the single-particle energy levels. The first term is the electrostatic energy stored in the dot capacitances, while the second is the sum of the single-particle confinement energies.

In this approximation, the electrochemical potential is:
\begin{equation}
\mu(N)=(N-N_0-\frac{1}{2})E_\mathrm{C}-\frac{E_\mathrm{C}}{e}(C_\mathrm{S}V_\mathrm{S}+C_\mathrm{D}V_\mathrm{D}+C_\mathrm{G}V_\mathrm{G})+E_N,
\label{chemicalpotential}
\end{equation}
where $E_\mathrm{C} = e^2/C$ is the charging energy. The electrochemical potential increases for successive values of $N$, forming a ladder of levels as shown in Fig.~\ref{CNTspectroscopy}. The separation between adjacent levels is the addition energy:
\begin{eqnarray}
\label{eqEadd}
E_\mathrm{add}(N) 	&=& \mu(N)-\mu(N-1)\notag \\
				&=& E_\mathrm{C} + \Delta E(N)
\end{eqnarray}
and includes both an electrostatic term $E_\mathrm{C}$ and the quantum energy level spacing $\Delta E(N) \equiv E_{N}-E_{N-1}$. From~Eq.~\eqref{chemicalpotential}, changing the gate voltage moves the entire ladder of electrochemical potentials up or down.

\subsection{Low-bias spectroscopy}
\label{ap_specHighBias}
The condition that both tunneling events be energetically favourable is equivalent to saying that the chemical potential must decrease at each step\footnote{Thermal excitations in the leads relax this constraint, broadening the Coulomb peaks. Here this effect is ignored, which is permissible if the temperature is less than both $\Delta E$ and $E_\mathrm{C}$.}. In other words, some $N$ must exist for which:
\begin{equation}
\mu_\mathrm{S} > \mu(N) > \mu_\mathrm{D}.
\label{CBcondition}
\end{equation}
Consider first the situation of low bias, where $eV_\mathrm{SD} \ll \Delta E, E_\mathrm{add}$. The corresponding level diagram in the blockaded case (Fig.~\ref{CNTspectroscopy}(a)), shows that no ground-state chemical potential satisfies Eq.~\eqref{CBcondition}, so no current flows. However, by increasing $V_\mathrm{G}$ to lower the ladder of electrochemical potentials, the blockade can be lifted (Fig.~\ref{CNTspectroscopy}(b)). As a function of $V_\mathrm{G}$, the current shows a series of Coulomb peaks, with each valley between the peaks corresponding to Coulomb blockade with a different fixed occupation $N$ (Fig.~\ref{CNTspectroscopy}(c)). For each $N$, the peak separation in $V_\mathrm{G}$ is equal to $E_\mathrm{add}(N)/e\alpha_\mathrm{G}$, where $\alpha_\mathrm{G}=C_\mathrm{G}/C$ is the lever arm that characterizes the coupling of the gate to the dot. By measuring the Coulomb peak positions in gate space as a function of some external parameter, the evolution of energy levels can be deduced.

\subsection{High-bias spectroscopy}

Low-bias transport is only sensitive to the ground-state energy of the device. With a bias larger than the single-particle level spacing, $e|V_\mathrm{SD}|>\Delta E$, the excited states can also be populated.  To interpret the resulting transport features, it is necessary to consider the corresponding electrochemical potentials. We define the first excited state electrochemical potential as
\begin{eqnarray}
\mu'(N)  &=& U'(N)-U(N-1) \\
&=& \mu(N)+\Delta E(N+1)
\end{eqnarray}
where $U'(N)=U(N)+\Delta E(N+1)$ is the first excited state energy of the $N$-electron dot. Higher excited states can be defined the same way~\cite{KouwenhovenRPP2001}.

As illustrated in Fig.~\ref{CNTspectroscopy}(f-h), transport can proceed via both the ground state and excited states within the bias window. As $V_\mathrm{G}$ is increased from the $N-1$ electron valley, the current first increases when $\mu(N)$ crosses~$\mu_\mathrm{S}$ (Fig.~\ref{CNTspectroscopy}(f)). With a further increase in $V_\mathrm{G}$, $\mu'(N)$ enters the bias window (Fig.~\ref{CNTspectroscopy}(g)). This allows transport via the first excited state, which continues until $\mu(N)$ crosses $\mu_\mathrm{D}$ (Fig.~\ref{CNTspectroscopy}(h)). For more positive values of~$V_\mathrm{G}$, Coulomb blockade is reestablished, and transport is blocked through both ground and excited states.

This series of resonances between dot and lead electrochemical potentials is usually seen by plotting the conductance $dI/dV_\mathrm{SD}$ as a function of $V_\mathrm{G}$, which results in a series of peaks as each transport channel is opened or closed (Fig.~\ref{CNTspectroscopy}(i))\footnote{Although the most common situation is for all peaks to have positive $dI/dV_\mathrm{SD}$ as drawn, the excited-state peaks can be negative if there is a strong difference of tunnel coupling between different states~\cite{WeinmannPRL1995}.}. From the peak spacings, the excited-state energies can be read off as shown. This technique also gives a convenient way to measure $\alpha_\mathrm{G}$; the gate voltage separation of the first and last peaks corresponds to shifting $\mu(N)$ from $\mu_\mathrm{S}$ to $\mu_\mathrm{D}$, and is therefore equal to $V_\mathrm{SD}/\alpha_\mathrm{G}$, where $V_\mathrm{SD}$ is set in the experiment\footnote{This discussion ignores processes whereby  tunneling \emph{off} the quantum dot leaves it in an $N-1$ electron excited state, which lead to additional conductance peaks not shown in Fig.~\ref{CNTspectroscopy}(g). These additional peaks, which are not relevant to data discussed in this Review, are clearly distinguished from those in~Fig.~\ref{CNTspectroscopy}(i) in a two-dimensional plot of conductance versus $V_\mathrm{SD}$ and $V_\mathrm{G}$, and can usually be eliminated by appropriate tuning of the tunnel barriers. For full discussion, see~\cite{KouwenhovenRPP2001}.}. These two complementary spectroscopy methods allow the energy levels of the quantum dot to be measured as a function of various experimental parameters.

\section{Theoretical background}
\label{ap_isospin}
\label{ap_SO}

This Appendix introduces the theoretical background of the review. The focus will be pedagogical and we will not attempt to refer to relevant theoretical papers. For this we refer to more theoretical reviews, and to the main text.
We derive the Dirac equation used extensively in the main text. Physical pictures of the two types of spin-orbit coupling (orbital- and Zeeman-like terms), as well as the curvature-induced band gap are given. Finally, we address the form of the single-particle and two-particle wave functions.

\subsection{Graphene band structure near the Dirac points}

\begin{figure}
\center
\includegraphics{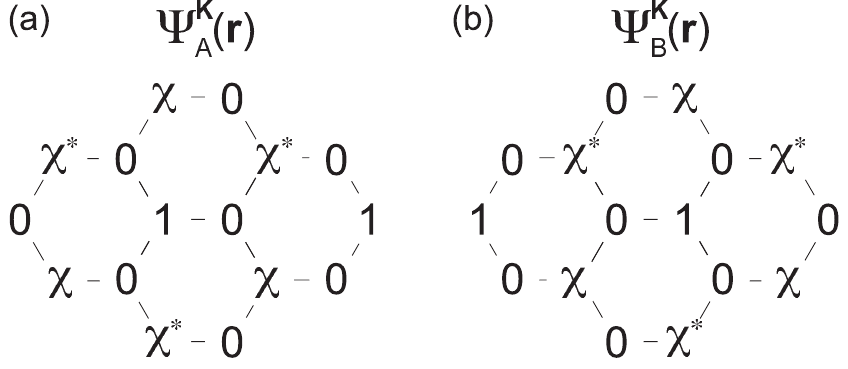}

\caption{Examples of orthonormal spatial eigenfunctions satisfying both Eqs.~\eqref{Translation} and \eqref{rotation} at the $K$ point for the minimal tight-binding model. The site coefficients for a one-orbital description of the eigenfunctions are shown, with $\chi=e^{i 2\pi/3}$. In the general case, the $\Psi_A^\mathbf{K}$ and $\Psi_B^\mathbf{K}$ wave functions have the same rotational symmetry as in the figure. The eigenfunctions at the $K'$ point are the same but with $\chi$ replaced by $\chi^*$.
\label{Diracwavefunction}}
\end{figure}

Since we are interested in the behavior close to the Fermi level, we focus on the bandstructure near the two Dirac points. This will be done in two ways: first a simple $\mathbf{k}\cdot\mathbf{p}$ calculation is applied to show that the spectrum can be derived from symmetry arguments alone. This is confirmed using a tight-binding calculation, which is also easier to generalize to the case with broken symmetry, as in a nanotube.

\subsubsection{The $\kkk\cdot\ppp$ derivation}

The $\kkk\cdot\ppp$ calculation for graphene \cite{DiVincenzoPRB1984,MarderBook2000} uses the fact that the potential has a unit cell of two carbon atoms and is invariant under translations $\mathbb{T}_{\mathbf{a}_{1,2}}$ by lattice vectors~$\mathbf{a}_{1,2}$ as well as under rotation $\mathbb{R}_{2\pi/3}$ by angle $2\pi/3$ about a lattice symmetry point ($C_3$ symmetry). At the Dirac points $\mathbf{K}$ and $\mathbf{K'}$, we define Bloch states, denoted $\Psi^{\KKK(\KKK')}_A(\boldr)$ and $\Psi^{\KKK(\KKK')}_B(\rrr)$, which are eigenstates of the translation operator:
\begin{equation}\label{Translation}
  \mathbb{T}_{\mathbf{a}_{1,2}}\Psi^{\KKK}_{A(B)}(\rrr)
  =e^{i\mathbf{K}\cdot\mathbf{a}_{1,2}}\Psi^{\KKK}_{A(B)}(\rrr),
\end{equation}
and likewise for $\KKK'$. They are degenerate because of the inversion symmetry (which interchanges $A$ and $B$ and $\kkk$ and $-\kkk$). The relative phase between two sites separated by the unit vector $\mathbf{a}_1$ is given by
\begin{equation}\label{phaseChi}
  e^{i\mathbf{K}\cdot\mathbf{a}_1}=e^{-i2\pi/3}=\chi^*,
\end{equation}
and by $\chi$ for $\KKKp$. Symmetry allows the functions $\Psi^{\KKK}_{A,B}(\boldr)$ to be chosen as eigenstates of the rotations $\mathbb{R}_{2\pi/3}$ around a center of a hexagon in the following way:
\begin{equation}\label{rotation}
\mathbb{R}_{2\pi /3}\Psi _{A}^{\mathbf{K}} =\chi \Psi _{A}^{\mathbf{K}},\quad \mathbb{R}_{2\pi /3}\Psi _{B}^{\mathbf{K}}=\chi ^{\ast}\Psi _{B}^{\mathbf{K}}.
\end{equation}
(At the $\mathbf{K}^{\prime }$ point, one should replace $\chi $ by $\chi ^{\ast}$.)

We now derive an effective Hamiltonian using $\kkk\cdot\ppp$ perturbation theory. Expanding the Bloch Hamiltonian~$H_{\boldk} $ around the Dirac points, we write $\boldk = \KKK+\bm{\kappa}$ and
\begin{equation}
\label{eq_effectiveHamiltonian}
H_{\boldk} \approx H_{\KKK} + \mathit{H}^\tau_{\bm{\kappa}},
\end{equation}
where $\tau=1(-1)$ for the $K(K')$ point. At $K$ the functions $\PsiAK$ and $\PsiBK$ are degenerate and we define our energy scale so that $\langle\PsiAK|H_{\KKK}|\PsiAK\rangle=0$. The correction to the Hamiltonian is
\begin{equation}\label{Hkappa}
H_{\bm{\kappa}}^\tau \equiv \frac{\hbar}{m}\bm{\kappa} \cdot \ppp,
\end{equation}
with $\ppp=-i \hbar \bm{\nabla}$ being the momentum operator. In the $\{\Psi^{\KKK}_A(\boldr),\Psi^{\KKK}_B(\boldr)\}$ basis, one can now find the matrix elements of the momentum operator  $\mathbf{p}_{CC'}^{\mathbf{K}}=\left\langle
\Psi _{C}^{\mathbf{K}}|\mathbf{p}|\Psi _{C'}^{\mathbf{K}}\right\rangle $, where $C,C'$ both take the values $A$ or $B$. Each matrix element $\mathbf{p}_{CC'}^{\mathbf{K}}$ is a vector in the $x$-$y$ plane of the graphene sheet. The $AB$ component follows from the rotational symmetry of the wave functions:
\begin{equation}\label{RpAB}
\mathbb{R}_{2\pi /3} \mathbf{p}_{AB}^{\mathbf{K}}=\chi^* \mathbf{p}_{AB}^{\mathbf{K}},
\quad
\mathbb{R}_{2\pi /3} \mathbf{p}_{AB}^{\mathbf{K'}}=\chi \mathbf{p}_{AB}^{\mathbf{K'}}.
\end{equation}
The eigenvectors in~Eq.~\eqref{RpAB} are
\begin{equation}\label{pAB}
\mathbf{p}_{AB}^{\mathbf{K}}\propto \left(
\begin{array}{c}
1 \\
-i
\end{array}
\right) ,\quad \mathbf{p}_{AB}^{\mathbf{K}^{\prime }}\propto \left(
\begin{array}{c}
1 \\
i%
\end{array}
\right) .
\end{equation}
The diagonal elements~$\mathbf{p}_{AA}^{\mathbf{K}}$ and~$\mathbf{p}_{BB}^{\mathbf{K}}$ vanish, since they obey an analogous equation to~Eq.~\eqref{RpAB}, but with $\chi$ replaced by 1, which only has zero solutions.

The low-energy Hamiltonian \eqref{Hkappa} is thus
\begin{equation}
H_\mathbf{\kappab}^\tau\propto\left(\begin{array}{cc}
0 & \tau \kappa_{x}-i\kappa_{y} \\
\tau \kappa_{x}+i\kappa_{y} & 0%
\end{array}
\right) =\hbar v_{F}( \tau \kappa_{x}\sigmaone+\kappa_{y}\sigmatwo),\label{HDirac}
\end{equation}
where the Fermi velocity $v_{F}$ was introduced as a phenomenological parameter, to match Eq.~(1).
Here $\sigmaone, \sigmatwo,$ are the usual Pauli matrices, now working in the $A/B$ sublattice.
To describe the Hilbert space spanned by the valley index $\tau$ and $A/B$ sublattice spinor some authors use the terms isospin and pseudospin.
However, there is no established convention as to which is which; see opposite definitions for example in \cite{KanePRL1997,McCannPRL2006}.

\subsubsection{The tight binding derivation}

In the tight binding approach the starting point is the Bloch functions of $A$ and $B$ sublattices:
\begin{equation}\label{PsiTB}
  \Psi_{A/B}^\kkk(\rrr)=\sum_n e^{i\kkk\cdot\mathbf{R}_n} \varphi_{A/B}(\rrr-\mathbf{R}_n),
\end{equation}
where $\varphi_{A/B}$ are the local basis functions\footnote{These functions are not orthogonal. The overlap matrix should therefore in principle be included when solving Schr\"odinger's equation \eqref{Hschr}. However, it turns out to be higher order in $\kappa$ and we can neglect it here.} for atomic orbitals at sites $A$ and $B$, respectively (see Fig.~\ref{Diracwavefunction}). Neighbouring orbitals hybridize via bonds along direction $\taub_i$ with overlap matrix elements $t_i$, with $i=1,2,3$. The Hamiltonian overlap of $\PsiAk$ and $\PsiBk$ is therefore
\begin{equation}\label{HABTB}
 \langle\PsiAk|H|\PsiBk\rangle = \sum_i t_i e^{i\kkk\cdot(\taub_i-\taub_1)}.
\end{equation}
For graphene, rotational symmetry means all bonds are wquivalent ($t_i=t$), and expanding around the $K$ point gives (using the definitions in Table I):
\begin{equation}\label{HABTB2}
 \langle\PsiAk|H|\PsiBk\rangle = t\sum_i  e^{i\kkk\cdot(\taub_i-\taub_1)}=\frac{it}{\sqrt3}(\kappa_x -i\kappa_y).
\end{equation}
To be consistent with the $\kkk\cdot\ppp$ result \eqref{HDirac}, we then make a transformation of the phase of the basis states so that $\PsiBk\rightarrow -i\PsiBk$, which recovers the Hamiltonian \eqref{HDirac}.

\subsubsection{Graphene wave functions}

Solving for the eigenenergies and eigenstates of the low-energy Hamiltonian \eqref{HDirac}
\begin{equation}\label{Hschr}
  H_\kappab^\tau \psi_{\kappab}^\tau=E_\kappab \psi_{\kappab}^\tau,
\end{equation}
one finds that the energies are $E_\kappab=\pm\hbar v_\mathrm{F}\sqrt{\kappa_x^2+\kappa_y^2}$, with the upper (lower) sign corresponding to the conduction (valence) band. The corresponding eigenstates are ``spinors" in $A/B$ space:
\begin{equation}\label{spinorsol}
\psi_{\bm{\kappa}}^\tau=\left(\begin{array}{c}
                                         F_A^\tau(\kappab)\\
                                         F_B^\tau(\kappab)
                                       \end{array}
                                     \right)=
                                     \frac{1}{\sqrt{2}}\left(
                                       \begin{array}{c}
                                         \pm\frac{\tau\kappa_x+i\kappa_y}{ \sqrt{\kappa_x^2+\kappa_y^2}}\\
                                         1 \\
                                       \end{array}
                                     \right).
\end{equation}
The spinor gives the weights of the sublattice Bloch states \eqref{PsiTB}. The wave function near $K$ is thus
\begin{equation}\label{PsiBloch}
\Psi_{\KKK+\kappab}(\rrr)= F_A^\KKK(\kappab)\Psi^{\KKK+\kappab}_A(\boldr) +  F_B^\KKK(\kappab)\Psi^{\KKK+\kappab}_B(\boldr),
\end{equation}
which can be separated into a fast and a slow part:
\begin{equation}\label{PsiBlochenv}
\Psi_{\KKK+\kappab}(\rrr)\approx e^{i\kappab\cdot\rrr}\left(F_A^\KKK(\kappab)\Psi^{\KKK}_A(\boldr) +  F_B^\KKK(\kappab)\Psi^{\KKK}_B(\boldr)\right),
\end{equation}
where the plane wave part $e^{i\kappab\cdot\rrr}$ is the so-called envelope function. Of course, a similar relation holds near $\KKK'$.

\subsection{Nanotubes: Graphene on a cylinder}

One must take care when transforming from the graphene coordinates to the nanotube cylindrical coordinates. Above we used the coordinate system defined in Table~I, \ie the horizontal bonds in Fig.~55 are along our $x$-direction. When changing to the coordinate system $(k_{\perp},k_{||})$, the coordinates are rotated as:
\begin{equation}\label{cylrotation}
\left(
\begin{array}{c}
\kappa_{x} \\
\kappa_{y}%
\end{array}%
\right) =\left(
\begin{array}{cc}
\cos \vartheta & \sin \vartheta \\
-\sin \vartheta & \cos \vartheta%
\end{array}%
\right) \left(
\begin{array}{c}
\kappa_{\perp } \\
\kappa_{\parallel }%
\end{array}%
\right) ,
\end{equation}%
where $\vartheta $ is the angle between $\mathbf{C}$ and $\mathbf{\hat x}$, or $\vartheta =\pi /6-\theta$ in terms of the chiral angle $\theta $. In these coordinates, Eq.~\eqref{HDirac} becomes
\begin{equation}
H_\kappab^\tau=\hbar v_{F}\left(
\begin{array}{cc}
0 & e^{i\tau \vartheta }(\tau \kappa_{\perp }-i\kappa_{\parallel }) \\
e^{-i\tau \vartheta }(\tau \kappa_{\perp }+i\kappa_{\parallel }) & 0%
\end{array}%
\right) .
\end{equation}
The coordinate rotation thus creates a phase factor $e^{i\tau \vartheta }$, which can be removed by a unitary transformation. Doing this, we get the nanotube Hamiltonian in cylindrical coordinates:
\begin{equation}\label{HCNT}
H_{\mathrm{CNT},\kappab}^\tau=U_{\vartheta }^{{}}HU_{\vartheta }^{\dagger} =\hbar v_{F}( \tau \kappa_\perp\sigmaone+\kappa_\parallel\sigmatwo) ,
\end{equation}%
where the unitary operator $U_\vartheta$ is
\begin{equation}
\quad U_{\vartheta }^{{}}=\left(
\begin{array}{cc}
1 & 0 \\
0 & e^{i\tau \vartheta }%
\end{array}%
\right) .  \label{UNI}
\end{equation}
The Hamiltonian \eqref{HCNT} has eigenenergies $E(\kappa_\parallel,\kappa_\perp)=\pm \sqrt{\kappa_\parallel^2+\kappa_\perp^2}$.

When graphene is rolled up to a nanotube (Fig.~\ref{structure}), the wave functions are restricted by periodic boundary conditions ($\Psi_{\mathbf{K}+\mathbf{\kappab}}(\mathbf{r})=\Psi_{\mathbf{K}+\mathbf{\kappab}}(\mathbf{r+C})$), implying that 
\begin{equation}\label{boundarycondition2}
  e^{i (\mathbf{K}+\bm{\kappa})\cdot\mathbf{C}} =1.
\end{equation}
The condition for a metallic nanotube is that the spectrum of the Hamiltonian \eqref{HCNT} is gapless, \ie $\kappa_\perp=0$. This happens when~$\mathbf{K}\cdot\mathbf{C}=2\pi M$, where $M$ is an integer. From the definitions in Table I, this is equivalent to $n-m=3M$~(Fig.~5).

\subsection{The curvature-induced gap}

Deformation of graphene due to nanotube curvature causes the overlap matrix elements to depend on direction, breaking the $C_3$ rotational symmetry. Because the effect on the Dirac cones is to shift them in momentum space~\cite{KanePRL1997}, graphene remains ungapped when curved. However, in a nominally metallic nanotube, this shift opens a band gap (Fig.~8). To understand these effects, we return to the degenerate subspace $\{\PsiAK,\PsiBK\}$ and calculate the correction to first order in the curvature perturbation $\Hcv$. First, we see the diagonal matrix elements are equal\footnote{This is because the combination of inversion and time-reversal symmetries rules out a $\sigma_z$ term in the Hamiltonian.}, $\langle \PsiAK|\Hcv|\PsiAK\rangle=\langle \PsiBK|\Hcv|\PsiBK\rangle$. Next, using the tight-binding wave functions, we find the off-diagonal elements are
\begin{equation}\label{C3broken}
  \langle \PsiAK|H_\mathrm{cv}|\PsiBK\rangle\propto \delta t_1+\delta t_2\chi+\delta t_3\chi^2,
\end{equation}
where $\delta t_i$ is the curvature correction to the hopping matrix elements for the three bonds. Since $\delta t$ is zero for a bond parallel to the nanotube axis and maximal for one perpendicular to it, the leading angular dependence is $\delta t_i\propto \cos^2\eta_i$, where $\eta_i$ is the angle between the bond $i$ and the chiral vector. Equation \eqref{C3broken} then evaluates to
\begin{equation}
\left\langle \Psi _{A}^{\mathbf{K}}|H_\mathrm{cv}|\Psi _{B}^{\mathbf{K}}\right\rangle \propto \sum_{p=-1}^{1}\cos ^{2}\left( \vartheta +2p\pi
/3\right) (\chi^*)^p=\frac{3}{4}e^{-i2\vartheta },
\end{equation}
with $\chi$ replaced by $\chi^*$ in the $K'$ valley. Applying the same transformation that led from Eq.~\eqref{HABTB2} to Eq.~\eqref{HDirac} and changing the phase as explained below Eq.~\eqref{HABTB2}, we obtain the curvature correction in cylindrical coordinates:
\begin{eqnarray}
H_{\mathrm{CNT,cv}}&=&\hbar v_{F}\Delta \kappa^\mathrm{cv}\left(
\begin{array}{cc}
0 & e^{i3\tau\theta } \\
e^{-i3\tau\theta } & 0
\end{array}\right)\notag \\
&=& \hbar v_{F}\Delta \kappa^\mathrm{cv}\left(\cos(3\theta)\sigmaone -\tau\sin(3\theta)\sigmatwo\right).\phantom{\sum}
\end{eqnarray}
When this is added to the Hamiltonian \eqref{HCNT}, it is clear that both $\kappa_\perp$ and $\kappa_\parallel$ are shifted. However, the shift in $\kappa_\parallel$ is unimportant and can be absorbed into the longitudinal momentum, whereas the shift in $\kappa_\perp$ (the coefficient to $\sigmaone$) gives a gap in the nanotube spectrum as shown in Fig.~8.

\subsection{Spin-orbit coupling}

We now include the effect of spin-orbit coupling. Special relativity tells us that an electric field is experienced by a moving electron as a magnetic field. The resulting spin-orbit interaction (SOI) Hamiltonian is given by
\begin{equation}\label{HSOI}
H_{\mathrm{SOI}}=\alpha \left( \mathbf{E}\times \mathbf{p}\right) \cdot
\mathbf{s,}
\end{equation}%
where $\mathbf{E}$ is the electric field and $\alpha $ is a constant derived from relativistic quantum mechanics. Each carbon ion contributes to $\mathbf{E}$, giving rise to matrix elements between $\pi$-orbitals and in-plane orbitals, which in turn are coupled by curvature. To describe this based on microscopic parameters, one must start from the known atomic spin-orbit coupling of carbon and the $sp^2$ tight-binding parameters. However, since we do not aim to determine the size of the effect, we use a simpler approach, namely to introduce phenomenological parameters for the SOI-induced coupling between $\pi$ orbitals in curved graphene. In this approach, the perturbation \eqref{HSOI} gives matrix elements in the Bloch basis \eqref{PsiBloch}.

As explained in Sec. III.F.1, broken symmetry in the nanotube allows the first-order matrix elements to be non-zero. Consider the simplest way to break mirror symmetry, namely a constant radial electric field~$\mathbf{E}_r$. We need to calculate the matrix elements of the spin-orbit Hamiltonian \eqref{HSOI} in the $\{\Psi^{\KKK}_A(\boldr),\Psi^{\KKK}_B(\boldr)\}$ basis. Taking the cross product in~Eq.~\eqref{HSOI} and projecting to the parallel direction $\mathbf{\hat{T}}$ gives
\begin{equation}
E_{r}\left( \mathbf{\hat{z}}\times \mathbf{p}^\mathrm{K}_{AB}\right) _{\parallel }= \hbar v_\mathrm{F}E_{r}\,\mathbf{\hat{T}}\cdot\left( i \mathbf{\hat{x}}+\mathbf{\hat{y}}\right) =\hbar v_\mathrm{F}E_{r}e^{i\vartheta },
\end{equation}
where we used the matrix element of $\ppp$ from Eq.~\eqref{pAB} and the rotation \eqref{cylrotation}. This also holds for the $\mathrm{K}'$ valley, if $e^{i\vartheta }$ is replaced by $e^{-i\vartheta }$. After the unitary transformation \eqref{UNI}, we obtain the form of the orbital-like spin-orbit interaction in cylindrical coordinates:
\begin{equation}
H_{\mathrm{SOI,orbital-like}}^{{}}=\Delta_{\mathrm{SO}}^{1}\left(
\begin{array}{cc}
0 & 1 \\
1 & 0%
\end{array}%
\right) \tau s=\Delta_{\mathrm{SO}}^{1}\sigmaone\tau s,
\end{equation}%
where $\Delta_{\mathrm{SO}}^{1}$ is a phenomenological parameter. As we said, this derivation is based only on symmetry and gives no information about the magnitude of $\Delta_{\mathrm{SO}}^{1}$, except that it is linear in the atomic SOI and inversely proportional to nanotube diameter. Thus the orbital-like SOI has an easy physical interpretation: it is caused by a Rashba effect, because it is proportional to the azimuthal momentum and the (mean) radial electric field. This is the SOI that was originally derived for graphene \cite{KanePRL1997} and for nanotubes \cite{AndoJPSJ2000}.

It was realized later that one more term is allowed by the reduced symmetry (see \ref{sec_SOdifferenttypes}). This term comes from the diagonal matrix elements:
\begin{equation}
B_{AA,\parallel }^{\mathbf{K}}=\left\langle \Psi _{A}^{\mathbf{K}}\right|\left(
\mathbf{E}\times \mathbf{p}\right) _{\parallel }\left|\Psi _{A}^{\mathbf{K}%
}\right\rangle,
\end{equation}
and the identical expression for the $B$ sublattice. For a constant electric field (or one obeying the $C_3$ symmetry)~$B_{AA,\parallel }^{\mathbf{K}}$ is zero. When the rotational symmetry is broken, it can be non-zero. It turns out that $B_{AA,\parallel }^{\mathbf{K}}$ depends on the chirality. We  study two special cases: armchair and zig-zag nanotubes.

Armchair nanotubes have mirror planes perpendicular to the nanotube axis through an $A$ atom. Therefore the curvature-induced electric field has the same symmetry, and so does  $\left( \mathbf{E}\times \mathbf{p}\right) _{\parallel }=E_{r}p_{\perp }$. Using that the wave function transforms under reflection as $\Psi_{A}^{\mathbf{K}}\rightarrow \left( \Psi _{A}^{\mathbf{K}}\right) ^{\ast }$, as evident from Fig.~\ref{Diracwavefunction}, we have for armchair nanotubes:
\begin{equation}
B_{AA,\parallel}^{\mathbf{K}}=\left\langle (\Psi _{A}^{\mathbf{K}})^{\ast }\right|\left( \mathbf{E}\times \mathbf{p}\right) _{\parallel
}\left|(\Psi _{A}^{\mathbf{K}})^{\ast }\right\rangle=-(B_{AA,\parallel}^{\mathbf{K}})^*,
\end{equation}
and therefore purely imaginary. On the other hand, since the operator $\left( \mathbf{E}\times \mathbf{p}\right) _{\parallel }$ is Hermitian, we must also have $B_{AA,\parallel }^{\mathbf{K}}=(B_{AA,\parallel }^{\mathbf{K}})^{\ast }$. Hence for an armchair nanotube $B_{AA,\parallel}^{\mathbf{K}}=0$.

Zig-zag nanotubes have no mirror plane through an atom and perpendicular to the axis. Instead, there is a mirror plane along the axis, and since $\left( \mathbf{E}\times \mathbf{p}\right) _{\parallel }$ changes sign under this symmetry, there is no cancellation as for the armchair case.

We conclude that there is a spin-orbit-interaction contribution diagonal in pseudospin, of the form
\begin{equation}\label{Zeemnanlike}
H_\mathrm{SOI, Zeeman-like}=\Delta_\mathrm{SO}^{0}\tau s,
\end{equation}%
where $\Delta_\mathrm{SO}^{0}$ depends on chirality and therefore has leading harmonic $\Delta_\mathrm{SO}^{0}\propto \cos 3\theta$. This is the Zeeman-like term of~Eq.~\eqref{eq_SOparam1}, which thus comes from intra-sublattice matrix elements, \ie from next-nearest neighbor couplings.

\subsection{Final form of the Hamiltonian}

The final low-energy Hamiltonian is the sum of the metallic nanotube Hamiltonian, the curvature term, and the two spin-orbit terms:
\begin{equation}\label{Hfinal}
H_{\bm{\kappa}}=\hbar v_{\mathrm{F}}\left(\kappa_{\parallel}\sigmatwo+\Delta \kappa_\perp\sigmaone\right) +\Delta_\mathrm{SO}^{0}\tau s+\Delta_\mathrm{SO}^{1}\sigmaone\tau s,
\end{equation}
where both $\Delta \kappa_\perp$ and $\Delta_\mathrm{SO}^{0}$ are proportional to $\cos3\theta$. In addition, a magnetic field gives rise to an Aharonov-Bohm phase, as well as the usual Zeeman term:
\begin{equation}\label{HBfinal}
H_B=\hbar v_{\mathrm{F}}\tau \Delta \kappa_{\perp}^B \sigmaone+\frac{1}{2} g_s \mu_{B}\mathbf{s}\cdot\mathbf{B}.
\end{equation}
with $\Delta\kappa_{\perp}^B = eDB_\parallel/4\hbar$. The spectrum can be found by simple diagonalization of the Hamiltonian, giving the expression in Eq.~\eqref{eq_Epmtaus} for $E^\pm_{\tau,s}(B_{||})$ for the case of~$\mathbf{B}$ directed along the nanotube. In that equation the gap is $E_\mathrm{G}^0=-2\hbar v_\mathrm{F}\Delta\kappa_\perp$ and the confinement energy is~$E_\mathrm{conf}= \hbar v_\mathrm{F}\kappa_\parallel$. The convention used here has for the time reversal operator $\mathcal{T}=i\tau_x s_y \mathcal{K}$, where $\mathcal{K}$ denotes complex conjugation. Time reversal therefore transforms $s\rightarrow-s$, $\tau\rightarrow-\tau$, $\sigmatwo\rightarrow-\sigmatwo$, and $k_\parallel\rightarrow-k_\parallel$.\footnote{Some authors use a convention where $\sigma_i\rightarrow \tau \sigma_i$, $i=1,2$. This is equivalent to a unitary transformation $H\rightarrow U HU^\dagger$, with $U=e^{i\pi(1-\tau)\sigma_3/4}$. This also changes the definition of the time-reversal operator to $\mathcal{T}\rightarrow i s_2 \sigma_3 \tau_1 \mathcal{K}$, such that time reversal corresponds to a sign change of $\sigma_1,s,\tau$ and $k_\parallel$. }

\subsection{Single-particle quantum dot states}
\label{appendixsingleparticlestates}

Next we discuss the single-particle states in a quantum dot. We assume a parallel magnetic field (making the spin projection $s$ along the axis a good quantum number) and a potential that is flat in the middle of the dot. The wave function is thus a superposition of right-moving and left-moving waves. In principle, both plane wave solutions close to $\KKK$ and $\KKK'$  should be included, because the dot terminations can mix the two valleys. However, we approximate the valley index a good quantum number, assuming a smooth confining potential, and later include mixing via a matrix element between the valley-polarized states, as explained in Sec.~III.B. The standing wave is then of the form
\begin{equation}\label{Psidot}
  \Psi_{\mathrm{dot},s}(t,c)= \left(Ce^{i\kappa_\parallel t}+De^{-i\kappa_\parallel t}\right) u^\KKK(t,c)\otimes|s\rangle,
\end{equation}
where $u^\KKK(t,c)$ is the periodic part of the Bloch wave function, and
where the relation between $C$ and $D$ is determined by the reflection coefficients $r_\mathrm{L}$ and $r_\mathrm{R}$ at the ends of the dot, because the condition for a bound state is $C=r_\mathrm{L}D=r_\mathrm{L}r_\mathrm{R}C$. Importantly, the envelope wave function depends on valley and spin, because $\kappa_{||}$ depends on both through the energy:
\begin{equation}
\hbar\vF\kappa_\parallel
=\pm\sqrt{(E-s\tau\Delta^0_\mathrm{SO}-E_Z)^2-(s\Delta^1_\mathrm{SO}- \tau E_\mathrm{G}^0/2+ E_B)^2}.
\end{equation}
(Here $E_B=e\vF DB_{||}/4$.) For $B=0$, we thus have two Kramers doublets separated by an energy $\Delta_\mathrm{SO}$, as explained in the main text. We emphasize that because of the spin-orbit induced difference in $\kappa_\parallel$, the two doublets cannot be exactly written as product states of longitudinal, valley, and spin components.

In the main text, we refer to the single-particle states as $|\nu\tau s\rangle$, where $\nu$ is a quantum number that labels the solutions to the above problem. Two-electron states are then built from the two-electron Slater determinants: $(|\nu\tau s\rangle_1|\nu'\tau' s'\rangle_2-|\nu\tau s\rangle_2|\nu'\tau' s'\rangle_1)/\sqrt{2}$ as in~Table~\ref{tab_symmetricantisymmtric}. These states do not take into account that the longitudinal wave functions are easily distorted by Coulomb interactions, resulting in highly correlated two-electron states (cf. Sec.~\ref{correlations}).

\subsection{Two-electron states and exchange interaction}
\label{twoelectronexchange}

In a two-electron quantum dot the question of interaction corrections to the simple single-particle filling arises. We will discuss these corrections using first-order perturbation theory. The two-electron matrix element between four states $\psi_a,\psi_b,\psi_c,$ and $\psi_d$ has the form:
\begin{equation}\label{generaltwoelectronmatrixelements}
  V_{abcd}=\langle a,b|V|c,d\rangle=\int d\rrr\int d\rrr'\,\rho_{ad}(\rrr)V(\rrr,\rrr')\rho_{bc}(\rrr')
\end{equation}
where
\begin{equation}\label{rhoab}
 \rho_{ab}(\rrr)=\langle \psi_a|\rrr\rangle\langle\rrr|\psi_b\rangle,
\end{equation}
and $V(\rrr,\rrr')$ is the Coulomb interaction potential. The state $|a,b\rangle$ is a simple product state: $|a,b\rangle=|a\rangle|b\rangle$ and not an antisymmetrized state as in the headers of Table~IV.
This matrix element between two Bloch states in Eq.~\eqref{PsiBloch} is
\begin{equation}
 \rho_{\kkk_1\kkk_2}(\rrr)=\frac{1}{N_A}\sum_{\RRR_n}\left(|\varphi_A(\rrr)|^2+|\varphi_B(\rrr)|^2\right)e^{-i(\kkk_1-\kkk_2)\cdot\RRR_n},
\end{equation}
where $N_A$ is a normalization equal to the number of unit cells and where we have neglected contributions from the overlap of neighboring atoms. The Coulomb matrix element obtained by inserting Eq.~\eqref{rhoab} into \eqref{generaltwoelectronmatrixelements} naturally separates into a short and a long-range part $ V= V^\mathrm{SR}+V^\mathrm{LR}$.  The two contributions are
\begin{subequations}
\label{shortlargerange}
\begin{eqnarray}
 V^\mathrm{SR}_{\kkk_1\kkk_2\kkk_3\kkk_4} &\approx& V^\mathrm{SR}_0\delta_{\kkk_1+\kkk_2-\kkk_3-\kkk_4} \\
  V^\mathrm{LR}_{\kkk_1\kkk_2\kkk_3\kkk_4} &\approx& \frac{1}{\mathcal{V}}
  \int_\mathcal{V} d\rrr\int_\mathcal{V} d\rrr'\,V(r-r')\nonumber\\
  &&\times e^{i(\kkk_1-\kkk_4)\cdot\rrr+i(\kkk_2-\kkk_3)\cdot\rrr'},
\end{eqnarray}
\end{subequations}
where $V^\mathrm{SR}_0=2U_0/N_A$ and where $\mathcal{V}$ is a normalization area of the nanotube surface.
For a quantum dot of length $L$ and diameter $D$, the short-range part is $V^\mathrm{SR} \approx U_0/(60 LD/[\mathrm{nm}]^2)$, which for parameters $L=300$~nm, $D=2$~nm, and $U_0=10$ eV gives $V^\mathrm{SR}_0=0.3$ meV.

\subsection{Exchange integrals due to long-range Coulomb interaction}
\label{exlongrange}

Rotational symmetry can be used to show that the long-range part vanishes except when $(\kkk_1+\kkk_2-\kkk_3-\kkk_4)\cdot\mathbf{C}=0$, which in terms of valley index means $\tau_1+\tau_2=\tau_3+\tau_4$ \cite{WeissPRB2010}. This valley selection rule contrasts with that for spin, which is $s_1=s_4$ and $s_2=s_3$.
Separating both the wave vectors and the coordinates into a transverse part and a longitudinal part, it becomes evident that the rapid oscillations of the Bloch wave functions make it a good approximation to ignore terms off-diagonal in valley indices, because
\begin{equation}\label{rapid}
  V^\mathrm{LR}_{\tau_1,\tau_2,\tau_3 ,\tau_4 }\propto \int d\varphi d\varphi' e^{-iM(\tau_1-\tau_4)(\varphi-\varphi)}
  V(\varphi-\varphi'),
\end{equation}
where the condition $\tau_1+\tau_2=\tau_3+\tau_4$ was used and where $M=\KKK\cdot\mathbf{C}/2\pi$ is an integer. Therefore, it is generally true that for narrow-gap nanotubes, which have $M\neq0$, the Coulomb matrix element is strongly suppressed.

Turning to the case of two electrons in a single dot, occupying single-particle states $|\nu\tau s\rangle$ in symmetric or antisymmetric combinations as discussed in Sec.~IV, the matrix elements between such orbitals therefore obey
\begin{equation}\label{12matrixelements}
  \langle 1\tau_1s_1,2\tau_2 s_2|V^\mathrm{LR}|3\tau_3 s_3,4\tau_4 s_4\rangle \propto  \delta_{s_1,s_4}  \delta_{s_2,s_3}\delta_{\tau_1,\tau_4}  \delta_{\tau_2,\tau_3}.
\end{equation}
This equation can be applied to calculate the long-range interaction energies of the states in Table~IV. For the $\SSS(0,2)$ states, we get four terms. The cross terms vanish, because they cannot have $\tau=\tau'$ and $s=s'$ at the same time. We therefore conclude that there are only diagonal, direct Coulomb interaction terms, \ie no exchange corrections. Assuming the charge distribution of both Kramers pairs to be approximately the same, we thus have
\begin{equation}\label{CoulombS02}
  \langle \psi |V^\mathrm{LR}|\psi \rangle=E_\mathrm{C}, \quad \mathrm{where } \quad \psi \in {S(0,2)}.
\end{equation}
Consider now the symmetric states containing different longitudinal modes, $\SSSprime(0,2)$. There are now eight terms in the interaction integral. The non-zero terms have the same valley and spin for the same electron label. If $\tau=\tau'$ and $s\neq s'$,  there are two positive cross terms. The same is true if $\tau\neq \tau$ and $s=s'$ (because of the fast oscillations that lead to Eq.~\eqref{12matrixelements}). We therefore conclude:
\begin{equation} 
  \langle S'(0,2)|V^\mathrm{LR}|S'(0,2)\rangle=E_\mathrm{C}+2\mathcal{C}.
\end{equation}
where
\begin{equation}\label{2Cdef}
  \mathcal{C}= \frac14 \langle 1Ks ,2K's|V^\mathrm{LR}|1K's,2Ks\rangle.
\end{equation}
is an exchange-like Coulomb integral.

The situation is similar for the antisymmetric states $\AS(0,2)$, except now the sign is opposite, \ie~there is an energy reduction due to exchange integrals.
These energy shifts can be absorbed into the definition of $\DeltaSAS$ and $\DeltaASSprime$, which is what is done in Table IV.

\subsection{Exchange integrals due to short-range Coulomb interaction}
\label{appendixexchangeshortrange}

For the short-range Coulomb interaction the valley selection rules in Eq.~\eqref{12matrixelements} do not apply. Within the~$S(0,2)$ multiplet we find that states with antiparallel spin are raised in energy, assuming a repulsive short-range interaction $U_0$.
This is because in such a state, two electrons can occupy the same atomic site, in contrast to the spin-polarized states where spatial antisymmetry forbids this.

The antisymmetric states $\AS(0,2)$ are almost unaltered by the short range interaction \cite{secchi2013inter}, because their longitudinal symmetries differ from those in $\SSS(0,2)$.
This can be intuitively understood by considering the limit $\Delta_\mathrm{SO} \rightarrow 0$, which allows us to separate each two-electron wave function into a longitudinal part and a spin-valley part. In this limit, $\AS$ states have zero amplitude to occupy the same cross section of nanotube and therefore the same atomic site.
As an example, consider the valley-polarized state $K{\downarrow},K{\uparrow}$.
For $\Delta_\mathrm{SO} \rightarrow 0$ it can be rewritten as  $(|\nu,\nu'\rangle - |\nu',\nu\rangle) \otimes |K,K\rangle \otimes (|{\downarrow},{\uparrow}\rangle+ |{\uparrow},{\downarrow}\rangle)$, with asymmetric longitudinal part as expected. In contrast, the analogous state in $\SSS(0,2)$ is longitudinally symmetric: $K{\downarrow},K{\uparrow}\rightarrow |\nu,\nu\rangle \otimes |K,K\rangle \otimes (|{\downarrow},{\uparrow}\rangle- |{\uparrow},{\downarrow}\rangle)$.


\subsection{Two-electron states and Pauli blockade}
\label{appendixtwoelectronstatesPauli}

In Sec. \ref{doubledots} we discussed Pauli blockade of two-electron states at zero magnetic field. We noted that for some of the blocked states Pauli blockade can be lifted by a dephasing mechanism only, whereas other blocked states require that the spin or valley quantum number within one of the dots be flipped. Here we give some examples.

As an example for dephasing, consider the blocked state $\AS(1,1)K{\downarrow},K'{\uparrow}$. From Table \ref{tab_symmetricantisymmtric} we can write out the wave function by identifying $\nu=1$(2) with the lowest longitudinal shell in the right(left) dot. After rearranging some terms this state has the form:
\begin{equation}
\begin{split}\label{eqFK2}
\AS(1,1)K{\downarrow}, K'{\uparrow}=|1K{\downarrow}\rangle_1|2K'{\uparrow}\rangle_2 - |2K'{\uparrow}\rangle_1|1K{\downarrow}\rangle_2  \\
+ |1K'{\uparrow}\rangle_1|2K{\downarrow}\rangle_2 - |2K{\downarrow}\rangle_1|1K'{\uparrow}\rangle_2.
\end{split}
\end{equation}
We are interested in the time evolution of this state, and model dephasing by assuming  an effective Kramers splitting in the left dot that is slightly larger than that in the right dot: $(E_\mathrm{1K'{\uparrow}}-E_\mathrm{1K{\downarrow}})=(E_\mathrm{2K'{\uparrow}}-E_\mathrm{2K{\downarrow}})+\delta$. Up to a trivial dynamical phase, the state evolves with time $t$ to:
\begin{equation}\label{eqFK3}
\begin{split}
|1K{\downarrow}\rangle_1|2K'{\uparrow}\rangle_2 - |2K'{\uparrow}\rangle_1|1K{\downarrow}\rangle_2\\
+ e^{-2i\delta t}\left(|1K'{\uparrow}\rangle_1|2K{\downarrow}\rangle_2 - |2K{\downarrow}\rangle_1|1K'{\uparrow}\rangle_2\right).
\end{split}
\end{equation}
When $e^{-2i\delta t}=-1$ the blocked state $\AS(1,1)K{\downarrow},K'{\uparrow}$ has dephased into the unblocked state $\SSSprime(1,1)K{\downarrow},K'{\uparrow}$. This is similar to the rapid dephasing from $T_0$ to $S$ due to an Overhauser field difference in a conventional double dot.

Next, consider the blocked state $\AS(1,1)K{\downarrow},K{\downarrow}$. We show that a valley flip in either dot will lift Pauli blockade. We can write out the state as
\begin{equation}\label{eqFK4}
\AS(1,1)K{\downarrow},K{\downarrow}\propto
|1K{\downarrow}\rangle_1|2K{\downarrow}\rangle_2 - |2K{\downarrow}\rangle_1|1K{\downarrow}\rangle_2.
\end{equation}
A valley flip in e.g.\ the left dot ($2K{\downarrow}\rightarrow2K'{\downarrow}$) leads to the state
\begin{equation}\label{eqFK5}
|1K{\downarrow}\rangle_1|2K'{\downarrow}\rangle_2 - |2K'{\downarrow}\rangle_1|1K{\downarrow}\rangle_2,
\end{equation}
which is a superposition of $\AS(1,1)K{\downarrow},K'{\downarrow}$ and the unblocked state  $\SSSprime(1,1)K{\downarrow},K'{\downarrow}$. Alternatively, a spin flip in the left dot results in a superposition of $\AS(1,1)K{\downarrow},K{\uparrow}$ and $\SSSprime(1,1)K{\downarrow},K{\uparrow}$, whereas a combined spin-valley flip results in a superposition of $\AS(1,1)K{\downarrow},K'{\uparrow}$ and $\SSS(1,1)K{\downarrow},K'{\uparrow}$. In all cases, Pauli blockade is circumvented by the admixture of unblocked states (\ie~longitudinal symmetric states).
However, because of spin-orbit coupling, a single flip in one of the dots requires emission or absorption of energy, whereas a spin-and-valley flip may be allowed by energy conservation. Therefore, ``spin-valley" blockade can occur not primarily due to spin and valley conservation, but due to energy conservation.
Thus energy conservation plays a central role in establishing Pauli blockade, just as in conventional quantum dots (cf. Fig. \ref{STblockade}). In nanotubes the relevant energy scale is the spin-orbit energy rather than the orbital level spacing. Therefore, Pauli blockade in nanotubes is generically weaker than singlet-triplet Pauli blockade in conventional semiconductors.

\bibliographystyle{apsrmp4-1}
\bibliography{Laird}

\end{document}